\documentclass[superscriptaddress,nofootinbib,notitlepage]{revtex4-1}
\pdfoutput=1

\usepackage{graphicx}
\usepackage{amsthm}
\usepackage{thmtools}
\usepackage{mathtools}
\usepackage{thm-restate}
\usepackage{amsmath}
\usepackage{algorithm}
\usepackage{algorithmic}
\usepackage{amssymb}
\usepackage{bm}
\usepackage{xcolor}
\usepackage{placeins}
\usepackage[normalem]{ulem} 
\usepackage[colorlinks]{hyperref}
\hypersetup{
	pdfstartview={FitH},
	pdfnewwindow=true,
	colorlinks=true,
	linkcolor=blue,
	citecolor=blue,
	filecolor=blue,
	urlcolor=blue}
\usepackage{tikz}
\usetikzlibrary{calc,backgrounds}
\usepackage{physics}
\usepackage{cancel}
\usepackage{multirow}
\usepackage{pgffor}
\usepackage{url}
\usepackage{hypcap}
\usepackage{dsfont}
\usepackage{soul}
\usepackage{inconsolata}

\newcommand{\eq}[1]{Eq.~\hyperref[eq:#1]{(\ref*{eq:#1})}}
\renewcommand{\sec}[1]{\hyperref[sec:#1]{Section~\ref*{sec:#1}}}
\newcommand{\app}[1]{\hyperref[app:#1]{Appendix~\ref*{app:#1}}}
\newcommand{\tab}[1]{\hyperref[tab:#1]{Table~\ref*{tab:#1}}}
\newcommand{\fig}[1]{\hyperref[fig:#1]{Figure~\ref*{fig:#1}}}
\newcommand{\figa}[2]{\hyperref[fig:#1]{Figure~\ref*{fig:#1}#2}}
\newcommand{\figx}[2]{\hyperref[fig:#1]{Figure~\ref*{fig:#1}(#2)}}
\newcommand{\thm}[1]{\hyperref[thm:#1]{Theorem~\ref*{thm:#1}}}
\newcommand{\lem}[1]{\hyperref[lem:#1]{Lemma~\ref*{lem:#1}}}
\newcommand{\cor}[1]{\hyperref[cor:#1]{Corollary~\ref*{cor:#1}}}
\newcommand{\defn}[1]{\hyperref[def:#1]{Definition~\ref*{def:#1}}}
\newcommand{\alg}[1]{\hyperref[alg:#1]{Algorithm~\ref*{alg:#1}}}

\newcommand{\sel}{\textsc{select}}
\newcommand{\prep}{\textsc{prepare}}

\def\avg#1{\mathinner{\langle{#1}\rangle}}
\def\bra#1{\mathinner{\langle{#1}|}}
\def\ket#1{\mathinner{|{#1}\rangle}}
\newcommand{\proj}[1]{\ket{#1}\!\!\bra{#1}}

\newcommand{\be}{\begin{equation}}
\newcommand{\ee}{\end{equation}}
\newcommand{\ba}{\begin{eqnarray}}
\newcommand{\ea}{\end{eqnarray}}

\newcommand{\nn}{\nonumber \\}

\newcommand{\inA}{\kappa}
\newcommand{\inB}{\lambda}
\newcommand{\mulAB}{\gamma}
\newcommand{\dimA}{{\ensuremath{d_A}}}
\newcommand{\dimB}{{\ensuremath{d_B}}}
\newcommand{\dimAB}{\ensuremath{d}} 
\newcommand{\mulError}{\varepsilon}
\newcommand{\gradbits}{{b_{\rm grad}}}
\newcommand{\brot}{b_{\rm rot}}
\newcommand{\gap}{\Delta}

\usepackage{color}
\definecolor{darkgreen}{rgb}{0.0, 0.5, 0.0}
\newcommand{\blu}{\color{blue}}

\newlength\replength
\newcommand\repfrac{.33}

\newcommand\rulewidth{.6pt}
\newcommand\tdashfill[1][\repfrac]{\cleaders\hbox to \replength{%
  \smash{\rule[\arraystretch\ht\strutbox]{\repfrac\replength}{\rulewidth}}}\hfill}

\newcommand\tdotfill[1][\repfrac]{\cleaders\hbox to \replength{%
  \smash{\raisebox{\arraystretch\dimexpr\ht\strutbox-.1ex\relax}{.}}}\hfill}

\newcommand{\chunk}{k} 

\input{Qcircuit}

\usepackage[capitalise]{cleveref}

\newcommand{\MQ}{\affiliation{%
Department of Physics and Astronomy,
Macquarie University, Sydney, NSW 2109, Australia} }

\newcommand{\EQUS}{\affiliation{%
ARC Centre of Excellence in Engineered Quantum System,
Macquarie University, Sydney, NSW 2109, Australia}}

\newcommand{\UW}{\affiliation{%
Department of Physics, University of Washington, Seattle, WA 18195, United States of America} }

\newcommand{\PNNL}{\affiliation{%
Pacific Northwest National Laboratory, Richland, WA 99354, United States of America} }

\newcommand{\Google}{\affiliation{%
Google Research, Venice, CA 90291, United States of America}}

\newcommand{\bdirect}{b_{\rm dir}}
\newcommand{\bdiff}{b_{\rm dif}}
\newcommand{\bphase}{b_{\rm pha}}
\newcommand{\bLCU}{b_{\rm LCU}}
\newcommand{\bfun}{b_{\rm fun}}
\newcommand{\bsmooth}{b_{\rm sm}}

\begin{document}

\title{Compilation of Fault-Tolerant Quantum Heuristics for Combinatorial Optimization}

\date{\today}
\author{Yuval R.~Sanders} \MQ \EQUS
\author{Dominic W.~Berry}
\email{corresponding author: dominic.berry@mq.edu.au} \MQ
\author{Pedro C.~S.~Costa} \MQ
\author{Louis W.~Tessler} \MQ
\author{Nathan Wiebe} \UW \PNNL \Google
\author{Craig Gidney} \Google
\author{Hartmut Neven} \Google
\author{Ryan Babbush}
\email{corresponding author: ryanbabbush@gmail.com} \Google

\begin{abstract}
Here we explore which heuristic quantum algorithms for combinatorial optimization might be most practical to try out on a small fault-tolerant quantum computer. We compile circuits for several variants of quantum accelerated simulated annealing including those using qubitization or Szegedy walks to quantize classical Markov chains and those simulating spectral gap amplified Hamiltonians encoding a Gibbs state. We also optimize fault-tolerant realizations of the adiabatic algorithm, quantum enhanced population transfer, the quantum approximate optimization algorithm, and other approaches. Many of these methods are bottlenecked by calls to the same subroutines; thus, optimized circuits for those primitives should be of interest regardless of which heuristic is most effective in practice. We compile these bottlenecks for several families of optimization problems and report for how long and for what size systems one can perform these heuristics in the surface code given a range of resource budgets. Our results discourage the notion that any quantum optimization heuristic realizing only a quadratic speedup will achieve an advantage over classical algorithms on modest superconducting qubit surface code processors without significant improvements in the implementation of the surface code. For instance, under quantum-favorable assumptions (e.g., that the quantum algorithm requires exactly quadratically fewer steps), our analysis suggests that quantum accelerated simulated annealing would require roughly a day and a million physical qubits to optimize spin glasses that could be solved by classical simulated annealing in about four CPU-minutes.
\end{abstract}

\maketitle

\tableofcontents

\makeatletter
\let\toc@pre\relax
\let\toc@post\relax
\makeatother 

\listoftables

\newpage

\section{Introduction}
\label{sec:introduction}

The prospect of quantum enhanced optimization has driven much interest in quantum technologies over the years. This is because discrete optimization problems are ubiquitous across many industries and faster solutions could potentially revolutionize fields as broad as logistics, finance, machine learning, and more. Since combinatorial optimization problems are often NP Hard, we do not expect that quantum computers can provide efficient solutions in the worst case. Rather, the hope is that there may exist ensembles of instances with structure that would enable a significant quantum speedup on average, or for which a quantum computer can approximate better solutions.

Among the most studied algorithms for quantum optimization are those that can function as heuristics. The objective of a heuristic algorithm is to produce a solution given a reasonable amount of  computational resources that is ``good enough'' (or at least the best one can afford) for solving the problem at hand. While heuristics are often able to efficiently find the exact solution, sometimes they might fail to do so and instead only approximate the exact solution (potentially in an uncontrolled fashion). But such techniques are still valuable because finding some usable result does not require a prohibitively long time. Accordingly, heuristics are often used without regard for rigorous bounds on their performance. Indeed, the NP Hardness of many combinatorial optimization problems makes heuristics the only viable option for many problems that need to be routinely solved in real-world applications.

While some heuristic algorithms have a strong theoretical basis, many of the most effective heuristics are based on intuitive principles and then honed empirically through data and experimentation. However, today, our ability to evaluate quantum heuristics through experimentation is limited since the only available quantum computers are small and noisy \cite{Arute2019}. We can perform numerics on small instances but extrapolation from those small system size numerics can be potentially misleading \cite{Farhi2001}. Still, it is reasonable to ask the question: what would be some of the most compelling quantum heuristics for optimization that we would want to attempt on a small fault-tolerant quantum computer, and how many resources would be required to implement their primitives?

There are many prominent approaches to combinatorial optimization on a quantum computer. These include variants of Grover's algorithm \cite{Grover1996,Durr1996AMinimum}, quantum annealing \cite{ray1989,Kadowaki1998}, adiabatic quantum computing \cite {farhi2000,Aharonov2007}, the shortest path algorithm \cite{Hastings2018}, quantum enhanced population transfer \cite{Kechedzhi2018,Smelyanskiy2018}, the quantum approximate optimization algorithm \cite{Farhi2014}, quantum versions of classical simulated annealing \cite{Somma2008b,Boixo2014a}, quantum versions of backtracking \cite{Montanaro2015QuantumAlgorithms,Campbell2018ApplyingProblems} as well as branch and bound techniques \cite{Montanaro2020QuantumAlgorithms}, among many others. While often these works focus on the asymptotic scaling of exact quantum optimization, in many cases one can use these algorithms heuristically through trivial modifications of the approach. For instance, the quantum adiabatic algorithm requires that one evolve the system for an amount of time scaling polynomially with the inverse of the minimum spectral gap of the adiabatic evolution. However, one can instead use this algorithm as a heuristic by choosing to evolve for a much shorter amount of time, and hoping for the best (this is similar to the strategy usually employed with quantum annealing).

What essentially all forms of quantum optimization have in common is the requirement that the quantum algorithm query some function of the cost function of interest. This is how the quantum computer accesses information about the energy landscape. For instance, if our cost function is $H$ and $H \ket{x} = E_x \ket{x}$ so that $E_x$ is the value of the cost function for bit string $\ket{x}$, then often we need to phase the computational basis by a function $f(\cdot)$ of $E_x$, e.g.,
\begin{equation}
\label{eq:phases}
\sum_{x} a_x \ket{x} \mapsto \sum_{x} e^{-i f\left(E_x\right)} a_x \ket{x}.
\end{equation}
For example, $f(E_x) \propto E_x$ is required to implement the quantum approximate optimization algorithm, quantum enhanced population transfer, digitized forms of quantum annealing and the shortest path algorithm. Alternatively, $f(E_x) \propto \arccos(E_x)$ would describe something related to the quantum walk forms of those algorithms. If $f(E_x) \propto (-1)^{(E_x \leq K)}$ this primitive would be the bottleneck subroutine for amplitude amplification to boost our support on energies less than $K$. In most quantum approaches to optimization, a unitary like this is interleaved with a much cheaper operation which does not commute with the operation in \eq{phases}. Some algorithms instead call for simultaneously evolving under a function of the cost function together with a simple non-commuting Hamiltonian, but still the bottleneck is usually the complexity of the cost function Hamiltonian. The difference between many of these algorithms often comes down to the choice of $f(\cdot)$ and the choice of the much cheaper non-commuting unitary. 

The quantum algorithms for simulated annealing (e.g.\ \cite{Somma2008b}) work slightly differently as those algorithms are based on making local updates to the wavefunction. For instance, the quantum version of a simulated annealing algorithm that updates with single bit flips requires
\begin{equation}
\label{eq:annealing_template}
\sum_{x} a_x \ket{k}\ket{x}\ket{0} \mapsto \sum_{x} a_x  \ket{k} \left(\sqrt{1-f\left(E_x, E_{x_k}\right)} \ket{x}\ket{0} + \sqrt{f\left(E_x, E_{x_k}\right)} \ket{x_k}\ket{1}\right)
\end{equation}
where $x_k$ is defined as the bit string $x$ with the $k^{\rm th}$ bit flipped, i.e.~$\ket{x_k} = \textsc{not}_k \ket{x}$, with $k=0$ corresponding to no bit flip. But again, these approaches are still typically bottlenecked by our ability to compute these functions of the cost function $f(\cdot)$.

This paper will \emph{not} address the important question of how well various heuristic quantum optimization approaches might perform in practice. Rather, our main motivation to is compile common bottleneck primitives for these approaches to quantum circuits suitable for execution on a small fault-tolerant quantum computer. In doing this, we will see that most contemporary approaches to quantum optimization are actually bottlenecked by the same subroutines (e.g., those required for \eq{phases} and \eq{annealing_template}), and thus improved strategies for realizing those subroutines are likely of interest regardless of which paradigm of quantum optimization is ultimately found to be most effective in practice. In essentially all heuristic approaches to quantum optimization there is a primitive that is repeated many times in order to perform the optimization. Instead of investigating how many times those primitives must be repeated, we focus on the best strategies for realizing those primitives within a fault-tolerant cost model. For all algorithms we consider, we report the constant factors in the leading order scaling of the Toffoli and ancilla complexity of these primitives.

For some algorithms studied, such as for the quantum algorithms for simulated annealing, this work is the first to give concrete implementations which determine constant factors in the scaling. In other cases our contribution is to optimize the scaling for certain problem Hamiltonians or improve details of the implementation. We focus on Toffoli complexity since we imagine realizing these algorithms in the surface code \cite{Kitaev1997Fault-tolerantAnyons,Fowler2012}, where non-Clifford gates such as Toffoli or T gates require considerably more time (and physical qubits) to implement than Clifford gates.

\subsection{Overview of results}
\label{sec:overview}

The goal of this paper is to estimate the performance of an early
universal quantum computer for key steps of combinatorial optimization.
To achieve this goal, we consider prominent heuristic-based methods for 
combinatorial optimization on a quantum computer and how their key steps 
could be executed on early hardware.
We consider the following heuristic-based methods:
amplitude amplification~\cite{Brassard2002} as a heuristic
for optimization and in combination with other approaches;
quantum approximate optimization algorithms (QAOA)~\cite{Farhi2014};
time-evolution approaches such as adiabatic algorithms~\cite{Farhi2001}
(including a variant incorporating a Zeno-like measurement~\cite{Boixo2009a}),
quantum enhanced population transfer~\cite{Smelyanskiy2018},
and ``shortest path'' optimization~\cite{Hastings2018};
and three quantum methods for simulated annealing (QSA), namely,
a Szegedy walk-based \cite{Szegedy2004} implementation of
Markov Chain Monte Carlo~\cite{Somma2008b},
a qubitized form of the Metropolis-Hastings approach~\cite{lemieux2019efficient},
and simulation of a spectral gap amplified Hamiltonian~\cite{Boixo2014a}.
We review existing approaches in detail and develop several new methods
or improvements. For each approach, we compile the primitive operations into quantum circuits optimized for execution in the surface code \cite{Fowler2012}.

For concreteness, we focus our analysis on four families of combinatorial optimization problems:
the $L$-term spin model, in which the Hamiltonian is specified
as a real linear combination of $L$ tensor products of
Pauli-$Z$ operators; Quadratic Unconstrained Binary Optimization (QUBO),
which is an NP-hard special case of a 2-local $L$-term spin model;
the Sherrington-Kirkpatrick (SK) model, which is a model of spin-glass
physics and an instance of QUBO that has been well-studied in the context of
simulated annealing~\cite{Kirkpatrick671};
and the Low Autocorrelation Binary Sequences (LABS) problem, which is
a problem with many terms but significant structure that is known to be extremely challenging in practice. For each of the above problems, we design several methods of calculating
the cost function on a quantum computer depending on how a given algorithmic
primitive is supposed to query and process the cost of a candidate solution. We present these methods in \sec{oracles}.

Our analysis has produced several novel techniques that yield improvements
over previous approaches. We recount the main ones here in order of appearance. In \sec{direct_oracle/SK}, we reduce by a logarithmic factor the cost of
calculating the Hamming weight of a bit string using our method from \cite{Kivlichan2019}.
This new technique leads to improvements in several other parts of our paper.
In \sec{functions}, we introduce a new technique for evaluating costly arithmetic
functions when computational cost matters more than accuracy.
Our new technique is based on approximating the function using linear
interpolation between classically precomputed points that can be
accessed using quantum read-only memory (QROM)~\cite{Babbush2018}, or a new variant of QROM designed
for sampling at exponentially growing spacings.

In \sec{qaoa}, we introduce a method of cost function evaluation
for QAOA based on amplitude estimation. This technique gives a quadratic improvement over the original approach.
In \sec{evolution}, we introduce a heuristic method for adiabatic optimization
that is likely to be computationally cheaper for some applications of
early quantum computers, although we do not expect an asymptotic advantage
over other state-of-the-art approaches.
The idea is to simulate the adiabatic path generated by the 
arccosine of the given Hamiltonian, not by the Hamiltonian directly,
by ``stroboscopically'' simulating time evolution with
short time steps produced by evolving under a qubitized walk.

In \sec{szegmain} we give a new method for constructing the Szegedy walk operator suggested in~\cite{Somma2008b}.
Our key technique is a state preparation circuit that avoids expensive
on-the-fly calculations by using the techniques introduced in~\cite{SLSB19}.
In \sec{QSA/qubitized}, we introduce an alternative method for executing
the controlled qubit-rotation step in the qubitized Metropolis-Hastings
approach introduced in~\cite{lemieux2019efficient}. Our approach is
preferable in cases where the Hamiltonian has a higher connectivity;
i.e.~when the probability of accepting a proposed transition depends on
many bits in the candidate solution. In those cases the approach of \cite{lemieux2019efficient} would have exponential complexity.
In \sec{spectral}, we give an explicit LCU-based oracle for the
spectral gap amplified Hamiltonian introduced in~\cite{Boixo2014a}.
This explicit oracle enables a cost analysis of the approach, which we provide.
Apart from assisting with our goal of estimating early quantum computer performance,
many of these innovations produce asymptotic improvements to the approaches
we consider.

Having compiled the primitive operations of our chosen approaches and
established how to query cost functions for our chosen problems,
we are able to numerically estimate the computational resources needed
to execute these primitives on a quantum computer.
Based on our assumption that the quantum computer will be built from
superconducting qubits and employ the surface code to protect the
computation from errors, we focus on minimizing the number of ancilla qubits and non-Clifford gates that would be required. This approach is founded on the
knowledge that non-Clifford operations are significantly harder
than Clifford operations to perform in the surface code.

We give an example of some of our ultimate findings in \tab{summary_of_results}.
In the table we provide the leading order scaling of the number of Toffoli gates
needed to perform an update using five of the heuristics that we consider for
two benchmark problems -- LABS and SK. These scalings are reproduced from
\tab{primitives} and presented in a simplified form where we assume that
the working precision for various calculations is a constant. We also reproduce key figures from \tab{sk_estimates} and \tab{LABS_estimates}
to show how we expect these estimated complexity scalings translate into
the runtime of an early quantum computer. In \tab{summary_of_results}
we show the estimated number of steps of the chosen algorithmic primitive that 
could be executed in a single day on a quantum computer for a problem size of
$N=256$, a relatively small problem size that would be reasonable to execute 
with only a single Toffoli factory as we assume in \tab{sk_estimates} and \tab{LABS_estimates}.
We also present the estimated number of physical qubits needed.

We find that, despite great efforts made to optimize our compiled quantum circuits,
the costs involved in implementing heuristics for combinatorial optimization will be taxing for early quantum computers. Not surprisingly, to implement problems between $N=64$ and $N=1024$ we find that hundreds of thousands of physical qubits are required when physical gate error rates are on the order of $10^{-4}$ and sometimes over a million are required for physical gate error rates on the order of $10^{-3}$. But even more concerning is that the number of updates that we can achieve in a day (given realistic cycle times for the error correcting codes) is relatively low, on the order of about ten thousand updates for the smallest instances considered of the cheapest cost functions. With such overheads, these heuristics would need to yield dramatically better improvements in the objective function per step than classical optimization heuristics. From this we conclude that, barring significant advances in the implementation of the surface code (e.g., much faster state distillation), quantum optimization algorithms offering only a quadratic speedup are unlikely to produce any quantum advantage on the first few generations of superconducting qubit surface code processors.

\begin{table*}[t]
\def\arraystretch{1.2}
\begin{tabular}{|c|c|c|c|c|}
\hline
\multirow{2}{*}{Problem}
  & \multirow{2}{*}{Algorithm Primitive}
    & \multicolumn{2}{c|}{(\tab{sk_estimates} and \tab{LABS_estimates})} 
    & (\tab{primitives}) \\ \cline{3-5}
  & 
    & steps per day
    & physical qubits
    & Toffoli count \\
\hline
\multirow{5}{*}{SK}
  & Amplitude Amplification \hfill (\S\,\ref{sec:amp_amp})
    & $4.8 \times 10^3$
    & $8.1 \times 10^5$
    & $\qquad 2 N^2 + N \hfill {\scriptstyle + \order{\log N}} \qquad$ \\
  & QAOA / $1^{\rm st}$ order Trotter \hfill (\S\,\ref{sec:qaoa})
    & $4.7 \times 10^3$
    & $8.6 \times 10^5$
    & $\qquad 2 N^2 + 4 N \hfill {\scriptstyle + \order{1}} \qquad$ \\
  & Hamiltonian Walk \hfill (\S\,\ref{sec:evolution})
    & $3.3 \times 10^5$
    & $8.0 \times 10^5$
    & $\qquad 6 N \hfill {\scriptstyle + \order{\log^2 N}} \qquad$ \\
  & QSA / Qubitized \hfill (\S\,\ref{sec:QSA/qubitized})
    & $3.3 \times 10^5$
    & $8.4 \times 10^5$
    & $\qquad 5 N \hfill {\scriptstyle + \order{\log N}} \qquad$ \\
  & QSA / Gap Amplification \hfill (\S\,\ref{sec:spectral})
    & $3.9 \times 10^5$
    & $8.4 \times 10^5$
    & $\qquad 5 N \hfill {\scriptstyle + \order{\log N}} \qquad$ \\
\hline
\multirow{5}{*}{LABS}
  & Amplitude Amplification \hfill (\S\,\ref{sec:amp_amp})
    & $3.3 \times 10^3$
    & $8.0 \times 10^5$
    & $\qquad 5 N^2/2 + 7 N / 2 \quad \hfill {\scriptstyle + \order{\log N}} \qquad$ \\
  & QAOA / $1^{\rm st}$ order Trotter \hfill (\S\,\ref{sec:qaoa})
    & $3.4 \times 10^3$
    & $8.4 \times 10^5$
    & $\qquad 5 N^2 / 2 \hfill {\scriptstyle + \order{N}} \qquad$ \\
  & Hamiltonian Walk \hfill (\S\,\ref{sec:evolution})
    & $4.9 \times 10^5$
    & $8.0 \times 10^5$
    & $\qquad 4 N \hfill {\scriptstyle + \order{\log N}} \qquad$ \\
  & QSA / Qubitized \hfill (\S\,\ref{sec:QSA/qubitized})
    & $1.7 \times 10^3$
    & $8.8 \times 10^5$
    & $\qquad 5 N^2  \hfill {\scriptstyle + \order{N}} \qquad$ \\
  & QSA / Gap Amplification \hfill (\S\,\ref{sec:spectral})
    & $1.7 \times 10^3$
    & $8.8 \times 10^5$
    & $\qquad 5 N^2  \hfill {\scriptstyle + \order{N}} \qquad$ \\
\hline
\end{tabular}
\caption[Summary of resource estimates]{\label{tab:summary_of_results}
We compare the cost of implementing various types of heuristics optimization primitives in a fault-toleration cost model. For concreteness, we give results for two problems: the Sherrington-Kirkpatrick model (SK) and
Low Autocorrelation Binary Sequences problem (LABS).
The numerical values from \tab{sk_estimates} and \tab{LABS_estimates} are based on
a problem size of $N=256$, a surface code cycle time of 1 {\textmu}s, and a physical gate error rate of $10^{-3}$ (there are other assumptions as well, covered in more detail in \sec{conclusion}). Note that depending how they would be used, it might be appropriate to scale the Hamiltonian walk steps by a factor of $\lambda$ which is roughly  $\lambda_{\rm SK} \approx N^2/2$ and $\lambda_{\rm LABS} \approx N^3/3$. We simplify the complexity scaling estimates from \tab{primitives} by treating as constant the bits of precision for numerical values.}
\end{table*}

\subsection{Organization of paper}
\label{sec:organization}

Our paper is divided into essentially two parts. In the first part (\sec{oracles}) we introduce and provide explicit compilations for a wide variety of subroutine or ``oracle'' circuits which perform operations related to specific problem Hamiltonians. In the second part of our paper (\sec{algorithms}) we describe a variety of heuristic algorithms for quantum optimization and discuss how the oracle circuits of \sec{oracles} can be called in order to implement these algorithms. We will see that the same ``oracle'' circuits are required by many algorithms. The results of \sec{algorithms} essentially provide query complexities to implement the primitives of common quantum optimization heuristics with respect to the oracles of \sec{oracles}. Thus, while the results of \sec{oracles} are adapted to particular problem Hamiltonians, the results of \sec{algorithms} are fairly general. We now describe our results in slightly more detail.

\sec{oracles} details strategies for realizing five straightforward oracle circuits which are detailed therein for each of four problem Hamiltonians in \tab{oracle_defs}. The specific problems we focus on are introduced at the beginning of \sec{oracles}. These five oracles correspond to: (\sec{direct_oracle}) the direct computation of a cost function into a quantum register, (\sec{diff}) the computation of the difference between the cost of two computational basis states which differ by a specific single bit, (\sec{phase_oracle}) an operation which phases the computational basis by an amount proportional to the cost, (\sec{lcu_oracle}) the realization of a qubitized quantum walk \cite{Low2016} which encodes eigenvalues of the cost function, and (\sec{functions}) the computation of arithmetic functions of an input value using QROM \cite{Babbush2018}. Our approach to computing arithmetic operations using QROM is likely useful in other contexts and is a new technique from this work. The culmination of \sec{oracles} is \tab{oracles} which gives leading order constants in the scaling of Toffoli, T and ancilla complexities for all five of these oracles and for all four of the problems. Even though the first two cost functions we introduce in \sec{oracles} have fairly general specifications, they do not capture exploitable structure in all optimization problems of interest. Still, we imagine that the motifs developed in \sec{oracles} will be helpful for any future work seeking to develop similar circuits for other cost functions.

\sec{algorithms} describes how the oracle circuits of \sec{oracles} are queried in order to realize the essential primitives of many fault-tolerant quantum heuristics for optimization. This section contains a mixture of new results and a review of established methods. \sec{amp_amp} reviews how one can use amplitude amplification \cite{Brassard2002} heuristically for optimization and also discusses how and why one might combine amplitude amplification with other algorithms in this section. \sec{qaoa} discusses strategies for executing QAOA \cite{Farhi2014} within fault-tolerant cost models. While most of this section is review, we also discuss the combination of QAOA with amplitude amplification based methods for more efficiently extracting the cost function value.

\sec{evolution} discusses several approaches to quantum optimization that are based on time evolution or quantum walks generated by a cost function and simple driver. First, we review the adiabatic algorithm \cite{Farhi2001} and well known methods for how it might be digitized using product formula type circuits. We then introduce a method of simulating the adiabatic algorithm based on qubitized quantum walks. Next, we review how the adiabatic algorithm can be combined with a Zeno-like measurement approach which corresponds to evolution under static Hamiltonians for random durations \cite{Boixo2009a}, and give some new results about how to optimally choose the distribution of those durations.

The remainder of \sec{algorithms} focuses on three approaches to a quantum algorithm which accelerates classical simulated annealing. In terms of implementation, these are the most complex algorithms studied in the paper. For the three variants of the quantum simulated annealing algorithms, we provide the first complete compilation of circuits which execute the heuristic primitive. In \sec{szegmain} we analyze and compile the original version \cite{Somma2008b} of these algorithms that is based on Szegedy quantum walks \cite{Szegedy2004}. As anticipated, this approach is the least efficient of the three studied. In \sec{QSA/qubitized} we focus on what is essentially a qubitized version of the Szegedy quantum walk. The primary characteristics of this approach were independently described in \cite{lemieux2019efficient} (a paper that came out during the preparation of our own) but we go beyond that work to determine (and in some ways improve upon)
constant factors in the scaling. 
Finally, in \sec{spectral} we compile the algorithm for quantum simulated annealing based on spectral gap amplification \cite{Boixo2014}, using an improvement based on qubitization. The results of \sec{algorithms} are summarized in \tab{query_complexity} and \tab{primitives}, which give the query complexities with respect to the oracles of \sec{oracles} and overall gate and ancilla complexities of all algorithms of \sec{algorithms} for all of the cost functions of \sec{oracles}.

Finally, we conclude in \sec{conclusion} with a discussion of these results. Our discussion includes an attempt to contextualize the ultimate cost of these heuristic primitives by giving the Toffoli count, ancilla count, and total number of physical qubits and wallclock time that would be required to realize these primitives given various resource budgets and assumptions in the surface code. These concrete resource estimates are given in \tab{sk_estimates} and \tab{LABS_estimates}. We then finish with a discussion of how these results lead to a fairly pessimistic outlook on the viability of obtaining quantum advantage for optimization by using a small quantum computer unless one is able to obtain significantly better than a quadratic speedup over classical alternatives.

\section{Oracles and Circuit Primitives for Specific Cost Functions}
\label{sec:oracles}

While many paradigms of quantum optimization require the same bottleneck subroutines for their implementation, aspects of these subroutines will always be specific to the particular problem that one intends to optimize. Thus, in order to give concrete implementations and develop a sense of how many resources would be required for steps of common quantum heuristics, aspects of our work are adapted to particular problem Hamiltonians (equivalently here, ``cost functions'') of interest. There are four main types of Hamiltonians that we consider in this paper. 

The first two types of Hamiltonians we will study are of interest because they are programmable instances of optimization problems that one might encounter in practical situations. The second two types of problems we will study are of interest more to those who study statistical physics and for different reasons: because they define ensembles of instances for which the average case has known and interesting properties. While solutions to specific instances of the latter two problems are probably not of much value, we anticipate they will be interesting problems on which to investigate the performance of a quantum computer. The four problems we study are described below.
\begin{enumerate}
    \item \textbf{$L$-term spin model:} The most general Hamiltonian we will consider is the one we will refer to simply as the ``$L$-term spin model''. This Hamiltonian is a linear combination of $L$ tensor products of Pauli-$Z$ operators,
\begin{equation}
\label{eq:mukL}
H_{L} = \sum_{\ell =1}^{L} w_{\ell} \prod_{i \in q_\ell} Z_i,
\end{equation}
where $w_\ell$ are real scalars, $Z_i$ is the Pauli-$Z$ operator on qubit $i$, $N$ is the number of qubits in the cost function, and $q_\ell$ is a unique set of up to $N$ integers which also take values between 1 and $N$ (it is a set of integers corresponding to the indices of qubits on which term $\ell$ acts). One might anticipate that it would be helpful to also specify this Hamiltonian in terms of its many-body order $k = \textrm{max} |\{q_\ell\}|$. However, perhaps surprisingly, none of the algorithms discussed in this paper have a Toffoli complexity that scales explicitly in $k$.

    \item \textbf{Quadratic Unconstrained Binary Optimization:} We will also consider an NP-Hard example of $H_{N^2/2}$ known as Quadratic Unconstrained Binary Optimization (QUBO). The QUBO Hamiltonian is expressed as
\begin{equation}
H_{\rm QUBO} = \sum_{i \leq j} w_{ij} \left(\frac{\openone - Z_i}{2} \right)\left(\frac{\openone - Z_j}{2} \right) = \sum_{i < j} J_{ij} Z_i Z_j + \sum_{i} h_i Z_i + K
\end{equation}
where $K$ is a constant term that we will ignore from this point forward as this never needs to be explicitly simulated or computed for the purposes of optimizing the model, and the coefficients $J_{ij}$ and $h_i$ can be computed from the $w_{ij}$. This form of the model is also known as the Ising model but we refer to it here as QUBO since the Ising model can also mean a model with more limited connectivity and regular coefficients in some contexts.

    \item \textbf{Sherrington-Kirkpatrick:} This problem corresponds to a widely studied model of spin glass physics  \cite{Kirkpatrick671}. The  Sherrington-Kirkpatrick (SK) model is an example of the following QUBO Hamiltonian:
\begin{equation}
H_{\rm SK} = \sum_{i < j} w_{ij} Z_i Z_j, \qquad \qquad w_{ij} \in \left\{-1, 1\right\}, \qquad \qquad \left\| H_{\rm SK} \right\| \leq N^2 /2,
\end{equation}
and the values of $w_{ij}$ are usually chosen at random. The SK model is the focus of many studies on heuristic optimization, especially ones focusing on variants of simulated annealing. There is also a variant of the SK model which has the same statistical properties where the coefficients are Gaussian distributed real numbers.

    \item \textbf{Low Autocorrelation Binary Sequences:} We think it would be interesting to use a quantum computer to attempt to optimize problems that are very challenging on average. One problem is the Low Autocorrelation Binary Sequences (LABS) problem, also known as the Bernasconi model in physics \cite{Bernasconi1987}:
\begin{equation}
\label{eq:labs}
H_{\rm LABS} = \sum_{k=0}^{N-1} H_k^2  \qquad \qquad H_k = \sum_{i=1}^{N-k} Z_i Z_{i + k}, \qquad \qquad \left\| H_{\rm LABS} \right\| \approx N^3/3,
\end{equation}
which is an instance of $H_{N^3}$. This model is known to be extremely difficult; in fact the best classical algorithm has scaling like $\Theta(1.73^N)$ and has only been run for problem sizes up to $N = 66$ \cite{Packebusch2016}. However, we note that the model is not really a ``problem'' in the usual computer science sense because there is only one instance defined for each problem size. A variant of the LABS problem that we will consider is when the squared operators are instead replaced with absolute values, as one can verify that the ordering of the low energy solutions would be unchanged by this modification, and it is sometimes less expensive to simulate with a quantum computer.
\end{enumerate}

\begin{table*}[t]
\begin{tabular}{c|l}
    symbol & meaning \\ \hline
    $x$ & bitstring corresponding to a candidate solution of the optimization problem \\
    $N$ & number of bits needed to specify a candidate solution \\
    $E_x$ & cost (a.k.a.~energy) of candidate solution $x$ as specified by a cost function \\
    $H_\text{cf}$ & Hamiltonian operator corresponding to a cost function ``cf'' \\
    $b$ & number of bits used to specify the precision of an oracle \\
    $L$ & number of terms in a spin model (type of cost function) \\
    $\lambda$ & the normalization parameter for LCU methods, related to the Hamiltonian 1-norm\\
    $\beta$ & inverse temperature in simulated annealing\\
    ${\cal C}$ & Toffoli or T cost of some oracle\\
    ${\cal A}$ & ancilla required to implement some oracle that must be kept\\
    ${\cal B}$ & temporary ancilla required to implement some oracle
\end{tabular}
\caption[List of symbols]{\label{tab:symbols}
A list of common symbols we use throughout this paper.}
\end{table*}

\begin{table*}[t]
\begin{tabular}{c|c|c}
oracle & oracle definition & precision definition \\
\hline\hline
&&\\[0.1ex]
$O^{\textrm{direct}}$ & $O^{\textrm{direct}} \sum_{x} \psi_x \ket{x} \ket{0}^{\otimes \bdirect} \mapsto \sum_{x} \psi_x \ket{x} \ket{\tilde{E}_x}$
& $\left |E_x - \tilde{E}_x \right| \leq 2^{-\bdirect} \max_x |E_x|$\\[1.2ex]
$O_{k}^{\textrm{diff}}$ & $O_{k}^{\textrm{diff}} \sum_{x} \psi_x \ket{x} \ket{0}^{\otimes \bdiff} \mapsto \sum_{x} \psi_x \ket{x} \ket{\widetilde{\delta E}_x^{(k)}}, \quad \ket{y} = X_k \ket{x}$&
$\left |\widetilde{\delta E}_x^{(k)} - E_x + E_y \right| \leq 2^{-\bdiff} \max_{x,y} |E_x-E_y|$\\[1.5ex]
$O^{\textrm{phase}}(\gamma) $ & $O^{\textrm{phase}}(\gamma) \sum_x \psi_x \ket{x} \mapsto  \sum_x e^{-i \widetilde{\gamma E}_x} \psi_x \ket{x}$ &
$\left| \widetilde{\gamma E}_x - \gamma E_x\right| \leq 2^{-\bphase}$ \\ [1.5ex]
$O^{\textrm{LCU}}$ & $\bra{0}^{\otimes \log L} O^{\textrm{LCU}} \ket{0}^{\otimes \log L} = \tilde{H} / \lambda, \quad \tilde{H} = \sum_{\ell=1}^L \tilde{w}_\ell U_\ell, \quad \lambda = \sum_{\ell=1}^{L} \left|w_\ell\right|$ & $\left|\sqrt{w_\ell} - \sqrt{\tilde{w}_\ell}\right| \leq 2^{-\bLCU}$\\[1.5ex]
$O^{\textrm{fun}}_{\beta}$ & $O^{\textrm{fun}}_{\beta}\ket{z}\ket{0}^{\otimes \bsmooth} \mapsto \ket{x}\ket{\tilde f(\beta z)}$ & $\left|f(\beta z)-\tilde f(\beta z)\right|\le 2^{-\bfun}$ \\[1.5ex]
\hline
\end{tabular}
\caption[Oracle definitions]{\label{tab:oracle_defs} Quick definitions of the most important ``oracle'' circuits discussed in this work. Here, we slightly abuse the term ``oracle'' to mean a circuit primitive which is repeatedly queried throughout an algorithm, usually revealing information about the problem we are solving. Throughout the paper we will use ${\cal C}$ to denote Toffoli (or occasionally T) complexity while ${\cal A}$ and ${\cal B}$ will denote persistent and temporary ancilla costs, respectively.
For some of these oracles there are different Toffoli costs when performing them in the forward and reverse directions.
We always pair a forward oracle with a reverse oracle, so will give the average cost. 
In some cases the computation may introduce ancilla qubits not shown here, that are erased in the inverse computation.
For the function evaluation oracle we incorporate multiplication by the inverse temperature $\beta$.
The approximation $\tilde f$ is given to $\bsmooth$ bits, but for generality we allow an error $2^{-\bfun}$ which may be larger than $2^{-\bsmooth}$.}
\end{table*}

The remainder of \sec{oracles} discusses concrete circuit realizations
for ``oracles'' which provide information about these cost functions of interest. Here we slightly abuse the term ``oracle'' to mean a circuit primitive which is repeatedly queried throughout an algorithm, usually revealing information about the problem we are solving. These oracles are used by multiple algorithms throughout our paper.
In \sec{direct_oracle}, we explain how to implement cost function oracles
that are required to return the cost of a specific candidate solution $x$.
We refer to such oracles as ``direct energy oracles''.
In \sec{diff}, we explain how to implement cost function oracles that
are required to return the difference in cost between two candidate solutions
that differ by exactly one bit.
In \sec{phase_oracle}, we explain how to implement cost function oracles
that are required to return the cost function as a phase, rather than as a value
written to a separate quantum register.
In \sec{lcu_oracle}, we explain how to implement cost function
oracles that are required to implement the cost function as a direct application
of the Hamiltonian onto a target quantum register.
Finally, in \sec{functions}, we consider the cost of evaluating functions
whose input is the difference in cost of candidate solutions as described in the other parts of
this section.

We summarize the content of this section using three tables.
In \tab{symbols} we give a list of the symbols we use for reporting our
computational complexity results.
This table aids in the interpretation of the following two tables.
In \tab{oracle_defs}, we summarize the definitions of the various different kinds of
oracles considered in this section.
Finally, in \tab{oracles}, we summarize the complexities of each of the sixteen
cost function oracles (four types of oracles for each of four types of cost functions)
as well as the complexity of calculating functions of those oracle outputs.
In these tables, and throughout the paper, we use $\log$ to indicate logarithms base $2$.

\begin{table*}[t]\resizebox{\textwidth}{!}{
\begin{tabular}{c|c|c|c|c}
cost function & oracle type & Toffoli (*or T) gate count ${\cal C}$ & persistent ancilla ${\cal A}$ & temporary ancilla ${\cal B}$ \\
\hline\hline
$L$-term Spin Model & direct energy & $L \, \bdirect$ \eqref{eq:HLdirtof} & $\bdirect$ \eqref{eq:HLdirper} & $\bdirect-1$ \eqref{eq:HLdirtem} \\
 $H_L$ & energy difference & $2\,L \, \bdiff + {\cal O}(1)$ \eqref{eq:HLdirtof} & $\bdiff$ \eqref{eq:HLdirper} & $\bdiff-1$ \eqref{eq:HLdirtem} \\
& direct phase* & $1.15 L (\bphase + \log L )
+ {\cal O}(\log L)$ \eqref{eq:HLphaT} & 0 \eqref{eq:HLphaper} & $1$ \eqref{eq:HLphatem} \\
& Hamiltonian walk & $3 \, L + 2 \, \bLCU + {\cal O}(\log L)$ \eqref{eq:HLLCUtof} & $2\log L + 2\,\bLCU + {\cal O}(1)$ \eqref{eq:HLLCUper} & $\log L + {\cal O}(1)$ \eqref{eq:HLLCUtem} \\
\hline
Quadratic & direct energy & $N^2 \bdirect / 2 + {\cal O}(N \bdirect)$ \eqref{eq:QUBOdirtof} & $\bdirect$ \eqref{eq:QUBOdirper} & $\bdirect-1$ \eqref{eq:QUBOdirtem} \\
  Unconstrained & energy difference & $N\, \bdiff$ \eqref{eq:QUBOdiftof} & $\bdiff$ \eqref{eq:QUBOdifper} & $\bdiff-1$ \eqref{eq:QUBOdiftem} \\
Binary Optimization & direct phase* & $0.575 \, N^2 (\bphase +  2 \log N)
+ {\cal O}(N^2)$ \eqref{eq:QUBOphaT} & $0$ \eqref{eq:QUBOphaper} & $1$ \eqref{eq:QUBOphatem} \\
$H_{\rm QUBO}$ & Hamiltonian walk & $ N(\bLCU+2\log N) + {\cal O}(N)$ \eqref{eq:QUBOLCUtof} & $2\bLCU+4 \log N + \order{1}$ \eqref{eq:QUBOLCUper} & $3\log N + {\cal O}(\log \bLCU)$ \eqref{eq:QUBOLCUtem} \\
\hline
Sherrington- & direct energy & $N^2$ \eqref{eq:SKdirtof} & $2\log N$ \eqref{eq:SKdirper} & $4\log N$ \eqref{eq:SKdirtem} \\
Kirkpatrick Model & energy difference & $2N$ \eqref{eq:SKdiftof} & $\log N + 1$ \eqref{eq:SKdifper} & $2\log N+\order{1}$ \eqref{eq:SKdiftem} \\
 $H_{\rm SK}$  & direct phase &  $2N^2+\bphase^2/2+\order{\bphase\log\bphase}$ \eqref{eq:SKphaT} & $2\log N+\bphase+\order{\log{\bphase}}$ \eqref{eq:SKphaper} & $4\log N$ \eqref{eq:SKphatem} \\
& Hamiltonian walk & $ 6\,N + {\cal O}(\log^2 N)$ \eqref{eq:SKLCUtof} & $2\log N +{\cal O}(1)$ \eqref{eq:SKLCUper} & $3\log N +{\cal O}(1)$ \eqref{eq:SKLCUtem} \\
\hline
Low & direct energy & $5N(N+1)/4$ \eqref{eq:LABSdirtof} & $2 \log N + 1$ \eqref{eq:LABSdirper} & $3 \log N + 3$ \eqref{eq:LABSdirtem} \\
 Autocorrelation  & energy difference & $5N(N+1)/2$ \eqref{eq:LABSdirtof} & $2 \log N + 1$ \eqref{eq:LABSdirper} & $3 \log N + 3$ \eqref{eq:LABSdirtem} \\
Binary Sequences& direct phase & $\tfrac 85 N^2 + \min\left(\tfrac 12 N \bphase^2,\tfrac 9{10}N^2\right) + \order{N\bphase\log\bphase}$ \eqref{eq:LABSphatof} & $\bphase + {\cal O}(\log\bphase)$ \eqref{eq:LABSphaper} &  $5\log N + {\cal O}(\log\bphase)$ \eqref{eq:LABSphatem} \\
$H_{\rm LABS}$  & Hamiltonian walk & $4\, N + {\cal O}(\log N)$ \eqref{eq:LABSLCUtof} & $3\log N + {\cal O}(1)$ \eqref{eq:LABSLCUper} & $2\log N + {\cal O}(1)$ \eqref{eq:LABSLCUtem} \\
\hline
& function evaluation &  $\bsmooth^2+\bdiff+\mathcal{O}(\bsmooth\log \bsmooth+2^{\bfun/2})$ \eqref{eq:funtof} & $2\bsmooth+\mathcal{O}(\log \bsmooth)$ \eqref{eq:funper} & $\bdiff-1$ \eqref{eq:funtem} \\
& arcsine evaluation & $(\bsmooth+\bfun)^2+\bdiff+\mathcal{O}(\bsmooth\log \bsmooth+2^{\bfun/2})$ \eqref{eq:ASfuntof} & $2\bsmooth+\bfun+\mathcal{O}(\log \bsmooth)$ \eqref{eq:ASfunper} &  $\bdiff-1$ \eqref{eq:ASfuntem} \\
\hline
\end{tabular}}
\caption[Oracle complexities]{\label{tab:oracles}
Summary of complexities for realizing oracles used throughout this paper. Next to the complexity entry is a number indicating the equation in the paper which gives the full expression in context.
The energy difference for $H_L$ and LABS just has twice the Toffoli cost and the same ancilla cost as the direct energy oracle, because it is found by evaluating the energy twice. These oracles and the meaning of their precision parameters $b$ are defined in \tab{oracle_defs}. The Toffoli count is reported except when the oracle type for that cost function is marked with (*), which indicates that T count is reported instead. Here we include only the main terms in the order expressions. We use these costings to determine the complexities in \tab{primitives}.}
\end{table*}

\subsection{Oracles for direct cost function evaluation}
\label{sec:direct_oracle}

Many of the algorithms considered in this work are formulated in terms of a query to an oracle which computes the value of the cost function $C$ (for instance, one of the Hamiltonians discussed above) in a binary register. For instance, if we have a wavefunction $\ket{\psi} = \sum_x \psi_x \ket{x}$ where the computational basis states $\ket{x}$ are eigenstates of $C$ such that $C \ket{x} = E_x \ket{x}$ then we define the direct energy evaluation oracle $O^{\textrm{direct}}$ as a circuit which  acts as
\begin{equation}
O^{\textrm{direct}} \sum_{x} \psi_x \ket{x} \ket{0}^{\otimes \bdirect} \mapsto \sum_{x} \psi_x \ket{x} \ket{\tilde{E}_x}
\end{equation}
where $\tilde{E}_x$ is a binary approximation to $E_x$ using $\bdirect$ bits. We provide some strategies for how to realize this oracle for specific problems with low Toffoli complexity. We will refer to the Toffoli complexity of this oracle as ${\cal C}^{\rm direct}$. However, first we will discuss an efficient method for performing reversible in-place addition of a constant. This routine will be critical to our implementation.

\subsubsection{Direct energy oracle for \texorpdfstring{$L$}{L}-term spin model and QUBO}
\label{sec:direnegor}

We will now explain how to implement the direct energy oracle for the $H_L$ Hamiltonian with low Toffoli complexity. We will represent the energy $\tilde{E}_x$ in the two's complement binary representation, as this encoding enables efficient methods for addition \cite{GidneyAdder}. In two's complement positive integers have a normal binary representation whereas negative integers are the complement of that representation minus one. For instance, in 4-bit two's complement $3_{10} = 0011_2$ whereas $-3_{10} = 1100_2 + 1 = 1101_2$. Zero still corresponds to all bits zero. The fact that we need to add one for negative numbers complicates our approach but this representation is still preferable for our purposes.

The main idea behind our approach will be to add or subtract the value of each term's coefficient $w_\ell$ to a $b$-bit output register based on the parity of the string $\prod_{i \in q_\ell} Z_i$.
To perform addition or subtraction controlled on a qubit, we use the fact that one can switch between addition and subtraction by applying \textsc{not} gates to the target register in two's complement representation.
That is, applying \textsc{not} gates to all qubits of a register will change $\ket{v}$ to $\ket{-v-1}$.
Adding $w$ to this register will give $\ket{w-v-1}$, then applying \textsc{not} gates to all qubits again will yield $\ket{v-w}$.
To perform addition or subtraction controlled on a qubit, one can use the procedure shown in Figure 4(a) of \cite{GidneyAdder} (see \app{multiplication/add_constant}).
The complete procedure to compute the energy is then as given in \alg{directenergyLtermQUBO}.

\begin{algorithm}[H]

\begin{algorithmic}[1]

\REQUIRE{A quantum state $\sum_x a_x\ket{x}$, a vector of weights $\{w_\ell\}$ that specifies the $L$-term spin model or QUBO Hamiltonian.}
\ENSURE{An output state of the form $\sum_x a_x \ket{x} \ket{\tilde E_x}$,
where $H$ is the relevant Hamiltonian and $\tilde E_x$ is the approximate eigenvalue of $H$ corresponding to $\ket{x}$.}

\STATE{Use Clifford gates (\textsc{cnot} gates) to compute the parity of the term $\prod_{i \in q_\ell} Z_i$ in-place in a single system qubit $\ket{\pi_\ell}$. Specifically, if $x_i$ is the $i^{\rm th}$ bit of computational basis state $x$ then we are using \textsc{cnot}s to compute $\pi_\ell = (\sum_{i \in q_\ell} x_i) \, {\rm mod}~ \,2$.}

\STATE{Controlled on $\ket{\pi_\ell}$, use more $\textsc{cnot}$ gates to negate every bit of the output register. We will refer to this output register as $\ket{v}$. Thus, after this step we will have the state $\ket{0}\ket{v}$ if the first bit holds $\pi_\ell = 0$ and we will have the state $\ket{1}\ket{-v-1}$ if the first bit holds $\pi_\ell = 1$.}

\STATE{Using the strategy described in \app{multiplication/add_constant} for the addition of a constant, add a $\bdirect$-bit binary approximation $\tilde{w}_\ell$ to the coefficient $w_\ell$ into the output register. This step has Toffoli complexity $\bdirect-2$ where $\bdirect$ is the size of the output register. After this step we will have the state $\ket{0}\ket{v + \tilde{w}_\ell}$ if $\pi_\ell = 0$ and we will have the state $\ket{1}\ket{\tilde{w}_\ell-v-1}$ if $\pi_\ell = 1$.}
    
\STATE{Negate the output register using \textsc{cnot} gates, controlled on $\ket{\pi_\ell}$. After this step we will have the state $\ket{0}\ket{v + \tilde{w}_\ell}$ if $\pi_\ell = 0$ and we will have the state $\ket{1}\ket{v-\tilde{w}_\ell}$ if $\pi_\ell = 1$.}

\STATE{Using Clifford gates uncompute the parity $\pi_\ell$}.
\end{algorithmic}

 \caption{ \label{alg:directenergyLtermQUBO} Energy evaluation for $L$-term spin model and QUBO
}
 
\end{algorithm}

After performing this for $L$ terms one can verify that this will produce the intended state $\ket{v} = \ket{\tilde{E}_x}$ in the output register.
Toffoli gates enter only through the adder in step 3. Thus, in total our approach has Toffoli complexity $C_{L}^{\textrm{direct}}$ and ancilla requirements ${\cal A}_{L}^{\textrm{direct}},{\cal B}_{L}^{\textrm{direct}}$ given by
\begin{align}\label{eq:HLdirtof}
{\cal C}_{L}^{\textrm{direct}} &= L \, (\bdirect -2) < L\,\bdirect, \\
\label{eq:HLdirper}
{\cal A}_{L}^{\textrm{direct}} &= \bdirect, \\
\label{eq:HLdirtem}
{\cal B}_{L}^{\textrm{direct}} &= \bdirect - 1 < \bdirect,
\end{align}
where the ancilla refer to the carry bits for the adder in addition to the $\bdirect$ bits required to output the energy.
We note that for this oracle these costs have no dependence on the many-body order of the Hamiltonian $H_L$ since this only affects the number of \textsc{cnot} gates used to compute the parity of the terms.

This exact same reasoning can be used to determine the complexity of computing the energies for the QUBO Hamiltonian.  Due to the relative lack of structure in QUBO, there is no obvious way to improve over this general complexity. There we have $L = N(N+1)/2$ terms and so from \eq{HLdirtof},  \eq{HLdirper} and  \eq{HLdirtem} we require a number of Toffolis and ancillas equal to
\begin{align}\label{eq:QUBOdirtof}
{\cal C}^{\textrm{direct}}_{\textrm{QUBO}} &= \frac{N^2 \bdirect}{2} + \frac{N \bdirect}{2} -N(N+1) = \frac{N^2 \bdirect}{2} + {\cal O}(N \bdirect),\\ \label{eq:QUBOdirper}
{\cal A}^{\rm direct}_{\textrm{QUBO}} &= \bdirect,\\ \label{eq:QUBOdirtem}
{\cal B}^{\rm direct}_{\textrm{QUBO}} &= \bdirect - 1 < \bdirect.
\end{align}

\subsubsection{Direct energy oracle for SK model}
\label{sec:direct_oracle/SK}

Here we show that the energy for the SK model can be computed with only $N^2$ Toffolis and a logarithmic number of ancillas.
The method we use is a sum of tree sums of bits.
It is also possible to just use a tree sum with a Toffoli cost of about $N^2/4$, but the drawback is that this method would need $N^2/2$ ancilla qubits, which is prohibitive.

For the SK model it is convenient to replace $-1$ with $0$, so the sum takes values between $0$ and $L$.
That corresponds to dividing the Hamiltonian by $2$ and shifting it, which does not change the optimization problem, but means we are only summing bits.
If we were to sum the bits in the obvious way, the Toffoli complexity would be approximately $N^2\log N$.
However, we can take advantage of the fact we are summing bits to reduce the complexity to $\order{N^2}$.

Our methods are based on tree sums of bits.
In \cite{Kivlichan2019} it was shown that it is possible to sum $L$ bits using $L-1$ Toffoli gates and $L-1$ ancilla qubits, and this sum can be uncomputed with no Toffoli cost.
As discussed in \cite{Kivlichan2019}, it is also possible to perform sums in approaches that reduce the number of ancilla at the price of increasing the number of Toffoli gates.
In particular, we can subdivide the bits we are summing into about $L/\log L$ groups of size $\log L$, start by using the tree sum approach to sum each of the groups, add it into a running sum, and uncompute it.
The number of ancillas needed is reduced to approximately $\log L$ for each of the tree sums.
There is also a cost of approximately $L$ for adding the tree sums, giving a total complexity of approximately $2L$.

To be more specific, taking into account that $L$ need not be a power of two, we can use $M=\lceil L/\lceil\log L\rceil \rceil-1$ groups of size $\lceil\log L\rceil$, except for a remaining group of size $J\le\lceil\log L\rceil$ such that $M\lceil\log L\rceil+J=L$.
That is, there are $\lceil L/\lceil\log L\rceil \rceil$ groups, and $J$ can be smaller than $\lceil\log L\rceil$.
The Toffoli cost of computing each of these sums is
\begin{align}
M\lceil \log L\rceil-M+J-1 &= L-M-1 
= L-\lceil L/\lceil\log L\rceil \rceil .\label{eq:treesum}
\end{align}

The cost of the additions is
\begin{align}\label{eq:treesumsum}
    \sum_{j=1}^M \left[\lceil \log(J+j\lceil\log L\rceil+1) \rceil -1 \right] &\le M(\lceil\log (L+1)\rceil -1) \nn
    &\le M\lceil\log L\rceil\nn
    &< (L/\lceil\log L\rceil) \lceil\log L\rceil=L.
\end{align}
We have assumed that $L>1$ and hence $\log L> 0$.
The first line of \eq{treesumsum} comes from starting with the sum over $J$ bits and then adding each of the sums over $\lceil\log L\rceil$ to it.
After adding $j$ of the sums over $\lceil\log L\rceil$ bits, the maximum value of the sum is $J+j\lceil\log L\rceil$, so the number of bits needed to store the result is $\lceil \log(J+j\lceil\log L\rceil+1) \rceil$, and the number of Toffolis needed for that sum is one less than that.
The inequality in the first line comes from the fact that the total number is never less than $L$, so the cost of the additions is never greater than $\lceil\log (L+1)\rceil -1$.
The inequality in the second line is because $\lceil\log (L+1)\rceil -1\le \log L$.
The inequality in the third line is using $M<L/\lceil\log L\rceil$.

Therefore, the total Toffoli cost is less than $2L$.
The ancilla cost of each tree sum is $\lceil\log L\rceil-1$, there are $\lceil \log (L+1)\rceil$ ancilla needed for the total, and $\lceil \log (L+1)\rceil-1$ temporary ancillas for the addition of the tree sum into the total.
Since the ancillas in the tree sum are uncomputed, they contribute to an overall temporary ancilla cost, meaning the temporary ancilla cost is $2\log L+\order{1}$ and the persistent ancilla cost (for the total) is $\log L+\order{1}$.

Since $L=N(N-1)/2$, if we were to use a tree sum the cost would be less than $N^2/2$, but the ancilla cost would be approximately $N^2/2$. The sum could be uncomputed without ancillas, giving an average (compute and uncompute) cost of $N^2/4$. We expect that the tradeoff is not worth it in this case. However, by using the sum of tree sums, we get a Toffoli cost less than $N^2$, and an ancilla cost that is logarithmic in $N$.
That gives costs for the SK model of
\begin{align}\label{eq:SKdirtof}
{\cal C}^{\textrm{direct}}_{\textrm{SK}} &< N^2,\\ \label{eq:SKdirper}
{\cal A}_{\rm SK}^{\rm direct} &\le 2 \log N, \\ \label{eq:SKdirtem}
{\cal B}_{\rm SK}^{\rm direct} &< 4 \log N.
\end{align}

\subsubsection{Direct energy oracle for LABS model}

Next we show that for the LABS problem it is possible to compute the energy with a Toffoli cost of $5N(N+1)/4$ for $N\ge 64$, with a logarithmic number of ancilla qubits.
We  improve over the application of our general technique by specializing the implementation to the LABS problem. Since the LABS problem has $L = {\cal O}(N^3)$ with maximum integer energy values of ${\cal O}(N^3)$, we would expect a complexity of ${\cal O}(N^3)$. Instead, we show that it is possible to perform the direct energy evaluation at cost ${\cal O}(N^2)$. We focus on the form of the LABS Hamiltonian that is expressed as $\sum_{k=1}^N | H_k |$ where $H_k$ is as defined in \eq{labs} (as we mentioned, this form of the problem has the same ordering of the low energy landscape). 

In the following we use $E_k$ to denote the eigenvalue of $H_k$.
It will be most efficient to use the sum of tree sums approach described above.
Here we need to find $E_k$ by using $+1$ and $-1$ rather than $+1$ and $0$, because we need to take the absolute value, so we need an extra bit for the sign.
Therefore, after summing bits, we will need to multiply by $2$ (which has no Toffoli cost), followed by subtracting the number of bits.
The overall approach is then as follows.
We will sum $k$ starting at $k=N-1$ and go down to zero, so the number of bits at each step is minimized.
For each value of $k$ we will perform \alg{LABS}.

\begin{algorithm}[H]

\begin{algorithmic}[1]

\REQUIRE{A quantum state $\sum_x a_x\ket{x}$, the set of all terms in the LABS Hamiltonian $\{H_k\}$.}
\ENSURE{An output state of the form $\sum_x a_x \ket{x} \ket{E_x}$.}

\STATE{Compute for computational basis vector $\ket{x}$ the value of $E_k$ in a scratch register $\ket{u}$ that will require $\lceil\log (N-k+1)\rceil+1$ ancilla to store (with $+1$ for the sign).}

\STATE{Controlled by the highest bit of $u$ (the sign bit in two's complement), use \textsc{cnot} gates to negate the value of the output register $\ket{v}$. At this point we have  $\ket{u}\ket{v}$ if $u \geq 0$ or $\ket{u}\ket{-v-1}$ if $u < 0$.}

\STATE{Add the scratch register into the output register. }
    
\STATE{Use \textsc{cnot}s to negate the output register controlled on the highest bit of the $\ket{u}$ register.}

\STATE{Uncompute $\ket{u}$.}
\end{algorithmic}
 
 \caption{Energy evaluation for LABS model \label{alg:LABS}}
\end{algorithm}
In Step 1, the Toffoli complexity computing each $E_k$ is approximately $2(N-k)$ plus the cost of subtracting $N-k$.
In two's complement we can determine whether the number is negative or positive by looking at the highest bit; if the highest bit is 1 then we know the value is negative.  This justifies the operations in Step 2.  Since Step 2 requires no non-Clifford operations, it can be neglected in our cost analysis.  In Step 3, the state is $\ket{u}\ket{u + v}$ if $u \geq 0$ or $\ket{u}\ket{v - u}$ if $u < 0$; equivalently we now have the state $\ket{u}\ket{v + |u|}$.  The output register will be of size $\lceil \log[(N-k)(N-k+1)/2+1] \rceil+1$ so the Toffoli cost is $\lceil \log[(N-k)(N-k+1)/2+1] \rceil$.
    
The output register is significantly larger than the scratch register.
However, with a slight modification of the procedure in \app{multiplication/add_constant} we can allow this register to be smaller with no additional Toffoli cost.
First, consider expanding the number of qubits $\ket{u}$ is encoded on.
This is of course trivial for positive numbers.
For negative $u$, for $n$ bits it is encoded as $2^n+u$.
Therefore, if we have a number that is negative and we need to map it to a negative number on some larger number of bits $n'$, then we need to map $2^n+u$ to $2^{n'}+u$, which means adding $2^{n'}-2^n=\sum_{j=n}^{n'-1} 2^j$.
This means that bits $n+1$ to $n'$ of the negative number encoded on the $n'$ bits need to be ones.
These can be set by using CNOTs controlled by bit $n$, which means no additional Toffoli cost is needed to encode the number into more qubits. A further simplification can be used to eliminate the need for those extra qubits.
First, rearrange the addition circuit as in \fig{adder2} so that the qubits of $\ket{u}$ are only used as controls and not changed.
Since all of the additional qubits for $\ket{u}$ contain the same value as the sign qubit of $\ket{u}$, we may use that sign qubit as the control instead of any of those additional qubits. Then the additional qubits are not used, and can be omitted.
    
There is an improvement that we can make when we take into account that each computation needs to be paired with an uncomputation.
This is because, in step 5, if we are computing an energy that we will later uncompute, then we can use the strategy of \cite{GidneyAdder} to erase $\ket{u}$ using $X$ measurements and no Toffoli cost.
A phase correction is required, but that can be done when we later uncompute the LABS energy.
This means that in step 5 we have a cost of $N-k$ in uncomputing the LABS energy, but no Toffoli cost in computing the LABS energy.
Because each computation is paired with an uncomputation, it is therefore convenient to give the average complexity of $N-k$. The largest temporary ancilla cost is when we need to uncompute the overall Hamiltonian, when it is $2\log(N-k)+\order{1}$.
That is still less than the temporary ancilla cost in step 3, so can be ignored.

\begin{figure}[t]
\centering
  \resizebox{.48\linewidth}{!}{\includegraphics{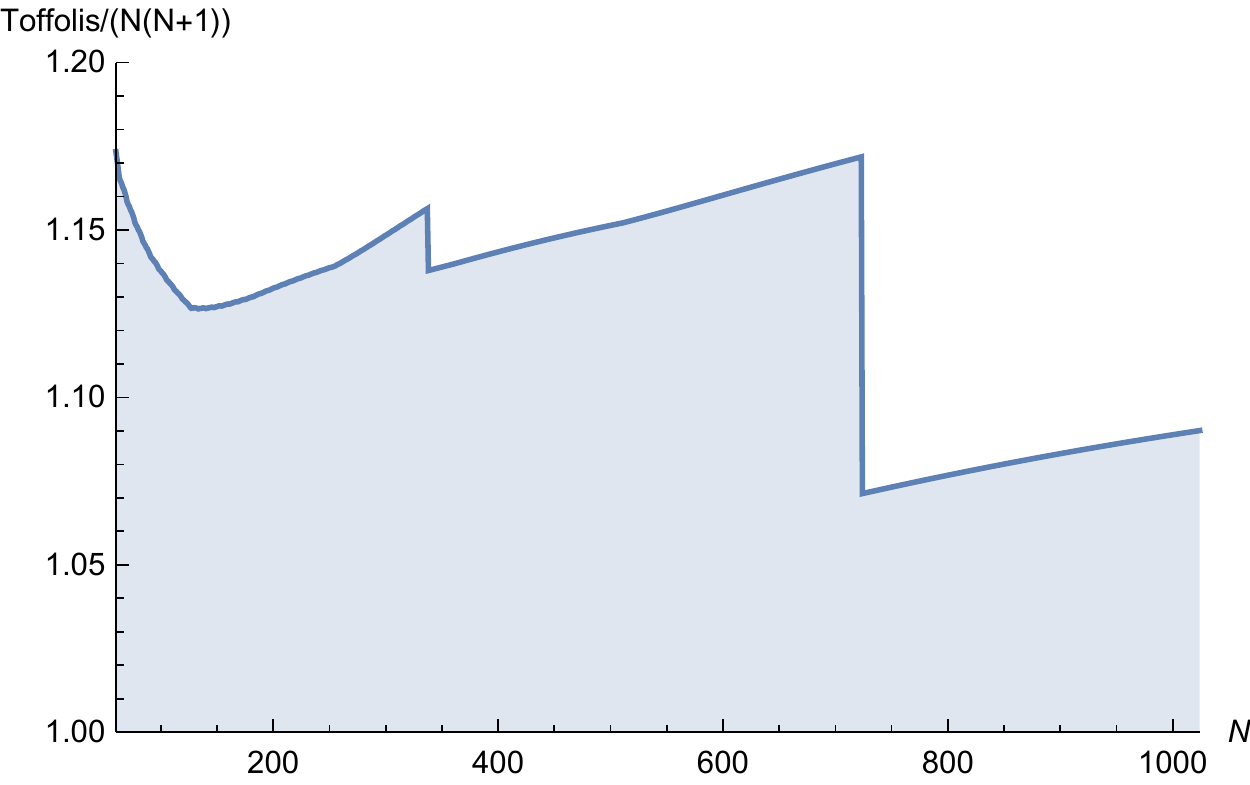}}\qquad
    \resizebox{.48\linewidth}{!}{\includegraphics{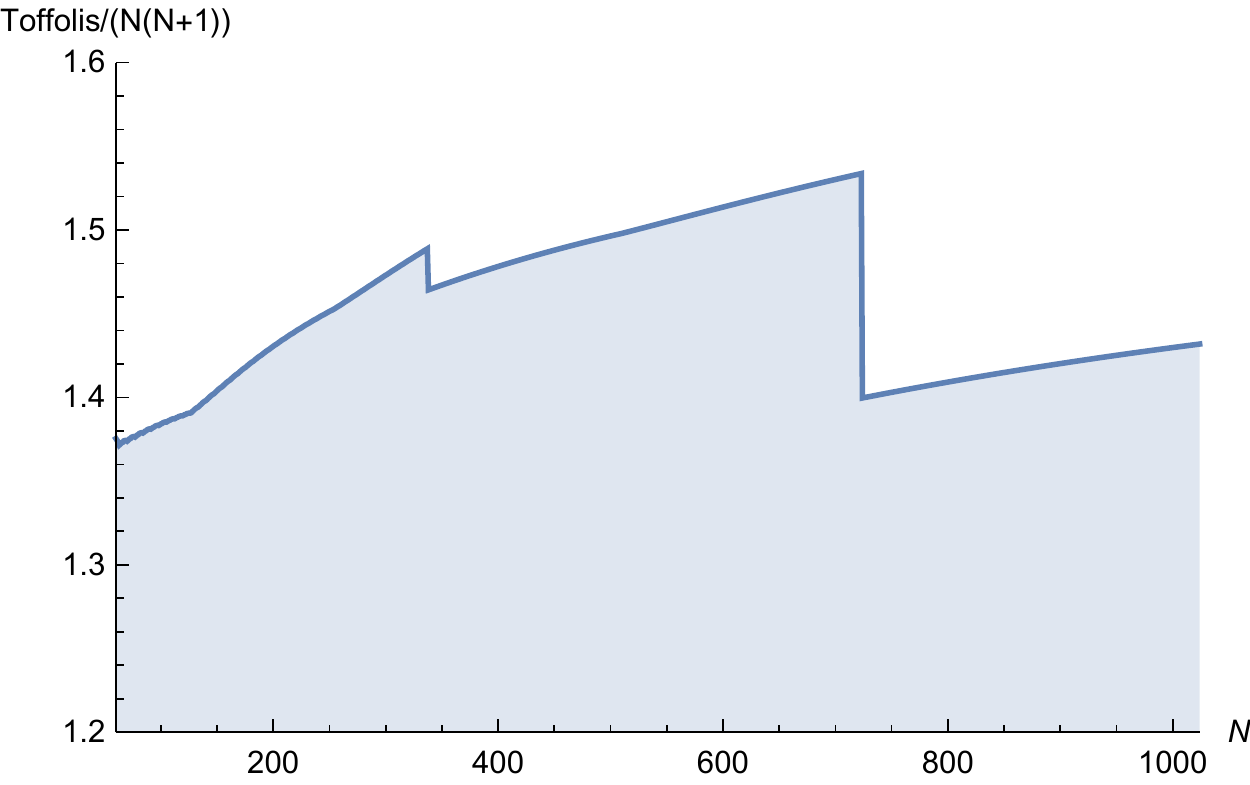}}
   \put(-388,150){(a)}
  \put(-125,150){(b)}
  \caption[Costs for LABS using tree sums]{\label{fig:LABSsum}
    The Toffoli costs for direct evaluation of the LABS Hamiltonian by computing $H_k$ with a sum of tree sums.
    The average cost when computing and uncomputing the Hamiltonian is shown in (a).
    The cost of just computing and uncomputing $H_k$ (omitting the cost of summing the absolute values), when we just compute the Hamiltonian, is given in (b).}
\end{figure}

After repeating this for the $N$ values of $k$ one can verify that the output register will contain the energy of the LABS Hamiltonian. Toffoli gates enter only through steps 1, 3 and 5. The primary contribution to the complexity is the computation of $E_k$ in steps 1 and 5.
Ignoring the complexity of subtracting $N-k$, the Toffoli complexity is
\begin{equation}
    \sum_{k=0}^{N-1}3(N-k) = 3N(N+1)/2.
\end{equation}
The cost of the subtractions as well as the additions in step 3 will increase the cost, but also $2(N-k)$ is an overestimate of the cost of adding $n-k$ bits.
In particular, we can use tree sums of as many as approximately $\log N$ bits, rather than just $\log(N-k)$, with no penalty in terms of the temporary ancilla cost.
The computed costs are shown in \figa{LABSsum}{(a)}, and it is found for the range of $N$ we are interested in (64 -- 1024), the constant factor on $N(N+1)$ is less than $1.2$, rather than $1.5$ (in fact, this bound is good for $N\ge 45$).
In particular, the constant factors for $N=64$, $128$, $256$ and $1024$ are $1.16466$, $1.12673$, $1.13945$, and $1.0901$, respectively.
To simplify the expressions we give the slightly looser bound in the table 
\begin{equation}\label{eq:LABSdirtof}
    {\cal C}^{\textrm{direct}}_{\textrm{LABS}} < 5N(N+1)/4 ,
\end{equation}
with the caveat that it is for $N\ge 45$.
The number of ancilla we will require is
\begin{align}\label{eq:LABSdirper}
{\cal A}^{\textrm{direct}}_{\textrm{LABS}} &= \lceil \log[N(N+1)/2+1] \rceil \le 2 \log N + 1, \\
\label{eq:LABSdirtem}
{\cal B}^{\textrm{direct}}_{\textrm{LABS}} &= \lceil \log[N(N+1)/2+1] \rceil+\lceil\log (N-k+1)\rceil+2  \le 3 \log N + 3.
\end{align}
The persistent ancilla are for the output value.
Approximately $2 \log N$ of the temporary ancilla are for carry bits in the addition and $\log N$ are for the scratch register.
We assume $N>1$ for the inequalities which omits the trivial case.
This example illustrates how taking advantage of problem structure can lead to advantages over the implementation of an oracle intended to handle a more general case.

\subsection{Energy difference oracles}
\label{sec:diff}

For some of the algorithms discussed in this work (specifically the quantum versions of simulated annealing) we often need the direct energy oracle only as means to compute a difference between the energies of two different states which differ in only one bit. The ultimate objective in that context is a circuit that performs the mapping
\begin{equation}
\label{eq:energy_difference}
O_k^{\textrm{diff}} \sum_{x} \psi_x \ket{x} \ket{0}^{\otimes \bdiff} \mapsto \sum_{x} \psi_x \ket{x} \ket{\widetilde{\delta E}_x^{(k)}}, \qquad \qquad \delta E_x^{(k)} = E_x - E_y, \qquad \qquad \ket{y} = X_k \ket{x},
\end{equation}
where (as usual) $X_k$ is the \textsc{not} operation on qubit $k$ and $\widetilde{\delta E}_x^{(k)}$ is a binary approximation to $\delta E_x^{(k)}$ using $\bdiff$ bits. Especially when the many-body order is 2-local, it is more efficient to consider a specialized implementation of $O_k^{\textrm{diff}}$ than to try to realize this operation using one call to $O^{\rm direct}$ and one call to $O^{\rm direct} X_k$.

First, we will discuss the energy difference oracle for QUBOs. In this case, $\delta E_x^{(k)}$ is the eigenvalue of the operator
\begin{equation}
\delta H^{(k)} = 2 h_k Z_k + 2 \sum_{i \neq k} J_{ik} Z_i Z_k .
\end{equation}
We see that $\delta H^{(k)}$ is itself a simple cost function which is an example of $H_{N}$ (the $L$-term spin model with $L=N$). Thus, to compute the eigenvalue of this operator (equivalent to implementing $O_k^{\textrm{diff}}$) we would require
\begin{align}
\label{eq:QUBOdiftof}
{\cal C}^{\textrm{diff}}_{\textrm{QUBO}} &= N \, (\bdiff-2) < N \,\bdiff,\\
\label{eq:QUBOdifper}
{\cal A}^{\textrm{diff}}_{\textrm{QUBO}} &= \bdiff ,\\
\label{eq:QUBOdiftem}
{\cal B}^{\textrm{diff}}_{\textrm{QUBO}} &= \bdiff-1 .
\end{align}
This scaling is much less than the $N^2 \bdiff + {\cal O}(N \bdiff)$ Toffoli gates that would be required by making two queries to the direct energy oracle for QUBO.

For the SK model we can simplify the QUBO result. We would then have the difference operator $2 \sum_{i \neq k} w_{ik} Z_i Z_k$, so we just need to sum $N-1$ bits, and can take $\bdiff = \lceil\log N\rceil$. We also need to subtract $N-1$ from the bit sum to obtain the energy difference, but the cost of that subtraction plus the cost of the bit sum is still no more than the upper bound of $2N$ we gave previously on the cost of the bit sum. Therefore the energy difference oracle has cost
\begin{align}\label{eq:SKdiftof}
{\cal C}^{\textrm{diff}}_{\textrm{SK}} &< 2N , \\
\label{eq:SKdifper}
{\cal A}^{\textrm{diff}}_{\textrm{SK}} &= \,\lceil\log N\rceil \le  \,\log N + 1 ,\\
\label{eq:SKdiftem}
{\cal B}^{\textrm{diff}}_{\textrm{SK}} &\le  \,2\log N +\order{1}.
\end{align}

For higher many-body order Hamiltonians like LABS or the $H_L$ model of many-body order greater than two, the best strategy will probably involve two applications of the direct energy oracle $O^{\textrm{direct}}$. However, rather than actually use two registers to output the energy and then perform subtraction one can instead just compute the energy of $x$ first and then in the same register compute the energy of $y$ while subtracting all of the terms instead of adding them. There is a slightly greater Toffoli cost because the subtraction is on a slightly larger number of qubits, but that cost is small enough to be ignored.
This leads to Toffoli complexity of $2\, {\cal C}^{\textrm{direct}}$ but requires no additional ancilla.

\subsection{Oracles for phasing by cost function}
\label{sec:phase_oracle}

In some contexts our goal will be to phase each computational state on which the wavefunction has support by an amount proportional to the energy of that computational basis state (this task is equivalent to performing evolution under a diagonal Hamiltonian for unit time). We will refer to circuits that achieve this task as a ``phase'' oracle and define them to act as
\begin{equation}
O^{\textrm{phase}} \! \left(\gamma \right) \sum_x \psi_x \ket{x} \mapsto  \sum_x e^{-i \widetilde{\gamma E_x}} \psi_x \ket{x} \qquad \qquad \left | \widetilde{\gamma E_x} - \gamma E_x \right | \leq 2^{-\bphase} .
\end{equation}
To simplify the following discussion, we assume that $E_x$ is shifted such that it is non-negative. Such a shift corresponds to an unobservable global phase.

To realize this oracle, one strategy would be to first approximately compute $E_x$ into a register using $O^{\textrm{direct}}$, then multiply by $\gamma$ and perform further logic to phase the system by the amount in the register. For instance,
\begin{equation}
 \left(O^{\textrm{direct}}\right)^\dagger \left(\openone \otimes U^{\rm phase}\left(\gamma\right)\right) O^{\textrm{direct}} \sum_{x} \psi_x \ket{x} \mapsto \sum_{x} e^{-i \widetilde{\gamma E_x}} \psi_x \ket{x} \end{equation}
where the phasing operation needed is
\begin{equation}
U^{\rm phase}\left(\gamma\right) = \!\!\! \sum_{k=0}^{2^{\bdirect}-1}\!\!\! \exp\left(\frac{2 \pi i k \tilde{\gamma}}{2^{\bdirect}}\right) \proj{k}.
\end{equation}
The value of $2\pi k\tilde{\gamma}/2^{\bdirect}$ would correspond to the approximation of $\gamma E_x$, with $k$ the integer approximating $E_x$ (so $k \approx 2^{\bdirect} E_x/E_{\max}$) and $\tilde{\gamma}=\gamma E_{\max}/(2\pi)$ is a scaled form of $\gamma$.
We will limit ourselves to simulations where the phase factor is no more than a factor of $2\pi$, so $\tilde{\gamma}\le 1$. Using the ``phase gradient'' trick of \cite{Kitaev2002,GidneyAdder}, it is possible to apply a phase by adding into a reusable ancilla register initialized to the state
\begin{equation}\label{eq:phasegrad}
\ket{\phi} = \frac{1}{\sqrt{2^{\gradbits}}} \sum_{\ell=0}^{2^{\gradbits} - 1} e^{-2 \pi i \ell / 2^{\gradbits}} \ket{\ell}.
\end{equation}
Here we use $\gradbits$ rather than $\bdirect$ in this state to allow for needing to use more bits to obtain the required precision in the phase.
For details see \app{rotations}.
In this case we need to multiply by the classically specified number $\tilde{\gamma}$ to obtain the required phase.
This number can be given by $\log\tilde{\gamma} + \bphase+\order{1}$ digits in order to obtain error $<2^{-\bphase}$.
There will be error due to the finite number of digits for $E_x$, the finite number of bits for $\tilde{\gamma}$, and the multiplication.

Rather than performing the multiplication by $\tilde{\gamma}$, adding into the phase gradient state, then uncomputing the multiplication, a more efficient method is to perform the multiplication by repeated addition into the phase gradient state.
For each non-zero bit of $\tilde{\gamma}$, we can add a bit-shifted copy of $k$ into the phase gradient state.
Each addition into the phase gradient state has cost $\gradbits-2$, and on average approximately half the bits of $\tilde{\gamma}$ will be zero, giving cost roughly $\gradbits(\log \tilde{\gamma} + \bphase)/2$.
To address cases where more bits of $\tilde{\gamma}$ are nonzero, we can write $\tilde{\gamma}$ as a sum of powers of $2$ with plus and minus signs.
In that case it is possible to use no more than $(\log \tilde{\gamma} + \bphase)/2+\order{1}$ additions, giving cost $\gradbits(\log \tilde{\gamma} + \bphase)/2+\order{\gradbits}$.
The error due to omission of bits in the multiplication is no more than approximately $2^{-\gradbits}(\log \tilde{\gamma} + \bphase)\pi$, so to obtain error $<2^{-\bphase}$ one should take $\gradbits=\bphase+\order{\log \bphase}$.
That gives an overall cost for the multiplication
\begin{equation}
    \frac{\bphase(\log \tilde{\gamma} + \bphase)}{2}+\order{\bphase\log \bphase}.
\end{equation}
For more details see \app{rotations}.
Note finally that the state $\ket{\phi}$ can be initialized prior to simulation and reused throughout, with a negligible additive one time cost scaling as ${\cal O}(b_{\rm grad}^2)$. This one time cost comes from synthesizing $\gradbits$ arbitrary rotations. However, since this is additive to the overall cost (whereas all other oracle costs are multiplicative with the number of queries), we expect this will be negligible.

For the $L$-term spin Hamiltonians and QUBOs, the cost of the multiplication by $\gamma$ can be eliminated by simply including it in the coefficients of the problem Hamiltonian.
However for these cases an even more efficient approach is to simulate each term explicitly in a Trotter-like fashion and perform rotation synthesis to decompose each rotation into a sequence of T gates. In that case, one would require a number of T gates equal to the number of terms times the cost of rotation synthesis, which gives a complexity of ${\cal O}(L (\bphase + \log L))$.
Using the repeat until success circuits of \cite{Bocharov2015},
this would give T gate and ancilla complexities of roughly
\begin{align}\label{eq:HLphaT}
{\cal C}^{\textrm{phase}}_{L} &= 1.15 \, L(\bphase + \log L) + 10.925 \, L + {\cal O}\left(1\right) = 1.15 \, L (\bphase + \log L) + {\cal O}(L),\\
\label{eq:HLphaper}
{\cal A}^{\textrm{phase}}_{L} &= 0,\\
\label{eq:HLphatem}
{\cal B}^{\textrm{phase}}_{L} &= 1 .
\end{align}
There is a single temporary ancilla qubit used by the repeat until success circuits.
The measure of error in \cite{Bocharov2015} is the Frobenius distance $d(U,V)=\sqrt{1-|{\rm Tr}(UV^\dagger)|/2}$.
A phase error of $2^{-\bphase}$ gives $|{\rm Tr}(UV^\dagger)|/2=|1+\exp(2^{-\bphase}i)|/2=\cos(2^{-\bphase}/2)$. Expanding in a series gives a Frobenius distance of $2^{-\bphase}/\sqrt 8+{\cal O}(2^{-3 \bphase})$.
That means the cost becomes $1.15\big(\bphase + \log (\sqrt 8 )\big)+9.2=1.15 \,\bphase +10.925$, which is why the second term above is different than in \cite{Bocharov2015}.
Because Toffoli gates require roughly twice the resources to distill as T gates \cite{Gidney2019}, this approach is likely to be more efficient in practice. This would give T and ancilla complexities for QUBO of
\begin{align}\label{eq:QUBOphaT}
{\cal C}^{\textrm{phase}} &= 0.575 \, N(N\!+\!1) [\bphase + \log(N(N\!+\!1))] + 4.9 \, N(N\!+\!1) + {\cal O}\left(1\right) = 0.575 \, N^2 (\bphase + 2 \log N) + {\cal O}(N^2),\\
\label{eq:QUBOphaper}
{\cal A}^{\textrm{phase}} &= 0 ,\\
\label{eq:QUBOphatem}
{\cal B}^{\textrm{phase}} &= 1 ,
\end{align}
assuming $N > \bphase$.

For the SK model it is better to compute the energy, add the energy into the phase gradient state, then uncompute the energy.
That has Toffoli complexity $2N^2$, with $2\log N$ persistent ancillas and $4\log N$ temporary ancillas.
The cost of the multiplication directly into the phase gradient state is $\bphase^2/2+\order{\bphase\log\bphase}$ (with $\bar\gamma\le 1$), with $\gradbits$ permanent ancillas for the phase gradient state and $\gradbits-1$ temporary ancillas for the addition.
That gives costs for SK of
\begin{align}\label{eq:SKphaT}
{\cal C}^{\textrm{phase}}_{\textrm{SK}}  &= 2N^2+\bphase^2/2+\order{\bphase\log\bphase},\\
\label{eq:SKphaper}
{\cal A}^{\textrm{phase}}_{\textrm{SK}} &= 2\log N+\bphase+\order{\log{\bphase}},\\
\label{eq:SKphatem}
{\cal B}^{\textrm{phase}}_{\textrm{SK}} &= \max\left(4\log N,\bphase+\order{\log{\bphase}}\right).
\end{align}
For the parameters we consider for examples of gate counts, $4\log N\ge \bphase$, so we give that in \tab{oracles}.

For the LABS model we still need to explicitly compute the partial sum for $H_k$ and then take the absolute value. Instead of adding the absolute value of that to an output register we can $\textsc{cnot}$ the highest bit (indicating the sign of the partial sum $u$) into a single ancilla. Then, we can negate the whole partial sum controlled on this ancilla so that we have the state $\ket{u}\ket{0}$ if $u \geq 0$ or $\ket{-u-1}\ket{1}$ if $u < 0$. Then, we can add this ancilla to the partial sum register giving us either $\ket{u}\ket{0}$ if $u \geq 0$ or $\ket{-u}\ket{1}$ if $u < 0$. At this point we can multiply by $\gamma$ and add the value of $u$ to the $\ket{\phi}$ register and perform phase kickback in order to phase the system by the absolute value of the partial sum. Then, we need to invert adding the sign qubit register to the sum register and uncompute $\ket{u}$ and the ancilla.

Using the sum of tree sums, we numerically find that the Toffoli cost to compute and uncompute the partial sums is no greater than $8N(N+1)/5$ for $N$ in the range $64$ to $1024$ that we consider.
The numerically computed ratios are shown in \figa{LABSsum}{(b)}, and for $64$, $128$, $256$ and $1024$ we obtain $1.35962$, $1.38507$, $1.45027$, and $1.43186$.
Multiplying by $\bar\gamma$ directly into the phase gradient state has cost $\bphase^2/2+\order{\bphase\log\bphase}$, giving a total cost
\begin{equation}\label{eq:LABSphatof_ineq}
{\cal C}^{\textrm{phase}}_{\textrm{LABS}} 
\le 8N(N+1)/5 + N \bphase^2/2 + \order{N\bphase\log\bphase}.
\end{equation}
The number of ancillas needed is $\gradbits$ persistent ancillas for the phase gradient state, $\gradbits-1$ temporary ancillas for the addition, $\log N+\order{1}$ for the temporary ancilla with the partial sum for $H_k$, and $2\log N+\order{1}$ for the temporary ancillas used for the sum of tree sums.
The ancillas for the partial sum for $H_k$ are needed at the same time as those for the addition into the phase gradient state, but the temporary ancillas for the sum of tree sums are not.
The temporary ancillas for the sum of tree sums will be less than those for the addition into the phase gradient state, so can be ignored.
That gives us a total of $\gradbits+\lceil\log (N+1)\rceil+1$ temporary ancillas for a total
\begin{align}\label{eq:LABSphaper}
{\cal A}^{\textrm{phase}}_{\textrm{LABS}} &=
\gradbits =
\bphase + {\cal O}(\log\bphase), \\
{\cal B}^{\textrm{phase}}_{\textrm{LABS}} &= 
\gradbits+\lceil\log (N+1)\rceil+1 = \bphase + \log N + {\cal O}(\log\bphase).
\end{align}

For $9N/5<\bphase^2$, 
it is more efficient to just compute the entire energy, multiply by $\bar\gamma$, then uncompute the energy, as explained above.
Then we obtain complexity 
\begin{equation}
{\cal C}^{\textrm{phase}}_{\textrm{LABS}} 
\le 5N^2/2 + \order{N\bphase\log\bphase},
\end{equation}
where the cost of multiplying by $\bar\gamma$ is absorbed into the order term.
Because this is smaller than that given above for $9N/5<\bphase^2$, we should give the cost as
the minimum of the two complexities
\begin{equation}\label{eq:LABSphatof}
{\cal C}^{\textrm{phase}}_{\textrm{LABS}} 
\le 8N(N+1)/5 +\min\left(N \bphase^2/2,\tfrac 9{10}N^2\right) + \order{N\bphase\log\bphase}.
\end{equation}
In that case we need $2\log N+\order{1}$ temporary ancillas for the energy, and $\gradbits-1$ temporary ancillas for the addition into the phase gradient state at the same time.  
There are also $3\log N+\order{1}$ temporary ancillas for computing the energy, which are not used at the same time as $\gradbits-1$ temporary ancillas.
That gives a number of temporary ancillas increased to
\begin{equation}
    \label{eq:LABSphatem}
{\cal B}^{\textrm{phase}}_{\textrm{LABS}} = \max(\bphase,3\log N) + 2\log N + {\cal O}(\log\bphase).
\end{equation}
We give this cost in \tab{oracles} to account for the possibility of using either method.
In the table we assume $3\log N\ge \bphase$, because that is true for most combinations of parameters we consider.

\subsection{Oracles for linear combinations of unitaries}
\label{sec:lcu_oracle}

A number of approaches to quantum simulation are based on accessing the Hamiltonian as a linear combination of unitaries. This so-called ``linear combination of unitaries'' (LCU) query model \cite{Childs2012} has been used for Taylor series simulation \cite{Berry2015}, interaction picture simulation \cite{Low2018}, and generalized to block encodings for ``qubitization'' \cite{Low2016}.
These approaches begin from the observation that any Hamiltonian can be decomposed as a linear combination of unitaries,
\begin{equation}
\label{eq:lcu}
H = \sum_{\ell = 1}^{L} w_\ell U_\ell
\end{equation}
where $w_\ell$ are real scalars and $U_\ell$ are unitary operators.

\begin{figure}[t!]
\centering
  \resizebox{.8\linewidth}{!}{\includegraphics{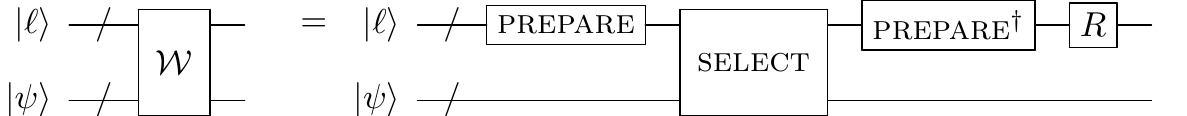}}
  \caption[Circuit realisation of qubitized quantum walk]{\label{fig:w}
    A circuit realizing the qubitized quantum walk operator ${\cal W}$ controlled on an ancilla qubit \cite{Low2016,Babbush2018}.
    Here $R$ is a reflection about the zero state for the entire $\ket{\ell}$ register, and therefore has Toffoli complexity $\log L+{\cal O}(1)$ where $\lceil\log L\rceil$ is the size of the $\ket{\ell}$ register. However, that overhead is negligible compared to the cost of the $\prep\,$ and $\sel\,$ operators in the constructions of this paper.}
\end{figure}

Here we consider an approach to forming quantum walks known as qubitization \cite{Low2016}. The quantum walk involves LCU using queries to two oracles, followed by a reflection operation as shown in \fig{w}. The first oracle circuit, the ``preparation oracle'', acts on an empty ancilla register of $\lceil\log L\rceil$ qubits and prepares a particular superposition state related to the notation of \eq{lcu},
\begin{equation}
\label{eq:prepare}
\prep \ket{0}^{\otimes \log L} \mapsto  \sum_{\ell = 1}^{L} \sqrt{\frac{w_\ell}{\lambda}} \ket{\ell},
\qquad \qquad
\lambda \equiv \sum_{\ell = 1}^{L} \left| w_\ell \right | \, .
\end{equation}
The quantity $\lambda$ has significant ramifications for the overall algorithm complexity; specifically, the qubitization oracles will need to be repeated a number of times proportional to $\lambda$ in order to realize the intended quantum walk.

The second oracle circuit we require acts on the ancilla register $\ket{\ell}$ as well as the system register $\ket{\psi}$ and directly applies one of the $U_\ell$ to the system, controlled on the ancilla register. For this reason, we refer to the ancilla register $\ket{\ell}$ as the ``selection register'' and name the second oracle the ``Hamiltonian selection oracle'',
\begin{equation}
\label{eq:select}
\sel \ket{\ell} \ket{\psi} \mapsto \ket{\ell} U_\ell \ket{\psi}.
\end{equation}

Using two queries to $\prep$ and a single query to $\sel$ we are able to implement a controlled quantum walk ${\cal W}$ which encodes the eigenvalues of $H$ as a function of its own eigenvalues \cite{Low2016}. Specifically, in a subspace this quantum walk has eigenvalues equal to the arccosine of the eigenvalues of the problem Hamiltonian divided by $\lambda$.
We now discuss the realization of this quantum walk for the problems discussed in \sec{oracles}.

\subsubsection{LCU oracles for \texorpdfstring{$L$}{L}-term Hamiltonian}

Using the strategy for unary iteration introduced in \cite{Babbush2018} we can implement $\sel\,$ for $H_{L}$ with Toffoli complexity of exactly $L-2$ and $\lceil\log L\rceil-1$ extra ancilla qubits (or $L-1$ and $\lceil\log L\rceil$ if the operation needs to be controlled by another ancilla, as it is in \cite{Babbush2018}). The circuit given there has $\lceil\log L\rceil$ ancilla. The other ancilla is just a control, it isn't needed for the iteration. If we don't want to make it controlled, then the number of ancilla needed is $\lceil\log L\rceil-1$. Also, the Toffoli cost is only $L-2$ if we don't need to make it controlled. The operator we are to implement is
\begin{equation}
\label{eq:unary_it}
\sel \ket{\ell}\ket{\psi} \mapsto \ket{\ell} \prod_{i \in q_\ell} Z_i \ket{\psi}.
\end{equation}
A simple way to understand the strategy would be to first map the binary representation of $\ket{\ell}$ to a one-hot unary register (a register that contains $L$ qubits which are all \textsc{off} except for qubit $\ell$ which is \textsc{on}). Then, one could control the application of the $Z_i$ associated with $i \in q_\ell$ on this qubit with only Clifford gates. This strategy would have low Toffoli complexity but would require $L$ ancilla. The basic insight of the unary iteration circuits in \cite{Babbush2018} is that one can stream through bits of this unary register using just $\lceil\log L\rceil-1$ extra ancilla.
A circuit primitive is repeated $L$ times and at iteration $j$, a particular ancilla is equal to \textsc{on} if and only if $\ell = j$. At that point in the circuit we can use Clifford gates to control the application of Hamiltonian terms like $Z_i Z_j Z_k$.

In \cite{Babbush2018} a strategy referred to therein as ``coherent alias sampling'' is introduced and explicit circuits are provided which allow one to realize $\prep$ for an arbitrary model with a Toffoli cost of $L + \bLCU + \log L+{\cal O}\left(1\right)$.
We need approximately $\log L$ ancillas for the state being prepared, $\log L$ for the alternate index values, and $\log L$ for the temporary ancillas in the QROM.  There are $\bLCU$ ancillas for the keep probabilities in the coherent alias sampling and $\bLCU$ for the equal superposition state.
Another temporary ancilla is used for the result of the inequality test. $\sel$ uses $L$ Toffolis and $\log L$ temporary ancilla, but these can be reused from the temporary ancilla used by $\prep$. Here, $\bLCU$ is a parameter that scales the precision of the cost function. In particular, this strategy will generate the state in \eq{prepare} but with approximate coefficients $\tilde{w}_\ell$ in place of the exact coefficients $w_\ell$ such that 
\begin{equation}
\left | \sqrt{w_\ell} - \sqrt{\tilde{w}}_\ell \right | \leq {2^{-\bLCU}}.
\end{equation}

Per the realization depicted in \fig{w},
the quantum walk of interest is realized using two queries to $\prep\,$ and one query to $\sel$. Thus, the strategy we have outlined requires Toffoli and ancilla counts of
\begin{align}
\label{eq:HLLCUtof}
{\cal C}^{\textrm{LCU}}_{L} &= 3\,L + 2\,\bLCU + 2\log L + {\cal O}\left(1\right),\\
\label{eq:HLLCUper}
{\cal A}^{\textrm{LCU}}_{L} &= 2\lceil\log L\rceil + 2 \,\bLCU + {\cal O}(1), \\
\label{eq:HLLCUtem}
{\cal B}^{\textrm{LCU}}_{L} &= \lceil\log L\rceil = \log L + \order{1}.
\end{align}

\subsubsection{LCU oracles for QUBO and using dirty ancilla}

In some cases, especially when there is some structure in the Hamiltonian terms and one is willing to reduce gate complexity at the cost of space complexity, another method of implementing $\prep$ might be appropriate. In particular, we can combine the coherent alias sampling ideas of \cite{Babbush2018} with the on-the-fly ``dirty QROAM'' of \cite{BGMMB19} (which is a concrete realization of an idea in \cite{Low2018a} which builds on the QROM idea of \cite{Babbush2018} and is named ``QROAM'' since it incorporates attributes of both QROM and QRAM). Using Theorem 1 of \cite{BGMMB19} in conjunction with the coherent alias sampling of \cite{Babbush2018} with cost $\bLCU+\mathcal{O}(\log N)$, we see that it is possible to implement $\prep\,$ with 
\begin{equation}
\frac{2 \,L}{\chunk} + 4 \, \bLCU \, \chunk + {\cal O}\left(\bLCU + \chunk\log L\right)
\end{equation}
Toffolis and $(\chunk-1) \bLCU$ dirty ancilla in addition to $2 \bLCU + \log(L / \chunk)  + {\cal O}(1)$ clean ancilla (not counting the selection register), where $\chunk \in [1, L]$ is a free parameter that must be a power of $2$. This sort of QROAM can be uncomputed faster than it can be computed \cite{BGMMB19}. 
Combining Theorem 3 in \cite{BGMMB19} with coherent alias sampling \cite{Babbush2018} leads us to the result that the Toffoli cost of uncomputing $\prep$ is less than the complexity quoted above by $4 (\bLCU-1) \chunk$
and can reuse the same ancilla. The number of dirty ancilla is reduced to $\chunk-1$, which means that the value of $\chunk$ can be taken to be larger, reducing the Toffoli complexity. See \tab{qrom_complexity} for detailed costs of various types of QROAM.

We will use this dirty QROAM strategy for the QUBO Hamiltonian. Our approach will involve indexing the terms and coefficients with two registers, each of size $\lceil\log N\rceil$ so that $\ket{\ell} = \ket{i}\ket{j}$. This makes applying $\sel$ particularly easy as we can use two applications of the unary iteration strategy that we discussed for implementing \eq{unary_it} to realize $\sel$ with Toffoli complexity $2 N -4$ and $\lceil\log N\rceil -1$ ancilla (again, not counting those in the selection register).
Because the QROAM strategy needs a single register that takes a contiguous set of values, we need to compute a new register for QUBO.
For QUBO where $i\le j$ one would calculate $j(j-1)/2+i$. (Note that this is with indexing starting from $1$, which we have done to simplify the sums, but $1$ would be represented in binary as $00\ldots00$, and so forth.)
We apply the QROAM to this register, then uncompute it afterwards. The cost of computing and uncomputing this register is $\mathcal{O}(\log^2N)$ due to the multiplications. Since $L = N(N+1) / 2$ for QUBO, the Toffoli cost of implementing $\sel$, in addition to implementing (and later uncomputing) $\prep$, will be
\begin{equation}
\frac{2 \, N^2}{\chunk} + 4 \, \bLCU \, \chunk + 2 \, N + {\cal O}\left(\bLCU + \log^2 N\right)
\end{equation}
and will require $\chunk \bLCU + {\cal O}(1)$ dirty ancilla and $2 \bLCU + 2 \log(N / \chunk) + 2\log N + {\cal O}(1)$ clean ancilla. For simplicity we are taking $\chunk$ to be the same for the computation and uncomputation here, though it is more efficient to take $\chunk$ larger for the uncomputation. Minimizing $\chunk$ by taking the derivative gives us
\begin{equation}
4 \bLCU - 2 N^2 / \chunk^2 = 0, \qquad \qquad \chunk = N / \sqrt{2 \bLCU} ,
\end{equation}
which leads to Toffoli complexity for the entire walk (including $\sel$) going like
\begin{align}
4\, N\sqrt{ 2\bLCU} + 2\,N +
{\cal O}\left(\bLCU + \log^2 N\right) = 4 N \sqrt{2 \bLCU} + {\cal O}(N)
\end{align}
and ancilla complexity for the entire walk going like
\begin{align}
N\sqrt{\bLCU/2} + 2\,\bLCU + 2 \log \bLCU + 2 \log N + {\cal O}(1) = N\sqrt{\bLCU/2}+ {\cal O}(\log (\bLCU N)),
\end{align}
where the first term in the ancilla scaling corresponds to the dirty ancilla, and thus can use the system qubits. For simplicity we have used the exact optimal value of $\chunk$ here; there will a slight increase to the complexity because $\chunk$ needs to be a power of $2$ so cannot be taken exactly equal to $N / \sqrt{2 \bLCU}$.

While this result optimizes the Toffoli complexity of our implementation it does so at a fairly high cost; we have increased the space complexity from $N + {\cal O}(\bLCU)$ to $N \sqrt{\bLCU / 2} + {\cal O}(\bLCU)$. In many cases this will not be a sensible tradeoff and one should instead choose a smaller $\chunk$ so that the total number of qubits is not increased. For instance, $\chunk = N / \bLCU$ will never increase the spatial complexity because we will always have $N$ system qubits available in the system register that are not acted upon while we apply $\prep$. In some cases (for instance, the quantum simulated annealing algorithm realized by Szegedy quantum walks) we will actually have $2N$ qubits available for use during $\prep$ and so we can safely take $\chunk = 2N / \bLCU$ without increasing the spatial complexity.

\begin{table*}[t]
\begin{tabular}{c|c|c|c|c}
type of ancilla & type of computation & Toffolis & clean ancilla & dirty ancilla \\
\hline\hline
clean & forward & $\lceil L/\chunk \rceil + M(\chunk-1)$ & $\lceil \log(L/\chunk) \rceil + M(\chunk-1)$ & 0 \\
dirty & forward & $2\lceil L/\chunk \rceil + 4M(\chunk-1)$ & $\lceil \log(L/\chunk) \rceil$ & $M(\chunk-1)$ \\
clean & reverse & $\lceil L/\chunk \rceil + \chunk$ & $\lceil \log(L/\chunk) \rceil + \chunk$ & 0 \\
dirty & reverse & $2\lceil L/\chunk \rceil + 4\chunk$ & $\lceil \log(L/\chunk) \rceil+1$ & $\chunk-1$ \\
\hline
\end{tabular}
\caption[QROAM complexities.]{\label{tab:qrom_complexity}
The QROAM complexities from \cite{BGMMB19}, where $L$ is the number of items, $\chunk$ is a power of $2$, and $M$ is the output size. This table omits the $\log L$ ancilla from the selection register and the $M$-qubit output.}
\end{table*}

Next we give a more detailed explanation of the costing. The QROAM costings are, for output size $M$, given in \tab{qrom_complexity}.
The value of $L$ is $L=N(N+1)/2$ for QUBO.
The output consists of $\bLCU$ qubits for the keep probability in the state preparation, plus $2\lceil\log N\rceil$ qubits for the alternate values of $i$ and $j$, so
\begin{equation}\label{eq:Mopt}
    M=\bLCU+2\lceil\log N\rceil.
\end{equation}
With clean ancilla qubits, the optimal value of $\chunk$ for preparation limited to powers of $2$ is
\begin{equation}\label{eq:kopt1}
\chunk_{c1} = 2^{{\rm round}(\log\sqrt{L/M})},
\end{equation}
and for inverse preparation is
\begin{equation}\label{eq:kopt2}
\chunk_{c2} = 2^{{\rm round}(\log\sqrt{L})},
\end{equation}
The other Toffoli costs in other parts of the LCU (beyond the QROAM) are as follows.
\begin{itemize}
    \item There is $\mathcal{O}(\log N)$ cost to prepare the equal superposition states over $i$ and $j$ with $i\le j$.
    \item There is $2(\bLCU+2\log N)+\mathcal{O}(1)$ Toffoli cost for the inequality test and controlled swaps for the state preparation and inverse preparation.
    \item The cost of the arithmetic for producing the contiguous ancilla is $\mathcal{O}(\log^2 N)$.
    \item The $\sel$ has a Toffoli cost of $2N-4$, or $2N-2$ if it needs to be made controlled.
\end{itemize}
Altogether these costs give a Toffoli cost with clean ancilla of
\begin{equation}
\frac{N(N+1)}{2\chunk_{c1}}+\frac{N(N+1)}{2\chunk_{c2}}+M(k_{c1}-1)+\chunk_{c2}+2\bLCU+2N+\mathcal{O}\left(\log^2 N\right),
\end{equation}
with the values of $M$, $\chunk_{c1}$, and $\chunk_{c2}$ in \eq{Mopt}, \eq{kopt1}, and \eq{kopt2}.
If we ignore the rounding in $\chunk_{c1}$ and $\chunk_{c2}$, then the Toffoli cost is
\begin{equation}
    \sqrt{2\bLCU}N + \mathcal{O}\left(N+\bLCU+\bLCU^{-1/2} N\log N\right).
\end{equation}
The rounding in $\chunk_{c1}$ and $\chunk_{c2}$ can potentially increase the cost by a factor of $1/\sqrt{2}+1/\sqrt{8}$, or about $6\%$.

In costing the total number of ancillas for the state preparation, we also need to account for the following (in addition to those in \tab{qrom_complexity}).
\begin{itemize}
\item There are $2\lceil\log N\rceil$ qubits needed for the prepared state.
\item There are $\bLCU$ qubits used for the register in equal superposition that we use to perform an inequality test with in the state preparation.
\item The $M$ output qubits.
\item There are $\lceil\log L\rceil$ temporary ancilla qubits used for the contiguous register.
\item There are $\bLCU-1$ temporary ancillas used in computing the inequality test for the state preparation.
\end{itemize}
There are also $\log N$ temporary registers needed for the $\sel$ step, but many of the qubits are only temporarily used by the QROAM, and these can be reused, so we do not get an additional ancilla cost for $\sel$.
The ancillas additional to those in \tab{qrom_complexity} can therefore be given as $2M$ persistent ancillas and $\max(\log L,\bLCU)+\mathcal{O}(1)$ temporary ancillas.
The ancillas in \tab{qrom_complexity} are temporary as well, and the $\bLCU$ qubits are not needed at the same time, giving a maximum of
\begin{equation}
\log(L/\chunk_{c1})+M(\chunk_{c1}-1)+\log L + \mathcal{O}(1),
\end{equation}
temporary ancillas. Ignoring the rounding in $\chunk_{c1}$ for simplicity gives the leading-order term as $N\sqrt{M/2}$ temporary ancillas. Next we consider the cost with $N$ dirty ancilla.
The optimal value of $\chunk$ for the QROAM computation is
\begin{equation}\label{eq:kvald1}
    \chunk_{d1} = 2^{\lfloor \log (N/M+1)\rfloor}.
\end{equation}
For the uncomputation cost it is optimal to take $\chunk_{d2}=\sqrt{L/2}$ which gives a cost of $4\sqrt{2L}$, ignoring rounding of $\chunk_{d2}$ to a power of $2$.
With $L=N(N + 1)/2$, the optimal $\chunk_{d2}$ is $\sqrt{N(N + 1)/4}<N$, so there are enough dirty ancilla available.
With rounding the value of $\chunk_{d2}$ for uncomputation would be
\begin{equation}\label{eq:kvald2}
    \chunk_{d2} = 2^{{\rm round} (\log \sqrt{N(N + 1)/4})}.
\end{equation}
Together with the additional Toffoli costs for the state preparation, the Toffoli cost for LCU is
\begin{equation}
    \frac{N(N+1)}{\chunk_{d1}}+\frac{N(N+1)}{\chunk_{d2}}+4M(\chunk_{d1}-1)+4\chunk_{d2}+2\bLCU+2N+\mathcal{O}(\log^2 N).
\end{equation}
To simplify the expression, we will use $N/M$ rather than $N/M+1$ in the expression for $\chunk_{d1}$, and not take into account rounding $k$ to a power of $2$.
Then we get a computation Toffoli cost of
\begin{equation}\label{eq:QUBOLCUtof}
{\cal C}^{\rm LCU}_{\rm QUBO} = N(\bLCU+2\log N) + {\cal O}(N).
\end{equation}

For the ancilla cost, the persistent ancilla cost is again $2M$, and the temporary ancilla cost loses the term $M(k-1)$ because dirty ancilla are used for that, so it does not increase the ancilla cost.
The temporary ancilla cost is
\begin{equation}
\max(\log(L/\chunk_{d1}) + \log L,\bLCU) + \mathcal{O}(1).
\end{equation}
Using $L=N(N + 1)/2$ and $\chunk_{d1}=N/M$ gives $\log(L/\chunk_{d1})=\log(N+ 1)+\log M-1$.
Then $\log(N + 1)=\log N+\mathcal{O}(1/N)$.
Using $M=\bLCU+2\log N+\mathcal{O}(1)$ then gives 
\begin{align}
\label{eq:QUBOLCUper}
{\cal A}^{\rm LCU}_{\rm QUBO} &= 2 \, \bLCU + 4\log N + \order{1},\\
\label{eq:QUBOLCUtem}
{\cal B}^{\rm LCU}_{\rm QUBO} &= \max(3\log N,\bLCU) +\mathcal{O}(\log \bLCU) .
\end{align}
In \tab{oracles} we just give $3\log N$ for the temporary ancilla cost, because it is true (or close to true) for the combinations of parameters we consider.

\subsubsection{LCU oracles for the SK model}

For the SK model we can considerably improve over the naive implementation. Because the SK model coefficients only need to give a sign, we just need to apply a sign to the terms in the superposition.
That corresponds to the phase fixup used for the QROAM uncomputation, and the cost is the same.
Another advantage of this approach is that we eliminate the $2(\bLCU+2\log N)+\mathcal{O}(1)$ cost for the inequality test and controlled swaps that would otherwise be needed for the coherent alias sampling.
Therefore the Toffoli cost with clean ancilla is
\begin{equation}
\frac{N(N-1)}{2\chunk_{c2}}+k_{c2}+2N+\mathcal{O}\left(\log^2 N\right).
\end{equation}
If we ignore the rounding in $\chunk_{c2}$ then we obtain the complexity
\begin{equation}
(2+\sqrt 2)N+\mathcal{O}\left(\log^2 N\right).
\end{equation}
Beyond the ancillas needed for the QROAM, we just need the $4\log N+\mathcal{O}(1)$ qubits for the $i$, $j$, and contiguous registers.
Again $\sel$ can use the same temporary ancillas as the QROAM and does not add to the ancilla cost.
Therefore the ancilla cost is
\begin{equation}
\log(L/\chunk_{c2})+\chunk_{c2}+4\log N +\mathcal{O}(1).
\end{equation}
Ignoring the rounding in $\chunk_{c2}$ for simplicity gives
\begin{equation}
N/\sqrt{2} + \mathcal{O}\left(\log N\right).
\end{equation}
If we are using dirty ancilla, then the Toffoli cost becomes
\begin{equation}
\frac{N(N-1)}{\chunk_{d2}}+4k_{d2}+2N+\mathcal{O}\left(\log^2 N\right).
\end{equation}
Ignoring the rounding in $\chunk_{d2}$ we obtain the complexity
\begin{equation}\label{eq:SKLCUtof}
{\cal C}^{\textrm{LCU}}_{\textrm{SK}} = 6N+\mathcal{O}\left(\log^2 N\right).
\end{equation}
The persistent ancilla cost is only $2\log N$ for the $i$ and $j$ registers, and there is temporary ancilla cost of $2\log N$ for the contiguous register and $\log(L/\chunk_{d2})\approx\log N$ from the QROAM.  The total ancilla costs are therefore
\begin{align}\label{eq:SKLCUper}
{\cal A}_{\rm SK}^{\rm LCU} &= 2\log N +\mathcal{O}(1), \\
\label{eq:SKLCUtem}
{\cal B}_{\rm SK}^{\rm LCU} &= 3\log N +\mathcal{O}(1).
\end{align}

\subsubsection{LCU oracles for the LABS model}

The LABS problem has $L = {\cal O}(N^3)$ terms in it, which would lead a high complexity quantum walk if our general strategy were applied. Fortunately, there is much structure in this problem. We start by rewriting \eq{labs} as
\begin{equation}
H_{\rm LABS} = \sum_{k=0}^{N-1} \sum_{j=1}^{N - k} \sum_{i=1}^{N - k} Z_i Z_{i + k} Z_j Z_{j + k}.
\end{equation}
Instead of linearly indexing all ${\cal O}(N^3)$ terms, we will use three registers, each of size $\log N$, which store the values of $i$, $j$ and $k$. Thus, our $\sel\,$ operation will act as
\begin{equation}
\sel \ket{i}\ket{j}\ket{k} \ket{\psi} \mapsto  \ket{i}\ket{j}\ket{k} Z_i Z_{i + k} Z_j Z_{j + k} \ket{\psi}.
\end{equation}
To accomplish this, we simply need 4 applications of the unary iteration primitive described in \cite{Babbush2018}. Each of these primitives require $N - 1$ Toffoli gates. The only nuance is that we will need to compute the values $i + k$ and $j + k$ before implementing the primitive to perform $Z_{i + k}$ and $Z_{j + k}$.

These additions can be performed in place (and then uncomputed) in the $i$ and $j$ registers and introduce a negligible additive $4\log N$ cost to the cost of unary iteration, where the cost of addition is $\lceil\log N\rceil-1\le \log N$. Thus, the total Toffoli cost of our $\sel\,$ implementation is $4 N + 4\log N$.
We require approximately $3\log N$ persistent ancilla for the $i$, $j$ and $k$ registers, another $\log N$ temporary ancilla for computing the $i+k$ and $j+k$ (since they are computed in place), and $\log N$ temporary ancilla for the addition. The unary iteration uses $\lceil\log N\rceil-2<\log N$ ancillas, which can be reused from the temporary ancillas for the addition so do not add to the cost.
Because all terms have the same coefficient, $\prep$ needs to initialize a superposition over a number of items that is not a power of $2$. The Toffoli cost is $\mathcal{O}(\log N)$. The only unfortunate aspect is that for the LABS problem the corresponding normalization $\lambda$, is quite large and this will enter into the complexity of our quantum walks as the number of times the quantum walk must be repeated to realize the intended unitary. In total then the cost to realize the quantum walk in \fig{w} is
\begin{align}\label{eq:LABSLCUtof}
{\cal C}^{\textrm{LCU}}_{\textrm{LABS}} &= 4\, N + {\cal O}(\log N),\\
\label{eq:LABSLCUper}
{\cal A}^{\textrm{LCU}}_{\textrm{LABS}} &= 3 \log N + {\cal O}(1),\\
\label{eq:LABSLCUtem}
{\cal B}^{\textrm{LCU}}_{\textrm{LABS}} &= 2 \log N + {\cal O}(1),\\
\lambda_{\textrm{LABS}} &\approx N^3/3.
\end{align}

\subsection{QROM-based function evaluation}
\label{sec:functions}

Now that we have explained how to implement oracles for various cost
functions of interest, we turn to the question of how to calculate
functions of the cost. This is important for several possible approaches
to heuristic-based combinatorial optimization. In simulated annealing,
for instance, the probability of moving from one candidate solution to
another is proportional to an exponential of the energy difference between
the two solutions, multiplied by an inverse temperature $\beta$. We would thus require the quantum computer to calculate
an exponential of the output of the relevant energy difference oracle.

Because we are implementing \emph{heuristic} approaches to combinatorial
optimization, we do not expect that the functions of the cost need to be
calculated to a high degree of accuracy so long as the functions we compute are still monotonic in the cost (to make sure that the energy landscape is not inverted in any way). We instead want to minimize the computational complexity of evaluating these functions given rather weak
requirements on the accuracy of the output. Here we describe a general
strategy for such cheap approximate function evaluation.

Our overall strategy is to approximate a function $f$ of a $b$-bit input
$z$ by a piecewise linear approximation, $\tilde f$. This approximation
$\tilde f$ is calculated based on a choice of \emph{sample} points
$z_0 < z_1 < \ldots < z_g$, where $z_0 \leq z < z_g$. These sample points
separate the interval $[z_0, z_L)$ into $g$ different sub-intervals of the
form $[z_\ell, z_{\ell+1})$ with $\ell = 0, 1, \ldots, g - 1$. The input $z$
belongs to exactly one of these sub-intervals, and so we find an $\ell$ such
that $z_\ell \leq z < z_{\ell + 1}$. Having found $\ell$, we use some data
that can be looked up in order to calculate $\tilde{f} (z) = \alpha f(z_\ell)
+ (1 - \alpha) f(z_{\ell + 1})$ for $\alpha=(z_{\ell+1}-z)/(z_{\ell+1}-z_\ell)$. That is,
the function $\tilde f$ is defined by interpolating between known values
$f(z_\ell)$ and $f(z_{\ell + 1})$ of the target function $f$.

QROM~\cite{Babbush2018} can be used to obtain the region that $z$ is in
(i.e.~the correct value of $\ell$ above),
and for that region the QROM outputs a slope and intercept for the linear approximation. The Toffoli cost of looking up one of $g$ different possible values in the scheme of \cite{Babbush2018} is
$g - 2$, or $g-1$ if the output is controlled by a qubit.
This Toffoli count relies on a technique from~\cite{GidneyAdder} in which certain naively expected Toffolis can be replaced with Clifford gates plus measurement.
Also note that the Toffoli count of QROM-based lookup is independent of the number of
bits of data output, meaning that we are free to choose any number of bits
to represent the slope and intercept without introducing a Toffoli cost from the QROM.
We choose the number of bits in order to obtain $\bsmooth$ bits for $\tilde f$.
That is, $\tilde f$ may be a rough approximation of $f$, but we give $\tilde f$ to more bits than needed by that approximation so $\tilde f$ has smooth behaviour.

We will not use QROM precisely as specified in~\cite{Babbush2018} but
rather a variant of it. To explain the distinction, we begin with some
terminology. QROM is a method for executing a quantum circuit that
operates on two registers, an \texttt{input} register and an 
\texttt{output} register. The \texttt{input} register has an initial
value of $\ell$ encoded into it and the
\texttt{output} register starts in the all-zero state. Each value of $\ell$
corresponds to some piece of data $d_\ell$ that has been specified 
classically before the quantum circuit was constructed.
The effect of the QROM is
\begin{equation}
    \text{QROM}:
    \ket{\ell}_\texttt{input} \ket{0}_\texttt{output} \mapsto
    \ket{\ell}_\texttt{input} \ket{d_\ell}_\texttt{output}.
\end{equation}

\begin{figure}[t]
\centering
  \includegraphics[height=.32\textheight]{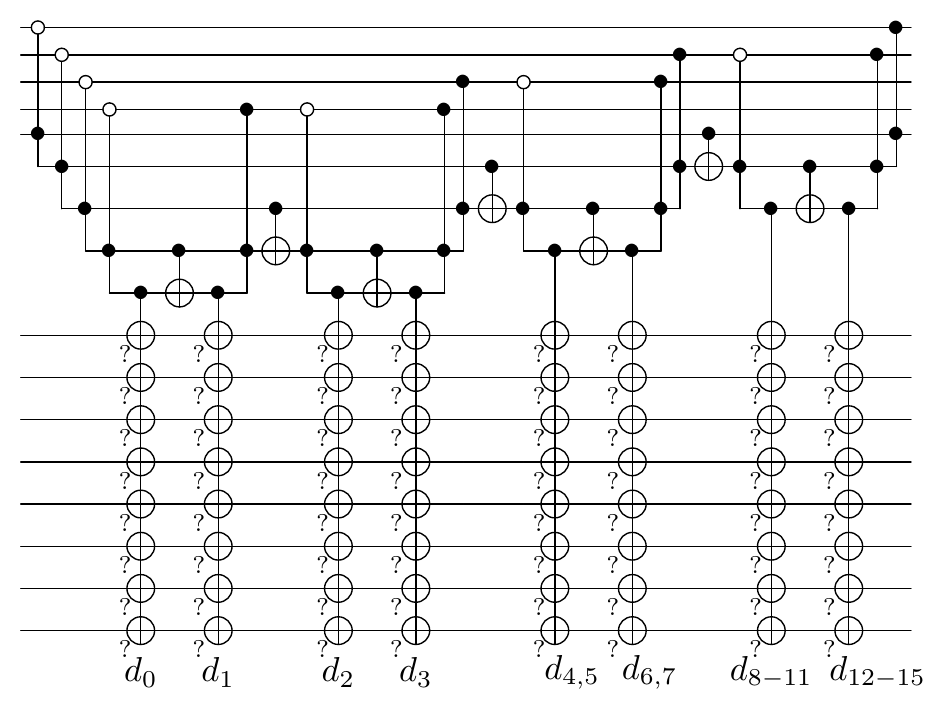} \quad
  \includegraphics[height=.32\textheight]{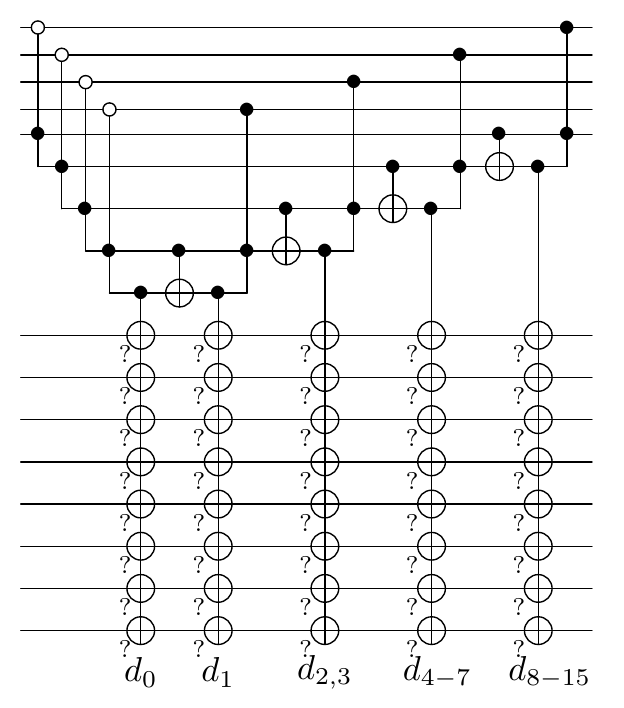}
  \put(-500,205){(a)}
  \put(-202,205){(b)}
  \caption[QROM with variable spacing.]{\label{fig:QROM}
    (a) This figure shows how to perform QROM with variable spacing for the example where there are 4 bits, and we aim to group the input numbers as $0$, $1$, $2$, $3$, $\{4,5\}$, $\{6,7\}$, $\{8,9,10,11\}$, $\{12,13,14,15\}$.
    That is, we output the same data for inputs of $4$ and $5$, and so forth.
    The first four lines are the four input bits and the fifth is a control register.
    There are $6$ Toffolis needed in this example for $8$ data points, with one more Toffoli for a control.
    (b) This figure shows how to perform QROM with variable spacing for the example where there are 4 bits, and we group the input numbers by powers of $2$ as $0$, $1$, $2-3$, $4-7$, and $8-15$.
    There are $3$ Toffolis needed in this example for $5$ data points, with one more Toffoli for a control.}
\end{figure}

Our variant of QROM is designed for the case in which there are
data collisions. That is to say, we consider the case where $d_\ell =
d_{\ell^\prime}$ for several different pairs $\ell$ and $\ell^\prime$.
In \figa{QROM}{(a)} we explain how this QROM variant works for $L = 16$
in the case where $d_4 = d_5$, $d_6 = d_7$, $d_8 = d_9 = d_{10} = d_{11}$,
and $d_{12} = d_{13} = d_{14} = d_{15}$. In this variant, we imagine that
we have distinct parts of the iteration: iterate by $\ell \mapsto \ell+1$,
iterate by $\ell \mapsto \ell + 2$, iterate by $\ell \mapsto \ell + 4$, 
and so on for each power of two. This variant of QROM is appropriate for our purposes because we want
to improve computational efficiency by spacing $z_\ell$ unevenly.
This is equivalent to treating many pieced of data $d_\ell$ as being equal, as the data is simply the information needed to calculate a linear function.
The total number of Toffoli gates is still $g-2$ for $g$ distinct regions, provided these regions correspond to ignoring bits of the input.
For example, we can use a region such as $\{4,5\}$, but not $\{3,4\}$, because $4\equiv 100$ and $5\equiv 101$, so grouping $4$ and $5$ corresponds to ignoring the least significant bit, but the least significant bit changes between $3$ and $4$.

A further subtlety is that all regions need to be a size corresponding to a power of $2$ for this cost.
In some cases we may wish to have a final region that is larger than half, so it is not a size that is a power of $2$.
That will occur because we can have a large energy difference, but the exponential will give a transition probability that will just be approximated as zero for a wide range of energies.
Then the cost can be larger.
For example, if we are distinguishing $0$ from $1-15$, then it will take $3$ Toffolis.
The cost can be seen from the diagram where the size of the regions increases in powers of $2$, shown in \figa{QROM}{(b)}.
There one can choose the numbers used for $d_{4-7}$ and $d_{8-15}$ to be equal, which gives a region for $4-15$. This choice corresponds to a situation where the gap between neighboring interpolation points $z_\ell$ grows exponentially.

For many of the piecewise approximations, we can obtain accurate approximations using just powers of $2$, as in \figa{QROM}{(b)}.
Two main types of function that we aim to approximate are the exponential and the arcsine of the exponential.
For the exponential the piecewise approximation can use points at argument values of $0$, $1/2$, $1$ and so on and achieve a piecewise linear approximation within about $0.03$.
The arcsine of the exponential is more difficult to approximate because the slope diverges at an argument of $0$, but using piecewise linear approximation points starting at $1/2^{11}$ and going up by powers of $2$ gives similar precision as for the exponential.

\begin{figure}[t]
\centering
  \resizebox{.413\linewidth}{!}{\includegraphics{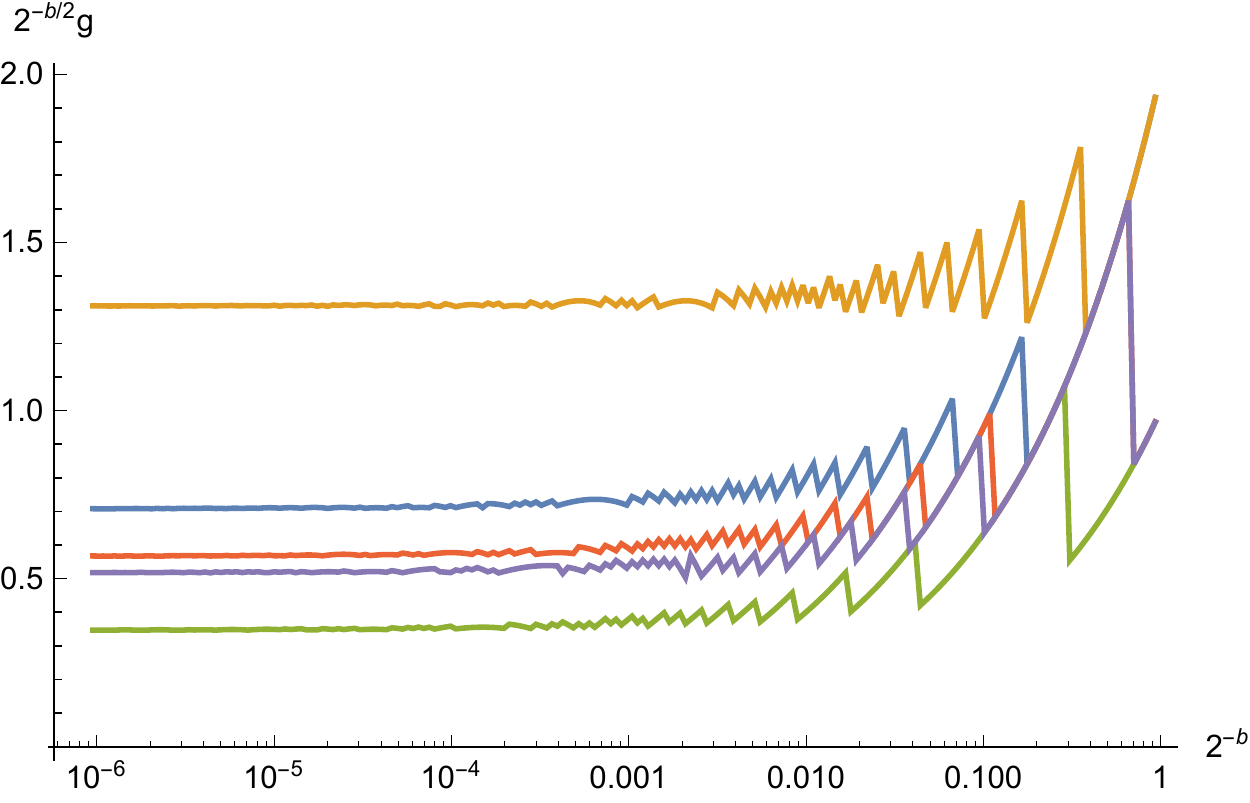}}\qquad
  \resizebox{.55\linewidth}{!}{\includegraphics{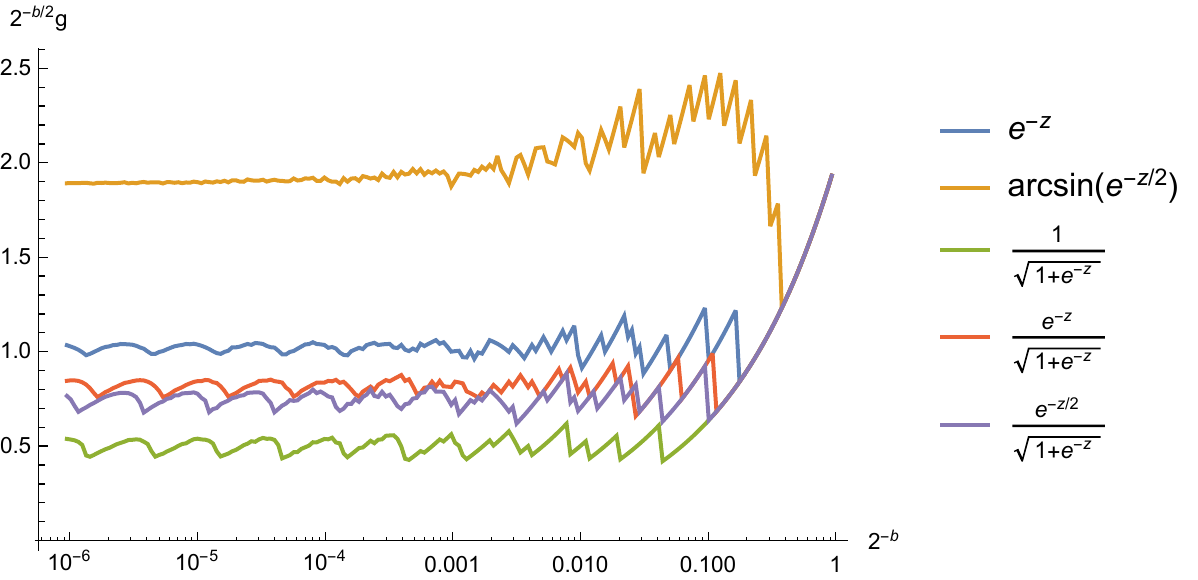}}
\put(-435,120){(a)}
  \put(-205,120){(b)}
  \caption[Numerical calculation of number of intervals for QROM.]{\label{fig:intervalsgen}
  The numbers of intervals multiplied by $2^{-\bfun/2}$ for the five functions we consider.
 In (a) we allow the intervals to have general endpoints, and in (b) we restrict the intervals to change by factors of $2$, to be consistent with the QROM method we use.
 This demonstrates that the number of intervals scales as $2^{\bfun/2}$ with a scaling constant around $1$.}
\end{figure}

To estimate the number of interpolation points needed for higher precision, note that the error of interpolation of function $f(z)$ is approximately
\begin{equation}
   \frac{(\delta z)^2}8 f''(z)\, ,
\end{equation}
where $\delta z$ is the width of the interval.
To obtain error no greater than $2^{-\bfun}$, we can therefore take
\begin{equation}
    \delta z = \frac {2^{-\bfun/2}\sqrt{8}}{\sqrt{f''(z)}} \, .
\end{equation}
We can therefore estimate the number of intervals needed to approximate the function by
\begin{equation}
    \frac {2^{\bfun/2}}{\sqrt{8}} \int_0^\infty dz \, \sqrt{f''(z)} \, .
\end{equation}
In the case where we are approximating $\arcsin(\exp(-z/2))$, then we would get $g \approx 1.31103\times 2^{\bfun/2}$,
and if we were approximating $\exp(-z)$, then we would have $g\approx 2^{(\bfun-1)/2}$.
For the three functions used for spectral gap amplification, $1/\sqrt{1+e^{-z}}$, $e^{-z}/\sqrt{1+e^{-z}}$, and $e^{-z/2}/\sqrt{1+e^{-z}}$ we get $2^{-\bfun/2}g$ of $0.346002$, $0.566302$, and $0.517075$ respectively.
The variation of $2^{-\bfun/2}g$ with $\bfun$ is shown in \figa{intervalsgen}{(a)}.
In practice, we need to limit the intervals to sizes that increase by factors of two as described above.
That increases the values of $2^{-\bfun/2}g$ to around
$1.0$, $1.9$, $0.5$, $0.8$, and $0.7$ for the five cases, as can be seen in \figa{intervalsgen}{(b)}, an increase of around $44\%$.
Nevertheless, it is reasonable to give the scaling of $g$ as $\mathcal{O}(2^{\bfun/2})$, with the constant factor somewhere between $0.5$ and $2$.

In the linear interpolation, the primary cost is that of multiplication of the argument times the slope.
This cost will depend on how many digits are used for the slope and the argument.
For simplicity, consider the case where bits of the argument can be divided between those before the decimal point and those after the decimal point.
The maximum value needed for the argument is $\order{\bsmooth}$, because beyond that the functions are within $1/2^{\bsmooth+1}$ of their asymptotic values.
That means only $\log\bsmooth +\order{1}$ bits would be needed before the decimal point.
The number of digits after the decimal point would depend on the maximum value of the slope.
In the case of the exponential the maximum slope is $1$, so only $\bsmooth$ bits would be needed.
Because the slope could be multiplied by an argument that is $\order{\bsmooth}$, it could need $\bsmooth+\log\bsmooth +\order{1}$ bits after the decimal point.
Both numbers would need approximately $\bsmooth+\log\bsmooth +\order{1}$ bits.
This gives a cost of multiplication of $\bsmooth^2+\mathcal{O}(\bsmooth\log \bsmooth)$ Toffoli gates.

The same result is obtained for all other functions we consider except the arcsine.
The arcsine has a slope that goes to infinity, but the linear interpolation will only use a finite slope.
The minimum interpolation point needs to be $\order{2^{-2\bfun}}$, which gives maximum slope of $\order{2^{\bfun}}$, so the argument would require another $\bfun+\order{1}$ bits after the decimal point.
The slope would need $\bfun+\order{1}$ bits before the decimal point, and $\bsmooth+\log\bsmooth+\order{1}$ bits after the decimal point to account for the maximum argument.
Then both numbers would need $\bfun+\bsmooth+\log\bsmooth+\order{1}$ bits.
We will take $\bfun$ similar to $\bsmooth$, giving a multiplication cost of $(\bsmooth+\bfun)^2+\mathcal{O}(\bsmooth\log \bsmooth)$ Toffoli gates.

\begin{figure}
    \resizebox{.8\linewidth}{!}{\includegraphics{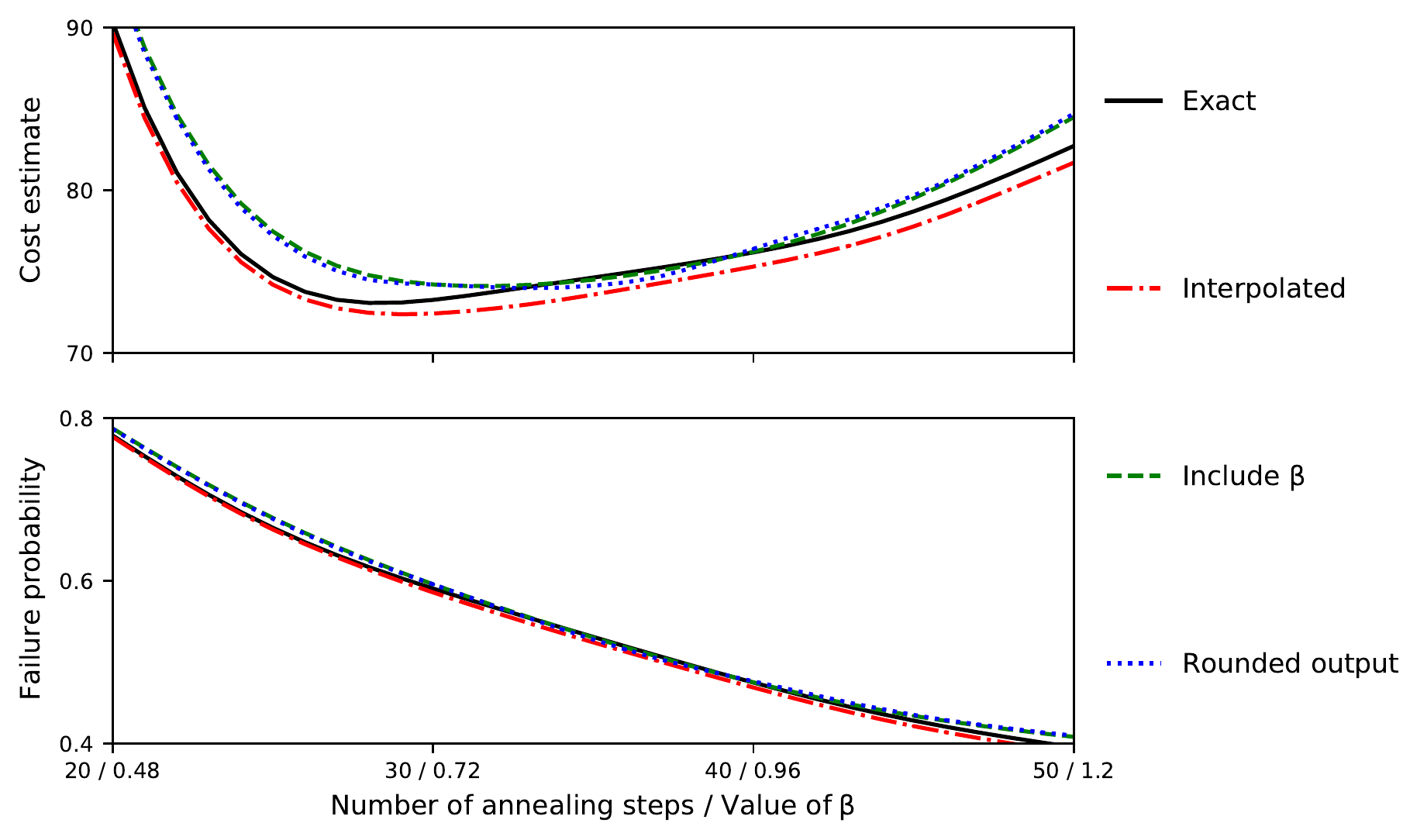}}
    \caption[Effect of approximations in qubitized quantum walk]{
    The effect of various methods of function approximation on optimization performance. We numerically simulate the quantum simulated annealing technique of \sec{QSA/qubitized}
    using various methods of approximating the transition probability.
    We consider the performance when the rotation angle
    is calculated to machine precision (``exact''), with piecewise linear
    approximation chosen to ensure the worst-case error does not exceed $0.01$ 
    (``interpolated''), incorporating the inverse temperature $\beta$ into the definition
    of the function so that we are interpolating $f(z)=\arcsin(\exp(-\beta z/2))$ rather 
    than $f(z)=\arcsin(\exp(-z/2))$ to avoid a multiplication (``include $\beta$''),
    and when we round off the output of the function to $7$ bits (``rounded output'').
    Each of these approximations builds upon the previous approximation,
    so we perform linear interpolation in all but the exact method.
    We simulate the performance averaging over $4096$ random SK instances on $16$ qubits, with $\beta$ linearly increasing over $50$ steps from $0$ to $1.2$.
    We report the average failure probability (bottom) as well as an estimate
    of the computational cost (top) in which we calculate the number of annealing steps 
    divided by the probability of success.
    We observe that the differences in performance are not meaningfully affected by
    the method of function approximation, suggesting that we can pick the computationally
    cheapest option for our cost analysis.
    \label{fig:comp_approx}}
\end{figure}

To estimate the numbers of bits needed, we have performed simulation of the technique of \sec{QSA/qubitized} with the SK Hamiltonian on 16 qubits, as shown in \fig{comp_approx}. In that technique, we need an approximation of the arcsine of the transition probability to control a qubit rotation, rather than the transition probability itself.
Choosing interpolation points such that the error in the approximation
of the rotation angle is no more than $0.01$, the success probabilities are almost unchanged.
So far we have assumed that the energy difference has been multiplied by the inverse temperature $\beta$ before being input to the procedure.
It is possible to bundle the multiplication by $\beta$ into the oracle, and as shown in \fig{comp_approx} that again has similar performance.
There is also the question of how many bits are needed in the function approximating the transition function.
We again find that low-precision approximations have very little impact on the success probability.

The overall complexity of the interpolation excluding the QROM is therefore $\bsmooth+\mathcal{O}(\bsmooth\log \bsmooth)$ or $(\bsmooth+\bfun)^2+\mathcal{O}(\bsmooth\log \bsmooth)$ when the arcsine is needed.
To estimate the QROM complexity, we need to account for the final region not being a size which is a power of $2$.
In the worst case the additional cost can be no larger than $\bdiff$, which is the total size of the input register.
We can therefore bound the QROM complexity as $\bdiff+\mathcal{O}(2^{\bfun/2})$, giving total interpolation complexity of
\begin{equation}\label{eq:funtof}
\mathcal{C}^{\rm fun} = \bsmooth^2+\bdiff+\mathcal{O}(\bsmooth\log \bsmooth+2^{\bfun/2}),
\end{equation}
or, for the case where the arcsine is needed,
\begin{equation}\label{eq:ASfuntof}
\mathcal{C}^{\rm fun} = (\bsmooth+\bfun)^2+\bdiff+\mathcal{O}(\bsmooth\log \bsmooth+2^{\bfun/2}).
\end{equation}

For the number of ancilla qubits needed, except for the arcsine case there are $2\bsmooth+\mathcal{O}(\log \bsmooth)$ needed for the slope and intercept, and $2\bsmooth+\mathcal{O}(\log \bsmooth)$ used as temporary ancillas for the arithmetic.
We need $\bdiff-1$ temporary ancillas for the QROM, which is more than the number used for the arithmetic. The output for the transition probability can be added into the slope, so does not increase the ancilla cost.
Therefore the ancilla costs are
\begin{align}\label{eq:funper}
    \mathcal{A}^{\rm fun} &= 2\bsmooth+\mathcal{O}(\log \bsmooth),\\
    \label{eq:funtem}
   \mathcal{B}^{\rm fun} &= \bdiff-1.
\end{align}
These considerations give the costs for function evaluation in \tab{oracles}.
For the arcsine case we need $2\bsmooth+\bfun+\mathcal{O}(\log \bsmooth)$ ancillas for the slope and intercept, because we need another $\bfun$ ancillas for the slope.
Again the temporary ancilla cost is primarily for the QROM, so the ancilla costs are
\begin{align}\label{eq:ASfunper}
    \mathcal{A}^{\rm fun} &= 2\bsmooth+\bfun+\mathcal{O}(\log \bsmooth),\\
    \label{eq:ASfuntem}
   \mathcal{B}^{\rm fun} &= \bdiff-1.
\end{align}

\section{Optimization Methods}
\label{sec:algorithms}

In this section we review proposals for heuristic quantum optimization algorithms and explain how those algorithms can be implemented in terms of the oracles we describe in \sec{oracles}. By this we include methods based on Hamiltonian walks, those based on time evolution, and methods related to simulated annealing. In most cases we will suggest improvements to these methods, but an important motivation for this section is to give a complete analysis of the complexity of these algorithms which includes constant factors so that we can estimate the resources required to realize them in the surface code in \sec{conclusion}. We describe the complexities of these methods in terms of the oracles from the previous section in \tab{query_complexity}, then give the complexity in terms of Toffoli or T gates in \tab{primitives}. 

As this section incorporates a wide variety of sophisticated techniques,
we begin with a brief summary of the approaches we are considering.
\begin{itemize}
    \item \emph{Amplitude amplification} (\sec{amp_amp}).    
    We start by considering amplitude amplification, which can be used to directly amplify the amplitude of the solution.
    Unlike the other methods, it takes no advantage of the structure of the solution, so is a useful reference point to compare to the other optimization approaches.
    Amplitude amplification can also be used in combination with the other optimization approaches, by performing amplitude amplification on the output of the optimization.
    \item \emph{The Quantum Approximate Optimization Algorithm} (\sec{qaoa}). The steps of this approach (QAOA) are equivalent to Trotter steps, so the costing for QAOA and Trotter steps is given in the same lines in \tab{query_complexity} and \tab{primitives}. Trotter steps can be used for adiabatic approaches, which are considered in the next subsection. But here we also focusing on strategies for efficiently estimating the QAOA objective value that are more appropriate for a fault-tolerant cost model than standard approaches.
    \item \emph{Adiabatic quantum optimization} (\sec{evolution}). We review the quantum adiabatic algorithm \cite{Farhi2001} and the most straightforward way of implementing that approach using a Trotter method that queries the phase oracles presented in \sec{oracles}. We will then suggest a strategy for implementing the adiabatic algorithm using the LCU oracles presented in \sec{oracles}. The LCU oracles have a different costing to Trotter/QAOA, so are given in separate lines in \tab{query_complexity} and \tab{primitives}.
    Next, we will review a method for digitizing the adiabatic algorithm while suppressing certain types of errors that is based on inducing quantum Zeno-effect-like projection to the ground state by randomizing phases \cite{Boixo2009a}. We will suggest how this approach can be improved by using carefully chosen probability distributions to eliminate the errors that would manifest from incorrect measurements in the Zeno approach. The time-evolution oracles used by these methods are also suitable for the quantum enhanced population transfer algorithm and the shortest path algorithm. Since we do not introduce new techniques for those algorithms, but rather review how our oracles can be queried within those frameworks, we discuss that content in \app{other_evolution_optimization_approaches}.
    
    \item \emph{Szegedy walk based quantum simulated annealing} (\sec{szegmain}).
    Simulated annealing is a classical algorithm that mimics a physical cooling process via Markov chain Monte Carlo techniques.
    The quantum algorithm of Somma \textit{et al.}~\cite{Somma2008b} is to replace the Markov chain with a corresponding Szegedy walk. If the spectral gap of the 
    Markov transition operator is  $\gap$, the  number of Szegedy walk 
    steps grows as ${\cal O}(1/\sqrt{\gap})$ in contrast with the best
    known bound on the worst case scaling of the number of Markov 
    transitions needed in the  classical approach,
    which goes like ${\cal O}(1/\gap)$.
    Thus, the result appears to be a quadratic speedup
    over simulated annealing. We note that the ${\cal O}(1/\gap)$
    scaling of classical simulated annealing is known to be a very loose bound for a 
    broad class of problems. Typically, simulated annealing is used 
    heuristically by lowering the temperature much faster than 
    would be suggested by this bound. Our results constitute the first complete cost analysis for this algorithm that involves constant factors in the complexity.
    
    \item \emph{LHPST qubitized walk based quantum simulated annealing} (\sec{QSA/qubitized}).
Lemieux, Heim, Poulin, Svore, and Troyer (LHPST)~\cite{lemieux2019efficient}
give a Metropolis-Hastings-like qubitized walk approach which is significantly more efficient than the direct Szegedy approach. We will refer to this method by their initials, but we provide an improved technique that is efficient for more complicated problem Hamiltonians with high connectivity.
LHPST consider a method that is efficient for simpler problem Hamiltonians with low connectivity, but would have exponential cost for the problem Hamiltonians considered here.
    
    \item \emph{Spectral gap amplification based quantum simulated annealing} (\sec{spectral}).
     In~\cite{Boixo2014a}, the authors construct an inverse-temperature-dependent Hamiltonian whose ground state in the zero-temperature limit is a superposition of solution  configurations. By performing spectral gap amplification on their Hamiltonian, they obtain a gap that is similar to that for the quantum walk approach, indicating a similar speedup.
     Our main purpose is to outline these techniques and summarize the work needed to execute such algorithms in general and for specific problems of interest as outlined in the Introduction. We will also suggest a variant of this algorithm where one can use qubitized quantum walks rather than time evolution for the adiabatic evolution. In both cases, our results provide the first constant factor bounds on the complexity of implementing these algorithms.
\end{itemize}

We summarize the outcomes of this section in \tab{query_complexity}.
The entries of \tab{query_complexity} show how the Toffoli complexity and ancilla
cost of each of the above named algorithm primitives depend on the relevant costs
of oracles presented in \tab{oracles}.  We can then use \tab{query_complexity} together with
\tab{oracles} to calculate the overall Toffoli complexity and ancilla cost of each
algorithm primitive for each type of cost function. The results of this analysis are
summarized in \tab{primitives}. In giving the complexities in this table, we assume $2^{\bfun/2} < \bsmooth \log \bsmooth < \brot$ to simplify the order terms, which is reasonable for the examples we consider in \sec{conclusion}.

\begin{table*}[t]\resizebox{\textwidth}{!}{
\begin{tabular}{c|c|c}
algorithm primitive & Toffoli complexity & ancilla complexity \\
\hline\hline
amplitude amplification step & $2 \,{\cal C}^{\textrm{direct}} + N+{\cal O}(\bdirect)$ \eqref{eq:amp_amp_step} & ${\cal A}^{\textrm{direct}}+{\cal B}^{\textrm{direct}} + {\cal O}(1)$ \\
QAOA/Trotter step & ${\cal C}^{\textrm{phase}}+ 4N+\bphase^2/2+\order{\bphase\log\bphase}$ \eqref{eq:suzuki} & ${\cal A}^{\textrm{phase}}+\max({\cal B}^{\textrm{phase}},3\log N)+\bphase+\order{\log\bphase}$\\
Hamiltonian walk step & ${\cal C}^{\textrm{LCU}} + {\cal O}(1)$ &  ${\cal A}^{\textrm{LCU}}+{\cal B}^{\textrm{LCU}} + {\cal O}(1)$\\
Szegedy walk annealing step & $\min(2(N+1)\mathcal{C}^{\textrm{direct}},2N \mathcal{C}^{\textrm{diff}}) + 2N\mathcal{C}^{\textrm{fun}} + 2N\log N + 8N\bsmooth+18\bsmooth^2 + \order{N}$ \eqref{eq:Szegtof} & $N\mathcal{A}^{\textrm{diff}}+ N \mathcal{A}^{\textrm{fun}}+ 5\bsmooth +\order{N}$ \eqref{eq:Szeganc} \\
LHPST walk annealing step & $2\mathcal{C}^{\textrm{diff}} + 2\mathcal{C}^{\textrm{fun}}+N+ 2\bdiff+9\log N+ \order{1}$  \eqref{eq:costqubtof} & $\mathcal{A}^{\textrm{diff}} + \mathcal{A}^{\textrm{fun}} + \mathcal{B}^{\textrm{diff}} + \log N + \bsmooth + \order{1}$ \eqref{eq:costqubanc} \\
gap amplified walk step & $2\mathcal{C}^{\textrm{diff}} + 2\mathcal{C}^{\textrm{fun}} + 2\bsmooth +N+14\log N+ \order{\brot}$ \eqref{eq:sgatof} & $\mathcal{A}^{\textrm{diff}} + \mathcal{A}^{\textrm{fun}}+ \mathcal{B}^{\textrm{diff}} + 2\log N + \bsmooth + \order{1}$ \eqref{eq:sgaanc} \\
\hline
\end{tabular}}
\caption[Query complexity of algorithm primitives]{\label{tab:query_complexity}
The Toffoli complexities and ancilla complexities needed to implement the basic primitives of various heuristic algorithms, reported in terms of the oracle costs from \tab{oracle_defs}. For both the LHPST walk step and the gap amplified walk step the cost is reduced by $8\log N$ when $N$ is a power of $2$.
By combining these scalings with the oracle costs given in \tab{oracles}, we arrive at the resource estimates in \tab{primitives}.
For order $\rho$ Suzuki (meaning that the error is $\order{\delta t^{\rho +1}}$) we multiply the QAOA cost by $2 \times 5^{\rho/2-1}$.
For the QAOA/Trotter step, the $\bphase$ ancillas are for a phase gradient state, and may be saved if those are already accounted for in ${\cal A}^{\textrm{phase}}$.
The quantity $\epsilon$ is an allowable error in synthesizing a rotation in the state preparation.}
\end{table*}

\begin{table*}[t]\resizebox{\textwidth}{!}{
\begin{tabular}{c|c|c|c}
cost function & algorithm primitive & Toffoli (* or T) count & total ancilla qubits \\
\hline\hline
$L$-term & amplitude amplification step & $2\,L \, \bdirect + N+{\cal O}(\bdirect)$ & $2 \, \bdirect + {\cal O}(1)$ \\
 Spin & QAOA/Trotter step* & $1.15 L (\bphase + \log L) + {\cal O}(N+\log L+\bphase^2)$ & $ 3\log N+\bphase+{\cal O}(\log\bphase)$\\
 Model  & Hamiltonian walk step & $3\,L + 2\bLCU + {\cal O}(\log L)$ &  $3\log L + 2\,\bLCU + {\cal O}(1)$\\
$H_L$ & Szegedy walk annealing step &  $2(N+1)L\bdirect + 2N( \bsmooth^2+\bdiff+\log N)  + \mathcal{O}(N\bsmooth\log \bsmooth)$ 
& $N\bdiff + 2N\bsmooth +\order{N\log \bsmooth}$
\\
& LHPST walk annealing step & $4L\bdiff+2(\bsmooth+\bfun)^2+ 2\bdiff+N+9\log N+\mathcal{O}(\bsmooth\log \bsmooth
)$ & $3\bsmooth+2\bdiff+\bfun+\log N+\order{\log \bsmooth}$ \\
& gap amplified walk step & $4L\bdiff+2 \bsmooth^2+2\bdiff+N+14\log N+\mathcal{O}( \brot
)$
& $3\bsmooth+2\bdiff+2\log N+\order{\log \bsmooth}$ \\
\hline
Quadratic & amplitude amplification & $N^2 \bdirect + {\cal O}(N \bdirect)$ & $2\,\bdirect + {\cal O}(1)$\\
 Unconstrained   & QAOA/Trotter step* & $ 0.575\, N^2 (\bphase + 2 \log N) + {\cal O}(N^2)$ & $3\log N+\bphase+{\cal O}(\log\bphase)$\\
 Binary & Hamiltonian walk step & $ N( \bLCU + 2\log N) + {\cal O}(N)$ & $ 7 \log N + 2\, \bLCU + {\cal O}(\log \bLCU)$\\
Optimization& Szegedy walk annealing step & $2N^2 \bdiff + 2N( \bsmooth^2+\bdiff+\log N)  + \mathcal{O}(N\bsmooth\log \bsmooth
)$ & $N\bdiff + 2N\bsmooth +\order{N\log \bsmooth}$ \\
$H_{\rm QUBO}$ & LHPST walk annealing step & $2N\bdiff+2(\bsmooth+\bfun)^2+ 2\bdiff+N+9\log N+\mathcal{O}(\bsmooth\log \bsmooth
)$ & $3\bsmooth+2\bdiff+\bfun+\log N+\order{\log \bsmooth}$ \\
& gap amplified walk step & $2N\bdiff+2 \bsmooth^2+2\bdiff+N+14\log N+\mathcal{O}(\brot
)$ & $3\bsmooth+2\bdiff+2\log N+\order{\log \bsmooth}$ \\
\hline
Sherrington-- & amplitude amplification step & $2\,N^2 + N +{\cal O}(\log N)$ 
& $6\log N + {\cal O}(1)$\\
Kirkpatrick   & QAOA/Trotter step & $2N^2+4N+\bphase^2+\order{\bphase\log\bphase}$ & $6\log N+\bphase+\order{\log{\bphase}}$ \\
Model  & Hamiltonian walk step & $6\, N + {\cal O}(\log^2 N)$ & $5 \log N + {\cal O}(1)$\\
$H_{\rm SK}$ & Szegedy walk annealing step & $4N^2 + 2N(\bsmooth^2+2\log N) + 8N\bsmooth+18\bsmooth^2 +\mathcal{O}(N\bsmooth\log \bsmooth
)$ & $N\log N + 2N\bsmooth+\order{N\log \bsmooth}$ \\
& LHPST walk annealing step & $5N+2(\bsmooth+\bfun)^2+ 11\log N+\mathcal{O}(\bsmooth\log \bsmooth
)$ &  $4\log N+3\bsmooth+\bfun+\order{\log \bsmooth}$ \\
& gap amplified walk step & $5N+2\bsmooth^2+ 16\log N+\mathcal{O}(\brot
)$ &  $5\log N+3\bsmooth+\order{\log \bsmooth}$ \\
\hline
Low & amplitude amplification step & $5\, N(N+1)/2 + N+{\cal O}(\log N)$ & $5 \log N + {\cal O}(1)$\\
 Autocorrelation & QAOA/Trotter step & $8 N^2 /5 + \min\left( N \bphase^2 / 2,9N^2/10\right) + \order{N\bphase\log\bphase}$ & $5\log N +\bphase + {\cal O}(\log\bphase)$\\
 Binary & Hamiltonian walk step & $4\, N + {\cal O}(\log N)$ & $5\log N + {\cal O}(1)$\\
 Sequences & Szegedy walk annealing step & $5N(N+1)^2/2 + 2N( \bsmooth^2+3\log N)+\mathcal{O}(N\bsmooth\log \bsmooth 
 )$ & $2N\log N + 2N\bsmooth+\order{N\log \bsmooth}$ \\
$H_{\rm LABS}$ & LHPST walk annealing step &  $5N^2+2(\bsmooth+\bfun)^2+6N+ 13\log N+\mathcal{O}(\bsmooth\log \bsmooth
)$ & $6\log N+3\bsmooth+ \bfun+ \order{\log \bsmooth}$ \\
& gap amplified walk step &  $5N^2+2 \bsmooth^2+6N+ 18\log N+\mathcal{O}( \brot
)$ & $7\log N+3\bsmooth+\order{\log \bsmooth}$ \\
\hline
\end{tabular}}
\caption[Resource estimates for optimization heuristic primitives]{ \label{tab:primitives}
Resource estimates for the various heuristic optimization primitives explored throughout this paper, applied to our four problems of interest.
For both the LHPST walk step and the gap amplified walk step the Toffoli count is reduced by $8\log N$ when $N$ is a power of $2$.
In all cases, these algorithms are refined by applying the primitive more times. The parameters used are as follows: $N$ is the number of bits on which our cost function is defined; $L$ is the numbers of terms in an $L$-term spin Hamiltonian; $\bphase$ is the number of bits we use to approximate phases in the implementation of our phase oracle; $\bdirect$ is the number of bits we use to approximate the value of energies; $\bLCU$ is the number of bits used to approximate the square root of Hamiltonian coefficients in LCU methods, and $\brot$ is the number of bits of precision used in rotations. The Trotter step and Hamiltonian walk steps can be used to realize the adiabatic algorithm, the Zeno phase randomization variant of the adiabatic algorithm, heuristic variants of the short path algorithm or quantum enhanced population transfer, and many other heuristics based on Hamiltonian time evolution. These scalings result from combining the query complexities in \tab{query_complexity} with the oracle costs in \tab{oracles}. When the algorithm type is decorated with (*) we report T complexity rather than Toffoli complexity.
We have only given the main terms in the order expressions to simplify them.}
\end{table*}

\subsection{Amplitude amplification}
\label{sec:amp_amp}

\subsubsection{Combining amplitude amplification with quantum optimization heuristics}
\label{sec:amp_combined}

All of the optimization heuristics discussed in this paper can be seen as methods of preparing a quantum state with overlap on a low energy subspace of interest. We will refer to the subspace of interest as ${\cal S}$. Sometimes this subspace of interest is actually the lowest energy state (or states) and other times it is any state with energy less than a certain threshold. Furthermore, all algorithms discussed in this paper are heuristics that can be systematically refined. Let us refer to an algorithm for quantum optimization that is run for duration $t$ as ${\cal U}(t)$. Let us assume that these algorithms always begin in the state $\ket{+}^{\otimes N}$ and denote the output state of the algorithm by $\ket{\psi(t)} = {\cal U}(t)\ket{+}^{\otimes N}$. Thus, after running our algorithm ${\cal U}(t)$ and sampling in the computational basis, the probability of measuring a state in the subspace of interest ${\cal S}$ is
\begin{equation}
p_0\!\left(t\right) = \sum_{x \in {\cal S}} \left | \braket{x}{\psi\!\left(t\right)} \right |^2.
\end{equation}

When we say that these heuristics can be systematically refined what we mean is that we can (on average) increase $p_0(t)$ by increasing $t$. This refinement will come at a cost ${\cal C}(t) > 0$ which we define as the complexity of implementing ${\cal U}(t)$. This complexity is greater than zero because preparing the initial state $\ket{+}^{\otimes N}$ requires nonzero time even if we do nothing further. We can also boost the probability of seeing a state in ${\cal S}$ by repeating ${\cal U}(t)$ more times and sampling. On average we will need to run our algorithm ${\cal U}(t)$ a number of times equal to $1 / p_0(t)$ in order to see a state in ${\cal S}$. Thus, on average the cost to sample a state ${\cal S}$ is given by
\begin{equation}
\label{eq:mean_tts}
\frac{{\cal C}\left(t\right)}{p_0\!\left(t\right)}.
\end{equation}

There is a compromise to be reached between the duration $t$ of the optimization heuristic ${\cal U}(t)$ and the success probability $p_0(t)$; heuristics run for more time can reach a higher success probability and therefore be repeated fewer times, but increasing $t$ beyond a certain point has a negligible impact on its success probability $p_0(t)$. While past work \cite{lemieux2019efficient} has discussed this dichotomy in terms of a minimum time to solution metric which is parameterized in terms of a target success probability, here we focus on the mean cost to succeed because this seems more reasonable to consider in a context where $p_0(t)$ is unknown. Still, given knowledge of $p_0(t)$ one could optimize this mean time by choosing $t$ to minimize \eq{mean_tts}. But rather than simply repeating the state preparation $1/p_0(t)$ times, one could instead boost the success probability with amplitude amplification.

Amplitude amplification is an idea which generalized Grover search and can be used to boost the probability of a marked state or subspace. For instance, we might define these marked states to be any state in ${\cal S}$. In this context, amplitude amplification would allow us to perform a series of $m$ reflections (involving two preparations of the state $\ket{\psi(t)}$) which boosts the probability of measuring the marked subspace to 
\begin{equation}
\label{eq:amp_amp}
p_m\!\left(t\right) = \sin^2\left(\left(2 m + 1\right) \arcsin\left(\sqrt{p_0\!\left(t\right)}\right)\right).
\end{equation}
For instance, if we hoped to boost the probability to $1$ then by using repeated sampling we would need roughly ${\cal O}(1 / p_0(t))$ repetitions. However, by using amplitude amplification we would need only
\begin{equation}
m \approx \frac{\pi}{4 \arcsin\left(\sqrt{p_0\!\left(t\right)}\right)} - 1 = {\cal O}\left(\frac{1}{\sqrt{p_0\!\left(t\right)}}\right)
\end{equation}
iterations if $p_0(t)$ is small (this is akin the usual quadratic Grover speedup).

For each round of amplitude amplification one needs to reflect about a qubit marking the subspace of interest ${\cal S}$. In our context the idea would be to amplify either a target energy (if a target energy, e.g.\ the ground state energy, is known) or to amplify all states with energy less than a certain threshold. To do this, one will need to compute the energy value into a register and perform either an equality or inequality test to determine whether we have reached a marked state. The energy can be computed simply by using the direct energy oracles introduced in \sec{direct_oracle}. However, both that step and the cost of the equality or inequality evaluation will typically have a negligible additive cost to the cost ${\cal C}(t)$ of actually running the quantum algorithm ${\cal U}(t)$. Moreover, the ancilla used for storing the value of the energy can be borrowed from ancilla used in other parts of the algorithm.

For amplitude amplification to be most effective one should have an estimate of the overlap $p_0(t)$ in order to avoid ``overshooting'' the peak of the function in \eq{amp_amp}. Unfortunately, a reliable estimate of $p_0(t)$ will not be known in advance in general. In some rare cases one might instead have a somewhat tight estimate of a lower bound to $p_0(t)$ and in those cases some advantages can be realized by using a variant of amplitude amplification known as fixed point amplitude amplification \cite{yoder}. However one can confirm that fixed point amplitude amplification will have no advantages in our context compared to the exponential search heuristic proposed in \cite{Brassard2002} when the best lower bound that is available is $p_0(t) > 0$. The idea behind the approach in \cite{Brassard2002} is to run amplitude amplification for $m = 2^j$ iterations for $j=0,1,2,3,...$ and so on until we sample a marked state. The cost of each iteration of amplitude amplification is $2\, {\cal C}(t)$ and so if we need to repeat this procedure until $m = 2^k$ it will have a total cost that goes like
\begin{equation}
    2\, {\cal C}\left(t\right) \sum_{j=0}^k 2^j = 2 \,{\cal C}\left(t\right) \left(2^k - 1\right).
\end{equation}
Therefore, since the probability of failure in a single run with $m = 2^j$ iterations is $1 - p_{2^j}(t)$, we see that the overall mean cost of the procedure is
\begin{equation}
\label{eq:mean_aa_tts}
2\, {\cal C}\left(t\right)\sum_{k=1}^\infty \left(2^k - 1\right) \prod_{j=1}^{k-1} \left(1 - p_{2^{j-1}}\!\left(t\right)\right) = {\cal O}\left(\frac{{\cal C}\left(t\right)}{\sqrt{p_0\!\left(t\right)}}\right).
\end{equation}
Though the left side of this expression cannot be simplified analytically, it converges quickly and can be easily numerically computed for any $p_0(t) > 0$.

Comparing \eq{mean_tts} to \eq{mean_aa_tts} we can see that there is a clear asymptotic advantage to using amplitude amplification over classical sampling and expect this advantage will be realizable in practice in many contexts of interest for us. Like with \eq{mean_tts}, if one has knowledge of $p_0(t)$ then one can minimize \eq{mean_aa_tts} with respect to $t$ to make the optimal tradeoff between running the algorithm ${\cal U}(t)$ for longer and using more rounds of amplitude amplification. In some cases it might actually be the case that the optimal choice is $t = 0$, which would correspond to using amplitude amplification directly as a heuristic for optimization. The only downside to using amplitude amplification in conjunction with other heuristic quantum algorithms for optimization is that we have traded incoherent repetitions of the primitive of ${\cal U}(t)$ for coherent repetitions of the primitive of ${\cal U}(t)$. In some cases this will mean that we need to target a higher error rate to make the calculation fault-tolerant by using an error-correcting code.

\subsubsection{Directly using amplitude amplification}
\label{sec:amp_sole}

In the prior section we described how amplitude amplification can be combined with any of the other optimization heuristics in this paper in order to boost overlap on a target low energy subspace of interest. However, one can also use amplitude amplification by itself as a heuristic for optimization. This heuristic provides an interesting point of comparison to other algorithms because it offers a quadratic advantage over classical brute force search without leveraging any structure that might be available in a particular optimization problem. Thus, it is asymptotically the optimal strategy for solving totally unstructured problems like those described by the typical Grover oracle (all computational basis states have energy zero except for a solution with energy $-1$) or the random energy model (all computational basis states have a unique, Gaussian distributed energy).

To use amplitude amplification on its own all one needs to do is to regard the algorithm ${\cal U}(t)$ as the preparation of the symmetric superposition state $\ket{+}^{\otimes N}$, which requires only Clifford gates. In the analysis of \sec{amp_combined} we assumed that the cost of directly computing the energy and then performing the comparison operation would be negligible compared to the cost of applying ${\cal U}(t)$ but that is not the case when we aim to directly apply amplitude amplification. Here, the main cost of a step will be the cost to compute (and then later uncompute) the energy. Following \eq{mean_aa_tts}, in this context we would find that the mean cost of applying amplitude amplification directly will then scale like
\begin{equation}
\label{eq:mean_aa}
\left(2\,{\cal C}^{\textrm{direct}} +N+ {\cal O}\left(\bdirect\right) \right)\!\sum_{k=1}^\infty \left(2^k \!-\! 1\right) \prod_{j=1}^{k-1}\! \left(1 \!-\! \sin^2\left(\!\left(2^j \!+\! 1\right) \arcsin\left(\!\sqrt{\frac{1}{2^N}}\right)\right)\right) = {\cal O}\left(\left({\cal C}^{\textrm{direct}} +N+ {\cal O}\left(\bdirect\right)\right)\sqrt{2^N}\right)
\end{equation}
where we have used $p_0(t) = 1/2^N$.
The cost $2\,{\cal C}^{\textrm{direct}} +N+ {\cal O}\left(\bdirect\right)$ comes from cost ${\cal C}^{\textrm{direct}}$ to directly compute the energy, cost ${\cal O}\left(\bdirect\right)$ to apply the inequality operator to determine whether the energy is below the target threshold, cost ${\cal C}^{\textrm{direct}}$ to uncompute the energy, and cost $N-2$ to reflect about the equal superposition state. Note that this procedure is exactly the heuristic approach introduced in \cite{Brassard2002}. When the subspace ${\cal S}$ contains only a single state this algorithm reduces exactly to standard Grover search \cite{Grover1996}. For later comparisons in this paper we will refer to the cost of a single step of amplitude amplification as having \begin{equation}
    \label{eq:amp_amp_step}
    2\,{\cal C}^{\textrm{direct}} + N+{\cal O}\left(\bdirect\right)
\end{equation}
Toffoli complexity and requiring ${\cal A}^{\textrm{direct}} + {\cal B}^{\textrm{direct}} + {\cal O}(1)$ ancilla.

\subsection{The Quantum Approximate Optimization Algorithm}
\label{sec:qaoa}

The quantum approximate optimization algorithm (QAOA) is another popular approach to quantum optimization, introduced in \cite{Farhi2014}. The QAOA initially attracted significant interest after it was shown to produce a better approximation ratio for a specific combinatorial optimization problem of bounded occurrence, Max E3LIN2, than any known efficient classical method \cite{Farhi2014}. While a more efficient classical algorithm was presented shortly afterwards \cite{Barak2015}, interest in QAOA has only increased since then. While bounds on the performance of QAOA are sometimes available, in most contexts it is studied as a heuristic in the sense that the intention is to use the algorithm without knowing how well it will perform in practice. Part of the appeal of QAOA has been that it is an easy-to-implement algorithm that can be tested on noisy intermediate scale quantum (NISQ) devices even before fault-tolerance is available \cite{Arute2020QuantumProcessor}. Nonetheless, QAOA would still be an interesting algorithm to perform within error-correction.

\newcommand{\gammavector}{\boldsymbol{\gamma}}
\newcommand{\betavector}{\boldsymbol{\beta}}

The QAOA is more straightforward than other algorithms discussed in this work. The QAOA consists of two components that are repeatedly applied. The first component is parameterized evolution under the diagonal problem Hamiltonian $C$,
\begin{equation}
\label{eq:qaoa_phase}
U_C\left(\gamma\right) = e^{-i \gamma C } = O^{\textrm{phase}}\left(\gamma\right),
\end{equation}
where in the last equality we emphasize that $U_C(\gamma)$ is equivalent to the phase oracle $O^{\textrm{phase}}(\gamma)$ that we introduce and provide explicit circuit constructions for in \sec{phase_oracle}.

The second component is parameterized evolution under a local transverse field driver Hamiltonian $B$,
\begin{equation}
U_B\left(\beta\right) = e^{-i \beta B} \qquad \qquad B = \sum_{j=1}^N X_j.
\end{equation}
The QAOA is a variational algorithm that uses repeated application of these unitaries to prepare a parameterized state that is then optimized. The depth of the variational algorithm is usually denoted as ``$p$'' in the QAOA literature. Specifically, for depth $p$ we prepare a state parameterized by $\gammavector = (\gamma_1, \dots, \gamma_p)$ and $\betavector = (\beta_1, \dots, \beta_p)$,
\begin{align}\label{eq:gamma-beta-state}
	\ket{\gammavector, \betavector} = U_B\left(\beta_p\right)  U_C\left(\gamma_p\right) \dots U_B\left(\beta_1
	\right) U_C\left(\gamma_1 \right) \ket{+}^{\otimes N}
\end{align}
where $\ket{+}^{\otimes N}$ is the symmetric superposition of all $2^N$ computational basis states.

For a given $p$, we attempt to find parameters that minimize the expectation value of the cost
\begin{align}
\avg{C} = \bra{\gammavector , \betavector} C \ket{\gammavector, \betavector}.
\end{align}
The QAOA proposes to use the quantum computer to estimate this expectation value and then to use a classical processor to perform a classical optimization, in a fashion similar to other variational algorithms \cite{Peruzzo2013,McClean2015}. In general finding the globally optimal values of $\gammavector$ and $\betavector$ could prove to be very challenging. However, QAOA is a heuristic algorithm and the idea is that even locally optimal parameter settings might provide good approximations.

The original implementation of QAOA suggested that one directly sample the cost function $C$ to estimate $\avg{C}$. Using this method, if one wishes to converge an unbiased estimator $\avg{\widetilde{C}}$ so that $|\avg{\widetilde{C}} - \avg{C} | \leq \Delta_C$ then the state $\ket{\gammavector, \betavector}$ must be prepared and sampled a number of times equal to
\begin{equation}
\label{eq:sampling_bound}
\sigma^2 / \Delta_C^2 \qquad {\rm where} \qquad \sigma^2 = \avg{C^2} - \avg{C}^2.
\end{equation}
While one will not know $\sigma^2$ in advance, one can obtain a reasonable estimate of $\sigma^2$ after only handful of measurements and use that to determine how many more measurements are required.

The cost of QAOA is always dominated by the number of times that one must repeat the unitary $U_C(\gamma)$; the cost to implement $U_B(\beta)$ is essentially free in comparison. Thus, if $J$ is the number of outer-loop optimization iterations which each require a query of the energy accurate to within $\Delta_C$ then in total we will require Toffoli complexity
\begin{equation}
\label{eq:qaoa_reps}
\frac{p \, J \, {\cal C}^{\textrm{phase}} \sigma^2}{\Delta_C^2}.
\end{equation}
It is difficult to say what an appropriate choice of the quantities $J$ and $\Delta_C$ should be as this depends on the problem, the choice of optimizer one is using, and how aggressively one is attempting to optimize. However, in many circumstances one might not need to perform the outer-loop optimization at all and can thus take $J=1$. This is the case when optimal (or ``good enough'') parameters can be inferred before running the algorithm. Such a situation often arises when running large instances of optimization problems that are characteristic of a well defined ensemble (for example, if one is running instances of the Sherrington-Kirkpatrick model). This is due to the observation that normalized energy landscapes (proportional to $\avg{C}$ as a function of $\gammavector$ and $\betavector$) concentrate to instance and size independent average values for large $N$ \cite{Brandao2018,Farhi2019}. Thus, surprisingly, it is possible to find the optimal values of $\gammavector$ and $\betavector$ by optimizing much smaller (presumably classically tractable) instances of these problems. Another possibility is that one simply use $\gammavector$ and $\betavector$ parameters that would be obtained from a Trotterization of the quantum adiabatic algorithm; in fact, there is evidence that these parameters become optimal as one increases $p$ \cite{Zhou2019}. Thus, for problems where it is appropriate to forgo the outer-loop optimization step of QAOA, we can approximate the Toffoli complexity as $p \, M\, {\cal C}^{\textrm{phase}}$ where $M$ is the number of samples we desire. The number of logical qubits required for its implementation will be $N$ (not counting any extra ancilla used for the phase oracle).

Within the context of NISQ computations it makes sense to use this method of sampling to estimate the cost function expectation value and then to perform the optimization on a classical computer. The reason is because both strategies minimize the size of each quantum circuit that must be executed, although potentially at a cost of needing a larger number of repetitions compared to other strategies. However, within cost models appropriate for fault-tolerance the primary resource to consider is the total number of gates required by the computation and no particular distinction is made whether those gates are involved in repeated applications of short quantum circuits or a single application of a longer quantum circuit. Thus, on a fault-tolerant quantum computer it may make sense to consider more elaborate versions of QAOA in which the expectation value estimation and potentially even the optimization is also performed on a quantum computer. For instance, perhaps the variational parameters $\gammavector$ and $\betavector$ can be stored in a quantum register on which the QAOA unitary is controlled. Such a scheme is considered in \cite{Gilyen2018} where it shown that such a method can enable quadratically faster resolution of the gradient than would be otherwise required, however with significant constant overhead. Similarly, by using the amplitude amplification based Monte Carlo techniques discussed in \cite{Montanaro2015} (see Theorem 5 therein) one can reduce the number of state preparations needed for an estimate of the cost function to ${\cal O}((\sigma / \Delta_C) \log^{3/2}(\sigma / \Delta_C) \log \log (\sigma / \Delta_C))$, an almost quadratic improvement over the naive sampling strategy. However, as that method requires a number of copies of the system register scaling as ${\cal O}( \log(\sigma / \Delta_C) \log \log (\sigma / \Delta_C))$, it might prove to be prohibitively expensive for realization on small fault-tolerant quantum computers. We now consider two alternative ways to measure the energy in QAOA which might prove more practical for small fault-tolerant quantum computers.

\subsubsection{Amplitude estimation based direct phase oracle evaluation}

Apart from sampling, the next most natural algorithm for estimating the energy is using amplitude estimation to compute the expectation value of each term in the cost function in sequence.  Let us assume that the cost function takes the form, $C=\sum_{\ell=1}^L w_\ell U_\ell$, where $U_\ell$ is a unitary operator (and will typically be a sum of diagonal Pauli operators), as in \eq{lcu}.  Further we will take $\lambda = \sum_\ell |w_\ell|$. The algorithm that we employ is simple, for each $\ell$ from $1$ to $L$ we compute the quantity $\bra{\psi} U_\ell \ket{\psi}$ within error $\Delta_C / (L|w_\ell|)$.  An unbiased estimate of the cost function is then given by $\sum_\ell w_\ell \bra{\psi} U_\ell \ket{\psi} $ and from the triangle inequality the error is at most $\Delta_C$.

\begin{figure}
\[
\centerline{
\Qcircuit @C=0.9em @R=.6em {
    \lstick{\ket{0}_A}& \gate{H}              &\ctrl{1}   &\gate{H}&\qw\\ 
   \lstick{\ket{0}}& \gate{{\cal U}_{\psi}} &\gate{U_\ell}&\qw&\qw
}
}
\]
\caption[Hadamard test]{\label{fig:hadamard} Hadamard test circuit for computing the expectation value of one of the terms in the cost-function.  Here ${\cal U}_\psi$ is a unitary operation that prepares the ansatz state: ${\cal U}_\psi \ket{0} = \ket{\psi}$.}
\end{figure}
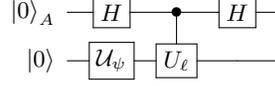

An estimate of $\bra{\psi} U_\ell \ket{\psi}$ can be obtained by performing the Hadamard test (as shown in \fig{hadamard}).  Specifically, the probability of measuring the ancillary qubit to be zero is $(1+{\rm Re}(\bra{\psi} U_\ell \ket{\psi}))/2$.  If Amplitude Amplification is used to mark the zero state for this circuit then the eigenphases of the resultant walk operator (within the two dimensional space spanned by the initial state and the marked state) is \cite{Brassard2002}
\begin{equation}
\phi=\pm 2\arcsin(\sqrt{P_0})= \pm  2\arcsin\left(\sqrt{\frac{1+{\rm Re}\left({\bra{\psi} U_\ell \ket{\psi}}\right)}{2}} \right).
\end{equation}
We then have that
\begin{equation}
    2\sin^2(\phi/2)-1 = {\rm Re}({\bra{\psi}U_\ell\ket{\psi}}).
\end{equation}
From calculus we then see that  
\begin{equation}
    \partial_\phi (2\sin^2(\phi/2)) = 2 \sin(\phi/2) \cos(\phi/2) \le \phi.
\end{equation}
Thus from Taylor's remainder theorem we have that for any $\delta\ge 0$
\begin{equation}
    |2\sin^2(\phi/ 2) - 2\sin^2((\phi+\delta)/2)| \le \delta
\end{equation}
and if $\phi\rightarrow \phi+\delta$ for some error $\delta$ we have that the uncertainty that propagates to the expectation value is at most
\begin{equation}
     |{\rm Re}(\bra{\psi} U_\ell \ket{\psi}) - {\rm Re}(\bra{\psi} U_\ell \ket{\psi})_{\rm est}| \le \delta.\label{eq:directest}
\end{equation}
Therefore if we wish to estimate the energy of a configuration within error $\epsilon$ it suffices to use phase estimation with an error of $\epsilon$ on the Grover operator.  Finally, as discussed above we take $\epsilon=\Delta_C/(L|w_\ell|)$ to ensure that the error sums up to $\Delta_C$ as required.

Using the QFT-based phase estimation algorithm in~\cite{Babbush2018} we find that, if we neglect the cost of the Quantum Fourier Transform and any additional costs due to additional precision required in the QROM then the number of queries to the Grover oracle needed is (for $\epsilon \le \pi$) $2^m \le 2\left\lfloor\frac{\pi}{\epsilon} \right\rfloor \le \frac{2\pi}{\epsilon} $. Here the factor of $2$ comes from the fact that the need to round to a power of $2$ leads to, in the worst case scenario, a factor of $2$ in the number of iterations required.

Next the Grover oracle requires two reflection operators, one that reflects about the state yielded by the Hadamard test circuit and another that reflects about the target space which is marked by the top qubit in~\fig{hadamard} being zero (i.e.\ $R_0=\openone - 2\ketbra{0}{0} \otimes \openone$).  The Grover walk operator is a product of these two operators $W=-R_1R_0$ and as a result, if we neglect the cost of the additional Hadamard and Toffoli gates needed to implement the conditional phase flip, the costs of this process are entirely due to the reflection about the initial state which requires two applications of the preparation of the initial state. 
We further will follow the assumption in the previous section that the cost of state preparation dwarfs the cost of applying prepare or select.  Thus under these assumptions, and taking the uncertainty in the objective function to be $\Delta_C$ the Toffoli complexity for the entire simulation is approximately
\begin{equation}
    \sum_{\ell=1}^L\frac{4p J\pi|w_\ell|L\mathcal{C}^{\rm phase}}{\Delta_C} =\frac{4 p J\pi\lambda L\mathcal{C}^{\rm phase}}{\Delta_C} . \label{eq:cost_direct}
\end{equation}
Thus, under these assumptions, direct energy evaluation yields an advantage over sampling if
\begin{equation}
    \sigma^2\ge 4 \pi \lambda \Delta_C L.
\end{equation}
We expect this to occur when the error tolerance is small and the number of terms is relatively modest.  On the other hand if the variance is small, target uncertainty is large, or $L$ is large then sampling will be preferable to the direct phase oracle evaluation process.

\subsubsection{Amplitude estimation based LCU evaluation}

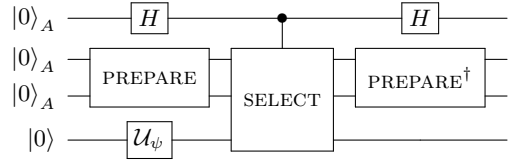
\begin{figure}
\[
\centerline{
\Qcircuit @C=0.9em @R=.6em {
  \lstick{\ket{0}_A}  & \gate{H}              &\ctrl{1}   &\gate{H}&\qw\\ 
  \lstick{\ket{0}_A} & \multigate{1}{\prep} &\multigate{2}{\sel}&\multigate{1}{\prep^\dagger}&\qw\\
    \lstick{\ket{0}_A}& \ghost{\prep}         &\ghost{\sel}&\ghost{\prep^\dagger}&\qw\\
\lstick{\ket{0}} & \gate{{\cal U}_{\psi}}                   &\ghost{\sel}&\qw&\qw
}
}
\]
\caption[Generalized Hadamard test]{\label{fig:hadamard_lcu} Generalized form of a Hadamard test that we will use for our QAOA implementation using LCU oracles.  Here ${\cal U}_\psi$ is a unitary operation that prepares the ansatz state: ${\cal U}_\psi \ket{0} = \ket{\psi}$.}
\end{figure}

One inexpensive approach that can be used to estimate the expectation value comes from combining the Hadamard test circuit and amplitude estimation~\cite{Brassard2002}.  
Here we used a slightly generalized form of a generalized Hadamard test circuit shown in \fig{hadamard_lcu}. The expectation value of the first qubit for the above circuit is $1/2+{\rm Re}(\bra{\psi} C \ket{\psi})/2$.
In order to see this, consider the following,
\begin{align}
    \ket{0}\ket{0}\ket{\psi} &\mapsto {\ket{0}} \left(\sum_\ell \sqrt{\frac{w_\ell}{\lambda}}\ket{\ell}  \right)\ket{\psi} \mapsto \frac{\ket{0} + \ket{1}}{\sqrt{2}} \left(\sum_\ell \sqrt{\frac{w_\ell}{\lambda}}\ket{\ell}  \right)\ket{\psi}\nonumber\\
    &\mapsto \frac{\ket{0}}{\sqrt{2}} \left(\sum_\ell \sqrt{\frac{w_\ell}{\lambda}}\ket{\ell}  \right)\ket{\psi}+\frac{\ket{1}}{\sqrt{2}} \left(\sum_\ell \sqrt{\frac{w_\ell}{\lambda}}\ket{\ell}  U_\ell\ket{\psi}\right)\nonumber\\
    &\mapsto \frac{\ket{0}}{{2}} \left(\left(\prep \ket{0}  \right)\ket{\psi}+ \sum_\ell \sqrt{\frac{w_\ell}{\lambda}}\ket{\ell}  U_\ell\ket{\psi}\right)+\frac{\ket{1}}{{2}}\left( \left(\prep \ket{0}  \right)\ket{\psi}- \sum_\ell \sqrt{\frac{w_\ell}{\lambda}}\ket{\ell}  U_\ell\ket{\psi}\right)\nonumber\\
    &\mapsto \frac{\ket{0}}{{2}} \left( \ket{0}  \ket{\psi}+ \prep^{\dagger}\sum_\ell \sqrt{\frac{w_\ell}{\lambda}}\ket{\ell}  U_\ell\ket{\psi}\right)+\frac{\ket{1}}{{2}} \left( \ket{0}  \ket{\psi}- \ket{{\rm junk}}\right) . \label{eq:chi}
\end{align}
Therefore, the probability of measuring $0$ in the top-most qubit in~\fig{hadamard_lcu} is
\begin{equation}
\frac{1}{4}\left(2 + \bra{0}\bra{\psi} \prep^\dagger \sum_\ell \sqrt{{\frac{w_\ell}{\lambda}}}\ket{\ell}  U_\ell\ket{\psi} +  \left(\bra{0}\bra{\psi} \prep^\dagger \sum_\ell \sqrt{{\frac{w_\ell}{\lambda}}}\ket{\ell}  U_\ell\ket{\psi} \right)^*\right)= \frac{1+{\rm Re}\left(\frac{\bra{\psi} C \ket{\psi}}{\lambda}\right)}{2}.
\end{equation}
If amplitude estimation is used, the number of invocations of $\prep$ and $\sel$ needed to estimate this probability within $\epsilon$ error is ${\cal O}(\lambda/\epsilon)$, which is a quadratic improvement over the sampling bound in \eq{sampling_bound}. 

Following the same reasoning used to derive \eq{directest} we find that the over all Toffoli count is then, under the assumptions that the Toffoli count is dominated by applications of the $\prep$, $\sel$ and phase circuit operations and further that the cost of adding an additional control to $\sel$ is negligible, given by
\begin{equation}
    \frac{4\pi p J \lambda (\mathcal{C}_{\rm phase} + \mathcal{C}_{\rm Sel} + 2 \mathcal{C}_{\rm Prep}) }{\Delta_C}.\label{eq:cost_lcuhadamard}
\end{equation}
Here $\mathcal{C}_{\rm Sel}$ and $\mathcal{C}_{\rm Prep}$ are the Toffoli counts for $\sel$ and $\prep$ respectively.

Equation \eq{cost_lcuhadamard} shows that the favorable scalings of the sampling approach and the direct phase evaluation methods can be combined together in a single method.  However, this advantage comes potentially at the price of a worse prefactor owing to the additional complexity of the $\prep$ and $\sel$ circuits.  In particular, we find that this approach will be preferable to sampling and direct phase estimation, respectively, when
\begin{align}
    \sigma^2&\ge 4\pi \lambda \Delta_C \left(\frac{\mathcal{C}_{\rm phase} + \mathcal{C}_{\rm Sel} + 2\mathcal{C}_{\rm Prep}}{\mathcal{C}_{\rm phase}} \right), \\
    L&\ge \frac{\mathcal{C}_{\rm phase} + \mathcal{C}_{\rm Sel} + 2\mathcal{C}_{\rm Prep}}{\mathcal{C}_{\rm Phase}}.
\end{align}
In general, we suspect that in fault-tolerant settings this this approach will be preferable to direct phase oracle evaluation because the costs of the prepare and select circuits will often be comparable, or less than, that of ${\cal U}_{\psi}$ as we will see in the following section where we provide explicit constructions for the $\prep$ and $\sel$ oracles.

\subsection{Adiabatic quantum optimization}
\label{sec:evolution}

\subsubsection{Background on the adiabatic algorithm}

The adiabatic algorithm \cite{Farhi2000a} works by initializing a system as an easy-to-prepare ground state of a known Hamiltonian, and then slowly (adiabatically) deforming that system Hamiltonian into the Hamiltonian whose ground state we wish to prepare. For instance, we might use a Hamiltonian parameterized by $s\in [0,1]$,
\begin{equation}
\label{eq:simple_adiabatic}
    H(s) = (1-s)H_0 + sH_1,
\end{equation}
where $H_0$ is a Hamiltonian with an easy-to-prepare ground state and $H_1$ is a Hamiltonian whose ground state we wish to prepare. We start the system in the ground state of $H(0) = H_0$ and then slowly deform the Hamiltonian by increasing $s$ from 0 to 1 until $H(1)=H_1$. If this is performed slowly enough, then the system will be in the ground state of $H_1$ at the end of the evolution.

The main challenge with the adiabatic algorithm is that we may need to turn $s$ on extremely slowly in order for the procedure to succeed. The rate at which we can turn on $s$ will depend on features of the spectrum of $H(s)$, including its derivatives and the minimum gap $\gap$ between the ground state eigenvalue and first excited state eigenvalue. It is often empirically observed that the total time of the evolution $T$ should scale as ${\cal O}(1/\gap^2)$. Indeed, this result has been proven using the so-called boundary adiabatic theorem. This result analyzes the adiabatic algorithm in terms of phase randomization between the different paths that describe quantum dynamics for a slowly varying time-dependent Hamiltonian. This randomization causes paths that lead different excitations to destructively interfere, which effects a mapping from the eigenvectors of an initial Hamiltonian to the corresponding eigenvectors of the target Hamiltonian in the limit of slow evolution relative to a relevant gap in the instantaneous eigenvalues of the time-dependent Hamiltonian. The boundary adiabatic theorem holds that if we let $\ket{\psi_k(s)}$ be the $k^{\rm th}$ instantaneous eigenvector of any Gevrey-class time-dependent $H(s)$ then we have that~\cite{elgart2012note}
\begin{equation}
    \left\|{\mathcal{T}} e^{-i\int_0^1 H(x)T \mathrm{d}x} \ket{\psi_k(0)}- \ket{\psi_k(1)}\right\| \in \widetilde{\cal O}\left(\frac{1}{\gap^2 T} \right),
\end{equation}
where $\gap$ is the minimum eigenvalue gap between the state $\ket{\psi_k(s)}$ and the remainder of the spectrum.  It then follows if we pick an appropriate value for  $T\in \order{1/\gap^2 \epsilon}$ then we can make the error less than $\epsilon$ for an arbitrary gapped adiabatic path.
Alternatively, if very high precision is required then the time required for adiabatic state preparation can also be improved for analytic Hamiltonians to  $\widetilde{\mathcal{O}}({\rm poly}(\|\dot{H}\|, \|\ddot{H}\|,\ldots)(1/\gap^2 +\log(1/\epsilon)/\gap))$ by adaptively choosing the adiabatic path to have obey $\|\partial_s^q H(0)\| = \|\partial_s^q H(1)\| = 0$ for all positive integers less than $Q(\epsilon)\in \order{\log(1/\epsilon)}$; however, this approach requires small error tolerance on the order of $\epsilon \in \order{\gap}$ in order to see the benefits of these improved adiabatic paths~\cite{lidar2009adiabatic,wiebe2012improved,kieferova2014power}.

Note that the boundary adiabatic theorem only tells us about the state at the end of the evolution, and does not actually tell us anything the state we would be in at the middle of the evolution. For that there are ``instantaneous'' adiabatic theorems which bound the probability of being in the ground state throughout the entire evolution. For instance, one such way to show this is based on the Zeno stabilized adiabatic evolutions described in \sec{zeno} \cite{Boixo2009a}.
These instantaneous adiabatic theorems have complexity $\order{L^2/(\epsilon\gap)}$, where
\begin{equation}
    L = \int_0^1 \| \dot\psi (s) \| ds
\end{equation}
is the path length.
In the case of simulated annealing, one can show that the path length is independent of $\gap$, whereas in general the worst-case bound is $L\le \|\dot H\|/\gap$, which yields $\order{\|\dot H\|^2/\gap^3}$ complexity \cite{Boixo2009a}. It is not completely clear which style of adiabatic evolution will give the best results when using the approach as a heuristic, and so we discuss both here. With either approach we typically take $H_1$ to be the cost function of interest and take $H_0$ to be a simple-to-implement Hamiltonian that does not commute, with an easy-to-prepare ground state. For instance, a common choice is to take $H_0 = \sum_{i=1}^N X_i$ where $X_i$ is the Pauli-$X$ operator, so that the initial state is $\ket{+}^{\otimes N}$. Other $H_0$ Hamiltonians (or more complicated adiabatic paths) are also possible.

The simplest way to use the adiabatic algorithm as a heuristic is to discretize the evolution using product formulas. For instance, if we assume the adiabatic schedule in \eq{simple_adiabatic} then we could attempt to prepare the ground state as
\begin{equation}
\prod_{k=1}^M \exp\left(-i \left(\frac{M-k}{M^2}\right) H_0 T\right) \exp\left(-i \left(\frac{k}{M^2}\right) H_1 T\right) \ket{\psi_0(0)},
\end{equation}
where $M$ is the number of first order Trotter steps used to discretize the adiabatic evolution. The idea of the heuristic is to choose $M$ based on available resources. $T$ will also need to be chosen heuristically rather than based on knowledge of the gap, which we do not expect to have in general. For fixed $M$, smaller $T$ will enable more precise approximation of the continuous-time algorithm, but smaller $T$ also means the system is less likely to stay adiabatic.

Of course, one can also easily extend this strategy to using higher-order product formulas, or to using either different adiabatic interpolations or adiabatic paths. For example, if we define
\begin{equation}
    U_2\left(\frac{k-1}{M},\frac{k}{M}\right) = \exp\left(-i \left(\frac{M-k-1/2}{2 M^2}\right) H_0 T\right) \exp\left(-i \left(\frac{k+1/2}{M^2}\right) H_1 T\right)\exp\left(-i \left(\frac{M-k-1/2}{2 M^2}\right) H_0 T\right),
\end{equation} 
then we have that  $\left\|\prod_{k=1}^MU_2\left(\frac{k-1}{M},\frac{k}{M}\right) - \mathcal{T}\exp(-i\int_0^T H(t) \mathrm{d} t)\right\| \in \mathcal{O}(T^3/M^2)$.
Higher-order versions of such integrators of order can be formed via Suzuki's recursive construction (for any $s\in [0,1]$):
\begin{align}
    U_{\rho}(s,s+\delta)&:= U_{\rho-2}(s+[1-\gamma_\rho]\delta,s+\delta)U_{\rho-2}(s+[1-2\gamma_\rho]\delta,s+[1-\gamma_\rho]\delta)\nonumber\\ &\quad \times U_{\rho -2}(s+2\gamma_\rho\delta,s+[1-2\gamma_\rho]\delta)U_{\rho-2}(s+\gamma_\rho\delta, s+2\gamma_\rho\delta)U_{\rho-2}(s, s+\gamma_\rho\delta).\label{eq:suzrecur}
\end{align}
Here $\gamma_\rho = (4-4^{1/(\rho-1)})^{-1}$ which approaches $1/3$ as the order of the formula, $\rho$, goes to infinity.  Furthermore we have that the error in the Trotter-Suzuki algorithm scales as as
\begin{equation}
    \left\|\prod_{k=1}^MU_{\rho}\left(\frac{k-1}{M},\frac{k}{M}\right) - \mathcal{T}\exp(-i\int_0^T H(t) \mathrm{d} t)\right\|\in \mathcal{O}\left(T^{\rho+1} / M^{\rho}\right),
\end{equation}
which results in near linear scaling with $T$ in the limit as $T$ approaches infinity.  

In practice, however, since the number of exponentials in the Trotter-Suzuki formula grows exponentially with the order there is in practice an optimal tradeoff in gate complexity that is satisfied by a finite order formula for a fixed $\epsilon$ and $T$.  For simplicity, we will assume that the time evolution under $H_0$ will be much cheaper to implement than the time evolution under $H_1$. As $H_1$ can be implemented using the phase oracles $O^{\rm phase}$ discussed in \sec{oracles}, the total cost of the procedure will be approximately $M \, {\cal C}^{\rm phase}$.  This implies that, for a finite value of $M$, the cost of performing the heuristic optimization using the above adiabatic sequence is approximately
\begin{equation}
    {\cal C}_{\rm adiabatic}=2M 5^{\rho/2-1} {\cal C}^{\textrm{phase}}.
\end{equation}
Again, assuming that our target error in the adiabatic sweep is $\epsilon$ and $\gap^2 \in {\cal O}(\epsilon)$ then it suffices to take $T\in \mathcal{O}(1/\gap^2\epsilon)$ and further after optimizing the cost by setting $M$ equal to $2^{\rho/2-1}$ we find that $M\in (T^{1+o(1)}/\epsilon^{o(1)})$.  Therefore the total cost obeys
\begin{equation}
    {\cal C}_{\rm adiabatic} \in \frac{{\cal C}_{\rm phase}}{(\epsilon \gap^{2})^{1+o(1)}}.
\end{equation}
Similarly, if we are interested in the limit where $\gap^2 \in \omega(\epsilon)$, then boundary cancellation methods~\cite{lidar2009adiabatic,wiebe2012improved} can be used to improve the number of gates needed to reach the global optimum to
\begin{equation}
    {\cal C}_{\rm adiabatic} \in \frac{{\cal C}_{\rm phase}\log^{1+o(1)}(1/\epsilon)}{\gap(\epsilon \gap)^{o(1)}}.
\end{equation}
These results show that, provided the eigenvalue gap is polynomial, we can use a simulation routine for $e^{-iH_1(t)}$ and $e^{-iH_0(t)}$ to find the local optimum in polynomial time.  However, in practice we will likely want to use such an algorithm in a heuristic fashion wherein the timesteps do not precisely conform to the adiabatic schedule.

To give the cost for a single step a little more precisely, we can also include the cost of implementing a transverse driving field.
Since that involves applying a phase to $\bphase$ bits to each of $N$ qubits,
using repeat-until-success circuits, this has cost $1.15 N\bphase+\order{N}$ in terms of T gates, with a single ancilla qubit.
It is also possible to sum the bits with Toffoli cost $N$, then phase by the sum with cost $\bphase^2/2+\mathcal{O}(\bphase\log\bphase)$ (accounting for multiplying the phase by a constant factor), though that has large ancilla cost.
Using the sum of tree sums approach would give complexity $4N+\bphase^2/2+\mathcal{O}(\bphase\log\bphase)$, with $3\log L+\order{1}$ temporary ancillas.
There will be $\gradbits=\bphase+\order{\log\bphase}$ persistent ancillas needed for a phase gradient state as well, but in many cases that state will be the same as in other steps of the procedure, so does not increase the ancilla cost.
Using this approach, and omitting the factor of $2\times 5^{\rho/2-1}$ for order $\rho$ Suzuki, gives Toffoli cost
\begin{equation}\label{eq:suzuki}
    {\cal C}^{\textrm{phase}} + 4N+\bphase^2/2+\order{\bphase\log\bphase}
\end{equation}
for a single step.

\subsubsection{Heuristic adiabatic optimization using quantum walks}\label{sec:heuwalk}

While the procedure we described for heuristically using the adiabatic algorithm with Trotter based methods is well known, it is less clear how one might heuristically use LCU methods with the adiabatic algorithm. One reason we might try doing this is because the qubitized quantum walks that we discuss in \sec{oracles} are sometimes cheaper to implement than Trotter steps for some problems. One approach to using LCU methods for adiabatic state preparation might be to directly attempt to simulate the time-dependent Hamiltonian evolution using a Dyson series approach, as was recently suggested for the purpose of adiabatic state preparation in \cite{wan2020fast}. However, this would require fairly complicated circuits due to the many time registers that one must keep to index the time-ordered exponential operator. In principle, we could always use quantum signal processing (or more generally quantum singular value transformations) to convert the walk operator at time $t$ into the form $e^{-iH(t)\delta}$ for some timestep $\delta$.

Instead, here we will suggest a strategy which is something of a combination between using qubitized quantum walks and using a product formula approximation. Our method is unlikely to be asymptotically optimal for this purpose but it is simple to implement and we suspect it would be cheaper than either a Dyson series approach or a Trotter approach for some applications on a small error-corrected quantum computer. The idea is to stroboscopically simulate time evolution as a short-time evolved ``qubitized'' walk. The result will be that we actually simulate the adiabatic path generated by the arccosine of the normalized Hamiltonian $H(s)$ rather than the adiabatic path generated directly by $H(s)$, but we expect that the relevant part of the eigenspectrum will be in the linear part of the arccosine, which means there will be not much effect on the dynamics. The main challenge in this approach will be to artificially shrink the effective duration of these quantum walk steps so that the method can be refined.

In the following we will assume that $\sel^2=1$ which is to say that every Hamiltonian in the decomposition is self-adjoint (consistent with the problem Hamiltonians we consider).  For every eigenvector $\ket{\psi_k(t)}$ of $H(t)$ with $H\ket{\psi_k(t)} = E_k(t)$, if we define $\ket{L} = \prep \ket{0}$ then we can write
\begin{equation}
    W= (I -2 I\otimes \ket{L}\!\bra{L}) \sel . \label{eq:W_no_t}
\end{equation}
The walk operator can be seen as a direct sum of two different walk operators,  $W= W_H \oplus W_\perp$, where $W_H$ is the portion of the walk operator that acts non-trivially on $\ket{\psi_k(t),L}=\ket{\psi_k(t)} \otimes \ket{L}$ and $W_\perp$ is the operator that acts on the remaining states.  Next, if for each $k$ and $t$ we define $\ket{\psi_k^\perp(t)}$ such that
\begin{equation}
    \ket{\psi_k^\perp(t)}=\frac{\left(W - \frac{E_k(t)}{\lambda(t)}\right)\ket{\psi_k(t),L}}{\sqrt{1 - \frac{E_k^2(t)}{\lambda^2(t)}}}=\frac{\left(W_H - \frac{E_k(t)}{\lambda(t)}\right)\ket{\psi_k(t),L}}{\sqrt{1 - \frac{E_k^2(t)}{\lambda^2(t)}}},\label{eq:perp}
\end{equation}
then we can express
\begin{equation}
    W_H(t) = \exp\left({-i\left(\sum_k i\ket{\psi_k^\perp(t)}\!\bra{\psi_k(t),L}-i \ket{\psi_k(t),L}\!\bra{\psi_k^\perp(t)}\right)\arccos\left(\frac{E_k(t)}{\lambda(t)} \right)}\right) .
\end{equation}
It may be unclear how to implement a time-step for $W(t)$ since the operation is only capable of applying unit-time evolutions.  Fortunately, we can address this by taking for any $r\ge 1$
\begin{equation}
    H(t) = \sum_k \lambda_k(t) U_k \mapsto \sum_k \lambda_k(t) U_k + \frac{(r-1)\lambda(t)}{2}\left( I - I\right) .
\end{equation}
In this case we can block encode the Hamiltonian using a unary encoding of the extra two operators via
\begin{equation}
    \ket{L(t,r)} = \sum_{k}\sqrt{\frac{\lambda_k(t)}{\lambda(t)r}}\ket{k}\ket{00}+ \sqrt{\frac{r-1}{2r}}\ket{0}\left(\ket{10} + \ket{11}\right) . \label{eq:L}
\end{equation}
The select oracles for this Hamiltonian require one additional control for each of the original terms in the Hamiltonian and the additional terms only need a single Pauli-$Z$ gate to implement.  We will define this operator to be $\sel'$.

With these two oracles defined, we can then describe the walk operator $W_r(t)$ for any fixed value of $t$ to be
\begin{equation}
    W_r(t) = (I -2 I\otimes \ket{L(t,r)}\!\bra{L(t,r)}) \sel' . \label{eq:W}
\end{equation}
This new Hamiltonian has exactly the same eigenvectors, however its value of $\lambda$ is greater by a factor of $r$.  In particular, we can express the walk operator (restricted to the eigenspace supported by the instantaneous eigenvectors of $H(t)$) is
\begin{align}
    W_{H,r}(t)
    &=\exp\left(\left({-i\left(\sum_k i\ket{\psi_k^\perp(t)}\!\bra{\psi_k(t),L(t,r)}-i \ket{\psi_k(t),L(t,r)}\!\bra{\psi_k^\perp(t)}\right)r\arccos\left(\frac{E_k(t)}{r\lambda(t)} \right)}\right)\frac{1}{r}\right) . \label{eq:WHR}
\end{align}
Using the fact that $\arccos(x) = \pi/2- \arcsin(x)$ we have that, up to an irrelevant global phase this operator can be written as
\begin{align}
    V_{H,r}(t) = \exp\left(\left({i\left(\sum_k i\ket{\psi_k^\perp(t)}\!\bra{\psi_k(t),L(t,r)}-i \ket{\psi_k(t),L(t,r)}\!\bra{\psi_k^\perp(t)}\right)r\arcsin\left(\frac{E_k(t)}{r\lambda(t)} \right)}\right)\frac{1}{r}\right).
\end{align}
Thus the operator $V_{H,r}(t)$ can be seen to generate a short time-step of duration $1/r$ for an effective Hamiltonian
\begin{equation}
    H_r(t) :=\left(\sum_k i\ket{\psi_k^\perp(t)}\!\bra{\psi_k(t),L(t,r)}-i \ket{\psi_k(t),L(t,r)}\!\bra{\psi_k^\perp(t)}\right)r\arcsin\left(\frac{E_k(t)}{r\lambda(t)} \right).
\end{equation}
Note that as $r\rightarrow \infty$ the eigenvalues of this Hamiltonian approach $\pm E_k(t)/\lambda(t)$ and more generally $$\left|r\sin^{-1}(E_k(t)/(r\lambda(t))) - E_k(t)/\lambda(t)\right| \in O(1/r^2).$$
For any fixed value of $r$ we can choose an adiabatic path between an initial Hamiltonian and a final Hamiltonian.  The accuracy of the adiabatic approximation depends strongly on how quickly we traverse this path so it is customary to introduce a dimensionless time $s=t/T$ which allows us to easily change the speed without altering the shape of the adiabatic path. In \app{adiabatic_walk} we are able to show that the adiabatic theorem then implies that the number of steps of the quantum walk required to achieve error $\epsilon$ in an adiabatic state preparation for a maximum rank Hamiltonian with gap $\gap$ is in
\begin{equation}
\widetilde{\mathcal{O}}\left( \frac{1}{\epsilon^{3/2}}\sqrt{\frac{ \max_s\left( \|\ddot{H}\| +|\ddot{\lambda}|\right)\max_s\left(|\dot{\lambda}| + \|\dot{H}\| \right)}{\min({\gap},\min_k |E_k|)^2}+ \frac{\lambda \max_s\left(|\dot{\lambda}| + \|\dot{H}\| \right)^3}{\min({\gap},\min_k |E_k|)^4 }} \right).
\end{equation}
The reason why this result depends on the minimum value of $E_k$ is an artifact of the fact that several of the eigenvalues of the walk operator can be mapped to $1$ under repeated application of $W_r$.  This potentially can alter the eigenvalue gaps for eigenvalues near zero which impacts the result.

The key point behind this scaling is that it shows that as the number of time slices increases this heuristic converges to the true adiabatic path.  Just as the intuition behind Trotterized adiabatic state preparation hinged on this fact, here this result shows that we can similarly use a programmable sequence of parameterizable walk operators to implement the dynamics.  The main advantage relative to Trotter methods is that the price that we have to pay using this technique does not depend strongly on the number of terms in the Hamiltonian which can lead to advantages in cases where the problem or driver Hamiltonians are complex.

This scaling can be improved by using higher-order splitting formulas for the time evolution~\cite{Wiebe2008} and by using boundary cancellation methods to improve the scaling of the error in adiabatic state preparation.  In general, if we assume that $\gap\in O(1)$ for the problem at hand then it is straightforward to see that we can improve the scaling from $O(1/\epsilon^{3/2})$ to $1/\epsilon^{o(1)}$~\cite{lidar2009adiabatic,wiebe2012improved,kieferova2014power}.  It is also worth noting that the bounds given above for the scaling with respect to the derivatives of the Hamiltonian and the coefficients of the Hamiltonian is expected to be quite loose owing to the many simplifying bounds used to make the expression easy to use.  On the other hand, the scaling with the gap and error tolerance is likely tighter.

\subsubsection{Zeno projection of adiabatic path via phase randomization}
\label{sec:zeno}


The principle of the Zeno approach is to increment the parameter for the Hamiltonian $s$ or $\beta$ by some small amount such that the overlap of the ground state of the new Hamiltonian with that of the previous Hamiltonian is small.
One can then perform phase estimation to ensure that the system is still in the ground state.
This approach was used in \cite{lemieux2019efficient,Lemieux2020ResourceComputer}, and combined with a rewind procedure to give a significant reduction in gate complexity compared to other approaches.
An alternative approach was proposed in \cite{Boixo2009a,Chiang2014}, where the measurement was replaced with phase randomization.
Here we summarize this method and show how to further optimize it.

When using phase estimation, if it verifies that the system is still in the ground state, one continues with incrementing the parameter.
If the ground state is not obtained from the phase estimation, one could abort, in which case no output is given and one needs to restart.
Because the probability of failure is low, one could just continue regardless, and check at the end.
That means that the result of the phase estimation is discarded.

The phase estimation is performed with control qubits controlling the time of the evolution, then an inverse quantum Fourier transform on the control qubits to give the phase.
But, if the result of the measurement is ignored, then one can simply ignore the inverse quantum Fourier transform, and regard the control qubits as being measured in the computational basis and the result discarded.
That is equivalent to randomly selecting values for these control qubits in the computational basis at the beginning.
But, if these qubits take random values in the computational basis, one can instead just classically randomly generate a time, and perform the evolution for that time.

In performing a phase measurement using control qubits, one uses a superposition state on those control qubits, and the error in the phase measurement corresponds to the Fourier transform of those amplitudes.
That is, with $b$ control qubits, we have a state of the form
\begin{equation}
\ket{\chi_\phi}=\sum_{z=0}^{2^b-1} e^{iz\phi} \chi_z \ket{z},
\end{equation}
where $\phi$ is a phase that would correspond to $-E\delta t$, the energy eigenvalue of the Hamiltonian times the shortest evolution time.
Then the phase measurement using the quantum inverse Fourier transform corresponds to the POVM $\ket{\hat\phi}\!\bra{\hat\phi}$, with
\begin{equation}
\ket{\hat\phi} = \frac 1{\sqrt{2\pi}}\sum_{z=0}^{2^b-1} e^{iz\hat\phi} \ket{z}.
\end{equation}
The probability distribution for the error $\delta\phi = \hat\phi-\phi$ is then given by
\begin{equation}
{\rm Pr}(\delta\phi) = \left|\langle{\hat\phi}|{\chi_\phi}\rangle\right|^2 = \frac 1{2\pi} \left| \sum_{z=0}^{2^b-1} e^{iz\delta\phi} \chi_z\right|.
\end{equation}
These measurements are equivalent to the theory of window functions in spectral analysis. A particularly useful window to choose is the Kaiser window, because it has exponential suppression of errors \cite{Kaiser}.

In the case where the evolution time is chosen classically, it can be given by a real number, and we do not need any bound on the evolution time. 
Then the the expected cost is the expectation value of $|t|$
\begin{equation}\label{eq:tcost}
  \langle |t| \rangle =  \int dt\, |t| p_{\rm time}(t) .
\end{equation}
Because there is no upper bound on $t$, we can obtain a probability distribution for the error that drops strictly to zero outside the given interval, rather than being exponentially suppressed.
Still considering a coherent superposition for the moment, the state is given by
\begin{equation}
\ket{\psi_E}=\int dt \, e^{-iEt} \chi_t \ket{t},
\end{equation}
where $E$ is the energy, $t$ is the evolution time, and $p_{\rm time}(t)=|\chi_t|^2$.
Then the POVM is $\ket{\hat E}\!\bra{\hat E}$ with
\begin{equation}
\ket{\hat E} = \frac 1{\sqrt{2\pi}}\int dt \, e^{-i\hat Et} \ket{t}.
\end{equation}
The probability distribution for the error in the measurement of $E$ is
\begin{equation}
{\rm Pr}(\delta E) = \frac 1{2\pi} \left| \int dt \, e^{it\delta E} \chi_t\right|^2.
\end{equation}
An alternative description is to describe the system as being in state
\begin{equation}
\ket{\psi} = \sum_j \braket{\psi_j}{\psi}\ket{\psi_j},
\end{equation}
where $\ket{\psi_j}$ is an eigenstate of the Hamiltonian with energy $E_j$.
Then evolving for time $t$ with probability $p_{\rm time}(t)$ gives the state
\begin{equation}
\sum_{j,k} \braket{\psi_{j}}{\psi} \braket{\psi}{\psi_{k}} \tilde p_{\rm time}(E_{j}-E_{k})\ket{\psi_{j}}\!\bra{\psi_{k}},
\end{equation}
where
\begin{equation}
    \tilde p_{\rm time}(E_{j}-E_{k}) = \int dt\, p_{\rm time}(t) e^{-i(E_{j}-E_{k})t}.
\end{equation}
If the width of the Fourier transform of the probability distribution $p_{\rm time}$ is less than the spectral gap $\gap$, then the state is
\begin{equation}
\sum_{j} |\!\braket{\psi_{j}}{\psi}\!|^2 \ket{\psi_{j}}\!\bra{\psi_{j}}.
\end{equation}
In comparison, if $\Pr(\delta E)$ is equal to zero for $|\delta E|\le E_{\rm max}$, then the same result will be obtained for $2E_{\max}=\gap$.
This is what would be expected, because if $p_{\rm time}(t)=\chi_t^2$, then the Fourier transform of $p_{\rm time}$ is the autocorrelation of the Fourier transform of $\chi_t$, and therefore has twice the width.

Next we consider appropriate probability distributions.
A probability distribution for $t$ that was suggested in \cite{Boixo2009a} was
\begin{equation}
p_{\rm time}(t) = \frac {8\pi\, {\rm sinc}^4(t \gap/4)}{3 \gap}.
\end{equation}
That gives $\langle |t| \rangle=12\ln 2/(\pi \gap)$, so $\langle |t| \rangle \gap\approx 2.648$.
There $\Pr(\delta E)$ is equivalent to the square of a triangle window, but greater performance can be obtained by using the triangle window
\begin{equation}
{\rm Pr}(\delta E) = \frac 2{\gap} (1-|2\delta E/\gap|).
\end{equation}
Then the corresponding $\psi_t$ is obtained from the Fourier transform of $\sqrt{{\rm Pr}(\delta E)}$ as
\begin{equation}
    \chi_t = \frac{\sin(\gap t/2)C(\sqrt{\gap t/\pi}) -\cos(\gap t/2)S(\sqrt{\gap t/\pi})}{(\gap t/2)^{3/2}},
\end{equation}
where $C$ and $S$ are Fresnel integral functions.
That gives $\langle |t| \rangle=7/(3\gap)$, so $\langle |t| \rangle \gap \approx 2.333$.

To find the optimal window, we can take
\begin{equation}
\frac 1{\sqrt{2\pi}} \int dt \, e^{itx} \chi_t=(1-x^2)\sum_{\ell} a_\ell x^{2\ell},
\end{equation}
for $x$ the difference in energy divided by $E_{\max}$.
We use only even orders, so it is symmetric, and the factor of $(1-x^2)$ ensures that it goes to zero at $\pm 1$.
Then
\begin{equation}
    \chi_t = \frac 1{\sqrt{2\pi}}\sum_{\ell} a_\ell \int_{-1}^1 dx \cos(xt) (1-x^2)x^{2\ell}.
\end{equation}
Then the expectation of the absolute value of the time is
\begin{equation}
\int dt \, |t|\, |\chi_t| = \frac 1{2\pi}\sum_{k,\ell}a_k a_\ell A_{k\ell},
\end{equation}
where
\begin{equation}
A_{k\ell} = \int dt \, |t| \left(\int_{-1}^1 dx \cos(xt) (1-x^2)x^{2k}\right)\left(\int_{-1}^1 dz \cos(zt) (1-z^2)z^{2\ell}\right).
\end{equation}
We also need, for normalization,
\begin{equation}
1=\sum_{k,\ell} a_k a_\ell \int_{-1}^1 dx \, (1-x^2)^2 x^{2(k+\ell)} = \sum_{k,\ell} a_k a_\ell B_{k\ell},
\end{equation}
where
\begin{equation}
B_{k\ell} = \frac {16}{[2(k+\ell)+1][2(k+\ell)+3][2(k+\ell)+5]}.
\end{equation}
Then defining $\vec b = B^{1/2}\vec a$, the normalisation corresponds to $\|\vec b\|=1$.
Then the minimum $\langle |t|\rangle$ corresponds to minimizing $\vec a^T A\vec a/\pi$, which is equivalent to minimizing $\vec a^T B^{-1/2}AB^{-1/2} \vec a/\pi$, so we need to find the minimum eigenvalue of $B^{-1/2}AB^{-1/2}$.
That gives $\langle |t| \rangle E_{\max}\approx 1.1580$ with terms up to $a_{22}$ (a 46th order polynomial).

This explanation is for the case where there is Hamiltonian evolution for a time $t$ which can take any real value. In the case of steps of a quantum walk with eigenvalues $e^{\pm i\arccos(H/\lambda)}$, the number of steps would take an integer value. For the Hamiltonian evolution it could be implemented by steps of a quantum walk as well but it is more efficient to simply use the steps of that quantum walk directly without signal processing. To obtain the corresponding probability distribution for a discrete number of steps, we simply take the probability distribution for $t$ at points separated by $1/\lambda$. That will yield a probability distribution for the error that is the same as for the continuous distribution, except with a periodicity of $\lambda$. That periodicity has no effect on the error, because it is beyond the range of possible values for the energy. The reason for this correspondence is that taking the probability distribution at a series of discrete points is like multiplying by a comb function, equivalent to convolving the error distribution with a comb function.

\subsection{Szegedy walk based quantum simulated annealing}\label{sec:szegmain}

In the remainder of \sec{algorithms} we consider quantum simulated annealing, where the goal is to prepare a coherent equivalent of a Gibbs state and cool to a low temperature.
More specifically, the coherent Gibbs state is
\begin{equation}
    \ket{\psi_\beta} :=
    \sum_{x \in \Sigma} \sqrt{\pi_\beta (x)} \ket{x},\quad
    \pi_\beta (x) \propto \exp(-\beta E_x),
\end{equation}
where $\beta$ is the inverse temperature.
For annealing, we have transition probabilities of obtaining $y$ from $x$ denoted $\Pr(y|x)$, which must satisfy the detailed balance condition
\begin{equation}
\Pr(y|x)\pi_\beta(x)=\Pr(x|y)\pi_\beta(y).
\end{equation}
The detailed balance condition ensures that $\pi_\beta$ is the equilibrium distribution with these transition probabilities.
For the costings in this work we take for $y$ differing from $x$ by a single bit flip,
\begin{equation}
    \Pr(y|x) :=
     \min\left\{ 1, \exp \left( \beta \left( E_{x} - E_{y} \right) \right)\right\}/N,
\end{equation}
and $\Pr(x|x)=1-\sum_{y\ne x} \Pr(y|x)$.
This choice is similar to that in \cite{lemieux2019efficient}.
Another choice, used in \cite{Boixo2014a}, is
$\Pr(y|x) = \chi\exp \left( \beta \left( E_{x} - E_{y} \right) \right)$ for $\chi$ chosen to prevent sums of probabilities greater than $1$.
If one were to construct a Hamiltonian as
\begin{equation}\label{eq:annealham}
    \bra{x} H_\beta \ket{y} =
    \delta_{x,y} - \sqrt{\Pr(x|y)\Pr(y|x)},
\end{equation}
then the detailed balance condition ensures that the ground state is $\ket{\psi_\beta}$ with eigenvalue zero.
One can then apply an adiabatic evolution under this Hamiltonian to gradually reduce the temperature (increase $\beta$).

In the approach of \cite{Somma2008b}, the method used is to instead construct a quantum walk where the quantum Gibbs state is an eigenstate.
One could change the value of $\beta$ between each step of the quantum walk similarly to the adiabatic algorithm for the Hamiltonian.
Alternatively, for each value of $\beta$ one can apply a measurement of the walk operator to project the state to $\ket{\psi_\beta}$ via the quantum Zeno effect.
Reference \cite{Somma2008b} also proposes using a random number of steps of the walk operator to achieve the same effect as the measurement.
The advantage of using the quantum walk is that the complexity scales as ${\cal O}(1/\sqrt{\delta})$, where $\delta$ is the spectral gap of $H_\beta$, rather than $\order{1/\delta}$, which is the best rigorous bound for the scaling of (classical) simulated annealing.

\begin{figure}[t]
\centering
  \resizebox{.8\linewidth}{!}{\includegraphics{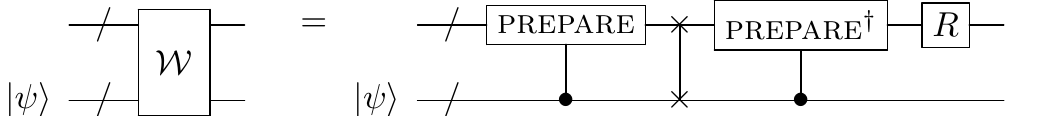}}
  \caption[Qubitized walk for Szegedy approach]{\label{fig:szeg}
    The qubitized quantum walk operator ${\cal W}$ using the Szegedy approach.}
  \vspace{5mm}
\centering
  \resizebox{.8\linewidth}{!}{\includegraphics{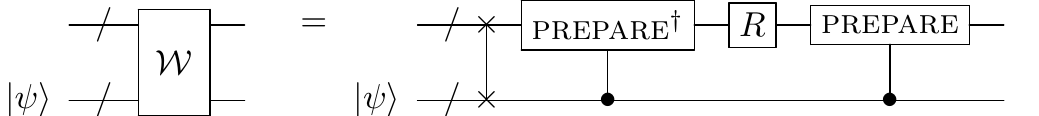}}
  \caption[Szegedy walk with controlled preparation shifted]{\label{fig:szeg2}
    The quantum walk operator using the Szegedy approach, where we have moved the controlled preparation to the end.}
\end{figure}

The quantum walk used in \cite{Somma2008b} is based on a Szegedy walk, which involves a controlled state preparation, a \textsc{swap} between the system and the ancilla, and inversion of the controlled state preparation.
Then a reflection on the ancilla is required.
The sequence of operations is as shown in \fig{szeg}.
The dimension of the ancilla needed is the same as the dimension as the system.
The reflection and \textsc{swap} have low cost, so the Toffoli cost is dominated by the cost of the controlled state preparation.

The Szegedy approach builds a quantum walk in a similar way as the LCU approach in \fig{w}, where there is a block encoded operation followed by a reflection \cite{Low2016}.
That is, preparation of the ancilla in the state $\ket{0}$, followed by unitary operations $U$ and projection onto $\ket{0}$ on the ancilla would yields the block encoded operator $A=\bra{0}U\ket{0}$.
Instead of performing a measurement on the ancilla, the reflection about $\ket{0}$ results in a joint operation that has eigenvalues related to the eigenvalues of $A$ as $e^{\pm i\arccos a}$, where $a$ is an eigenvalue of $A$.

Here the controlled state preparation is of the form
\begin{equation}
\label{eq:cprep_defn}
    \textsc{cprep}\ket{x} \ket{0} =
    \sum_{y} 
    \sqrt{\Pr(y|x)}
    \ket{x} \ket{y} \equiv \ket{\alpha_x},
\end{equation}
where the sum is taken over all $y$ that differ from $x$ by at most one bit.
As a result, the block-encoded operation is
\begin{equation}
    \bra{0}\textsc{cprep}^\dagger\, \textsc{swap}\, \textsc{cprep} \ket{0} =
    \sum_{x,y} 
    \sqrt{\Pr(x|y)\Pr(y|x)}
    \ketbra{y}{x} . 
\end{equation}
Thus the block-encoded operation has a matrix representation of the form $\sqrt{\Pr(x|y)\Pr(y|x)}$, which is equivalent to $\openone-H_\beta$.
Therefore the quantum Gibbs state $\ket{\psi_\beta}$ is an eigenstate of this operation with eigenvalue $1$.
Combining this operation with the reflection gives a step of a quantum walk with eigenvalues corresponding to the arccosine of the block-encoded operator \cite{BerryNPJ18,PoulinPRL18}. It is this arccosine that causes a square root improvement in the scaling with the spectral gap.
This is because if the block-encoded operation has gap $\delta E$ from the eigenvalue of 1 for the target state, taking the arccosine yields a gap of approximately $\sqrt{2\delta E}$ for the quantum walk.
This gap governs the complexity of the algorithm based on the quantum walk.

In implementing the step of the walk, the state preparation requires calculation of each of the $\Pr(y|x)$ for a given $x$.
In turn these require computing the energy difference under a bit flip, and the exponential.
The probability $\Pr(x|x)$ is computed from the formula $\Pr(x|x) \equiv
1 - \sum_{y \neq x} \Pr(y|x)$ required for normalization of the probabilities. To prepare the state one can first prepare a state of the form
\begin{equation}
\label{eq:main_Vec}
\ket{\psi_x}=
    \sum_{k} 
    \sqrt{\Pr(x_k|x)}
    \ket{x} \ket{k},
\end{equation}
where $x_k$ indicates that bit $k$ of $x$ has been flipped with $k=0$ indicating no bit flip, and $\ket{k}$ is encoded in one-hot unary.
The state $\ket{\alpha_x}$ can then be prepared by applying \textsc{cnot}s between the respective bits of the two registers.

In order to prepare the state $\ket{\psi_x}$ in unary, an obvious method is to perform a sequence of controlled rotations depending on the transition probabilities.
However, that tends to be expensive because our method of performing rotations involves multiplications, and high precision is required because the error in each rotation accumulates.
A better method can be obtained by noting that the amplitudes for $k>0$ are limited.
We can then perform the state preparation by the following method.
\begin{enumerate}
\item Compute $N\Pr(x_k|x)$ for all $N$ bit flips, and subtract those values from $N$ to obtain $N\Pr(x|x)$.
Note that $N\Pr(x_k|x)\le 1$, and we compute this value to $\bsmooth$ bits.
The value of $N\Pr(x|x)$ will need $\lceil \log N \rceil + \bsmooth$ bits, but only the leading $\bsmooth$ bits can be regarded as reliable.
The complexity of the subtractions is $N(\lceil \log N \rceil + \bsmooth)$.
\item We have $N$ qubits in the target system we need to prepare the state and five ancillas,
\begin{equation}
\ket{0}_{\texttt{A}}\ket{0}_{\texttt{K}}\ket{0}_{\texttt{Z}}\ket{0}_{\texttt{ZZ}}\ket{0}_{\texttt{B}}\ket{0}_{\texttt{C}},
\end{equation}
where $\texttt{K}$ is the target system, $\texttt{A}$, $\texttt{B}$, and $\texttt{C}$ are single-qubit ancillas, and $\texttt{Z}$ and $\texttt{ZZ}$ are $s$-qubit ancillas.
Apply Hadamards to the ancillas to give equal superpositions on all except $\texttt{ZZ}$ and $\texttt{B}$.
\begin{equation}
\ket{+}_{\texttt{A}}\ket{0}_{\texttt{K}}\frac 1{2^{s/2}}\sum_{z=0}^{2^s-1}\ket{z}_{\texttt{Z}}\ket{0}_{\texttt{ZZ}}\ket{0}_{\texttt{B}}\ket{+}_{\texttt{C}}.
\end{equation}
\item Controlled on ancilla $A$, prepare an equal superposition state on $\lceil\log N\rceil$ qubits of $\texttt{K}$.
If $N$ is a power of $2$, then it can be performed with $\log N$ controlled Hadamards, each of which can be performed with two $T$ gates.
It is also possible to prepare an equal superposition for $N$ not a power of $2$ with complexity $\order{\log N}$.
For more details see \sec{QSA/qubitized/V}.
\item We can map the binary to unary in place, with cost no more than $N-\log N$ (see \app{b2u}), to give
\begin{equation}
\frac 1{2^{s/2}\sqrt{2}}\left( \ket{0}_{\texttt{A}}\ket{0}_{\texttt{K}} + \frac 1{\sqrt N}\ket{1}_{\texttt{A}}\sum_{k=1}^{N}\ket{k}_{\texttt{K}}\right)\sum_{z=0}^{2^s-1}\ket{z}_{\texttt{Z}}\ket{0}_{\texttt{ZZ}}\ket{0}_{\texttt{B}}\ket{+}_{\texttt{C}},
\end{equation}
where $\ket{k}_{\texttt{K}}$ is a value in one-hot unary.
\item Compute the approximate square of $z$, denoted $\tilde z^2$, placing the result in register $\texttt{ZZ}$, to give
\begin{equation}
\frac 1{2^{s/2}\sqrt{2}}\left( \ket{0}_{\texttt{A}}\ket{0}_{\texttt{K}} + \frac 1{\sqrt N}\ket{1}_{\texttt{A}}\sum_{k=1}^{N}\ket{k}_{\texttt{K}}\right)\sum_{z=0}^{2^s-1}\ket{z}_{\texttt{Z}}\ket{\tilde z^2}_{\texttt{ZZ}}\ket{0}_{\texttt{B}}\ket{+}_{\texttt{C}}.
\end{equation}
The complexity is no greater than $s^2/2$, as discussed in \app{multiplication/real_square}.
To obtain $\bsmooth$ bits of precision in the square, we need to take $s=\bsmooth+\order{\log \bsmooth}$, giving complexity $\bsmooth^2/2+\order{\bsmooth\log \bsmooth}$.
\item For each $k=1,\ldots,N$, perform an inequality test between $N\Pr(x_k|x)$ and $z^2$ in the $\texttt{ZZ}$ register, controlled by qubit $k$ in $\texttt{K}$, placing the result in $\texttt{B}$.
This has cost $N\bsmooth$ Toffolis.
\item Controlled on ancilla $\texttt{A}$ being zero, perform an inequality test between $N\Pr(x|x)$ and $N z^2$, with the output in $\texttt{B}$.
The inequality test has complexity $\bsmooth$.
In the case where $N$ is not a power of $2$, multiplying by $N$ has complexity approximately $\bsmooth^2+\order{\bsmooth\log\bsmooth}$ to obtain $\bsmooth$ bits,
and we incur this cost twice, once for computation and once for uncomputation.
If $N$ is a power of $2$ the multiplication by $N$ has no cost.
We obtain the state
\begin{align}
&\frac 1{2^{s/2}\sqrt{2}}\left( \ket{0}_{\texttt{A}}\ket{0}_{\texttt{K}}\sum_{z=0}^{2^s \sqrt{\widetilde{\Pr}(x|x)} -1}\ket{z}_{\texttt{Z}}\ket{\tilde z^2}_{\texttt{ZZ}}\ket{0}_{\texttt{B}}+\ket{0}_{\texttt{A}}\ket{0}_{\texttt{K}}\sum_{z=2^s \sqrt{\widetilde{\Pr}(x|x)}}^{2^s-1}\ket{z}_{\texttt{Z}}\ket{z^2}_{\texttt{ZZ}}\ket{1}_{\texttt{B}} \right.
\\
& \left. + \frac 1{\sqrt N}\ket{1}_{\texttt{A}}\sum_{k=1}^{N}\ket{k}_{\texttt{K}}\sum_{z=0}^{2^s \sqrt{N\widetilde{\Pr}(x_k|x)}-1}\ket{z}_{\texttt{Z}}\ket{\tilde z^2}_{\texttt{ZZ}}\ket{0}_{\texttt{B}} +\frac 1{\sqrt N}\ket{1}_{\texttt{A}}\sum_{k=1}^{N}\ket{k}_{\texttt{K}}\sum_{z=2^s \sqrt{N\widetilde{\Pr}(x_k|x)}}^{2^s-1}\ket{z}_{\texttt{Z}}\ket{\tilde z^2}_{\texttt{ZZ}}\ket{1}_{\texttt{B}}\right)\ket{+}_{\texttt{C}},
\end{align}
where $\widetilde{\Pr}$ indicates an approximation of the probability, with the imprecision primarily due to imprecise squaring of $z$.
\item Uncompute $z^2$ in register $\texttt{ZZ}$ with complexity no more than $s^2/2$.
\item Use a sequence of CNOTs with the $N$ qubits of $\texttt{K}$ as controls and ancilla $\texttt{A}$ as target.
This will reset $\texttt{A}$ to zero.
\item Perform Hadamards on the qubits of $\texttt{K}$, giving a state of the form
\begin{align}
&\frac 1{2}\ket{0}_{\texttt{A}}\left( \sqrt{\widetilde{\Pr}(x|x)} \ket{0}_{\texttt{K}}+\frac 1{\sqrt N}\sum_{k=1}^{N} \sqrt{\widetilde{\Pr}(x_k|x)}  \ket{k}_{\texttt{K}}\right)\ket{0}_{\texttt{Z}}\ket{0}_{\texttt{ZZ}}\ket{0}_{\texttt{B}}\ket{0}_{\texttt{C}} +\ket{\psi^\perp},
\end{align}
where $\ket{\psi^\perp}$ is the component of the state perpendicular to zero states on $Z$, $B$, and $C$.
\item Now conditioned on $\ket{0}_{\texttt{Z}}\ket{0}_{\texttt{B}}\ket{0}_{\texttt{C}}$, we have the correct state with amplitude approximately $1/2$.
We simply need to perform one round of amplitude amplification.
We reflect about $\ket{0}_{\texttt{Z}}\ket{0}_{\texttt{B}}\ket{0}_{\texttt{C}}$, invert steps 10 to 2, reflect about zero, then perform steps 2 to 10 again.
In the limit of large $s$ we then have the correct state.
As well as incurring three times the cost of steps 2 to 10, we have a cost of $N+\order{\bsmooth}$ for the reflection.
\end{enumerate}

The overall Toffoli complexity of this procedure, excluding the computation of $\Pr(x_k|x)$, is
\begin{equation}\label{eq:prepcom}
    N(\lceil \log N \rceil + \bsmooth)+N+3\left[N+\bsmooth^2+2\bsmooth^2+(N+1)\bsmooth\right] + \order{\log N+\bsmooth\log\bsmooth}.
\end{equation}
Here is first term is for the subtractions in step 1, the second term $N$ is for the reflection, then the terms inside the square brackets are from steps 2 to 10.
In the square brackets $N$ is for the binary to unary conversion, $\bsmooth^2$ is for computation and inverse computation of $z^2$, $2\bsmooth^2$ is for multiplication by $N$ (computation and uncomputation), which is only needed for $N$ not a power of two, and $(N+1)\bsmooth$ is for the $N+1$ inequality tests.
The cost $\log N$ in the order term is for the controlled preparation of an equal superposition state, and $\bsmooth\log\bsmooth$ is the order term for the squaring and multiplication.

Note that the preparation will not be performed perfectly, because the initial amplitude is not exactly $1/2$.
We will use a flag qubit to indicate success, which will control the \textsc{swap}.
To see the effect of this procedure, suppose the system is in basis state $x$.
Then the state that is prepared is
\begin{equation}
\textsc{cprep}\ket{0}\ket{x}\ket{0}= \mu_x \ket{1} \ket{x} \sum_{y} 
    \sqrt{\Pr(y|x)}
    \ket{y} + \nu_x \ket{0} \ket{x}\ket{\phi_x}
\end{equation}
where the first qubit flags success, $\mu_y$ is an amplitude for success, 
$\nu_x$ is an amplitude for failure,
and $\phi_x$ is some state that is prepared in the case of failure and can depend on $x$.
Here we have ignored the imperfect approximation of $\Pr(y|x)$, and are focusing just on the imperfect success probability.
Then the $\textsc{swap}$ is only performed in the case of success, which gives
\begin{equation}
\textsc{swap}\, \textsc{cprep}\ket{0}\ket{x}\ket{0}= \mu_x \ket{1} \sum_{y} 
    \sqrt{\Pr(y|x)}
    \ket{y} \ket{x} + \nu_x \ket{0} \ket{x}\ket{\phi_x}.
\end{equation}
Then we can write
\begin{equation}
\bra{0}\bra{y}\bra{0}\textsc{cprep}^\dagger= \mu_y \ket{1} \sum_{x} 
    \sqrt{\Pr(x|y)}
    \bra{y} \bra{x} + \nu_y \bra{0} \bra{y}\bra{\phi_y},
\end{equation}
so
\begin{align}\label{eq:qubfail}
    \bra{0}\bra{y}\bra{0}\textsc{cprep}^\dagger\textsc{swap}\, \textsc{cprep}\ket{0}\ket{x}\ket{0}&=\mu_x\mu_y \sqrt{\Pr(y|x)\Pr(x|y)} + \delta_{x,y}\nu_x^2\nonumber \\
    &= \sqrt{{\Pr}'(y|x){\Pr}'(x|y)},
\end{align}
where we define
\begin{equation}\label{eq:newprob}
{\Pr}'(x|y) = \begin{cases}
\mu_y^2 \Pr(x|y), & x\ne y \\
1-\mu_y^2 \sum_{z\ne y} \Pr(z|y), & x=y.
\end{cases}
\end{equation}
That is, the effect of the imperfect preparation is that the qubitized step corresponds to a slightly lower probability of transitions, which should have only a minor effect on the optimization.

The cost of the quantum walk in this approach is primarily in computing all transition probabilities $N\Pr(x_k|x)$.
If we were only concerned with the inequality tests for $k>0$, then we could incur that cost only once with a simple modification of the above scheme.
The problem is that we also need $N\Pr(x|x)$, which requires computing all $N\Pr(x_k|x)$.
The steps of computing each $N\Pr(x_k|x)$ are as follows.
\begin{enumerate}
  \item Query the energy difference oracle to find the energy difference 
  $\delta E$ of a proposed transition to $\bdiff$ bits,
  \item Calculate $\exp(-\beta \delta E)$ to $\bsmooth$ bits using the 
  QROM/interpolation method from
  \sec{functions}.
\end{enumerate}
The costs for the energy difference oracles were discussed in \sec{direct_oracle}, and are as in \tab{oracles}.
In this table, the costs for the energy difference oracles for the $L$-term spin model and LABS problem are obtained by evaluating the energy twice. Computing $N$ values of the energy difference would suggest we multiply this cost by $N$, but we can save computation cost by just calculating the energy for $x$ once, and computing the energy for each of the $x_k$.
That means the cost for these problem Hamiltonians can be given as the cost for a single energy evaluation multiplied by $N+1$.
For QUBO and the SK model it is considerably more efficient to compute the energy difference than the energy, so in these cases we simply compute the energy difference $N$ times.
The number of output registers is increased by a factor of $N$ in all cases.
For the cases where we compute the starting energy and the $N$ energies under bit flips, we can compute the starting energy first, copy it into the $N$ outputs, and subtract the energy under the bit flip from each of the output registers.
In summary, the complexity can be given as the minimum of $N+1$ times the cost of the energy oracle, and $N$ times the cost of energy difference oracle.

To perform the state preparation, we need to compute the energy differences, use those to compute the transition probabilities, prepare the state, then uncompute the transition probabilities and energy differences.
In each step of the Szegedy walk as shown in \fig{szeg}, we need to do the controlled preparation and inverse preparation, which means that the energy differences and need to be computed four times for each step.
That would give a cost of
\begin{equation}
    \min(4(N+1)\mathcal{C}^{\textrm{direct}},4N \mathcal{C}^{\textrm{diff}}) + 4N\mathcal{C}^{\textrm{fun}}.
\end{equation}

However, we can save a factor of two by taking the controlled preparation and moving it to the end of the step, as shown in \fig{szeg2}.
The reason why we can save a factor of two is that then, in between the controlled inverse preparation and preparation, there is a reflection on the target, but the control is not changed.
That means we can keep the values of the energy differences and transition probabilities computed in the controlled inverse preparation without uncomputing them, then only uncompute them after the controlled preparation.

This approach does not change the effect of a sequence of steps if $\beta$ is kept constant.
However, if $\beta$ is changed between steps, then the procedure as shown in \fig{szeg} will be different to that taking the controlled preparation and moving it to the end of the state.
That is, the value of $\beta$ is changed at the swap operation, rather than the reflection. Because there is only a factor of 2 rather than 4, the resulting cost is
\begin{equation}
    \min\left(2(N+1)\mathcal{C}^{\textrm{direct}}, 2N \mathcal{C}^{\textrm{diff}}\right) + 2N\mathcal{C}^{\textrm{fun}}.
\end{equation}
Now adding twice the complexity of the state preparation from \eq{prepcom} gives complexity
\begin{equation}\label{eq:Szegtof}
    \min\left(2(N+1)\mathcal{C}^{\textrm{direct}},2N \mathcal{C}^{\textrm{diff}}\right) + 2N\mathcal{C}^{\textrm{fun}} + 2N\log N + 8N\bsmooth+18\bsmooth^2 + \order{N}.
\end{equation}
Here we have omitted $\bsmooth\log\bsmooth$ in the order term because it is smaller than $N$ for the parameter ranges we are interested in.
The term $9\bsmooth^2$ includes $3\bsmooth^2$ from squaring and $6\bsmooth^2$ from multiplication.
In the case where $N$ is a power of $2$ the cost of $6\bsmooth^2$ can be omitted.

To evaluate the numbers of ancillas needed, we need to distinguish between the persistent ancillas and temporary ancillas in \tab{oracles}.
This is because the persistent ancillas need to be multiplied by $N$, whereas the temporary ancillas are reused, so we only need to take the maximum.
Considering the persistent ancillas first, the ancilla costs are as follows.
\begin{enumerate}
    \item The $N$ qubits for the Szegedy walk for the copy of the system.
    \item $N$ times the ancilla cost for the energy evaluation.
    \item $N$ times the ancilla cost for the function evaluation.
    \item The ancillas $\texttt{Z}$, $\texttt{A}$, $\texttt{B}$, $\texttt{C}$ in the state preparation use $\bsmooth+\order{\log\bsmooth}$ qubits.
\end{enumerate}
For the temporary ancillas, we have contributions from the energy difference evaluation, the function evaluation, and the state preparation.
Since these operations are not done concurrently, we can take the maximum of the costs.
The most significant will be that for the state preparation.
In the state preparation we have costs
\begin{enumerate}
    \item Ancilla $\texttt{ZZ}$ has $\bsmooth+\order{\log\bsmooth}$ qubits, and it is temporary because it is uncomputed.
    \item If $N$ is not a power of $2$ then we need another $\bsmooth+\order{\log\bsmooth}$ qubits for an ancilla with $Nz^2$.
    \item We use $\bsmooth+\order{\log\bsmooth}$ qubits for squaring, or $2\bsmooth+\order{\log\bsmooth}$ qubits if we are performing the multiplication by $N$.
\end{enumerate}
As a result, the temporary ancilla cost is $2\bsmooth+\order{\log\bsmooth}$ qubits if $N$ is a power of $2$, or $4\bsmooth+\order{\log\bsmooth}$ otherwise.
Considering the worst-case that $N$ is not a power of $2$,
this temporary ancilla cost is larger than that for the difference function evaluation, giving a total ancilla cost
\begin{equation}\label{eq:Szeganc}
    N\mathcal{A}^{\textrm{diff}}+ N \mathcal{A}^{\textrm{fun}} +5\bsmooth+\order{\log\bsmooth}.
\end{equation}

\subsection{LHPST qubitized walk based quantum simulated annealing}
\label{sec:QSA/qubitized}

The same quantum walk approach to quantum simulated annealing can be achieved using an improved form of quantum walk given by Lemieux, Heim, Poulin, Svore, and Troyer (LHPST) \cite{lemieux2019efficient} that requires only computation of a \emph{single} transition probability for each step.
Here we provide an improved implementation of that quantum walk that can be efficiently achieved for more general types of cost Hamiltonians than considered in \cite{lemieux2019efficient}.
The operations used to achieve the step of the walk are
\begin{equation}
\tilde U_W =  R V^\dagger B^\dagger FBV
\end{equation} 
where 
\begin{align}
V :&\ \ket 0_M \rightarrow \frac 1{\sqrt{N}}\sum_j \ket j_M, \label{eq:move}\\
B :&\ \ket x_S \ket j_M \ket 0_C 
\rightarrow \ket x_S \ket j_M  \left( \sqrt{1-p_{x,x_j}}\ket 0_C + \sqrt{p_{x,x_j}}\ket 1_C\right), \label{eq:coin}\\
F :&\
  \begin{array}{l}
    \ket x_S\ket j_M\ket 0_C \rightarrow \ket{x}_S\ket j_M \ket 0_C,  \\
    \ket x_S\ket j_M\ket 1_C \rightarrow \ket{x_j}_S\ket j_M \ket 1_C,
  \end{array} \label{eq:flip} \\
R :&\ 
  \begin{array}{l}
    \ket 0_M\ket 0_C \rightarrow -\ket 0_M \ket 0_C, \\
    \ket j_M \ket c_C \rightarrow \ket j_M \ket c_C \ {\rm for}\ (j,c) \neq (0,0).
  \end{array} \label{eq:reflection}
\end{align}
Here $p_{x,y}=N \Pr(y|x)$ in the notation used above, and we have specialized to an equal superposition over $j$ and only single bit flips.

This walk is equivalent to the Szegedy approach of \cite{Somma2008b} because it yields the same block-encoded operation. That is,
$\bra{0}V^\dagger B^\dagger FBV\ket{0}$ has matrix representation $\sqrt{\Pr(x|y)\Pr(y|x)}$. To show this fact, the sequence of operations gives
\begin{align}
V\ket{0}_M\ket{0}_C &= \frac 1{\sqrt{N}}\sum_{j=1}^N \ket{j}_M\ket{0}_C, \\
BV\ket{0}_M\ket{0}_C &= \frac 1{\sqrt{N}}\sum_x \sum_{j=1}^N \ketbra{x}{x} \otimes \ket{j}\left( \sqrt{1-p_{x,x_j}}\ket 0_C + \sqrt{p_{x,x_j}}\ket 1_C \right), \\
FBV\ket{0}_M\ket{0}_C &= \frac 1{\sqrt{N}}\sum_x\sum_{j=1}^N \ketbra{x}{x}\otimes \ket{j}\sqrt{1-p_{x,x_j}}\ket 0_C +
\frac 1{\sqrt{N}}\sum_x \sum_{j=1}^N \ketbra{x_j}{x}\otimes \ket{j}
\sqrt{p_{x,x_j}}\ket 1_C  \nonumber \\
&= \frac 1{\sqrt{N}}\sum_x\sum_{j=1}^N \ketbra{x}{x}\otimes \ket{j}\sqrt{1-p_{x,x_j}}\ket 0_C +
\frac 1{\sqrt{N}}\sum_x \sum_{j=1}^N \ketbra{x}{x_j}\otimes \ket{j}
\sqrt{p_{x_j,x}}\ket 1_C , \\
_M\!\bra{0}_C\!\bra{0}V^\dagger B^\dagger FBV\ket{0}_M\ket{0}_C &=
\frac 1N \sum_x \sum_{j=1}^N\ketbra{x}{x} (1-p_{x,x_j})
+\frac 1N \sum_x \sum_{j=1}^N\ketbra{x}{x_j} \sqrt{p_{x,x_j}p_{x_j,x}}\nonumber \\
&=
\sum_x \ketbra{x}{x} \left(1-\frac 1N \sum_{j=1}^Np_{x,x_j}\right)
+\frac 1N \sum_x \sum_{j=1}^N\ketbra{x_j}{x} \sqrt{p_{x,x_j}p_{x_j,x}}\nonumber \\
&= \sum_{x,y} \ketbra{y}{x} \sqrt{\Pr(y|x)\Pr(x|y)}\, .
\end{align}

Just as with the Szegedy approach, most operations are trivial to perform, and the key difficulty is in the operation $B$ which depends on the transition probability.
However, $B$ only depends on \emph{one} transition probability, whereas the Szegedy approach requires computing all the transition probabilities for a state preparation.
Lemieux \textit{et al.}\ \cite{lemieux2019efficient} propose a method for the $B$ operation that is not useful for the cost Hamiltonians considered here, but is useful for Hamiltonians with low connectivity.
Instead of computing the energy difference then the exponential, they consider an approach where the required angle of rotation is found from a database.

That is, one considers the qubits that the transition probability for the move (here a bit flip) depends on, and classically precomputes the rotation angle for each basis state on those qubits.
For each value of $j$, one sequentially performs a multiply-controlled Toffoli for each computational basis state for these qubits, and performs the required rotation on the ancilla qubit $C$.
The complexity that is given by \cite{lemieux2019efficient} is $\mathcal{O}(2^{|\mathcal{N}_j|}{|\mathcal{N}_j|}\log(1/\epsilon))$, where $|\mathcal{N}_j|$ is the number of qubits that the transition probability for move $j$ depends on.
That complexity is a slight overestimate, because each multiply controlled Toffoli has a cost of $|\mathcal{N}_j|$, then the cost of the rotation synthesis is $\mathcal{O}(\log(1/\epsilon))$.
It should also be noted that this is the cost for each value of $j$, and there are $N$ values of $j$, giving an overall cost
$\mathcal{O}(N 2^{|\mathcal{N}_j|}[{|\mathcal{N}_j|}+\log(1/\epsilon)])$.

To improve the complexity, one can divide this procedure into two parts, where first a QROM is used to output the desired rotation in an ancilla, and then those qubits are used to control a rotation.
Using the QROM procedure of \cite{Babbush2018} to output the required rotation, the cost in terms of Toffoli gates is $\mathcal{O}(N 2^{|\mathcal{N}_j|})$.
Then one can apply rotations using the phase gradient state, which was discussed above in \sec{phase_oracle}.
Addition of the register containing the rotation to an ancilla with state $\ket{\phi}$ from \eq{phasegrad} results in a phase rotation.
To rotate the qubit, simply make the addition controlled by this qubit, and use Clifford gates before and after so that the rotation is in the $y$-direction.
The cost of this rotation is $\mathcal{O}(\log(1/\epsilon))$ Toffolis; for more details see \app{rotations}.
With that improvement the complexity is reduced to $\mathcal{O}(N 2^{|\mathcal{N}_j|}+\log(1/\epsilon))$.

Even with that improvement, any procedure of that type is exponential in the number of qubits that the energy difference depends on, $|\mathcal{N}_j|$.
That is acceptable for the types of Hamiltonians considered in \cite{lemieux2019efficient}, but here we consider Hamiltonians typical of real world problems where the energy difference will depend on most of the system qubits, because the Hamiltonians have high connectivity.
We thus propose alternative procedures to achieve the rotation $B$.

\subsubsection{Rotation \texorpdfstring{$B$}{B}}
\label{sec:QSA/qubitized/B}

We propose a completely different method to perform the rotation $B$ than that of LHPST \cite{lemieux2019efficient}.
We can first compute the energy difference $E_x-E_{x_j}$, then the rotation $\arcsin\sqrt{p_{x,x_j}}$ with the result put in an ancilla register.
The rotation of the qubit ancilla $C$ is controlled on the value of this ancilla as explained above, then the value of $\arcsin\sqrt{p_{x,x_j}}$ is uncomputed.
There are many possible approaches to the computation of $\arcsin\sqrt{p_{x,x_j}}$, for example that of \cite{hner2018optimizing}.
For the purposes of quantum optimization, we expect that we will not need to compute this function to high precision as long as the function we compute is still monotonic in the actual energy, so there is the opportunity to use methods that are very efficient for low precision but would not be suitable for high precision.
We propose a method based on using a piecewise linear approximation, with the coefficients output by a QROM, as described in \sec{functions}.

One could then apply the controlled rotation with cost $\bsmooth$ Toffolis using the phase gradient state in \eq{phasegrad}, as described in detail in \app{rotations}.
Then after uncomputing the rotation angle we would have implemented $B$.
That approach would then mean that a single step of the walk has four times the cost of computing $\arcsin\sqrt{p_{x,x_j}}$, because it needs to be computed and uncomputed for $B$, and the operation $B$ is applied twice in each step.

It is possible to halve that cost by only computing and uncomputing once in a step, and retaining the value of $\arcsin\sqrt{p_{x,x_j}}$ during the $F$ operation.
Because $F$ is a controlled flip of bit $j$ of $x$, it would reverse the role of $x$ and $x_j$, and the sign of $E_x-E_{x_j}$ would be flipped.
In more detail, the procedure is as follows.
\begin{enumerate}
    \item Compute the energy difference between $x$ and $x_j$, $E_x-E_{x_j}$.
    \item Compute $\arcsin\sqrt{p_{x,x_j}}$ based on $|E_x-E_{x_j}|$.
    \item If $E_{x_j}<E_x$ then perform an $X$ operation on the qubit $C$.
    That can be achieved with a \textsc{cnot} (Clifford) controlled by the sign bit of $E_x-E_{x_j}$.
    \item The remaining rotations for the case of $E_{x_j}>E_x$ need to be controlled on $-1$ for the sign bit.
    \item When we apply $F$, as well as applying the Toffolis to change $x$ to $x_j$, we need to flip the sign bit on $E_x-E_{x_j}$ controlled on qubit $C$.
    This is another \textsc{cnot}, with no non-Clifford cost.
    \item Then at the end we uncompute $\arcsin\sqrt{p_{x,x_j}}$ and $E_x-E_{x_j}$.
\end{enumerate}
This procedure assumes that $E_x-E_{x_j}$ is represented as a signed integer.
The computation of $E_x-E_{x_j}$ uses two's complement, so there will be an additional cost of $\bdiff$ to switch to a signed integer.
Because there is only a factor of two instead of four, the overall cost will then be $2\mathcal{C}^{\textrm{diff}}+2\mathcal{C}^{\textrm{fun}}+2\bdiff+\mathcal{O}(1)$.
Next we consider the other (simpler) operations used in the step of the quantum walk.

\subsubsection{Equal superposition \texorpdfstring{$V$}{V}}
\label{sec:QSA/qubitized/V}

The operation $V$ generates the equal superposition starting from a zero state
\begin{equation}
    V : \ket 0_M \rightarrow \frac 1{\sqrt{N}}\sum_j \ket j_M.
\end{equation}
In the case where $N$ is a power of $2$, then we can create the equal superposition over binary by using Hadamards (and no Toffolis).
More generally, if we wish to create an equal superposition where the number of items is \emph{not} a power of 2, we can rotate an ancilla qubit such that the net amplitude is $1/2$ for $\ket{1}\ket{1}$ on the result of the inequality test and the ancilla qubit.
We can then perform a single step of amplitude amplification to obtain the superposition state.
Our procedure is explained below and gives a complexity of $4\log N+\order{1}$ Toffolis.

Our method for $V$ is also very different that of LHPST \cite{lemieux2019efficient}.
There they proposed encoding the $M$ register in unary, whereas here we use binary which greatly reduces the ancilla cost (which is sublinear in $N$).
Moreover, LHPST did not consider using equal superpositions in cases where $N$ is not a power of $2$, and instead just allowed for a constant factor overhead in the complexity.

Our procedure to create an equal superposition over $N<2^k$ items is as follows.
With Hadamards we prepare
\be
\frac 1{\sqrt{2^k}} \sum_{j=0}^{2^k-1} \ket{j}.
\ee
Then we have an inequality test between $j$ and $N$ to give
\be
\frac 1{\sqrt{2^k}} \sum_{j=0}^{2^k-1} \ket{j} \ket{1}
+\frac 1{\sqrt{2^k}} \sum_{j=N}^{2^k-1} \ket{j}\ket{0}.
\ee
This is an inequality test on $k$ bits, and since it is an inequality test with a \emph{constant} we save a Toffoli gate.  The cost is therefore $k-2$ Toffolis as per the explanation in \cite{BGMMB19}.
We would have an amplitude of $\sqrt{N/2^k}$ for success, and would aim to multiply it by another amplitude of approximately $\frac 12 \sqrt{2^k/N}$ so the amplitude is $1/2$ and we can use a single step of amplitude amplification.
For an amplitude of $\frac 12 \sqrt{2^k/N}$, we can rotate another register 
according to the procedure in \app{rotations} to give
\begin{equation}
\cos\theta \ket{0} + \sin\theta \ket{1} .
\end{equation}
We can then perform a single step of amplitude amplification for $\ket{1}$ on both this qubit and the result of the inequality test.

The steps needed in the amplitude amplification and their costs are as follows.
If we use the procedure for the rotation with $s$ bits, it would take $s-3$ Toffolis because the angle of rotation is given classically.
\begin{enumerate}
\item A first inequality test ($k-2$ Toffolis) and a rotation on a qubit (cost $s-3$).
\item A first reflection on the rotated qubit and the result of the inequality test. This just needs a controlled phase (a Clifford gate).
\item Inverting the rotation and inequality test has cost $k+s-5$.
\item Hadamards then reflection of the $k$ qubits and the single qubit ancilla about zero ($k-1$ Toffolis).
\item Applying the inequality test ($k-2$ Toffolis).
\end{enumerate}
The total cost is $4k+2s-13$ Toffolis.

\begin{figure}
  \centering
    \includegraphics[width=0.5\textwidth]{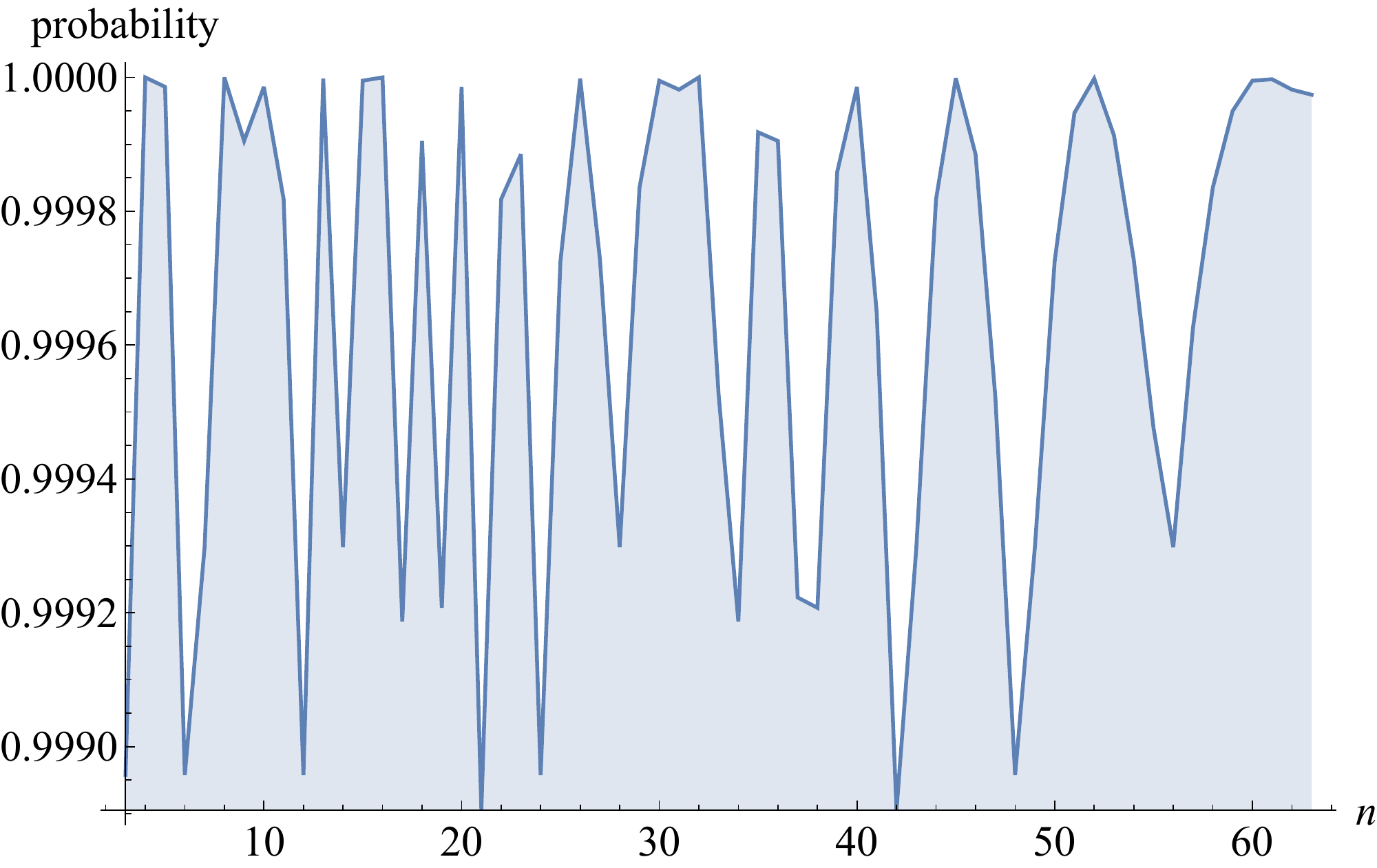}
  \caption[Probability of success for single-qubit rotation in qubitized walk approach.]{\label{ampfig}The probability for success using a rotation of the form $2\pi/2^s$ with $s=7$.}
\end{figure}

Conditioned on success of the inequality test, the state is
\begin{equation}
\frac {1}{\sqrt{2^k}}\sum_{j=0}^{N-1} \ket{j} \left[
\left( 1-
\frac {4N\sin^2\theta}{{2^k}}\right)
\ket{0}+
2\sin\theta
(\sin\theta \ket{0}+\cos\theta\ket{1})\right]\ket{1}.
\end{equation}
The probability for success is then given by the normalization
\begin{equation}
\frac {N}{2^k} \left[
\left( 1-
\frac {4N\sin^2\theta}{{2^k}}+2\sin^2\theta\right)^2
+
4\sin^2\theta\cos^2\theta\right].
\end{equation}
It is found that highly accurate results are obtained for $s=7$, as shown in Fig.~\ref{ampfig}.
This procedure enables construction of equal superposition states flagged by an ancilla qubit for $N$ not a power of $2$.
If we take $s=7$, then the cost is $4k+1$.

\subsubsection{Controlled bit flip \texorpdfstring{$F$}{F}}
\label{sec:QSA/qubitized/F}

We also need to modify the operation $F$ compared to that in LHPST to account for the $M$ register being encoded in binary.
This operation flips bit $j$ on $x$ for the control qubit $C$ being in the state $\ket{1}$,
\begin{equation}
    F :\ 
    \begin{array}{l}
      \ket x_S\ket j_M\ket 0_C \rightarrow \ket{x}_S\ket j_M \ket 0_C, \\
      \ket x_S\ket j_M\ket 1_C \rightarrow \ket{x_j}_S\ket j_M \ket 1_C.
    \end{array}
\end{equation}
This operation can be achieved using the iteration procedure of \cite{Babbush2018} with Toffoli complexity $N$, which allows us to perform the operation with register $M$ encoded in binary.

A complication is that, in the case where $N$ is not a power of $2$, there is a nonzero cost of the state preparation in $V$ failing.
We should only perform the operation $F$ in the case where we have success of the state preparation.
We include another two Toffolis to create and erase a register that gives a control qubit that flags whether the $C$ register is in the state $\ket{1}$ and there is success of the state preparation.
Because the other operations are inverted, in the case that the state preparation does not work the net operation performed is the identity.

To be more specific, $V$ prepares a state of the form 
\begin{equation}
V\ket{0}=\ket{1}\ket{\psi_1}+\ket{0}\ket{\psi_0} ,
\end{equation}
with the first register flagging success.
Since we only perform $F$ for the flag qubit in the state $\ket{1}$, we obtain
\begin{equation}
B^\dagger FB V \ket{0} = B^\dagger FB\ket{1}\ket{\psi_1}+\openone\ket{0}\ket{\psi_0}.
\end{equation}
To determine the block encoded operation
\begin{equation}
\bra{0} V^\dagger B^\dagger FBV \ket{0},
\end{equation}
we note that
\begin{equation}
\bra{0} V^\dagger = \bra{1}\bra{\psi_1}+\bra{0}\bra{\psi_0},
\end{equation}
so
\begin{equation}
\bra{0} V^\dagger B^\dagger FBV \ket{0} = \bra{\psi_1} B^\dagger FB \ket{\psi_1} + \openone\braket{\psi_0}{\psi_0},
\end{equation}
where $\openone$ indicates the identity on the target system.
The first term is the desired operation we would obtain if the equal superposition state was obtained exactly (with a multiplying factor corresponding to the probability of success), and the second term is proportional to the identity.
This small offset by the identity just gives a trivial shift to the eigenvalues.

\subsubsection{Reflection \texorpdfstring{$R$}{R}}
\label{sec:QSA/qubitized/R}

This operation applies a phase flip for zero on the ancillas as
\begin{align}
    R :&\ \ket 0_M\ket 0_C \rightarrow -\ket 0_M \ket 0_C, \nonumber \\
&\  \ket j_M \ket c_C \rightarrow \ket j_M \ket c_C \ {\rm for}\ (j,c) \neq (0,0).
\end{align}
As well as the ancillas $M$ and $C$, this reflection also needs to be on any ancilla qubits used to encode the ancilla for the preparation of the equal superposition state, and the flag qubit.
There are $\lceil \log N \rceil$ qubits used to encode $j$, one qubit for $C$, and one ancilla used for the rotation, for a total of $\lceil \log N \rceil+2$ qubits.
Therefore the number of Toffolis needed for reflection about zero is $\lceil \log N \rceil$.

\subsubsection{Total costs}\label{sectotcostqub}
The total Toffoli costs of implementing $\tilde U_W =  R V^\dagger B^\dagger FBV$ are as follows.
\begin{enumerate}
    \item The cost of $V$ and $V^\dagger$ is $8\log N+\order{1}$.
    \item The cost of $F$ is $N$ Toffolis.
    \item The cost of $R$ is $\lceil \log N\rceil$.
    \item The cost of two applications of $B$ is $2\mathcal{C}^{\textrm{diff}}+2\mathcal{C}^{\textrm{fun}}+2\bdiff+\mathcal{O}(1)$.
\end{enumerate}
The total cost of a step is then
\begin{equation}\label{eq:costqubtof}
    2\mathcal{C}^{\textrm{diff}} + 2\mathcal{C}^{\textrm{fun}}+N+2\bdiff+9\log N+ \order{1}.
\end{equation}
Note that $8\log N$ of this cost is for preparing equal superposition states, and can be omitted if $N$ is a power of $2$. The ancilla qubits needed are as follows.
\begin{enumerate}
    \item The ancilla registers $M$ and $C$ need $\lceil \log N\rceil +1$ qubits.
    \item The resource state used to implement the controlled rotations needs $\bsmooth$ qubits.
    \item The ancilla requirements of the energy difference and function evaluation oracles.
\end{enumerate}

For the temporary ancilla cost, we need to take the maximum of that for the energy difference and function evaluation, giving the total ancilla cost of
\begin{equation}\label{eq:costqubanc}
    \mathcal{A}^{\textrm{diff}} + \mathcal{A}^{\textrm{fun}}+ \max(\mathcal{B}^{\textrm{diff}},\mathcal{B}^{\textrm{fun}}) + \log N + \bsmooth + \order{1}.
\end{equation}

\subsection{Spectral gap amplification based quantum simulated annealing}
\label{sec:spectral}

An alternative, and potentially simpler, approach to preparing a low-temperature thermal state is given by~\cite{Boixo2014a}.  The idea behind this approach is to construct a Hamiltonian whose ground state is a purification of the Gibbs state.
Similarly to the case with the quantum walk, one can start with an equal superposition state corresponding to infinite temperature, and simulate the Hamiltonian evolution starting from $\beta=0$ and gradually increase $\beta$.
This approach can correspond to using an adibatic approach on this Hamiltonian, or one can also apply a quantum Zeno approach by phase measurements on the Hamiltonian evolution, or apply Hamiltonian evolutions for randomly chosen times.

A simple choice of Hamiltonian is similar to the block-encoded operation for the quantum walks, so has a small spectral gap.
In order to obtain a speedup, one needs to construct a new Hamiltonian with the square root of the spectral gap of the original Hamiltonian, thus yielding the same speedup as the quantum walks.
That procedure, from \cite{Boixo2014a}, is called spectral gap amplification.
Simulating the time-dependent Hamiltonian, for example using a Dyson series, has significant complexity.

To avoid that complexity, here we suggest that one instead construct a step of a quantum walk using a linear combination of unitaries.
Such a quantum walk could be used to simulate the Hamiltonian evolution, but as discussed in \cite{BerryNPJ18,PoulinPRL18} one can instead just perform steps of the quantum walk which has eigenvalues that are the exponential of the arccosine of those for the Hamiltonian.
By applying the steps of the quantum walk we can obtain the advantage of the spectral gap amplification, without the difficulty of needing to simulate a time-dependent Hamiltonian.
Unlike the quantum walks in the previous subsections, the arccosine does not yield a further square-root amplification of the spectral gap, because the relevant eigenvalue for the amplified Hamiltonian is not at $1$. However, it potentially gives other scaling advantages (for instance, in avoiding the need for quantum signal processing when using certain oracles) compared to other proposals in the literature for realizing quantum simulated annealing via spectral gap amplification.

\subsubsection{The spectral gap amplification Hamiltonian}

Here we summarise the method of spectral gap amplification from \cite{Boixo2014a}, but specialise to the case where only single bit flips are allowed to make the method clearer.
As discussed above, one can use a Hamiltonian simulation approach with Hamiltonian $H_\beta$ given in \eq{annealham} with ground state corresponding to the quantum Gibbs state $\ket{\psi_\beta}$.
Because the complexity depends on the spectral gap, it is advantageous to increase the spectral gap as much as possible, which was done via a quantum walk in the previous subsections.
The proposal in \cite{Boixo2014a} is to construct a different Hamiltonian whose spectral gap
has been amplified relative to $H_\beta$. To define this new Hamiltonian,
they introduce states equivalent to
\begin{equation}
    \ket{\lambda_{x,y}} := 
    \sqrt{\frac{p_{x,y}}{p_{x,y} + p_{y,x}}} \ket{y} -
    \sqrt{\frac{p_{y,x}}{p_{x,y} + p_{y,x}}} \ket{x},
\end{equation}
where as before $p_{x,y}=N \Pr(y|x)$.
This is the normalised form of the unnormalised kets
$\ket{\mu_\beta^{\sigma_i,\sigma_j}}$ presented in Eq.~(21)
of~\cite{Boixo2014a}. One can then write
\begin{equation}
    H_\beta = \frac 1{2N}
    \sum_{x,y} \left(p_{x,y} + p_{y,x}\right)
    \ketbra{\lambda_{x,y}}{\lambda_{x,y}}.
\end{equation}
In this work we consider only transitions with single bit flips, so the coefficient $\left(p_{x,y} + p_{y,x}\right)$
is non-zero only if $x$ and $y$ differ by exactly one bit.
We have included a factor of $1/2$ to account for the symmetry between $x$ and $y$.
We can use this condition to express $H_\beta$ as a sum of $2$-sparse matrices.
To do so, recall that each $x$ is an $N$-bit string.
Then for each $k=1,\ldots,n$ we define
\begin{equation}
    H_{\beta, k} := \frac 1{2N} \sum_x \left(p_{x,x_k} + p_{x_k,x}\right)
    \ketbra{\lambda_{x,x_k}}{\lambda_{x,x_k}},
\end{equation}
where $x_k = \textsc{not}_k (x)$, the result of flipping the 
$k^\text{th}$ bit of $x$. Then $H_\beta = \sum_k H_{\beta, k}$.
The operators $H_{\beta, k}$ here are equivalent to $O_{\beta, k}$ in \cite{Boixo2014a}, except we have specialized to the case where only transitions with single bit flips are allowed.

One can then define a new Hamiltonian (Eq.~(25) in \cite{Boixo2014a})
\begin{equation}
    A_\beta :=
    \sum_{k=1}^N \sqrt{H_{\beta,k}} \otimes (\ketbra{k}{0} + \ketbra{0}{k}).
\end{equation}
The projector structure of the Hamiltonian allows the square root to be easily implemented via
\begin{equation}
    \sqrt{H_{\beta, k}} = \frac 1{2\sqrt{N}} \sum_x \sqrt{p_{x,x_k} + p_{x_k,x}}
    \ketbra{\lambda_{x,x_k}}{\lambda_{x,x_k}}.
\end{equation}
Here the $1/2$ is still included to account for the symmetry between $i$ and $i^{(k)}$.
Following Eq.~(32) in~\cite{Boixo2014a}, a coherent Gibbs distribution can be seen to be the ground state of the following
Hamiltonian
\begin{equation}\label{eq:hamdef}
    \tilde{H}_\beta :=
    A_\beta + \sqrt{\Delta_\beta} (\openone - \ketbra{0}{0}),
\end{equation}
where $\Delta_\beta$ is a lower bound for the spectral gap of $H_\beta$.  This means that by preparing the minimum energy configuration of this Hamiltonian one, in effect, is capable of drawing a sample from the distribution that would be seen by running a simulated annealing procedure for sufficient time.

\subsubsection{Implementing the Hamiltonian}

In order to implement the Hamiltonian, we will use a linear combination of unitaries.
We can rewrite the square root of the Hamiltonian as
\begin{equation}
\sqrt{H_{\beta, k}} = \frac 1{2\sqrt{N}} \sum_x \left(p_{x,x_k} + p_{x_k,x}\right)^{-1/2} \left[
{p_{x,x_k}}\ketbra{x_k}{x_k}+{p_{x_k,x}}\ketbra{x}{x}
-\sqrt{p_{x,x_k}p_{x_k,x}}\left( \ketbra{x}{x_k}+\ketbra{x_k}{x}\right)
\right].
\end{equation}
This is a 2-sparse Hamiltonian, then summing over $k$ to obtain $A_\beta$ gives a $2N$-sparse Hamiltonian.
To express $A_\beta$ as a linear combination of unitaries, we can express $\sqrt{H_{\beta, k}}$ as
\begin{align}
\sqrt{H_{\beta, k}} &= \frac 1{\sqrt{N}}\sum_x q_{xk} \ketbra{x}{x}  -\frac 1{2\sqrt {2N}} \sum_x r_{xk}
\left( \ketbra{x}{x_k}+\ketbra{x_k}{x}\right) \nonumber \\
&= \frac 1{\sqrt N} \sum_x \int_0^1 dz \, (-1)^{2z> 1+q_{xk}} \ketbra{x}{x} 
-\frac 1{2\sqrt{2N}} \sum_x \int_0^1 dz \, (-1)^{2z> 1+r_{xk}}
\left( \ketbra{x}{x_k}+\ketbra{x_k}{x}\right) ,
\end{align}
where
\begin{align}
q_{xk} = \frac{p_{x_k,x}}{\sqrt{p_{x,x_k} + p_{x_k,x}}}, \\
r_{xk} = \sqrt{\frac{2 p_{x_k,x}p_{x,x_k}}{p_{x,x_k} + p_{x_k,x}}},
\end{align}
and we are taking the inequality test to yield a numerical value of $0$ for false and $1$ for true.
Note that with these definitions, $q_{xk}$ and $r_{xk}$ can take values in the range $[0,1]$.
We use the procedure from \cite{BerrySTOC14} (Lemma 4.3) to obtain a linear combination of unitaries.
The operator is then approximated as a sum
\begin{align}
\sqrt{H_{\beta, k}} &\approx
\frac 1{2^{s}\sqrt N} \sum_{z=0}^{2^s-1} \sum_x (-1)^{z/2^{s-1}> 1+q_{xk}} \ketbra{x}{x} 
-\frac 1{2^{s+1}\sqrt{2N}} \sum_{z=0}^{2^s-1} \sum_x (-1)^{z/2^{s-1}> 1+r_{xk}}
\left( \ketbra{x}{x_k}+\ketbra{x_k}{x}\right) .
\end{align}
The operator $A_\beta$ is then approximated by
\begin{align}
    A_\beta &\approx \frac 1{\sqrt N}
    \sum_{k=1}^N \left\{\left[\frac 1{2^{s}} \sum_{z=0}^{2^s-1} \sum_x (-1)^{z/2^{s-1}> 1+q_{xk}} \ketbra{x}{x} \right.\right. \nonumber \\
&\quad \left.\left. -\frac 1{2^{s+1}\sqrt{2}} \sum_{z=0}^{2^s-1} \sum_x (-1)^{z/2^{s-1}> 1+r_{xk}}
\left( \ketbra{x}{x_k}+\ketbra{x_k}{x}\right) \right] \otimes \left( \ketbra{k}{0} + \ketbra{0}{k}\right)  \right\}.
\end{align}
For the part $\sqrt{\Delta_\beta} (\openone - \ketbra{0}{0})$, we can write it as
\begin{align}
    \sqrt{\Delta_\beta} (\openone - \ketbra{0}{0}) &= 
     \frac{\sqrt{2N\Delta_\beta}}{(N-1)(\sqrt{2}-1)}\frac 1{\sqrt{N}}\left( 1-\frac 1{\sqrt 2}\right)(N-1)(\openone - \ketbra{0}{0})\nn
     &= \frac {\delta_\beta}{\sqrt{N}}
     \sum_{k=1}^N \left( 1-\frac 1{\sqrt 2}\right) \sum_{\ell>0, \ell\ne k} \ketbra{\ell}{\ell}\nn
     &= \frac 1{\sqrt{N}}\frac 1{2^{s}} \sum_{z=0}^{2^s-1} (-1)^{z/2^{s-1}> 1+\delta_\beta}
     \sum_{k=1}^N \left( 1-\frac 1{\sqrt 2}\right) \sum_{\ell>0, \ell\ne k} \ketbra{\ell}{\ell}
\end{align}
where
\begin{equation}
   \delta_\beta := \frac{\sqrt{2N\Delta_\beta}}{(N-1)(\sqrt{2}-1)} .
\end{equation}
Therefore the complete approximation of the Hamiltonian with spectral gap amplification is
\begin{align}
    \tilde{H}_\beta &\approx \frac 1{2^s\sqrt{N}}\sum_{k=1}^N \sum_{z=0}^{2^s-1} \left\{ \left[\sum_x (-1)^{z/2^{s-1}> 1+q_{xk}} \ketbra{x}{x} \otimes \left( \ketbra{k}{0} + \ketbra{0}{k}\right) + (-1)^{z/2^{s-1}> 1+\delta_\beta}\openone \otimes \sum_{\ell>0, \ell\ne k} \ketbra{\ell}{\ell}\right]
    \right. \nonumber \\
&\quad \left. - \frac 1{\sqrt{2}}\left[\frac 1{2} \sum_x (-1)^{z/2^{s-1}> 1+r_{xk}}
\left( \ketbra{x}{x_k}+\ketbra{x_k}{x}\right)  \otimes \left( \ketbra{k}{0} + \ketbra{0}{k}\right) + (-1)^{z/2^{s-1}> 1+\delta_\beta}\openone \otimes \sum_{\ell>0, \ell\ne k} \ketbra{\ell}{\ell}\right] \right\}.\label{eq:Htilde}
\end{align}
Here we have grouped the terms such that the operations in square brackets are unitaries.
Summing the coefficients in the sums gives a $\lambda$-value of
\begin{equation}\label{eq:sgalam}
    \lambda = \left( 1 + \frac 1{\sqrt{2}}\right) \sqrt{N}.
\end{equation}
To implement the operator by a linear combination of unitaries, we need two single qubit ancillas, a register with $z$ and a register with $k$.
The \textsc{prepare} operation is trivial, and just needs to prepare the state
\begin{equation}
   \frac 1{\sqrt{\lambda 2^{s}}}
   \sum_{k=1}^N \ket{k} \sum_{z=0}^{2^s-1}\ket{z}\left(\ket{0}_F+\frac 1{2^{1/4}}\ket{1}_F\right).
\end{equation}
The roles of these registers are as follows.
\begin{enumerate}
    \item The register with $k$ selects terms in the sum over $k$ in \eq{Htilde}.
    \item The register with $z$ selects terms in the sum over $z$ in \eq{Htilde}.
    \item The $F$ register selects between the terms in square brackets in the first and second lines of \eq{Htilde}.
\end{enumerate}
There are registers containing $k$ for both this prepared control state and the target state.  We will call these the control and target $K$ registers.
In the \textsc{prepare} operation, creating the superposition over $z$ can be trivially achieved with Hadamards.
The superposition over $N$ can be achieved similarly if $N$ is a power of $2$, but otherwise the procedure outlined in \sec{QSA/qubitized/V} can be used with cost $4\log N+\mathcal{O}(1)$.
The rotation on qubit $F$ can be achieved with precision $\epsilon$ using $1.15\brot+\mathcal{O}(1)$ T operations, where $\brot=\log(1/\epsilon)$.

The \textsc{select} procedure for the linear combinations of unitaries may be performed as follows.

\begin{enumerate}
\item Perform a test of whether the target system $K$-register is in the space $\{\ket{0},\ket{k}\}$, placing the result in an ancilla qubit, call this qubit $E$.
    \item Controlled on $E$ being $\ket{1}$ and $F$, compute $q_{xk}$ or $r_{xk}$.
    \item Controlled on $E$ being $\ket{0}$, place the value $\delta_\beta$ into the output register also used for $q_{xk}$ or $r_{xk}$.
    \item Perform the inequality test between $z/2^{s-1}$ and $1+q_{xk}$, $1+r_{xk}$, or $1+\delta_\beta$.
    \item Apply a $Z$ gate to the output of the inequality test.
    \item Controlled on the $E$ register being $\ket{1}$ \emph{and} the register $F$ being $\ket{1}$, apply $X$ to qubit $k$ of the target system.
    \item Apply a \textsc{not} between $\ket{0}$ and $\ket{k}$ for the target system. That gives $\ketbra{k}{0} + \ketbra{0}{k}$.
    \item Invert the inequality test from step 4.
    \item Invert step 3.
    \item Invert step 2 uncomputing $q_{xk}$ or $r_{xk}$.
    \item Invert step 1.
    \item Apply a $Z$ gate to $F$ to introduce the $-1$ sign.
\end{enumerate}
Here we call the register that would carry $\ket{k}$ for the target system the $K$-register.
The cost of these steps may be quantified as follows, ignoring $O(1)$ costs.\\
\textbf{Steps 1 and 11.}
We need an equality test between the $K$-register for the ancilla and the $K$-register for the system, with cost $\log N+\mathcal{O}(1)$.
We also test if the system has $0$ in its $K$-register, with cost $\log N+\mathcal{O}(1)$, and perform an OR on the results of the two comparisons with cost $1$.
Since the comparisons needs to be computed and uncomputed, there is cost $4\log N+\mathcal{O}(1)$ for the two steps.\\
\textbf{Steps 2 and 10.}
Computing $q_{xk}$ and $r_{xk}$ may be performed by first computing the energy difference, then using a QROM to output coefficients for linear interpolation.
The cost estimation is as given in \sec{functions}, and we pay the QROM lookup cost twice for $q_{xk}$ and $r_{xk}$, but we pay the multiplication cost only once.
Since that is the dominant cost, the cost may be regarded as that of a single function oracle.
The computation and uncomputation in the two steps means we pay twice the cost of the energy difference and function oracles.
Note that $q_{xk}$ and $r_{xk}$ are unchanged under the bit flip in step 6 (since there is no bit flip for $q_{xk}$ and $r_{xk}$ is symmetric under the bit flip).
There is $\mathcal{O}(1)$ cost to making the computation controlled on the ancilla in $E$.
\\
\textbf{Steps 3 and 9.} Outputting $\delta_\beta$ controlled on a single ancilla may be performed with CNOTs (no Toffoli cost) because $\delta_\beta$ is classically computed.\\
\textbf{Steps 4 and 8.} The inequality test is simply performed in the form $z< 2^{s-1}(1+q_{xk})$ and similarly for $r$.
There are no multiplications involved, because $q_{xk}$ and $r_{xk}$ are output as integer approximations.
The inequality test has cost $s$ Toffolis, so computation and uncomputation for the two steps has cost $2s$.\\
\textbf{Step 5.} This is just a $Z$ gate with no Toffoli cost.\\
\textbf{Step 6.} The cost is two Toffolis to prepare a control qubit that flags whether the conditions required are satisfied.
Then this qubit is used as a control register for the QROM on the value of $k$ to apply a $X$ operation to the target system.
That QROM has complexity $N$.\\
\textbf{Step 7.} Controlled on the system $K$-register being equal to $k$, we subtract $k$ from it, and controlled on the system $K$-register being $0$ we add $k$ to it.
We then swap the registers with the results of these two equality tests.
Since we still have the qubits with the results of the equality tests from step 1, we have no additional cost for that here.
The cost of the two additions is $2\log N+\mathcal{O}(1)$.\\

The Toffoli cost of the steps is therefore $2s+N+6\log N + \mathcal{O}(1)$, plus two times the cost of the function evaluation  and energy difference oracles.
Note that we pay four times the cost of the QROM lookup within the function evaluation oracle, but we are regarding the cost as two function oracles because the QROM lookup cost is a smaller cost given in an order term.
The cost of the preparation and inverse preparation is $8\log N+\mathcal{O}(1)$ Toffolis and $2.3\brot+\mathcal{O}(1)$ T gates, or just $2.3\brot+\mathcal{O}(1)$ T gates if $N$ is a power of $2$.
Taking $s=\bsmooth+\order{1}$, that gives total cost
\begin{equation}\label{eq:sgatof}
    2\mathcal{C}^{\textrm{diff}} + 2\mathcal{C}^{\textrm{fun}} + 2\bsmooth +N +14\log N+ \order{\brot},
\end{equation}
where we have put the $T$ cost in the order term.
The ancilla cost is as follows.
\begin{enumerate}
    \item Two qubits for the $E$ and $F$ ancillae.
    \item Two qubits from the results of the two equality tests for the system $K$-register.
    \item The register with $k$ for the control ancilla and that with $k$ for the system each need $\lceil \log N\rceil$ qubits.
    \item The register with $z$ for the control ancilla needs $s$ qubits.
    \item The ancillas for the energy difference oracle.
    \item The ancillas for the function evaluation oracle.
\end{enumerate}
The number of qubits $s$ used for $z$ can be taken to be within $\mathcal{O}(1)$ of the number of qubits $c$ used for $q_{xk}$ or $r_{xk}$.
We need temporary qubits for working, but the same working qubits as for the oracles can be used, so we will not count these ancilla costs again.
The function evaluation oracle may use more or less temporary ancilla than the energy difference, so we need to take the maximum of these two costs.
That gives an ancilla cost of $2\log N+\bsmooth+\mathcal{O}(1)$ plus the ancilla costs of the two oracles, or
\begin{equation}\label{eq:sgaanc}
    \mathcal{A}^{\textrm{diff}} + \mathcal{A}^{\textrm{fun}} + \max(\mathcal{B}^{\textrm{diff}},\mathcal{B}^{\textrm{fun}}) + 2\log N + \bsmooth + \order{1}.
\end{equation}

\section{Error-Correction Analysis and Discussion}
\label{sec:conclusion}

Previous sections of this paper have discussed and optimized the compilation of various heuristic approaches to quantum optimization into cost models appropriate for quantum error-correction. Specifically, we focused on reducing the Toffoli (and in some cases T) complexity of these algorithms while also keeping the number of ancilla qubits reasonable. This cost model is motivated by our desire to assess the viability of these heuristics within the surface code (the most practical error-correcting code suitable for a 2D array of physical qubits) \cite{Bravyi1998,Dennis2002,Raussendorf2007,Fowler2012}. T gates and Toffoli gates cannot be implemented transversely within practical implementations of the surface code. Instead, one must implement these gates by first distilling resource states. In particular, to implement a T gate one requires a T state ($\ket{\rm T}$ = ${\rm T}\ket{+}$) and to implement a Toffoli gate one requires a CCZ state ($\ket{\rm CCZ} = {\rm CCZ}\ket{+++}$); in both cases these states are consumed during the implementation of the associated gates. Distilling T or CCZ states requires a substantial amount of both time and hardware. 

\begin{table*}[t]
\begin{tabular}{c|c|c|c|c|c|c|c}
\multicolumn{4}{c|} {} & \multicolumn{2}{c} { one hour runtime} & \multicolumn{2}{|c} {one day runtime} \\
\hline
algorithm applied to & problem & logical & Toffolis & maximum & physical & maximum & physical \\
Sherrington-Kirkpatrick model & size, $N$ & qubits & per step & steps & qubits & steps & qubits  \\
\hline\hline
  & 64 & 100 & $6.3\!\times\! 10^3$ & $3.3\!\times\! 10^3$ & $3.1\!\times\! 10^5$ $(1.8\!\times\! 10^5)$ & $7.9\!\times\! 10^4$ & $3.7\!\times\! 10^5$ $(2.0\!\times\! 10^5)$ \\ 
  & 128 & 170 & $2.6\!\times\! 10^4$ & $7.9\!\times\! 10^2$ & $4.2\!\times\! 10^5$ $(2.1\!\times\! 10^5)$ & $1.9\!\times\! 10^4$ & $5.2\!\times\! 10^5$ $(2.3\!\times\! 10^5)$ \\ 
amplitude amplification & 256 & 304 & $1.0\!\times\! 10^5$ & $2.0\!\times\! 10^2$ & $7.2\!\times\! 10^5$ $(3.0\!\times\! 10^5)$ & $4.8\!\times\! 10^3$ & $8.1\!\times\! 10^5$ $(3.0\!\times\! 10^5)$ \\ 
  & 512 & 566 & $4.6\!\times\! 10^5$ & $4.5\!\times\! 10^1$ & $1.2\!\times\! 10^6$ $(4.3\!\times\! 10^5)$ & $1.1\!\times\! 10^3$ & $1.4\!\times\! 10^6$ $(4.3\!\times\! 10^5)$ \\ 
  & 1024 & 1084 & $1.8\!\times\! 10^6$ & $1.1\!\times\! 10^1$ & $2.2\!\times\! 10^6$ $(7.0\!\times\! 10^5)$ & $2.7\!\times\! 10^2$ & $2.9\!\times\! 10^6$ $(8.8\!\times\! 10^5)$ \\ 
\hline   & 64 & 120 & $6.8\!\times\! 10^3$ & $3.1\!\times\! 10^3$ & $3.4\!\times\! 10^5$ $(1.9\!\times\! 10^5)$ & $7.3\!\times\! 10^4$ & $4.1\!\times\! 10^5$ $(2.1\!\times\! 10^5)$ \\ 
QAOA / $1^{\rm st}$ order Trotter & 128 & 190 & $2.7\!\times\! 10^4$ & $7.7\!\times\! 10^2$ & $5.0\!\times\! 10^5$ $(2.4\!\times\! 10^5)$ & $1.9\!\times\! 10^4$ & $5.6\!\times\! 10^5$ $(2.4\!\times\! 10^5)$ \\ 
e.g., for population transfer & 256 & 324 & $1.1\!\times\! 10^5$ & $2.0\!\times\! 10^2$ & $7.6\!\times\! 10^5$ $(3.1\!\times\! 10^5)$ & $4.7\!\times\! 10^3$ & $8.6\!\times\! 10^5$ $(3.1\!\times\! 10^5)$ \\ 
or adiabatic algorithm & 512 & 586 & $4.6\!\times\! 10^5$ & $4.5\!\times\! 10^1$ & $1.2\!\times\! 10^6$ $(4.5\!\times\! 10^5)$ & $1.1\!\times\! 10^3$ & $1.4\!\times\! 10^6$ $(4.5\!\times\! 10^5)$ \\ 
  & 1024 & 1104 & $1.8\!\times\! 10^6$ & $1.1\!\times\! 10^1$ & $2.2\!\times\! 10^6$ $(7.1\!\times\! 10^5)$ & $2.7\!\times\! 10^2$ & $2.9\!\times\! 10^6$ $(8.9\!\times\! 10^5)$ \\ 
\hline   & 64 & 94 & $3.8\!\times\! 10^2$ & $5.4\!\times\! 10^4$ & $3.0\!\times\! 10^5$ $(1.8\!\times\! 10^5)$ & $1.3\!\times\! 10^6$ & $3.5\!\times\! 10^5$ $(2.0\!\times\! 10^5)$ \\ 
Hamiltonian walk & 128 & 163 & $7.7\!\times\! 10^2$ & $2.7\!\times\! 10^4$ & $4.1\!\times\! 10^5$ $(2.1\!\times\! 10^5)$ & $6.5\!\times\! 10^5$ & $5.0\!\times\! 10^5$ $(2.3\!\times\! 10^5)$ \\ 
e.g., for population transfer & 256 & 296 & $1.5\!\times\! 10^3$ & $1.4\!\times\! 10^4$ & $7.0\!\times\! 10^5$ $(3.0\!\times\! 10^5)$ & $3.3\!\times\! 10^5$ & $8.0\!\times\! 10^5$ $(3.0\!\times\! 10^5)$ \\ 
or adiabatic algorithm & 512 & 557 & $3.1\!\times\! 10^3$ & $6.8\!\times\! 10^3$ & $1.2\!\times\! 10^6$ $(4.3\!\times\! 10^5)$ & $1.6\!\times\! 10^5$ & $1.4\!\times\! 10^6$ $(4.3\!\times\! 10^5)$ \\ 
  & 1024 & 1074 & $6.1\!\times\! 10^3$ & $3.4\!\times\! 10^3$ & $2.2\!\times\! 10^6$ $(6.9\!\times\! 10^5)$ & $8.1\!\times\! 10^4$ & $2.9\!\times\! 10^6$ $(8.7\!\times\! 10^5)$ \\ 
\hline   & 64 & 117 & $6.7\!\times\! 10^2$ & $3.1\!\times\! 10^4$ & $3.3\!\times\! 10^5$ $(1.9\!\times\! 10^5)$ & $7.5\!\times\! 10^5$ & $4.0\!\times\! 10^5$ $(2.1\!\times\! 10^5)$ \\ 
LHPST walk & 128 & 185 & $9.0\!\times\! 10^2$ & $2.3\!\times\! 10^4$ & $4.4\!\times\! 10^5$ $(2.2\!\times\! 10^5)$ & $5.6\!\times\! 10^5$ & $5.5\!\times\! 10^5$ $(2.4\!\times\! 10^5)$ \\ 
quantum simulated annealing & 256 & 317 & $1.5\!\times\! 10^3$ & $1.4\!\times\! 10^4$ & $7.4\!\times\! 10^5$ $(3.1\!\times\! 10^5)$ & $3.3\!\times\! 10^5$ & $8.4\!\times\! 10^5$ $(3.1\!\times\! 10^5)$ \\ 
  & 512 & 577 & $2.6\!\times\! 10^3$ & $8.1\!\times\! 10^3$ & $1.2\!\times\! 10^6$ $(4.4\!\times\! 10^5)$ & $2.0\!\times\! 10^5$ & $1.4\!\times\! 10^6$ $(4.4\!\times\! 10^5)$ \\ 
  & 1024 & 1093 & $4.8\!\times\! 10^3$ & $4.4\!\times\! 10^3$ & $2.2\!\times\! 10^6$ $(7.0\!\times\! 10^5)$ & $1.0\!\times\! 10^5$ & $2.9\!\times\! 10^6$ $(8.9\!\times\! 10^5)$ \\ 
\hline   & 64 & 116 & $4.0\!\times\! 10^2$ & $5.2\!\times\! 10^4$ & $3.3\!\times\! 10^5$ $(1.9\!\times\! 10^5)$ & $1.2\!\times\! 10^6$ & $4.0\!\times\! 10^5$ $(2.1\!\times\! 10^5)$ \\ 
spectral gap amplified & 128 & 185 & $6.4\!\times\! 10^2$ & $3.3\!\times\! 10^4$ & $4.4\!\times\! 10^5$ $(2.2\!\times\! 10^5)$ & $7.8\!\times\! 10^5$ & $5.5\!\times\! 10^5$ $(2.4\!\times\! 10^5)$ \\ 
walk based quantum & 256 & 318 & $1.3\!\times\! 10^3$ & $1.6\!\times\! 10^4$ & $7.4\!\times\! 10^5$ $(3.1\!\times\! 10^5)$ & $3.9\!\times\! 10^5$ & $8.4\!\times\! 10^5$ $(3.1\!\times\! 10^5)$ \\ 
simulated annealing & 512 & 579 & $2.3\!\times\! 10^3$ & $9.0\!\times\! 10^3$ & $1.2\!\times\! 10^6$ $(4.4\!\times\! 10^5)$ & $2.2\!\times\! 10^5$ & $1.4\!\times\! 10^6$ $(4.4\!\times\! 10^5)$ \\ 
  & 1024 & 1096 & $4.5\!\times\! 10^3$ & $4.6\!\times\! 10^3$ & $2.2\!\times\! 10^6$ $(7.0\!\times\! 10^5)$ & $1.1\!\times\! 10^5$ & $2.9\!\times\! 10^6$ $(8.9\!\times\! 10^5)$ \\ 
\hline 
\end{tabular}
\caption[Resource estimates for the SK problem]{\label{tab:sk_estimates}
Estimates of resources required to implement steps of various heuristic algorithms for the Sherrington-Kirkpatrick (SK) model within the surface code. All error-correction overheads are reported assuming a single Toffoli factory using state distillation constructions from \cite{Gidney2019}. Surface code overheads in parenthesis assume a physical error rate of $10^{-4}$ whereas the overheads not in parenthesis assume a physical error rate of $10^{-3}$. The target success probability is $0.9$. These estimates are based on \tab{primitives} where we somewhat arbitrarily choose to set all values of the parameter quantifying the number of bits of precision ($b$) that appear in the table to 20 
except for $\bfun$ and $\bsmooth$ which can be smaller so we take $\bfun=\bsmooth=7$.}
\end{table*}

Here, we will analyze the cost to implement our various heuristic optimization primitives using the constructions of \cite{Gidney2019} which are based on applying the lattice surgery constructions of \cite{Fowler2018} to the fault-tolerant Toffoli protocols of \cite{Jones2012,Eastin2013}. We will further assume a correlated-error minimum weight perfect matching decoder capable of keeping pace with 1 {\textmu}s rounds of surface code error detection \cite{Fowler2013}, and capable of performing feedforward in about 15 {\textmu}s. We will assume that our physical hardware gates have error rates of either $10^{-3}$ or $10^{-4}$, the former consistent with the best error rates demonstrated in state-of-the-art intermediate scale superconducting qubit platforms \cite{Arute2019} and the latter consistent with improvements in the technology that we hope would be feasible in the next decade. Under these assumptions the spacetime volume required to implement one Toffoli gate or two T gates with two levels of state distillation and code distance $d=31$ (which is safely sufficient for the computations we analyze here) is equal to roughly 26 qubitseconds \cite{Gidney2019}. For instance, to distill one CCZ state using the approach in \cite{Gidney2019} requires $5.5 d + {\cal O}(1)$ cycles using a factory with a data qubit footprint of about $12 d \times 6 d$ (the total qubit count includes measurement qubits, and so is roughly double this figure). Specifically, in our estimates we will assume that executing a Toffoli gate requires about 170 microseconds and 150,000 physical qubits (see the resource estimation spreadsheet included in the supplementary materials of \cite{Gidney2019} for more detailed assumptions). Due to this large overhead we will focus on estimates assuming that we distill CCZ states in series, which is likely how we would operate the first generation of fault-tolerant surface code computers.

In \tab{sk_estimates} and \tab{LABS_estimates} we estimate the resources that would be required to implement various heuristic optimization primitives within the surface code (given the assumptions of the the prior paragraphs) for the Sherrington-Kirkpatrick and Low Autocorrelation Binary Sequences problems, respectively. We perform this analysis for the primitives of amplitude amplification, a first order Trotter step (which can be used for QAOA, population transfer, the adiabatic algorithm, etc.), a qubitized Hamiltonian walk realized from the linear combinations of unitaries query model (which can be used for measuring energies in QAOA, performing population transfer, the adiabatic algorithm, etc.), the qubitized quantum walk approach to quantum simulated annealing (``LHPST walk'') and the spectral gap amplified approach to quantum simulated annealing. The only primitive discussed in this paper omitted from these tables is the Szegedy walk approach to quantum simulated annealing. This is because we can see from \tab{primitives} that the Szegedy walk approach is strictly less efficient than the qubitized variant, and would require so many ancilla that analyzing it under the assumption of serial state distillation seems unreasonable. Because we do not know how many times one would need to repeat these primitives to solve the various optimization problems, in \tab{sk_estimates} and \tab{LABS_estimates} we report how many times one would be able to implement these primitives for various system sizes, assuming maximum run times of one hour or one day (24 hours). We also report how many physical qubits would be required to realize these computations assuming physical gate error rates of $10^{-3}$ or ($10^{-4}$).

\begin{table*}[t]
\begin{tabular}{c|c|c|c|c|c|c|c}
\multicolumn{4}{c|} {} & \multicolumn{2}{c} { one hour runtime} & \multicolumn{2}{|c} {one day runtime}\\
\hline
algorithm applied to & problem & logical & Toffolis & maximum & physical & maximum & physical\\
LABS problem & size, $N$ & qubits & per step & steps & qubits & steps & qubits\\
\hline\hline
  & 64 & 98 & $9.8\!\times\! 10^3$ & $2.1\!\times\! 10^3$ & $3.0\!\times\! 10^5$ $(1.8\!\times\! 10^5)$ & $5.1\!\times\! 10^4$ & $3.6\!\times\! 10^5$ $(2.0\!\times\! 10^5)$ \\ 
  & 128 & 167 & $3.7\!\times\! 10^4$ & $5.6\!\times\! 10^2$ & $4.1\!\times\! 10^5$ $(2.1\!\times\! 10^5)$ & $1.3\!\times\! 10^4$ & $5.1\!\times\! 10^5$ $(2.3\!\times\! 10^5)$ \\ 
amplitude amplification & 256 & 300 & $1.5\!\times\! 10^5$ & $1.4\!\times\! 10^2$ & $7.1\!\times\! 10^5$ $(3.0\!\times\! 10^5)$ & $3.3\!\times\! 10^3$ & $8.0\!\times\! 10^5$ $(3.0\!\times\! 10^5)$ \\ 
  & 512 & 561 & $6.1\!\times\! 10^5$ & $3.4\!\times\! 10^1$ & $1.2\!\times\! 10^6$ $(4.3\!\times\! 10^5)$ & $8.2\!\times\! 10^2$ & $1.4\!\times\! 10^6$ $(4.3\!\times\! 10^5)$ \\ 
  & 1024 & 1078 & $2.3\!\times\! 10^6$ & $9.0\!\times\! 10^0$ & $2.2\!\times\! 10^6$ $(6.9\!\times\! 10^5)$ & $2.2\!\times\! 10^2$ & $2.9\!\times\! 10^6$ $(8.8\!\times\! 10^5)$ \\ 
\hline   & 64 & 114 & $1.0\!\times\! 10^4$ & $2.1\!\times\! 10^3$ & $3.3\!\times\! 10^5$ $(1.9\!\times\! 10^5)$ & $5.0\!\times\! 10^4$ & $4.0\!\times\! 10^5$ $(2.1\!\times\! 10^5)$ \\ 
QAOA / $1^{\rm st}$ order Trotter & 128 & 183 & $3.8\!\times\! 10^4$ & $5.5\!\times\! 10^2$ & $4.4\!\times\! 10^5$ $(2.1\!\times\! 10^5)$ & $1.3\!\times\! 10^4$ & $5.5\!\times\! 10^5$ $(2.4\!\times\! 10^5)$ \\ 
e.g., for population transfer & 256 & 316 & $1.5\!\times\! 10^5$ & $1.4\!\times\! 10^2$ & $7.4\!\times\! 10^5$ $(3.1\!\times\! 10^5)$ & $3.4\!\times\! 10^3$ & $8.4\!\times\! 10^5$ $(3.1\!\times\! 10^5)$ \\ 
or adiabatic algorithm & 512 & 577 & $5.0\!\times\! 10^5$ & $4.2\!\times\! 10^1$ & $1.2\!\times\! 10^6$ $(4.4\!\times\! 10^5)$ & $1.0\!\times\! 10^3$ & $1.4\!\times\! 10^6$ $(4.4\!\times\! 10^5)$ \\ 
  & 1024 & 1094 & $1.7\!\times\! 10^6$ & $1.2\!\times\! 10^1$ & $2.2\!\times\! 10^6$ $(7.0\!\times\! 10^5)$ & $2.9\!\times\! 10^2$ & $2.9\!\times\! 10^6$ $(8.9\!\times\! 10^5)$ \\ 
\hline   & 64 & 94 & $2.6\!\times\! 10^2$ & $8.1\!\times\! 10^4$ & $3.0\!\times\! 10^5$ $(1.8\!\times\! 10^5)$ & $2.0\!\times\! 10^6$ & $3.5\!\times\! 10^5$ $(2.0\!\times\! 10^5)$ \\ 
Hamiltonian walk & 128 & 163 & $5.1\!\times\! 10^2$ & $4.1\!\times\! 10^4$ & $4.1\!\times\! 10^5$ $(2.1\!\times\! 10^5)$ & $9.8\!\times\! 10^5$ & $5.0\!\times\! 10^5$ $(2.3\!\times\! 10^5)$ \\ 
e.g., for population transfer & 256 & 296 & $1.0\!\times\! 10^3$ & $2.0\!\times\! 10^4$ & $7.0\!\times\! 10^5$ $(3.0\!\times\! 10^5)$ & $4.9\!\times\! 10^5$ & $8.0\!\times\! 10^5$ $(3.0\!\times\! 10^5)$ \\ 
or adiabatic algorithm & 512 & 557 & $2.0\!\times\! 10^3$ & $1.0\!\times\! 10^4$ & $1.2\!\times\! 10^6$ $(4.3\!\times\! 10^5)$ & $2.4\!\times\! 10^5$ & $1.4\!\times\! 10^6$ $(4.3\!\times\! 10^5)$ \\ 
  & 1024 & 1074 & $4.1\!\times\! 10^3$ & $5.1\!\times\! 10^3$ & $2.2\!\times\! 10^6$ $(6.9\!\times\! 10^5)$ & $1.2\!\times\! 10^5$ & $2.9\!\times\! 10^6$ $(8.7\!\times\! 10^5)$ \\ 
\hline   & 64 & 132 & $2.0\!\times\! 10^4$ & $1.0\!\times\! 10^3$ & $3.6\!\times\! 10^5$ $(2.0\!\times\! 10^5)$ & $2.5\!\times\! 10^4$ & $4.4\!\times\! 10^5$ $(2.1\!\times\! 10^5)$ \\ 
LHPST walk & 128 & 202 & $7.5\!\times\! 10^4$ & $2.8\!\times\! 10^2$ & $5.3\!\times\! 10^5$ $(2.5\!\times\! 10^5)$ & $6.7\!\times\! 10^3$ & $5.9\!\times\! 10^5$ $(2.5\!\times\! 10^5)$ \\ 
quantum simulated annealing & 256 & 336 & $3.0\!\times\! 10^5$ & $6.9\!\times\! 10^1$ & $7.8\!\times\! 10^5$ $(3.2\!\times\! 10^5)$ & $1.7\!\times\! 10^3$ & $8.8\!\times\! 10^5$ $(3.2\!\times\! 10^5)$ \\ 
  & 512 & 598 & $1.2\!\times\! 10^6$ & $1.7\!\times\! 10^1$ & $1.3\!\times\! 10^6$ $(4.5\!\times\! 10^5)$ & $4.1\!\times\! 10^2$ & $1.5\!\times\! 10^6$ $(4.5\!\times\! 10^5)$ \\ 
  & 1024 & 1116 & $4.6\!\times\! 10^6$ & $5.0\!\times\! 10^0$ & $2.2\!\times\! 10^6$ $(7.1\!\times\! 10^5)$ & $1.1\!\times\! 10^2$ & $3.0\!\times\! 10^6$ $(9.0\!\times\! 10^5)$ \\ 
\hline   & 64 & 131 & $2.0\!\times\! 10^4$ & $1.1\!\times\! 10^3$ & $3.6\!\times\! 10^5$ $(2.0\!\times\! 10^5)$ & $2.5\!\times\! 10^4$ & $4.3\!\times\! 10^5$ $(2.1\!\times\! 10^5)$ \\ 
spectral gap amplified & 128 & 202 & $7.5\!\times\! 10^4$ & $2.8\!\times\! 10^2$ & $5.3\!\times\! 10^5$ $(2.5\!\times\! 10^5)$ & $6.7\!\times\! 10^3$ & $5.9\!\times\! 10^5$ $(2.5\!\times\! 10^5)$ \\ 
walk based quantum & 256 & 337 & $3.0\!\times\! 10^5$ & $6.9\!\times\! 10^1$ & $7.8\!\times\! 10^5$ $(3.2\!\times\! 10^5)$ & $1.7\!\times\! 10^3$ & $8.8\!\times\! 10^5$ $(3.2\!\times\! 10^5)$ \\ 
simulated annealing & 512 & 600 & $1.2\!\times\! 10^6$ & $1.7\!\times\! 10^1$ & $1.3\!\times\! 10^6$ $(4.5\!\times\! 10^5)$ & $4.1\!\times\! 10^2$ & $1.5\!\times\! 10^6$ $(4.5\!\times\! 10^5)$ \\ 
  & 1024 & 1119 & $4.6\!\times\! 10^6$ & $5.0\!\times\! 10^0$ & $2.2\!\times\! 10^6$ $(7.2\!\times\! 10^5)$ & $1.1\!\times\! 10^2$ & $3.0\!\times\! 10^6$ $(9.0\!\times\! 10^5)$ \\ 
\hline
\end{tabular}
\caption[Resource estimates for the LABS problem]{\label{tab:LABS_estimates}
Estimates of resources required to implement steps of various heuristic algorithms for the Low Autocorrelation Binary Sequence (LABS) problem within the surface code. All overheads are reported assuming a single Toffoli factory using state distillation constructions from \cite{Gidney2019}. Surface code overheads in parenthesis assume a physical error rate of $10^{-4}$ whereas the overheads not in parenthesis assume a physical error rate of $10^{-3}$. Target success probability is 0.9. These estimates are based on \tab{primitives} where we somewhat arbitrarily choose to set all values of the parameter quantifying the number of bits of precision ($b$) that appear in the table to 20 except for $\bfun$ and $\bsmooth$ which can be smaller, so we take $\bfun=\bsmooth=7$.}
\end{table*}

We focus on the SK and LABS cost functions primarily for concreteness. As seen in \tab{oracles} and \tab{primitives}, the choice to focus on these specific problems rather than QUBO or the $H_L$ model means that we do not need to choose a precision parameter in some cases. For example, with amplitude amplification we know how many bits of precision we should compute the energy to since SK and LABS both have integer valued energies in a well defined range. However, in order to produce specific numerical estimates for other primitives it is necessary to assume values for the precision parameters $b$ appearing \tab{oracles} (defined in \tab{oracle_defs}); e.g., for the Trotter steps one must realize time evolutions of non-integer duration so that the phase is accurate to within some precision $\bphase$ which we must choose independently of the particular problem. Thus, in order to produce actual numerical estimates, in \tab{sk_estimates} and \tab{LABS_estimates} we choose to set many variants of the free precision parameter $b$ to 20; thus, $b=20$ bits of precision. However, as discussed in previous sections, the parameters $\bfun$ and $\bsmooth$ can generally be chosen to be smaller than the other values of $b$ without compromising precision; here we take $\bfun=\bsmooth=7$.

It is tempting to directly compare the costs of the various primitives shown in \tab{sk_estimates} and \tab{LABS_estimates}. While comparisons of the same primitives between the two problem types are straightforward (e.g., SK is more efficient than LABS in most, but not all, cases), comparisons between the different primitive types are challenging. Quantum simulated annealing, amplitude amplification, QAOA, population transfer, and the adiabatic algorithm are simply different algorithms so it is difficult to compare the relative values of a step of these algorithms.

It seems more reasonable to compare the Trotter steps to the qubitized Hamiltonian walk steps since these primitives can be used for the same ends (e.g., population transfer or the adiabatic algorithm). But first, the choice of $b=20$ means something different for these two algorithms. And second, while the Hamiltonian walks are capable of more precise evolutions (scaling as ${\cal O}(\log 1/\epsilon)$ in terms of precision compared to the ${\cal O}(\rm{poly}(1/\epsilon))$ scaling of fixed order Trotter based methods), for heuristic optimization the evolution does not necessarily need to be precise, so the Trotter approach may be more efficient by using large steps. The Trotter steps can be made arbitrarily large without increasing gate count (although at a cost of less precision), whereas the Hamiltonian walk effectively simulates time of at most $1/\lambda$ where $\lambda_{\rm SK} \approx N^2/2$ and $\lambda_{\rm LABS} \approx N^3/3$ (but it does so quite precisely). Thus, although the Hamiltonian walk steps require the fewest Toffolis in \tab{LABS_estimates}, they 
may still be less efficient than other approaches.

For the various forms of quantum simulated annealing, the number of steps needed is governed by the spectral gap.
The qubitized annealing (LHPST) and Szegedy approaches are directly comparable because they have the same gap, which means the same number of steps should be sufficient.
This means that the smaller step cost of LHPST means that it is more efficient.
The spectral gap amplified approach has a similar gap as the LHPST and Szegedy approaches, because it provides a similar square-root improvement. The problem is that the Hamiltonian has a $\lambda$-value proportional to $\sqrt{N}$, as shown in \eq{sgalam}.
This increases the cost of implementing the Hamiltonian by a factor of $\sqrt{N}$, so the cost given for a single step should be multiplied by $\sqrt{N}$ for a fair comparison with the other simulated annealing approaches.
When that is taken into account, the spectral gap amplified approach is less efficient.

With these caveats and context, we believe that \tab{sk_estimates} and \tab{LABS_estimates} give a rough sense for the feasibility of implementing these various heuristic optimization primitives on a small fault-tolerant surface code quantum processor. In most cases one can attempt these algorithms up to roughly a thousand bits with around a million physical qubits or less (especially given $10^{-4}$ error rates). However, we can see that the significant overheads of state distillation make the execution of these algorithms painfully slow. The quantum simulated annealing steps are often more efficient to implement than most other steps.
The one exception is the Hamiltonian walk steps, which are highly efficient.
But again, there it is likely that the large value of $\lambda$ means that many more Hamiltonian walk steps would be required.

We see that for SK model problem sizes between $N = 64$ and $N = 1024$ one can perform between about $4\!\times\!10^3$ and $3\!\times\!10^4$ quantum simulated annealing updates per hour. As a comparison, the work of \cite{Isakov2015} discusses the implementation of a very performant classical simulated annealing code for optimizing sparse spin glasses. This same code deployed to an $N=512$ spin instance of SK is capable of performing a simulated annealing update step in an average of 7 CPU-nanoseconds \cite{SergeiEmail} (this average accounts for the fact that most updates for the Sherrington-Kirkpatrick model are rejected). This works out to about $6\! \times\! 10^{11}$ attempted updates per core-hour, or about one-hundred million times more steps than the quantum computer can implement in that same amount of time for an $N=512$ spin SK model. The state produced after the $2\!\times\!10^5$ quantum simulated annealing steps that our quantum computer can make in one day for the $N=512$ spin SK model could be produced by a single classical core in about four CPU-minutes, assuming that the classical algorithm would require exactly quadratically more ($4\!\times\!10^{10}$) steps. The comparison is even less favorable for quantum computing if we consider larger problem sizes. Furthermore, given the high costs of quantum computing, it is unclear why we should restrict the classical competition to one core rather than to millions of cores.

The quantum computer must give a speedup for a sufficiently difficult problem if we assume a quadratic speedup in the number of annealing steps required. For the $N=512$ spin SK model, by comparing the number of steps that the classical algorithm from \cite{Isakov2015} can make in one hour  $(5\!\times\!10^{11})$ to the number of steps that the quantum algorithm can make in one hour $(8 \!\times\! 10^3)$, we can estimate a crossover point. In particular, solving $M / (8 \!\times \!10^3) = M^2 / (5\! \times\! 10^{11})$ yields $M \approx 7 \!\times\! 10^7$ as the minimum number of steps that would be required for the quantum algorithm to give an advantage. Unfortunately, this would mean the quantum computer would need to run for a number of hours that is $7\! \times\! 10^7 / (8\! \times \!10^3)$, which works out to about one year. Moreover, this analysis is very favorable to the quantum computer in that (1) it does not adjust the surface code distance (and thus, resource overheads) for runtimes longer than an hour, (2) it compares to a single classical core and (3) it assumes that $N=512$ is a large enough instance to warrant this many steps in some cases. Of course, most $N=512$ instances of the SK model can be solved with much less than a CPU year of simulated annealing run time, thus precluding the possibility of a quantum speedup for most instances at that size under the assumptions of our analysis.

Comparisons for amplitude amplification are similarly discouraging. For these two problems one can perform between about ten and three thousand steps of amplitude amplification using between about one-hundred thousand and one-million qubithours of state distillation. In the same amount of time one could conservatively check hundreds of billions of solutions on even a single core of a classical computer. Assuming the quantum computer would require quadratically fewer steps of amplitude amplification (still at least a hundred thousand steps) compared to random classical energy queries, we would still need roughly billions of qubithours of state distillation in order to compete with what a single core of a classical computer can do in one hour. Once again, if we instead make our comparisons to a classical supercomputing cluster rather than to a single classical core, the overheads appear even more daunting.

The LABS problem is an example where the scaling of the best known classical algorithm is worse than ${\cal O}(2^{N/2})$ and thus, an approach based on amplitude amplification would have better scaling. In particular, the best scaling method in the literature goes as $\Theta(1.73^N)$ \cite{Packebusch2016}. That scaling is obtained for a branch-and-bound type method that queries the effect of local spin flips (and thus, not the entire objective function). Each of these queries is slightly faster than requiring 7 CPU-microseconds with an optimized classical implementation for $N=64$ (about $5 \! \times \! 10^{8}$ steps per hour). If we were to compete with this approach using amplitude amplification on a quantum computer (where we can perform about $2\! \times \! 10^{3}$ steps per hour at $N=64$) then we can approximate the crossover point as $2^{M/2} / (2\! \times \! 10^{3}) = 1.73^M /(5 \! \times \! 10^{8})$ so long as we remember that these numbers are only valid in the vicinity of $M\approx N = 64$. Coincidentally, that is the case as we find that $M=62$, which corresponds to about $2\!\times\!10^9$ queries, which would take about 116 years. Once again, here we are being generous to the quantum computer by making comparisons to a single core and not adjusting the code distance for long runtimes. Still, we again see that a small error-corrected quantum computer cannot compete with classical methods under such a modest scaling advantage.

The heuristics based on Trotter steps or qubitized walk LCU queries are more difficult to compare to classical competition since algorithms such as QAOA, the adiabatic algorithm, or population transfer lack a clear classical analog. In that sense, it is not straightforward to predict what being able to perform a few hundred Trotter steps or a few thousand qubitized walk steps in an hour might buy us, but it is clear that these would be able to perform only very short quantum walks or time evolutions, or very inaccurate time evolutions. Eventually, it will at least be possible to find out by using our constructions to realize these algorithms on a small fault-tolerant quantum computer and experimentally discovering what happens. We note that for these algorithms the number of steps should be interpreted as the product of the number of repetitions of the primitive and the total number of times the algorithm is repeated. For instance, we see that for either the SK model or LABS at $N=256$, slightly more than 100 Trotter steps can be implemented in an hour. In the context of QAOA, this could mean that we run QAOA at $p=100$ and draw one sample, or we run QAOA at $p=10$ and draw ten samples or we run QAOA at $p=1$ and draw one-hundred samples, etc. However, as we have explained in \sec{amp_amp} and \sec{qaoa} one is probably better off using coherent repetitions in the context of an amplitude-amplification like scheme rather than making classical repetitions.

Although we have tried to optimize the realization of these heuristic primitives for the cost functions considered in this paper, clever improvements to our approaches might further reduce the resources required. However, we would expect the complexity of these primitives to be no better than $N$. In particular, LCU-based methods require a minimum of $N-1$ Toffolis just to access $N$ qubits in a controlled way.
For Trotter step methods, evolution under the problem Hamiltonian could be below $N$ for a particularly simple problem Hamiltonian, but then the evolution under the transverse field driver Hamiltonian would be the dominant cost and require $\order{N}$ non-Clifford gates.
For amplitude amplification, one could again have a small cost for a particularly simple problem Hamiltonian, but amplitude amplification requires a reflection on at least $N$ qubits, with cost at least $N-2$ Toffolis.

We are already at about $5N$ for the LHPST walk with SK, so we would not expect more than about a factor of 5 improvement even for the easiest problem. If we were to use the sum of bits directly as in \cite{Kivlichan2019}, then the complexity would be about $2N$, but another $N$ ancilla qubits would be needed. One could also propose to use a larger fault-tolerant quantum computer and distill more Toffoli states in parallel. But even if this strategy is pursued to the fullest extent possible (requiring tens or hundreds of millions of physical qubits) and parallelized near-optimally, the surface code will then be bottlenecked by Clifford gates (or the overhead of routing) which are, at best, only about a hundred to a thousand times faster to implement.

In conclusion, we have optimized and compiled the basic primitives required for many popular heuristic algorithms for quantum optimization to a cost model appropriate for practical quantum error-correction schemes. This allowed us to assess and compare the cost of several quantum algorithms that have not previously been compiled in such detail. We focused on doing this for only a subset of the possible cost function structures that one might hope to algorithmically exploit for more efficient implementations, but our constructions led to the development of various methodologies which we expect will be useful in a more general context. For instance, we expect that work outside the context of quantum optimization might benefit from our strategy of interpolating arithmetic functions using an adaptive QROM. However, despite our attempts at optimization, the concrete resource estimates from \tab{sk_estimates} and \tab{LABS_estimates} are predictably discouraging. The essential reason for this is the substantial constant factor slowdown between error-corrected quantum computation and classical computation. Based on these numbers we strongly suspect that in order for early fault-tolerant quantum computers to have a meaningful impact on combinatorial optimization, we will either need quantum optimization algorithms that afford speedups which are much better than quadratic, or we will need significant improvements in the way that we realize error-correction.

\section*{Acknowledgements}

The authors thank Sergio Boixo, Austin Fowler, Sergei Isakov, Kostyantyn Kechedzhi, M\'aria Kieferov\'a, Jessica Lemieux, Jarrod McClean, John Platt, and Vadim Smelyanskiy for helpful discussions. Y.R.S., D.W.B., P.C.S.C, and N.W.~acknowledge funding for this work from a grant from Google Quantum. D.W.B. is also funded by an Australian Research Council Discovery Project DP190102633.\\

\bibliographystyle{apsrev4-1}
\bibliography{references_new}

\begin{thebibliography}{75}%
\makeatletter
\providecommand \@ifxundefined [1]{%
 \@ifx{#1\undefined}
}%
\providecommand \@ifnum [1]{%
 \ifnum #1\expandafter \@firstoftwo
 \else \expandafter \@secondoftwo
 \fi
}%
\providecommand \@ifx [1]{%
 \ifx #1\expandafter \@firstoftwo
 \else \expandafter \@secondoftwo
 \fi
}%
\providecommand \natexlab [1]{#1}%
\providecommand \enquote  [1]{``#1''}%
\providecommand \bibnamefont  [1]{#1}%
\providecommand \bibfnamefont [1]{#1}%
\providecommand \citenamefont [1]{#1}%
\providecommand \href@noop [0]{\@secondoftwo}%
\providecommand \href [0]{\begingroup \@sanitize@url \@href}%
\providecommand \@href[1]{\@@startlink{#1}\@@href}%
\providecommand \@@href[1]{\endgroup#1\@@endlink}%
\providecommand \@sanitize@url [0]{\catcode `\\12\catcode `\$12\catcode
  `\&12\catcode `\#12\catcode `\^12\catcode `\_12\catcode `\%12\relax}%
\providecommand \@@startlink[1]{}%
\providecommand \@@endlink[0]{}%
\providecommand \url  [0]{\begingroup\@sanitize@url \@url }%
\providecommand \@url [1]{\endgroup\@href {#1}{\urlprefix }}%
\providecommand \urlprefix  [0]{URL }%
\providecommand \Eprint [0]{\href }%
\providecommand \doibase [0]{http://dx.doi.org/}%
\providecommand \selectlanguage [0]{\@gobble}%
\providecommand \bibinfo  [0]{\@secondoftwo}%
\providecommand \bibfield  [0]{\@secondoftwo}%
\providecommand \translation [1]{[#1]}%
\providecommand \BibitemOpen [0]{}%
\providecommand \bibitemStop [0]{}%
\providecommand \bibitemNoStop [0]{.\EOS\space}%
\providecommand \EOS [0]{\spacefactor3000\relax}%
\providecommand \BibitemShut  [1]{\csname bibitem#1\endcsname}%
\let\auto@bib@innerbib\@empty
\bibitem [{\citenamefont {Arute}\ \emph {et~al.}(2019)\citenamefont {Arute},
  \citenamefont {Arya}, \citenamefont {Babbush}, \citenamefont {Bacon},
  \citenamefont {Bardin}, \citenamefont {Barends}, \citenamefont {Biswas},
  \citenamefont {Boixo}, \citenamefont {Brandao}, \citenamefont {Buell},
  \citenamefont {Burkett}, \citenamefont {Chen}, \citenamefont {Chen},
  \citenamefont {Chiaro}, \citenamefont {Collins}, \citenamefont {Courtney},
  \citenamefont {Dunsworth}, \citenamefont {Farhi}, \citenamefont {Foxen},
  \citenamefont {Fowler}, \citenamefont {Gidney}, \citenamefont {Giustina},
  \citenamefont {Graff}, \citenamefont {Guerin}, \citenamefont {Habegger},
  \citenamefont {Harrigan}, \citenamefont {Hartmann}, \citenamefont {Ho},
  \citenamefont {Hoffmann}, \citenamefont {Huang}, \citenamefont {Humble},
  \citenamefont {Isakov}, \citenamefont {Jeffrey}, \citenamefont {Jiang},
  \citenamefont {Kafri}, \citenamefont {Kechedzhi}, \citenamefont {Kelly},
  \citenamefont {Klimov}, \citenamefont {Knysh}, \citenamefont {Korotkov},
  \citenamefont {Kostritsa}, \citenamefont {Landhuis}, \citenamefont
  {Lindmark}, \citenamefont {Lucero}, \citenamefont {Lyakh}, \citenamefont
  {Mandr{\`{a}}}, \citenamefont {McClean}, \citenamefont {McEwen},
  \citenamefont {Megrant}, \citenamefont {Mi}, \citenamefont {Michielsen},
  \citenamefont {Mohseni}, \citenamefont {Mutus}, \citenamefont {Naaman},
  \citenamefont {Neeley}, \citenamefont {Neill}, \citenamefont {Niu},
  \citenamefont {Ostby}, \citenamefont {Petukhov}, \citenamefont {Platt},
  \citenamefont {Quintana}, \citenamefont {Rieffel}, \citenamefont {Roushan},
  \citenamefont {Rubin}, \citenamefont {Sank}, \citenamefont {Satzinger},
  \citenamefont {Smelyanskiy}, \citenamefont {Sung}, \citenamefont
  {Trevithick}, \citenamefont {Vainsencher}, \citenamefont {Villalonga},
  \citenamefont {White}, \citenamefont {Yao}, \citenamefont {Yeh},
  \citenamefont {Zalcman}, \citenamefont {Neven},\ and\ \citenamefont
  {Martinis}}]{Arute2019}%
  \BibitemOpen
  \bibfield  {author} {\bibinfo {author} {\bibfnamefont {F.}~\bibnamefont
  {Arute}}, \bibinfo {author} {\bibfnamefont {K.}~\bibnamefont {Arya}},
  \bibinfo {author} {\bibfnamefont {R.}~\bibnamefont {Babbush}}, \bibinfo
  {author} {\bibfnamefont {D.}~\bibnamefont {Bacon}}, \bibinfo {author}
  {\bibfnamefont {J.~C.}\ \bibnamefont {Bardin}}, \bibinfo {author}
  {\bibfnamefont {R.}~\bibnamefont {Barends}}, \bibinfo {author} {\bibfnamefont
  {R.}~\bibnamefont {Biswas}}, \bibinfo {author} {\bibfnamefont
  {S.}~\bibnamefont {Boixo}}, \bibinfo {author} {\bibfnamefont {F.~G. S.~L.}\
  \bibnamefont {Brandao}}, \bibinfo {author} {\bibfnamefont {D.~A.}\
  \bibnamefont {Buell}}, \bibinfo {author} {\bibfnamefont {B.}~\bibnamefont
  {Burkett}}, \bibinfo {author} {\bibfnamefont {Y.}~\bibnamefont {Chen}},
  \bibinfo {author} {\bibfnamefont {Z.}~\bibnamefont {Chen}}, \bibinfo {author}
  {\bibfnamefont {B.}~\bibnamefont {Chiaro}}, \bibinfo {author} {\bibfnamefont
  {R.}~\bibnamefont {Collins}}, \bibinfo {author} {\bibfnamefont
  {W.}~\bibnamefont {Courtney}}, \bibinfo {author} {\bibfnamefont
  {A.}~\bibnamefont {Dunsworth}}, \bibinfo {author} {\bibfnamefont
  {E.}~\bibnamefont {Farhi}}, \bibinfo {author} {\bibfnamefont
  {B.}~\bibnamefont {Foxen}}, \bibinfo {author} {\bibfnamefont
  {A.}~\bibnamefont {Fowler}}, \bibinfo {author} {\bibfnamefont
  {C.}~\bibnamefont {Gidney}}, \bibinfo {author} {\bibfnamefont
  {M.}~\bibnamefont {Giustina}}, \bibinfo {author} {\bibfnamefont
  {R.}~\bibnamefont {Graff}}, \bibinfo {author} {\bibfnamefont
  {K.}~\bibnamefont {Guerin}}, \bibinfo {author} {\bibfnamefont
  {S.}~\bibnamefont {Habegger}}, \bibinfo {author} {\bibfnamefont {M.~P.}\
  \bibnamefont {Harrigan}}, \bibinfo {author} {\bibfnamefont {M.~J.}\
  \bibnamefont {Hartmann}}, \bibinfo {author} {\bibfnamefont {A.}~\bibnamefont
  {Ho}}, \bibinfo {author} {\bibfnamefont {M.}~\bibnamefont {Hoffmann}},
  \bibinfo {author} {\bibfnamefont {T.}~\bibnamefont {Huang}}, \bibinfo
  {author} {\bibfnamefont {T.~S.}\ \bibnamefont {Humble}}, \bibinfo {author}
  {\bibfnamefont {S.~V.}\ \bibnamefont {Isakov}}, \bibinfo {author}
  {\bibfnamefont {E.}~\bibnamefont {Jeffrey}}, \bibinfo {author} {\bibfnamefont
  {Z.}~\bibnamefont {Jiang}}, \bibinfo {author} {\bibfnamefont
  {D.}~\bibnamefont {Kafri}}, \bibinfo {author} {\bibfnamefont
  {K.}~\bibnamefont {Kechedzhi}}, \bibinfo {author} {\bibfnamefont
  {J.}~\bibnamefont {Kelly}}, \bibinfo {author} {\bibfnamefont {P.~V.}\
  \bibnamefont {Klimov}}, \bibinfo {author} {\bibfnamefont {S.}~\bibnamefont
  {Knysh}}, \bibinfo {author} {\bibfnamefont {A.}~\bibnamefont {Korotkov}},
  \bibinfo {author} {\bibfnamefont {F.}~\bibnamefont {Kostritsa}}, \bibinfo
  {author} {\bibfnamefont {D.}~\bibnamefont {Landhuis}}, \bibinfo {author}
  {\bibfnamefont {M.}~\bibnamefont {Lindmark}}, \bibinfo {author}
  {\bibfnamefont {E.}~\bibnamefont {Lucero}}, \bibinfo {author} {\bibfnamefont
  {D.}~\bibnamefont {Lyakh}}, \bibinfo {author} {\bibfnamefont
  {S.}~\bibnamefont {Mandr{\`{a}}}}, \bibinfo {author} {\bibfnamefont {J.~R.}\
  \bibnamefont {McClean}}, \bibinfo {author} {\bibfnamefont {M.}~\bibnamefont
  {McEwen}}, \bibinfo {author} {\bibfnamefont {A.}~\bibnamefont {Megrant}},
  \bibinfo {author} {\bibfnamefont {X.}~\bibnamefont {Mi}}, \bibinfo {author}
  {\bibfnamefont {K.}~\bibnamefont {Michielsen}}, \bibinfo {author}
  {\bibfnamefont {M.}~\bibnamefont {Mohseni}}, \bibinfo {author} {\bibfnamefont
  {J.}~\bibnamefont {Mutus}}, \bibinfo {author} {\bibfnamefont
  {O.}~\bibnamefont {Naaman}}, \bibinfo {author} {\bibfnamefont
  {M.}~\bibnamefont {Neeley}}, \bibinfo {author} {\bibfnamefont
  {C.}~\bibnamefont {Neill}}, \bibinfo {author} {\bibfnamefont {M.~Y.}\
  \bibnamefont {Niu}}, \bibinfo {author} {\bibfnamefont {E.}~\bibnamefont
  {Ostby}}, \bibinfo {author} {\bibfnamefont {A.}~\bibnamefont {Petukhov}},
  \bibinfo {author} {\bibfnamefont {J.~C.}\ \bibnamefont {Platt}}, \bibinfo
  {author} {\bibfnamefont {C.}~\bibnamefont {Quintana}}, \bibinfo {author}
  {\bibfnamefont {E.~G.}\ \bibnamefont {Rieffel}}, \bibinfo {author}
  {\bibfnamefont {P.}~\bibnamefont {Roushan}}, \bibinfo {author} {\bibfnamefont
  {N.~C.}\ \bibnamefont {Rubin}}, \bibinfo {author} {\bibfnamefont
  {D.}~\bibnamefont {Sank}}, \bibinfo {author} {\bibfnamefont {K.~J.}\
  \bibnamefont {Satzinger}}, \bibinfo {author} {\bibfnamefont {V.}~\bibnamefont
  {Smelyanskiy}}, \bibinfo {author} {\bibfnamefont {K.~J.}\ \bibnamefont
  {Sung}}, \bibinfo {author} {\bibfnamefont {M.~D.}\ \bibnamefont
  {Trevithick}}, \bibinfo {author} {\bibfnamefont {A.}~\bibnamefont
  {Vainsencher}}, \bibinfo {author} {\bibfnamefont {B.}~\bibnamefont
  {Villalonga}}, \bibinfo {author} {\bibfnamefont {T.}~\bibnamefont {White}},
  \bibinfo {author} {\bibfnamefont {Z.~J.}\ \bibnamefont {Yao}}, \bibinfo
  {author} {\bibfnamefont {P.}~\bibnamefont {Yeh}}, \bibinfo {author}
  {\bibfnamefont {A.}~\bibnamefont {Zalcman}}, \bibinfo {author} {\bibfnamefont
  {H.}~\bibnamefont {Neven}}, \ and\ \bibinfo {author} {\bibfnamefont {J.~M.}\
  \bibnamefont {Martinis}},\ }\href {\doibase 10.1038/s41586-019-1666-5}
  {\bibfield  {journal} {\bibinfo  {journal} {Nature}\ }\textbf {\bibinfo
  {volume} {574}},\ \bibinfo {pages} {505} (\bibinfo {year}
  {2019})}\BibitemShut {NoStop}%
\bibitem [{\citenamefont {Farhi}\ \emph {et~al.}(2001)\citenamefont {Farhi},
  \citenamefont {Goldstone}, \citenamefont {Gutmann}, \citenamefont {Lapan},
  \citenamefont {Lundgren},\ and\ \citenamefont {Preda}}]{Farhi2001}%
  \BibitemOpen
  \bibfield  {author} {\bibinfo {author} {\bibfnamefont {E.}~\bibnamefont
  {Farhi}}, \bibinfo {author} {\bibfnamefont {J.}~\bibnamefont {Goldstone}},
  \bibinfo {author} {\bibfnamefont {S.}~\bibnamefont {Gutmann}}, \bibinfo
  {author} {\bibfnamefont {J.}~\bibnamefont {Lapan}}, \bibinfo {author}
  {\bibfnamefont {A.}~\bibnamefont {Lundgren}}, \ and\ \bibinfo {author}
  {\bibfnamefont {D.}~\bibnamefont {Preda}},\ }\href
  {https://doi.org/10.1126/science.1057726} {\bibfield  {journal} {\bibinfo
  {journal} {Science}\ }\textbf {\bibinfo {volume} {292}},\ \bibinfo {pages}
  {472} (\bibinfo {year} {2001})}\BibitemShut {NoStop}%
\bibitem [{\citenamefont {Grover}(1996)}]{Grover1996}%
  \BibitemOpen
  \bibfield  {author} {\bibinfo {author} {\bibfnamefont {L.~K.}\ \bibnamefont
  {Grover}},\ }in\ \href {\doibase 10.1145/237814.237866} {\emph {\bibinfo
  {booktitle} {Proceedings of the Twenty-Eighth Annual ACM Symposium on Theory
  of Computing}}},\ \bibinfo {series and number} {STOC ’96}\ (\bibinfo
  {publisher} {Association for Computing Machinery},\ \bibinfo {address} {New
  York, NY, USA},\ \bibinfo {year} {1996})\ p.\ \bibinfo {pages}
  {212–219}\BibitemShut {NoStop}%
\bibitem [{\citenamefont {Durr}\ and\ \citenamefont
  {Hoyer}(1996)}]{Durr1996AMinimum}%
  \BibitemOpen
  \bibfield  {author} {\bibinfo {author} {\bibfnamefont {C.}~\bibnamefont
  {Durr}}\ and\ \bibinfo {author} {\bibfnamefont {P.}~\bibnamefont {Hoyer}},\
  }\href@noop {} {\enquote {\bibinfo {title} {A quantum algorithm for finding
  the minimum},}\ } (\bibinfo {year} {1996}),\ \Eprint
  {http://arxiv.org/abs/quant-ph/9607014} {arXiv:quant-ph/9607014} \BibitemShut
  {NoStop}%
\bibitem [{\citenamefont {Ray}\ \emph {et~al.}(1989)\citenamefont {Ray},
  \citenamefont {Chakrabarti},\ and\ \citenamefont {Chakrabarti}}]{ray1989}%
  \BibitemOpen
  \bibfield  {author} {\bibinfo {author} {\bibfnamefont {P.}~\bibnamefont
  {Ray}}, \bibinfo {author} {\bibfnamefont {B.~K.}\ \bibnamefont
  {Chakrabarti}}, \ and\ \bibinfo {author} {\bibfnamefont {A.}~\bibnamefont
  {Chakrabarti}},\ }\href {\doibase 10.1103/physrevb.39.11828} {\bibfield
  {journal} {\bibinfo  {journal} {Physical Review B}\ }\textbf {\bibinfo
  {volume} {39}},\ \bibinfo {pages} {11828} (\bibinfo {year}
  {1989})}\BibitemShut {NoStop}%
\bibitem [{\citenamefont {Kadowaki}\ and\ \citenamefont
  {Nishimori}(1998)}]{Kadowaki1998}%
  \BibitemOpen
  \bibfield  {author} {\bibinfo {author} {\bibfnamefont {T.}~\bibnamefont
  {Kadowaki}}\ and\ \bibinfo {author} {\bibfnamefont {H.}~\bibnamefont
  {Nishimori}},\ }\href {\doibase 10.1103/physreve.58.5355} {\bibfield
  {journal} {\bibinfo  {journal} {Physical Review E}\ }\textbf {\bibinfo
  {volume} {58}},\ \bibinfo {pages} {5355} (\bibinfo {year}
  {1998})}\BibitemShut {NoStop}%
\bibitem [{\citenamefont {Farhi}\ \emph
  {et~al.}(2000{\natexlab{a}})\citenamefont {Farhi}, \citenamefont {Goldstone},
  \citenamefont {Gutmann},\ and\ \citenamefont {Sipser}}]{farhi2000}%
  \BibitemOpen
  \bibfield  {author} {\bibinfo {author} {\bibfnamefont {E.}~\bibnamefont
  {Farhi}}, \bibinfo {author} {\bibfnamefont {J.}~\bibnamefont {Goldstone}},
  \bibinfo {author} {\bibfnamefont {S.}~\bibnamefont {Gutmann}}, \ and\
  \bibinfo {author} {\bibfnamefont {M.}~\bibnamefont {Sipser}},\ }\href@noop {}
  {\enquote {\bibinfo {title} {Quantum computation by adiabatic evolution},}\ }
  (\bibinfo {year} {2000}{\natexlab{a}}),\ \Eprint
  {http://arxiv.org/abs/quant-ph/0001106} {arXiv:quant-ph/0001106} \BibitemShut
  {NoStop}%
\bibitem [{\citenamefont {Aharonov}\ \emph {et~al.}(2007)\citenamefont
  {Aharonov}, \citenamefont {van Dam}, \citenamefont {Kempe}, \citenamefont
  {Landau}, \citenamefont {Lloyd},\ and\ \citenamefont {Regev}}]{Aharonov2007}%
  \BibitemOpen
  \bibfield  {author} {\bibinfo {author} {\bibfnamefont {D.}~\bibnamefont
  {Aharonov}}, \bibinfo {author} {\bibfnamefont {W.}~\bibnamefont {van Dam}},
  \bibinfo {author} {\bibfnamefont {J.}~\bibnamefont {Kempe}}, \bibinfo
  {author} {\bibfnamefont {Z.}~\bibnamefont {Landau}}, \bibinfo {author}
  {\bibfnamefont {S.}~\bibnamefont {Lloyd}}, \ and\ \bibinfo {author}
  {\bibfnamefont {O.}~\bibnamefont {Regev}},\ }\href
  {https://doi.org/10.1137/s0097539705447323} {\bibfield  {journal} {\bibinfo
  {journal} {{SIAM} Journal on Computing}\ }\textbf {\bibinfo {volume} {37}},\
  \bibinfo {pages} {166} (\bibinfo {year} {2007})}\BibitemShut {NoStop}%
\bibitem [{\citenamefont {Hastings}(2018{\natexlab{a}})}]{Hastings2018}%
  \BibitemOpen
  \bibfield  {author} {\bibinfo {author} {\bibfnamefont {M.~B.}\ \bibnamefont
  {Hastings}},\ }\href {\doibase 10.22331/q-2018-07-26-78} {\bibfield
  {journal} {\bibinfo  {journal} {Quantum}\ }\textbf {\bibinfo {volume} {2}},\
  \bibinfo {pages} {78} (\bibinfo {year} {2018}{\natexlab{a}})}\BibitemShut
  {NoStop}%
\bibitem [{\citenamefont {Kechedzhi}\ \emph {et~al.}(2018)\citenamefont
  {Kechedzhi}, \citenamefont {Smelyanskiy}, \citenamefont {McClean},
  \citenamefont {Denchev}, \citenamefont {Mohseni}, \citenamefont {Isakov},
  \citenamefont {Boixo}, \citenamefont {Altshuler},\ and\ \citenamefont
  {Neven}}]{Kechedzhi2018}%
  \BibitemOpen
  \bibfield  {author} {\bibinfo {author} {\bibfnamefont {K.}~\bibnamefont
  {Kechedzhi}}, \bibinfo {author} {\bibfnamefont {V.}~\bibnamefont
  {Smelyanskiy}}, \bibinfo {author} {\bibfnamefont {J.~R.}\ \bibnamefont
  {McClean}}, \bibinfo {author} {\bibfnamefont {V.~S.}\ \bibnamefont
  {Denchev}}, \bibinfo {author} {\bibfnamefont {M.}~\bibnamefont {Mohseni}},
  \bibinfo {author} {\bibfnamefont {S.}~\bibnamefont {Isakov}}, \bibinfo
  {author} {\bibfnamefont {S.}~\bibnamefont {Boixo}}, \bibinfo {author}
  {\bibfnamefont {B.}~\bibnamefont {Altshuler}}, \ and\ \bibinfo {author}
  {\bibfnamefont {H.}~\bibnamefont {Neven}},\ }in\ \href
  {https://doi.org/10.4230/LIPIcs.TQC.2018.9} {\emph {\bibinfo {booktitle}
  {13th Conference on the Theory of Quantum Computation, Communication and
  Cryptography (TQC 2018)}}},\ \bibinfo {series} {Leibniz International
  Proceedings in Informatics (LIPIcs)}, Vol.\ \bibinfo {volume} {111},\
  \bibinfo {editor} {edited by\ \bibinfo {editor} {\bibfnamefont
  {S.}~\bibnamefont {Jeffery}}}\ (\bibinfo  {publisher} {Schloss
  Dagstuhl--Leibniz-Zentrum fuer Informatik},\ \bibinfo {address} {Dagstuhl,
  Germany},\ \bibinfo {year} {2018})\ pp.\ \bibinfo {pages}
  {9:1--9:16}\BibitemShut {NoStop}%
\bibitem [{\citenamefont {Smelyanskiy}\ \emph {et~al.}(2020)\citenamefont
  {Smelyanskiy}, \citenamefont {Kechedzhi}, \citenamefont {Boixo},
  \citenamefont {Isakov}, \citenamefont {Neven},\ and\ \citenamefont
  {Altshuler}}]{Smelyanskiy2018}%
  \BibitemOpen
  \bibfield  {author} {\bibinfo {author} {\bibfnamefont {V.~N.}\ \bibnamefont
  {Smelyanskiy}}, \bibinfo {author} {\bibfnamefont {K.}~\bibnamefont
  {Kechedzhi}}, \bibinfo {author} {\bibfnamefont {S.}~\bibnamefont {Boixo}},
  \bibinfo {author} {\bibfnamefont {S.~V.}\ \bibnamefont {Isakov}}, \bibinfo
  {author} {\bibfnamefont {H.}~\bibnamefont {Neven}}, \ and\ \bibinfo {author}
  {\bibfnamefont {B.}~\bibnamefont {Altshuler}},\ }\href
  {https://doi.org/10.1103/physrevx.10.011017} {\bibfield  {journal} {\bibinfo
  {journal} {Physical Review X}\ }\textbf {\bibinfo {volume} {10}},\ \bibinfo
  {pages} {011017} (\bibinfo {year} {2020})}\BibitemShut {NoStop}%
\bibitem [{\citenamefont {Farhi}\ \emph {et~al.}(2014)\citenamefont {Farhi},
  \citenamefont {Goldstone},\ and\ \citenamefont {Gutmann}}]{Farhi2014}%
  \BibitemOpen
  \bibfield  {author} {\bibinfo {author} {\bibfnamefont {E.}~\bibnamefont
  {Farhi}}, \bibinfo {author} {\bibfnamefont {J.}~\bibnamefont {Goldstone}}, \
  and\ \bibinfo {author} {\bibfnamefont {S.}~\bibnamefont {Gutmann}},\
  }\href@noop {} {\enquote {\bibinfo {title} {A quantum approximate
  optimization algorithm},}\ } (\bibinfo {year} {2014}),\ \Eprint
  {http://arxiv.org/abs/1411.4028} {arXiv:1411.4028} \BibitemShut {NoStop}%
\bibitem [{\citenamefont {Somma}\ \emph {et~al.}(2008)\citenamefont {Somma},
  \citenamefont {Boixo}, \citenamefont {Barnum},\ and\ \citenamefont
  {Knill}}]{Somma2008b}%
  \BibitemOpen
  \bibfield  {author} {\bibinfo {author} {\bibfnamefont {R.~D.}\ \bibnamefont
  {Somma}}, \bibinfo {author} {\bibfnamefont {S.}~\bibnamefont {Boixo}},
  \bibinfo {author} {\bibfnamefont {H.}~\bibnamefont {Barnum}}, \ and\ \bibinfo
  {author} {\bibfnamefont {E.}~\bibnamefont {Knill}},\ }\href
  {https://doi.org/10.1103/PhysRevLett.101.130504} {\bibfield  {journal}
  {\bibinfo  {journal} {Physical Review Letters}\ }\textbf {\bibinfo {volume}
  {101}},\ \bibinfo {pages} {130504} (\bibinfo {year} {2008})}\BibitemShut
  {NoStop}%
\bibitem [{\citenamefont {Boixo}\ \emph {et~al.}(2015)\citenamefont {Boixo},
  \citenamefont {Ortiz},\ and\ \citenamefont {Somma}}]{Boixo2014a}%
  \BibitemOpen
  \bibfield  {author} {\bibinfo {author} {\bibfnamefont {S.}~\bibnamefont
  {Boixo}}, \bibinfo {author} {\bibfnamefont {G.}~\bibnamefont {Ortiz}}, \ and\
  \bibinfo {author} {\bibfnamefont {R.}~\bibnamefont {Somma}},\ }\href
  {\doibase 10.1140/epjst/e2015-02341-5} {\bibfield  {journal} {\bibinfo
  {journal} {The European Physical Journal Special Topics}\ }\textbf {\bibinfo
  {volume} {224}},\ \bibinfo {pages} {35} (\bibinfo {year} {2015})}\BibitemShut
  {NoStop}%
\bibitem [{\citenamefont
  {Montanaro}(2015{\natexlab{a}})}]{Montanaro2015QuantumAlgorithms}%
  \BibitemOpen
  \bibfield  {author} {\bibinfo {author} {\bibfnamefont {A.}~\bibnamefont
  {Montanaro}},\ }\href@noop {} {\enquote {\bibinfo {title} {Quantum walk
  speedup of backtracking algorithms},}\ } (\bibinfo {year}
  {2015}{\natexlab{a}}),\ \Eprint {http://arxiv.org/abs/1509.02374}
  {arXiv:1509.02374} \BibitemShut {NoStop}%
\bibitem [{\citenamefont {Campbell}\ \emph {et~al.}(2019)\citenamefont
  {Campbell}, \citenamefont {Khurana},\ and\ \citenamefont
  {Montanaro}}]{Campbell2018ApplyingProblems}%
  \BibitemOpen
  \bibfield  {author} {\bibinfo {author} {\bibfnamefont {E.}~\bibnamefont
  {Campbell}}, \bibinfo {author} {\bibfnamefont {A.}~\bibnamefont {Khurana}}, \
  and\ \bibinfo {author} {\bibfnamefont {A.}~\bibnamefont {Montanaro}},\ }\href
  {\doibase 10.22331/q-2019-07-18-167} {\bibfield  {journal} {\bibinfo
  {journal} {Quantum}\ }\textbf {\bibinfo {volume} {3}},\ \bibinfo {pages}
  {167} (\bibinfo {year} {2019})}\BibitemShut {NoStop}%
\bibitem [{\citenamefont {Montanaro}(2020)}]{Montanaro2020QuantumAlgorithms}%
  \BibitemOpen
  \bibfield  {author} {\bibinfo {author} {\bibfnamefont {A.}~\bibnamefont
  {Montanaro}},\ }\href {\doibase 10.1103/physrevresearch.2.013056} {\bibfield
  {journal} {\bibinfo  {journal} {Physical Review Research}\ }\textbf {\bibinfo
  {volume} {2}},\ \bibinfo {pages} {013056} (\bibinfo {year}
  {2020})}\BibitemShut {NoStop}%
\bibitem [{\citenamefont {Kitaev}(2003)}]{Kitaev1997Fault-tolerantAnyons}%
  \BibitemOpen
  \bibfield  {author} {\bibinfo {author} {\bibfnamefont {A.}~\bibnamefont
  {Kitaev}},\ }\href {\doibase 10.1016/s0003-4916(02)00018-0} {\bibfield
  {journal} {\bibinfo  {journal} {Annals of Physics}\ }\textbf {\bibinfo
  {volume} {303}},\ \bibinfo {pages} {2} (\bibinfo {year} {2003})}\BibitemShut
  {NoStop}%
\bibitem [{\citenamefont {Fowler}\ \emph {et~al.}(2012)\citenamefont {Fowler},
  \citenamefont {Mariantoni}, \citenamefont {Martinis},\ and\ \citenamefont
  {Cleland}}]{Fowler2012}%
  \BibitemOpen
  \bibfield  {author} {\bibinfo {author} {\bibfnamefont {A.~G.}\ \bibnamefont
  {Fowler}}, \bibinfo {author} {\bibfnamefont {M.}~\bibnamefont {Mariantoni}},
  \bibinfo {author} {\bibfnamefont {J.~M.}\ \bibnamefont {Martinis}}, \ and\
  \bibinfo {author} {\bibfnamefont {A.~N.}\ \bibnamefont {Cleland}},\ }\href
  {\doibase 10.1103/physreva.86.032324} {\bibfield  {journal} {\bibinfo
  {journal} {Physical Review A}\ }\textbf {\bibinfo {volume} {86}},\ \bibinfo
  {pages} {032324} (\bibinfo {year} {2012})}\BibitemShut {NoStop}%
\bibitem [{\citenamefont {Brassard}\ \emph {et~al.}(2002)\citenamefont
  {Brassard}, \citenamefont {H{\o}yer}, \citenamefont {Mosca},\ and\
  \citenamefont {Tapp}}]{Brassard2002}%
  \BibitemOpen
  \bibfield  {author} {\bibinfo {author} {\bibfnamefont {G.}~\bibnamefont
  {Brassard}}, \bibinfo {author} {\bibfnamefont {P.}~\bibnamefont {H{\o}yer}},
  \bibinfo {author} {\bibfnamefont {M.}~\bibnamefont {Mosca}}, \ and\ \bibinfo
  {author} {\bibfnamefont {A.}~\bibnamefont {Tapp}},\ }in\ \href {\doibase
  10.1090/conm/305/05215} {\emph {\bibinfo {booktitle} {Quantum Computation and
  Information}}},\ \bibinfo {editor} {edited by\ \bibinfo {editor}
  {\bibnamefont {{Vitaly I Voloshin}}}, \bibinfo {editor} {\bibnamefont
  {{Samuel J. Lomonaco}}}, \ and\ \bibinfo {editor} {\bibnamefont {{Howard E.
  Brandt}}}}\ (\bibinfo  {publisher} {American Mathematical Society},\ \bibinfo
  {address} {Washington D.C.},\ \bibinfo {year} {2002})\ Chap.~\bibinfo
  {chapter} {3}, pp.\ \bibinfo {pages} {53--74}\BibitemShut {NoStop}%
\bibitem [{\citenamefont {Boixo}\ \emph {et~al.}(2009)\citenamefont {Boixo},
  \citenamefont {Knill},\ and\ \citenamefont {Somma}}]{Boixo2009a}%
  \BibitemOpen
  \bibfield  {author} {\bibinfo {author} {\bibfnamefont {S.}~\bibnamefont
  {Boixo}}, \bibinfo {author} {\bibfnamefont {E.}~\bibnamefont {Knill}}, \ and\
  \bibinfo {author} {\bibfnamefont {R.}~\bibnamefont {Somma}},\ }\href
  {https://arxiv.org/abs/0903.1652} {\bibfield  {journal} {\bibinfo  {journal}
  {Quantum Information {\&} Computation}\ }\textbf {\bibinfo {volume} {9}},\
  \bibinfo {pages} {0833} (\bibinfo {year} {2009})}\BibitemShut {NoStop}%
\bibitem [{\citenamefont {{Szegedy}}(2004)}]{Szegedy2004}%
  \BibitemOpen
  \bibfield  {author} {\bibinfo {author} {\bibfnamefont {M.}~\bibnamefont
  {{Szegedy}}},\ }in\ \href {https://doi.org/10.1109/focs.2004.53} {\emph
  {\bibinfo {booktitle} {45th Annual IEEE Symposium on Foundations of Computer
  Science}}}\ (\bibinfo {year} {2004})\ pp.\ \bibinfo {pages}
  {32--41}\BibitemShut {NoStop}%
\bibitem [{\citenamefont {Lemieux}\ \emph
  {et~al.}(2020{\natexlab{a}})\citenamefont {Lemieux}, \citenamefont {Heim},
  \citenamefont {Poulin}, \citenamefont {Svore},\ and\ \citenamefont
  {Troyer}}]{lemieux2019efficient}%
  \BibitemOpen
  \bibfield  {author} {\bibinfo {author} {\bibfnamefont {J.}~\bibnamefont
  {Lemieux}}, \bibinfo {author} {\bibfnamefont {B.}~\bibnamefont {Heim}},
  \bibinfo {author} {\bibfnamefont {D.}~\bibnamefont {Poulin}}, \bibinfo
  {author} {\bibfnamefont {K.}~\bibnamefont {Svore}}, \ and\ \bibinfo {author}
  {\bibfnamefont {M.}~\bibnamefont {Troyer}},\ }\href {\doibase
  10.22331/q-2020-06-29-287} {\bibfield  {journal} {\bibinfo  {journal}
  {{Quantum}}\ }\textbf {\bibinfo {volume} {4}},\ \bibinfo {pages} {287}
  (\bibinfo {year} {2020}{\natexlab{a}})}\BibitemShut {NoStop}%
\bibitem [{\citenamefont {Kirkpatrick}\ \emph {et~al.}(1983)\citenamefont
  {Kirkpatrick}, \citenamefont {Gelatt},\ and\ \citenamefont
  {Vecchi}}]{Kirkpatrick671}%
  \BibitemOpen
  \bibfield  {author} {\bibinfo {author} {\bibfnamefont {S.}~\bibnamefont
  {Kirkpatrick}}, \bibinfo {author} {\bibfnamefont {C.~D.}\ \bibnamefont
  {Gelatt}}, \ and\ \bibinfo {author} {\bibfnamefont {M.~P.}\ \bibnamefont
  {Vecchi}},\ }\href {https://doi.org/10.1126/science.220.4598.671} {\bibfield
  {journal} {\bibinfo  {journal} {Science}\ }\textbf {\bibinfo {volume}
  {220}},\ \bibinfo {pages} {671} (\bibinfo {year} {1983})}\BibitemShut
  {NoStop}%
\bibitem [{\citenamefont {Kivlichan}\ \emph {et~al.}(2020)\citenamefont
  {Kivlichan}, \citenamefont {Gidney}, \citenamefont {Berry}, \citenamefont
  {Wiebe}, \citenamefont {McClean}, \citenamefont {Sun}, \citenamefont {Jiang},
  \citenamefont {Rubin}, \citenamefont {Fowler}, \citenamefont {Aspuru-Guzik},
  \citenamefont {Neven},\ and\ \citenamefont {Babbush}}]{Kivlichan2019}%
  \BibitemOpen
  \bibfield  {author} {\bibinfo {author} {\bibfnamefont {I.~D.}\ \bibnamefont
  {Kivlichan}}, \bibinfo {author} {\bibfnamefont {C.}~\bibnamefont {Gidney}},
  \bibinfo {author} {\bibfnamefont {D.~W.}\ \bibnamefont {Berry}}, \bibinfo
  {author} {\bibfnamefont {N.}~\bibnamefont {Wiebe}}, \bibinfo {author}
  {\bibfnamefont {J.}~\bibnamefont {McClean}}, \bibinfo {author} {\bibfnamefont
  {W.}~\bibnamefont {Sun}}, \bibinfo {author} {\bibfnamefont {Z.}~\bibnamefont
  {Jiang}}, \bibinfo {author} {\bibfnamefont {N.}~\bibnamefont {Rubin}},
  \bibinfo {author} {\bibfnamefont {A.}~\bibnamefont {Fowler}}, \bibinfo
  {author} {\bibfnamefont {A.}~\bibnamefont {Aspuru-Guzik}}, \bibinfo {author}
  {\bibfnamefont {H.}~\bibnamefont {Neven}}, \ and\ \bibinfo {author}
  {\bibfnamefont {R.}~\bibnamefont {Babbush}},\ }\href
  {https://quantum-journal.org/papers/q-2020-07-16-296/} {\bibfield  {journal}
  {\bibinfo  {journal} {Quantum}\ }\textbf {\bibinfo {volume} {4}},\ \bibinfo
  {pages} {296} (\bibinfo {year} {2020})}\BibitemShut {NoStop}%
\bibitem [{\citenamefont {Babbush}\ \emph {et~al.}(2018)\citenamefont
  {Babbush}, \citenamefont {Gidney}, \citenamefont {Berry}, \citenamefont
  {Wiebe}, \citenamefont {McClean}, \citenamefont {Paler}, \citenamefont
  {Fowler},\ and\ \citenamefont {Neven}}]{Babbush2018}%
  \BibitemOpen
  \bibfield  {author} {\bibinfo {author} {\bibfnamefont {R.}~\bibnamefont
  {Babbush}}, \bibinfo {author} {\bibfnamefont {C.}~\bibnamefont {Gidney}},
  \bibinfo {author} {\bibfnamefont {D.~W.}\ \bibnamefont {Berry}}, \bibinfo
  {author} {\bibfnamefont {N.}~\bibnamefont {Wiebe}}, \bibinfo {author}
  {\bibfnamefont {J.}~\bibnamefont {McClean}}, \bibinfo {author} {\bibfnamefont
  {A.}~\bibnamefont {Paler}}, \bibinfo {author} {\bibfnamefont
  {A.}~\bibnamefont {Fowler}}, \ and\ \bibinfo {author} {\bibfnamefont
  {H.}~\bibnamefont {Neven}},\ }\href {\doibase 10.1103/physrevx.8.041015}
  {\bibfield  {journal} {\bibinfo  {journal} {Physical Review X}\ }\textbf
  {\bibinfo {volume} {8}},\ \bibinfo {pages} {041015} (\bibinfo {year}
  {2018})}\BibitemShut {NoStop}%
\bibitem [{\citenamefont {Sanders}\ \emph {et~al.}(2019)\citenamefont
  {Sanders}, \citenamefont {Low}, \citenamefont {Scherer},\ and\ \citenamefont
  {Berry}}]{SLSB19}%
  \BibitemOpen
  \bibfield  {author} {\bibinfo {author} {\bibfnamefont {Y.~R.}\ \bibnamefont
  {Sanders}}, \bibinfo {author} {\bibfnamefont {G.~H.}\ \bibnamefont {Low}},
  \bibinfo {author} {\bibfnamefont {A.}~\bibnamefont {Scherer}}, \ and\
  \bibinfo {author} {\bibfnamefont {D.~W.}\ \bibnamefont {Berry}},\ }\href
  {\doibase 10.1103/physrevlett.122.020502} {\bibfield  {journal} {\bibinfo
  {journal} {Physical Review Letters}\ }\textbf {\bibinfo {volume} {122}},\
  \bibinfo {pages} {020502} (\bibinfo {year} {2019})}\BibitemShut {NoStop}%
\bibitem [{\citenamefont {Low}\ and\ \citenamefont {Chuang}(2019)}]{Low2016}%
  \BibitemOpen
  \bibfield  {author} {\bibinfo {author} {\bibfnamefont {G.~H.}\ \bibnamefont
  {Low}}\ and\ \bibinfo {author} {\bibfnamefont {I.~L.}\ \bibnamefont
  {Chuang}},\ }\href {\doibase 10.22331/q-2019-07-12-163} {\bibfield  {journal}
  {\bibinfo  {journal} {Quantum}\ }\textbf {\bibinfo {volume} {3}},\ \bibinfo
  {pages} {163} (\bibinfo {year} {2019})}\BibitemShut {NoStop}%
\bibitem [{\citenamefont {Boixo}\ \emph {et~al.}(2014)\citenamefont {Boixo},
  \citenamefont {Ronnow}, \citenamefont {Isakov}, \citenamefont {Wang},
  \citenamefont {Wecker}, \citenamefont {Lidar}, \citenamefont {Martinis},\
  and\ \citenamefont {Troyer}}]{Boixo2014}%
  \BibitemOpen
  \bibfield  {author} {\bibinfo {author} {\bibfnamefont {S.}~\bibnamefont
  {Boixo}}, \bibinfo {author} {\bibfnamefont {T.~F.}\ \bibnamefont {Ronnow}},
  \bibinfo {author} {\bibfnamefont {S.~V.}\ \bibnamefont {Isakov}}, \bibinfo
  {author} {\bibfnamefont {Z.}~\bibnamefont {Wang}}, \bibinfo {author}
  {\bibfnamefont {D.}~\bibnamefont {Wecker}}, \bibinfo {author} {\bibfnamefont
  {D.~A.}\ \bibnamefont {Lidar}}, \bibinfo {author} {\bibfnamefont {J.~M.}\
  \bibnamefont {Martinis}}, \ and\ \bibinfo {author} {\bibfnamefont
  {M.}~\bibnamefont {Troyer}},\ }\href {\doibase 10.1038/nphys2900} {\bibfield
  {journal} {\bibinfo  {journal} {Nature Physics}\ }\textbf {\bibinfo {volume}
  {10}},\ \bibinfo {pages} {218} (\bibinfo {year} {2014})}\BibitemShut
  {NoStop}%
\bibitem [{\citenamefont {Bernasconi}(1987)}]{Bernasconi1987}%
  \BibitemOpen
  \bibfield  {author} {\bibinfo {author} {\bibfnamefont {J.}~\bibnamefont
  {Bernasconi}},\ }\href {\doibase 10.1051/jphys:01987004804055900} {\bibfield
  {journal} {\bibinfo  {journal} {Journal de Physique}\ }\textbf {\bibinfo
  {volume} {48}},\ \bibinfo {pages} {559} (\bibinfo {year} {1987})}\BibitemShut
  {NoStop}%
\bibitem [{\citenamefont {Packebusch}\ and\ \citenamefont
  {Mertens}(2016)}]{Packebusch2016}%
  \BibitemOpen
  \bibfield  {author} {\bibinfo {author} {\bibfnamefont {T.}~\bibnamefont
  {Packebusch}}\ and\ \bibinfo {author} {\bibfnamefont {S.}~\bibnamefont
  {Mertens}},\ }\href {\doibase 10.1088/1751-8113/49/16/165001} {\bibfield
  {journal} {\bibinfo  {journal} {Journal of Physics A: Mathematical and
  Theoretical}\ }\textbf {\bibinfo {volume} {49}},\ \bibinfo {pages} {165001}
  (\bibinfo {year} {2016})}\BibitemShut {NoStop}%
\bibitem [{\citenamefont {Gidney}(2018)}]{GidneyAdder}%
  \BibitemOpen
  \bibfield  {author} {\bibinfo {author} {\bibfnamefont {C.}~\bibnamefont
  {Gidney}},\ }\href {\doibase 10.22331/q-2018-06-18-74} {\bibfield  {journal}
  {\bibinfo  {journal} {Quantum}\ }\textbf {\bibinfo {volume} {2}},\ \bibinfo
  {pages} {74} (\bibinfo {year} {2018})}\BibitemShut {NoStop}%
\bibitem [{\citenamefont {Kitaev}\ \emph {et~al.}(2002)\citenamefont {Kitaev},
  \citenamefont {Shen},\ and\ \citenamefont {Vyalyi}}]{Kitaev2002}%
  \BibitemOpen
  \bibfield  {author} {\bibinfo {author} {\bibfnamefont {A.~Y.}\ \bibnamefont
  {Kitaev}}, \bibinfo {author} {\bibfnamefont {A.~H.}\ \bibnamefont {Shen}}, \
  and\ \bibinfo {author} {\bibfnamefont {M.~N.}\ \bibnamefont {Vyalyi}},\
  }\href@noop {} {\emph {\bibinfo {title} {Graduate Studies in Mathematics}}},\
  Vol.~\bibinfo {volume} {47}\ (\bibinfo  {publisher} {American Mathematical
  Society},\ \bibinfo {address} {Providence, Rhode Island},\ \bibinfo {year}
  {2002})\BibitemShut {NoStop}%
\bibitem [{\citenamefont {Bocharov}\ \emph {et~al.}(2015)\citenamefont
  {Bocharov}, \citenamefont {Roetteler},\ and\ \citenamefont
  {Svore}}]{Bocharov2015}%
  \BibitemOpen
  \bibfield  {author} {\bibinfo {author} {\bibfnamefont {A.}~\bibnamefont
  {Bocharov}}, \bibinfo {author} {\bibfnamefont {M.}~\bibnamefont {Roetteler}},
  \ and\ \bibinfo {author} {\bibfnamefont {K.~M.}\ \bibnamefont {Svore}},\
  }\href {\doibase 10.1103/physrevlett.114.080502} {\bibfield  {journal}
  {\bibinfo  {journal} {Physical Review Letters}\ }\textbf {\bibinfo {volume}
  {114}},\ \bibinfo {pages} {080502} (\bibinfo {year} {2015})}\BibitemShut
  {NoStop}%
\bibitem [{\citenamefont {Gidney}\ and\ \citenamefont
  {Fowler}(2019)}]{Gidney2019}%
  \BibitemOpen
  \bibfield  {author} {\bibinfo {author} {\bibfnamefont {C.}~\bibnamefont
  {Gidney}}\ and\ \bibinfo {author} {\bibfnamefont {A.~G.}\ \bibnamefont
  {Fowler}},\ }\href {\doibase 10.22331/q-2019-04-30-135} {\bibfield  {journal}
  {\bibinfo  {journal} {Quantum}\ }\textbf {\bibinfo {volume} {3}},\ \bibinfo
  {pages} {135} (\bibinfo {year} {2019})}\BibitemShut {NoStop}%
\bibitem [{\citenamefont {Childs}\ and\ \citenamefont
  {Wiebe}(2012)}]{Childs2012}%
  \BibitemOpen
  \bibfield  {author} {\bibinfo {author} {\bibfnamefont {A.~M.}\ \bibnamefont
  {Childs}}\ and\ \bibinfo {author} {\bibfnamefont {N.}~\bibnamefont {Wiebe}},\
  }\href {https://dl.acm.org/citation.cfm?id=2481570} {\bibfield  {journal}
  {\bibinfo  {journal} {Quantum Information {\&} Computation}\ }\textbf
  {\bibinfo {volume} {12}},\ \bibinfo {pages} {901} (\bibinfo {year}
  {2012})}\BibitemShut {NoStop}%
\bibitem [{\citenamefont {Berry}\ \emph {et~al.}(2015)\citenamefont {Berry},
  \citenamefont {Childs}, \citenamefont {Cleve}, \citenamefont {Kothari},\ and\
  \citenamefont {Somma}}]{Berry2015}%
  \BibitemOpen
  \bibfield  {author} {\bibinfo {author} {\bibfnamefont {D.~W.}\ \bibnamefont
  {Berry}}, \bibinfo {author} {\bibfnamefont {A.~M.}\ \bibnamefont {Childs}},
  \bibinfo {author} {\bibfnamefont {R.}~\bibnamefont {Cleve}}, \bibinfo
  {author} {\bibfnamefont {R.}~\bibnamefont {Kothari}}, \ and\ \bibinfo
  {author} {\bibfnamefont {R.~D.}\ \bibnamefont {Somma}},\ }\href
  {http://link.aps.org/doi/10.1103/PhysRevLett.114.090502} {\bibfield
  {journal} {\bibinfo  {journal} {Physical Review Letters}\ }\textbf {\bibinfo
  {volume} {114}},\ \bibinfo {pages} {090502} (\bibinfo {year}
  {2015})}\BibitemShut {NoStop}%
\bibitem [{\citenamefont {Low}\ and\ \citenamefont {Wiebe}(2018)}]{Low2018}%
  \BibitemOpen
  \bibfield  {author} {\bibinfo {author} {\bibfnamefont {G.~H.}\ \bibnamefont
  {Low}}\ and\ \bibinfo {author} {\bibfnamefont {N.}~\bibnamefont {Wiebe}},\
  }\href@noop {} {\enquote {\bibinfo {title} {Hamiltonian simulation in the
  interaction picture},}\ } (\bibinfo {year} {2018}),\ \Eprint
  {http://arxiv.org/abs/1805.00675} {1805.00675} \BibitemShut {NoStop}%
\bibitem [{\citenamefont {Berry}\ \emph {et~al.}(2019)\citenamefont {Berry},
  \citenamefont {Gidney}, \citenamefont {Motta}, \citenamefont {McClean},\ and\
  \citenamefont {Babbush}}]{BGMMB19}%
  \BibitemOpen
  \bibfield  {author} {\bibinfo {author} {\bibfnamefont {D.~W.}\ \bibnamefont
  {Berry}}, \bibinfo {author} {\bibfnamefont {C.}~\bibnamefont {Gidney}},
  \bibinfo {author} {\bibfnamefont {M.}~\bibnamefont {Motta}}, \bibinfo
  {author} {\bibfnamefont {J.~R.}\ \bibnamefont {McClean}}, \ and\ \bibinfo
  {author} {\bibfnamefont {R.}~\bibnamefont {Babbush}},\ }\href
  {https://doi.org/10.22331/q-2019-12-02-208} {\bibfield  {journal} {\bibinfo
  {journal} {Quantum}\ }\textbf {\bibinfo {volume} {3}},\ \bibinfo {pages}
  {208} (\bibinfo {year} {2019})}\BibitemShut {NoStop}%
\bibitem [{\citenamefont {Low}\ \emph {et~al.}(2018)\citenamefont {Low},
  \citenamefont {Kliuchnikov},\ and\ \citenamefont {Schaeffer}}]{Low2018a}%
  \BibitemOpen
  \bibfield  {author} {\bibinfo {author} {\bibfnamefont {G.~H.}\ \bibnamefont
  {Low}}, \bibinfo {author} {\bibfnamefont {V.}~\bibnamefont {Kliuchnikov}}, \
  and\ \bibinfo {author} {\bibfnamefont {L.}~\bibnamefont {Schaeffer}},\
  }\href@noop {} {\enquote {\bibinfo {title} {Trading {T}-gates for dirty
  qubits in state preparation and unitary synthesis},}\ } (\bibinfo {year}
  {2018}),\ \Eprint {http://arxiv.org/abs/1812.00954} {arXiv:1812.00954}
  \BibitemShut {NoStop}%
\bibitem [{\citenamefont {Yoder}\ \emph {et~al.}(2014)\citenamefont {Yoder},
  \citenamefont {Low},\ and\ \citenamefont {Chuang}}]{yoder}%
  \BibitemOpen
  \bibfield  {author} {\bibinfo {author} {\bibfnamefont {T.~J.}\ \bibnamefont
  {Yoder}}, \bibinfo {author} {\bibfnamefont {G.~H.}\ \bibnamefont {Low}}, \
  and\ \bibinfo {author} {\bibfnamefont {I.~L.}\ \bibnamefont {Chuang}},\
  }\href {\doibase 10.1103/physrevlett.113.210501} {\bibfield  {journal}
  {\bibinfo  {journal} {Physical Review Letters}\ }\textbf {\bibinfo {volume}
  {113}},\ \bibinfo {pages} {210501} (\bibinfo {year} {2014})}\BibitemShut
  {NoStop}%
\bibitem [{\citenamefont {Barak}\ \emph {et~al.}(2015)\citenamefont {Barak},
  \citenamefont {Moitra}, \citenamefont {O'Donnell}, \citenamefont
  {Raghavendra}, \citenamefont {Regev}, \citenamefont {Steurer}, \citenamefont
  {Trevisan}, \citenamefont {Vijayaraghavan}, \citenamefont {Witmer},\ and\
  \citenamefont {Wright}}]{Barak2015}%
  \BibitemOpen
  \bibfield  {author} {\bibinfo {author} {\bibfnamefont {B.}~\bibnamefont
  {Barak}}, \bibinfo {author} {\bibfnamefont {A.}~\bibnamefont {Moitra}},
  \bibinfo {author} {\bibfnamefont {R.}~\bibnamefont {O'Donnell}}, \bibinfo
  {author} {\bibfnamefont {P.}~\bibnamefont {Raghavendra}}, \bibinfo {author}
  {\bibfnamefont {O.}~\bibnamefont {Regev}}, \bibinfo {author} {\bibfnamefont
  {D.}~\bibnamefont {Steurer}}, \bibinfo {author} {\bibfnamefont
  {L.}~\bibnamefont {Trevisan}}, \bibinfo {author} {\bibfnamefont
  {A.}~\bibnamefont {Vijayaraghavan}}, \bibinfo {author} {\bibfnamefont
  {D.}~\bibnamefont {Witmer}}, \ and\ \bibinfo {author} {\bibfnamefont
  {J.}~\bibnamefont {Wright}},\ }in\ \href
  {https://doi.org/10.4230/LIPIcs.APPROX-RANDOM.2015.110} {\emph {\bibinfo
  {booktitle} {Approximation, Randomization, and Combinatorial Optimization.
  Algorithms and Techniques, {APPROX/RANDOM} 2015, August 24-26, 2015,
  Princeton, NJ, {USA}}}},\ \bibinfo {series} {LIPIcs}, Vol.~\bibinfo {volume}
  {40},\ \bibinfo {editor} {edited by\ \bibinfo {editor} {\bibfnamefont
  {N.}~\bibnamefont {Garg}}, \bibinfo {editor} {\bibfnamefont {K.}~\bibnamefont
  {Jansen}}, \bibinfo {editor} {\bibfnamefont {A.}~\bibnamefont {Rao}}, \ and\
  \bibinfo {editor} {\bibfnamefont {J.~D.~P.}\ \bibnamefont {Rolim}}}\
  (\bibinfo  {publisher} {Schloss Dagstuhl - Leibniz-Zentrum f{\"{u}}r
  Informatik},\ \bibinfo {year} {2015})\ pp.\ \bibinfo {pages}
  {110--123}\BibitemShut {NoStop}%
\bibitem [{\citenamefont {Arute}\ \emph {et~al.}(2020)\citenamefont {Arute},
  \citenamefont {Arya}, \citenamefont {Babbush}, \citenamefont {Bacon},
  \citenamefont {Bardin}, \citenamefont {Barends}, \citenamefont {Boixo},
  \citenamefont {Broughton}, \citenamefont {Buckley}, \citenamefont {Buell},
  \citenamefont {Burkett}, \citenamefont {Bushnell}, \citenamefont {Chen},
  \citenamefont {Chen}, \citenamefont {Chiaro}, \citenamefont {Collins},
  \citenamefont {Courtney}, \citenamefont {Demura}, \citenamefont {Dunsworth},
  \citenamefont {Farhi}, \citenamefont {Fowler}, \citenamefont {Foxen},
  \citenamefont {Gidney}, \citenamefont {Giustina}, \citenamefont {Graff},
  \citenamefont {Habegger}, \citenamefont {Harrigan}, \citenamefont {Ho},
  \citenamefont {Hong}, \citenamefont {Huang}, \citenamefont {Ioffe},
  \citenamefont {Isakov}, \citenamefont {Jeffrey}, \citenamefont {Jiang},
  \citenamefont {Jones}, \citenamefont {Kafri}, \citenamefont {Kechedzhi},
  \citenamefont {Kelly}, \citenamefont {Kim}, \citenamefont {Klimov},
  \citenamefont {Korotkov}, \citenamefont {Kostritsa}, \citenamefont
  {Landhuis}, \citenamefont {Laptev}, \citenamefont {Lindmark}, \citenamefont
  {Leib}, \citenamefont {Lucero}, \citenamefont {Martin}, \citenamefont
  {Martinis}, \citenamefont {McClean}, \citenamefont {McEwen}, \citenamefont
  {Megrant}, \citenamefont {Mi}, \citenamefont {Mohseni}, \citenamefont
  {Mruczkiewicz}, \citenamefont {Mutus}, \citenamefont {Naaman}, \citenamefont
  {Neeley}, \citenamefont {Neill}, \citenamefont {Neukart}, \citenamefont
  {Neven}, \citenamefont {Niu}, \citenamefont {O'Brien}, \citenamefont
  {O'Gorman}, \citenamefont {Ostby}, \citenamefont {Petukhov}, \citenamefont
  {Putterman}, \citenamefont {Quintana}, \citenamefont {Roushan}, \citenamefont
  {Rubin}, \citenamefont {Sank}, \citenamefont {Satzinger}, \citenamefont
  {Skolik}, \citenamefont {Smelyanskiy}, \citenamefont {Strain}, \citenamefont
  {Streif}, \citenamefont {Sung}, \citenamefont {Szalay}, \citenamefont
  {Vainsencher}, \citenamefont {White}, \citenamefont {Yao}, \citenamefont
  {Yeh}, \citenamefont {Zalcman},\ and\ \citenamefont
  {Zhou}}]{Arute2020QuantumProcessor}%
  \BibitemOpen
  \bibfield  {author} {\bibinfo {author} {\bibfnamefont {F.}~\bibnamefont
  {Arute}}, \bibinfo {author} {\bibfnamefont {K.}~\bibnamefont {Arya}},
  \bibinfo {author} {\bibfnamefont {R.}~\bibnamefont {Babbush}}, \bibinfo
  {author} {\bibfnamefont {D.}~\bibnamefont {Bacon}}, \bibinfo {author}
  {\bibfnamefont {J.~C.}\ \bibnamefont {Bardin}}, \bibinfo {author}
  {\bibfnamefont {R.}~\bibnamefont {Barends}}, \bibinfo {author} {\bibfnamefont
  {S.}~\bibnamefont {Boixo}}, \bibinfo {author} {\bibfnamefont
  {M.}~\bibnamefont {Broughton}}, \bibinfo {author} {\bibfnamefont {B.~B.}\
  \bibnamefont {Buckley}}, \bibinfo {author} {\bibfnamefont {D.~A.}\
  \bibnamefont {Buell}}, \bibinfo {author} {\bibfnamefont {B.}~\bibnamefont
  {Burkett}}, \bibinfo {author} {\bibfnamefont {N.}~\bibnamefont {Bushnell}},
  \bibinfo {author} {\bibfnamefont {Y.}~\bibnamefont {Chen}}, \bibinfo {author}
  {\bibfnamefont {Z.}~\bibnamefont {Chen}}, \bibinfo {author} {\bibfnamefont
  {B.}~\bibnamefont {Chiaro}}, \bibinfo {author} {\bibfnamefont
  {R.}~\bibnamefont {Collins}}, \bibinfo {author} {\bibfnamefont
  {W.}~\bibnamefont {Courtney}}, \bibinfo {author} {\bibfnamefont
  {S.}~\bibnamefont {Demura}}, \bibinfo {author} {\bibfnamefont
  {A.}~\bibnamefont {Dunsworth}}, \bibinfo {author} {\bibfnamefont
  {E.}~\bibnamefont {Farhi}}, \bibinfo {author} {\bibfnamefont
  {A.}~\bibnamefont {Fowler}}, \bibinfo {author} {\bibfnamefont
  {B.}~\bibnamefont {Foxen}}, \bibinfo {author} {\bibfnamefont
  {C.}~\bibnamefont {Gidney}}, \bibinfo {author} {\bibfnamefont
  {M.}~\bibnamefont {Giustina}}, \bibinfo {author} {\bibfnamefont
  {R.}~\bibnamefont {Graff}}, \bibinfo {author} {\bibfnamefont
  {S.}~\bibnamefont {Habegger}}, \bibinfo {author} {\bibfnamefont {M.~P.}\
  \bibnamefont {Harrigan}}, \bibinfo {author} {\bibfnamefont {A.}~\bibnamefont
  {Ho}}, \bibinfo {author} {\bibfnamefont {S.}~\bibnamefont {Hong}}, \bibinfo
  {author} {\bibfnamefont {T.}~\bibnamefont {Huang}}, \bibinfo {author}
  {\bibfnamefont {L.~B.}\ \bibnamefont {Ioffe}}, \bibinfo {author}
  {\bibfnamefont {S.~V.}\ \bibnamefont {Isakov}}, \bibinfo {author}
  {\bibfnamefont {E.}~\bibnamefont {Jeffrey}}, \bibinfo {author} {\bibfnamefont
  {Z.}~\bibnamefont {Jiang}}, \bibinfo {author} {\bibfnamefont
  {C.}~\bibnamefont {Jones}}, \bibinfo {author} {\bibfnamefont
  {D.}~\bibnamefont {Kafri}}, \bibinfo {author} {\bibfnamefont
  {K.}~\bibnamefont {Kechedzhi}}, \bibinfo {author} {\bibfnamefont
  {J.}~\bibnamefont {Kelly}}, \bibinfo {author} {\bibfnamefont
  {S.}~\bibnamefont {Kim}}, \bibinfo {author} {\bibfnamefont {P.~V.}\
  \bibnamefont {Klimov}}, \bibinfo {author} {\bibfnamefont {A.~N.}\
  \bibnamefont {Korotkov}}, \bibinfo {author} {\bibfnamefont {F.}~\bibnamefont
  {Kostritsa}}, \bibinfo {author} {\bibfnamefont {D.}~\bibnamefont {Landhuis}},
  \bibinfo {author} {\bibfnamefont {P.}~\bibnamefont {Laptev}}, \bibinfo
  {author} {\bibfnamefont {M.}~\bibnamefont {Lindmark}}, \bibinfo {author}
  {\bibfnamefont {M.}~\bibnamefont {Leib}}, \bibinfo {author} {\bibfnamefont
  {E.}~\bibnamefont {Lucero}}, \bibinfo {author} {\bibfnamefont
  {O.}~\bibnamefont {Martin}}, \bibinfo {author} {\bibfnamefont {J.~M.}\
  \bibnamefont {Martinis}}, \bibinfo {author} {\bibfnamefont {J.~R.}\
  \bibnamefont {McClean}}, \bibinfo {author} {\bibfnamefont {M.}~\bibnamefont
  {McEwen}}, \bibinfo {author} {\bibfnamefont {A.}~\bibnamefont {Megrant}},
  \bibinfo {author} {\bibfnamefont {X.}~\bibnamefont {Mi}}, \bibinfo {author}
  {\bibfnamefont {M.}~\bibnamefont {Mohseni}}, \bibinfo {author} {\bibfnamefont
  {W.}~\bibnamefont {Mruczkiewicz}}, \bibinfo {author} {\bibfnamefont
  {J.}~\bibnamefont {Mutus}}, \bibinfo {author} {\bibfnamefont
  {O.}~\bibnamefont {Naaman}}, \bibinfo {author} {\bibfnamefont
  {M.}~\bibnamefont {Neeley}}, \bibinfo {author} {\bibfnamefont
  {C.}~\bibnamefont {Neill}}, \bibinfo {author} {\bibfnamefont
  {F.}~\bibnamefont {Neukart}}, \bibinfo {author} {\bibfnamefont
  {H.}~\bibnamefont {Neven}}, \bibinfo {author} {\bibfnamefont {M.~Y.}\
  \bibnamefont {Niu}}, \bibinfo {author} {\bibfnamefont {T.~E.}\ \bibnamefont
  {O'Brien}}, \bibinfo {author} {\bibfnamefont {B.}~\bibnamefont {O'Gorman}},
  \bibinfo {author} {\bibfnamefont {E.}~\bibnamefont {Ostby}}, \bibinfo
  {author} {\bibfnamefont {A.}~\bibnamefont {Petukhov}}, \bibinfo {author}
  {\bibfnamefont {H.}~\bibnamefont {Putterman}}, \bibinfo {author}
  {\bibfnamefont {C.}~\bibnamefont {Quintana}}, \bibinfo {author}
  {\bibfnamefont {P.}~\bibnamefont {Roushan}}, \bibinfo {author} {\bibfnamefont
  {N.~C.}\ \bibnamefont {Rubin}}, \bibinfo {author} {\bibfnamefont
  {D.}~\bibnamefont {Sank}}, \bibinfo {author} {\bibfnamefont {K.~J.}\
  \bibnamefont {Satzinger}}, \bibinfo {author} {\bibfnamefont {A.}~\bibnamefont
  {Skolik}}, \bibinfo {author} {\bibfnamefont {V.}~\bibnamefont {Smelyanskiy}},
  \bibinfo {author} {\bibfnamefont {D.}~\bibnamefont {Strain}}, \bibinfo
  {author} {\bibfnamefont {M.}~\bibnamefont {Streif}}, \bibinfo {author}
  {\bibfnamefont {K.~J.}\ \bibnamefont {Sung}}, \bibinfo {author}
  {\bibfnamefont {M.}~\bibnamefont {Szalay}}, \bibinfo {author} {\bibfnamefont
  {A.}~\bibnamefont {Vainsencher}}, \bibinfo {author} {\bibfnamefont
  {T.}~\bibnamefont {White}}, \bibinfo {author} {\bibfnamefont {Z.~J.}\
  \bibnamefont {Yao}}, \bibinfo {author} {\bibfnamefont {P.}~\bibnamefont
  {Yeh}}, \bibinfo {author} {\bibfnamefont {A.}~\bibnamefont {Zalcman}}, \ and\
  \bibinfo {author} {\bibfnamefont {L.}~\bibnamefont {Zhou}},\ }\href@noop {}
  {\enquote {\bibinfo {title} {Quantum approximate optimization of non-planar
  graph problems on a planar superconducting processor},}\ } (\bibinfo {year}
  {2020}),\ \Eprint {http://arxiv.org/abs/2004.04197} {arXiv:2004.04197}
  \BibitemShut {NoStop}%
\bibitem [{\citenamefont {Peruzzo}\ \emph {et~al.}(2014)\citenamefont
  {Peruzzo}, \citenamefont {McClean}, \citenamefont {Shadbolt}, \citenamefont
  {Yung}, \citenamefont {Zhou}, \citenamefont {Love}, \citenamefont
  {Aspuru-Guzik},\ and\ \citenamefont {O'Brien}}]{Peruzzo2013}%
  \BibitemOpen
  \bibfield  {author} {\bibinfo {author} {\bibfnamefont {A.}~\bibnamefont
  {Peruzzo}}, \bibinfo {author} {\bibfnamefont {J.}~\bibnamefont {McClean}},
  \bibinfo {author} {\bibfnamefont {P.}~\bibnamefont {Shadbolt}}, \bibinfo
  {author} {\bibfnamefont {M.-H.}\ \bibnamefont {Yung}}, \bibinfo {author}
  {\bibfnamefont {X.-Q.}\ \bibnamefont {Zhou}}, \bibinfo {author}
  {\bibfnamefont {P.~J.}\ \bibnamefont {Love}}, \bibinfo {author}
  {\bibfnamefont {A.}~\bibnamefont {Aspuru-Guzik}}, \ and\ \bibinfo {author}
  {\bibfnamefont {J.~L.}\ \bibnamefont {O'Brien}},\ }\href
  {https://doi.org/10.1038/ncomms5213} {\bibfield  {journal} {\bibinfo
  {journal} {Nature Communications}\ }\textbf {\bibinfo {volume} {5}},\
  \bibinfo {pages} {4213} (\bibinfo {year} {2014})}\BibitemShut {NoStop}%
\bibitem [{\citenamefont {McClean}\ \emph {et~al.}(2016)\citenamefont
  {McClean}, \citenamefont {Romero}, \citenamefont {Babbush},\ and\
  \citenamefont {Aspuru-Guzik}}]{McClean2015}%
  \BibitemOpen
  \bibfield  {author} {\bibinfo {author} {\bibfnamefont {J.~R.}\ \bibnamefont
  {McClean}}, \bibinfo {author} {\bibfnamefont {J.}~\bibnamefont {Romero}},
  \bibinfo {author} {\bibfnamefont {R.}~\bibnamefont {Babbush}}, \ and\
  \bibinfo {author} {\bibfnamefont {A.}~\bibnamefont {Aspuru-Guzik}},\ }\href
  {\doibase 10.1088/1367-2630/18/2/023023} {\bibfield  {journal} {\bibinfo
  {journal} {New Journal of Physics}\ }\textbf {\bibinfo {volume} {18}},\
  \bibinfo {pages} {023023} (\bibinfo {year} {2016})}\BibitemShut {NoStop}%
\bibitem [{\citenamefont {Brandao}\ \emph {et~al.}(2018)\citenamefont
  {Brandao}, \citenamefont {Broughton}, \citenamefont {Farhi}, \citenamefont
  {Gutmann},\ and\ \citenamefont {Neven}}]{Brandao2018}%
  \BibitemOpen
  \bibfield  {author} {\bibinfo {author} {\bibfnamefont {F.~G. S.~L.}\
  \bibnamefont {Brandao}}, \bibinfo {author} {\bibfnamefont {M.}~\bibnamefont
  {Broughton}}, \bibinfo {author} {\bibfnamefont {E.}~\bibnamefont {Farhi}},
  \bibinfo {author} {\bibfnamefont {S.}~\bibnamefont {Gutmann}}, \ and\
  \bibinfo {author} {\bibfnamefont {H.}~\bibnamefont {Neven}},\ }\href@noop {}
  {\enquote {\bibinfo {title} {For fixed control parameters the quantum
  approximate optimization algorithm's objective function value concentrates
  for typical instances},}\ } (\bibinfo {year} {2018}),\ \Eprint
  {http://arxiv.org/abs/1812.04170} {arXiv:1812.04170} \BibitemShut {NoStop}%
\bibitem [{\citenamefont {Farhi}\ \emph {et~al.}(2019)\citenamefont {Farhi},
  \citenamefont {Goldstone}, \citenamefont {Gutmann},\ and\ \citenamefont
  {Zhou}}]{Farhi2019}%
  \BibitemOpen
  \bibfield  {author} {\bibinfo {author} {\bibfnamefont {E.}~\bibnamefont
  {Farhi}}, \bibinfo {author} {\bibfnamefont {J.}~\bibnamefont {Goldstone}},
  \bibinfo {author} {\bibfnamefont {S.}~\bibnamefont {Gutmann}}, \ and\
  \bibinfo {author} {\bibfnamefont {L.}~\bibnamefont {Zhou}},\ }\href@noop {}
  {\enquote {\bibinfo {title} {The quantum approximate optimization algorithm
  and the {Sherrington-Kirkpatrick} model at infinite size},}\ } (\bibinfo
  {year} {2019}),\ \Eprint {http://arxiv.org/abs/1910.08187} {arXiv:1910.08187}
  \BibitemShut {NoStop}%
\bibitem [{\citenamefont {Zhou}\ \emph {et~al.}(2020)\citenamefont {Zhou},
  \citenamefont {Wang}, \citenamefont {Choi}, \citenamefont {Pichler},\ and\
  \citenamefont {Lukin}}]{Zhou2019}%
  \BibitemOpen
  \bibfield  {author} {\bibinfo {author} {\bibfnamefont {L.}~\bibnamefont
  {Zhou}}, \bibinfo {author} {\bibfnamefont {S.-T.}\ \bibnamefont {Wang}},
  \bibinfo {author} {\bibfnamefont {S.}~\bibnamefont {Choi}}, \bibinfo {author}
  {\bibfnamefont {H.}~\bibnamefont {Pichler}}, \ and\ \bibinfo {author}
  {\bibfnamefont {M.~D.}\ \bibnamefont {Lukin}},\ }\href
  {https://link.aps.org/doi/10.1103/PhysRevX.10.021067} {\bibfield  {journal}
  {\bibinfo  {journal} {Physical Review X}\ }\textbf {\bibinfo {volume} {10}},\
  \bibinfo {pages} {021067} (\bibinfo {year} {2020})}\BibitemShut {NoStop}%
\bibitem [{\citenamefont {Gily{\'{e}}n}\ \emph {et~al.}(2019)\citenamefont
  {Gily{\'{e}}n}, \citenamefont {Arunachalam},\ and\ \citenamefont
  {Wiebe}}]{Gilyen2018}%
  \BibitemOpen
  \bibfield  {author} {\bibinfo {author} {\bibfnamefont {A.}~\bibnamefont
  {Gily{\'{e}}n}}, \bibinfo {author} {\bibfnamefont {S.}~\bibnamefont
  {Arunachalam}}, \ and\ \bibinfo {author} {\bibfnamefont {N.}~\bibnamefont
  {Wiebe}},\ }in\ \href {\doibase 10.1137/1.9781611975482.87} {\emph {\bibinfo
  {booktitle} {Proceedings of the Thirtieth Annual {ACM}-{SIAM} Symposium on
  Discrete Algorithms}}}\ (\bibinfo  {publisher} {Society for Industrial and
  Applied Mathematics},\ \bibinfo {year} {2019})\ pp.\ \bibinfo {pages}
  {1425--1444}\BibitemShut {NoStop}%
\bibitem [{\citenamefont {Montanaro}(2015{\natexlab{b}})}]{Montanaro2015}%
  \BibitemOpen
  \bibfield  {author} {\bibinfo {author} {\bibfnamefont {A.}~\bibnamefont
  {Montanaro}},\ }\href
  {https://royalsocietypublishing.org/doi/10.1098/rspa.2015.0301} {\bibfield
  {journal} {\bibinfo  {journal} {Proceedings of the Royal Society A}\ }\textbf
  {\bibinfo {volume} {471}},\ \bibinfo {pages} {20150301} (\bibinfo {year}
  {2015}{\natexlab{b}})}\BibitemShut {NoStop}%
\bibitem [{\citenamefont {Farhi}\ \emph
  {et~al.}(2000{\natexlab{b}})\citenamefont {Farhi}, \citenamefont
  {Goldstone},\ and\ \citenamefont {Gutmann}}]{Farhi2000a}%
  \BibitemOpen
  \bibfield  {author} {\bibinfo {author} {\bibfnamefont {E.}~\bibnamefont
  {Farhi}}, \bibinfo {author} {\bibfnamefont {J.}~\bibnamefont {Goldstone}}, \
  and\ \bibinfo {author} {\bibfnamefont {S.}~\bibnamefont {Gutmann}},\
  }\href@noop {} {\enquote {\bibinfo {title} {A numerical study of the
  performance of a quantum adiabatic evolution algorithm for satisfiability},}\
  } (\bibinfo {year} {2000}{\natexlab{b}}),\ \Eprint
  {http://arxiv.org/abs/quant-ph/0007071} {arXiv:quant-ph/0007071} \BibitemShut
  {NoStop}%
\bibitem [{\citenamefont {Elgart}\ and\ \citenamefont
  {Hagedorn}(2012)}]{elgart2012note}%
  \BibitemOpen
  \bibfield  {author} {\bibinfo {author} {\bibfnamefont {A.}~\bibnamefont
  {Elgart}}\ and\ \bibinfo {author} {\bibfnamefont {G.~A.}\ \bibnamefont
  {Hagedorn}},\ }\href {\doibase 10.1063/1.4748968} {\bibfield  {journal}
  {\bibinfo  {journal} {Journal of Mathematical Physics}\ }\textbf {\bibinfo
  {volume} {53}},\ \bibinfo {pages} {102202} (\bibinfo {year}
  {2012})}\BibitemShut {NoStop}%
\bibitem [{\citenamefont {Lidar}\ \emph {et~al.}(2009)\citenamefont {Lidar},
  \citenamefont {Rezakhani},\ and\ \citenamefont {Hamma}}]{lidar2009adiabatic}%
  \BibitemOpen
  \bibfield  {author} {\bibinfo {author} {\bibfnamefont {D.~A.}\ \bibnamefont
  {Lidar}}, \bibinfo {author} {\bibfnamefont {A.~T.}\ \bibnamefont
  {Rezakhani}}, \ and\ \bibinfo {author} {\bibfnamefont {A.}~\bibnamefont
  {Hamma}},\ }\href {\doibase 10.1063/1.3236685} {\bibfield  {journal}
  {\bibinfo  {journal} {Journal of Mathematical Physics}\ }\textbf {\bibinfo
  {volume} {50}},\ \bibinfo {pages} {102106} (\bibinfo {year}
  {2009})}\BibitemShut {NoStop}%
\bibitem [{\citenamefont {Wiebe}\ and\ \citenamefont
  {Babcock}(2012)}]{wiebe2012improved}%
  \BibitemOpen
  \bibfield  {author} {\bibinfo {author} {\bibfnamefont {N.}~\bibnamefont
  {Wiebe}}\ and\ \bibinfo {author} {\bibfnamefont {N.~S.}\ \bibnamefont
  {Babcock}},\ }\href {\doibase 10.1088/1367-2630/14/1/013024} {\bibfield
  {journal} {\bibinfo  {journal} {New Journal of Physics}\ }\textbf {\bibinfo
  {volume} {14}},\ \bibinfo {pages} {013024} (\bibinfo {year}
  {2012})}\BibitemShut {NoStop}%
\bibitem [{\citenamefont {Kieferov{\'{a}}}\ and\ \citenamefont
  {Wiebe}(2014)}]{kieferova2014power}%
  \BibitemOpen
  \bibfield  {author} {\bibinfo {author} {\bibfnamefont {M.}~\bibnamefont
  {Kieferov{\'{a}}}}\ and\ \bibinfo {author} {\bibfnamefont {N.}~\bibnamefont
  {Wiebe}},\ }\href {https://doi.org/10.1088/1367-2630/16/12/123034} {\bibfield
   {journal} {\bibinfo  {journal} {New Journal of Physics}\ }\textbf {\bibinfo
  {volume} {16}},\ \bibinfo {pages} {123034} (\bibinfo {year}
  {2014})}\BibitemShut {NoStop}%
\bibitem [{\citenamefont {Wan}\ and\ \citenamefont {Kim}(2020)}]{wan2020fast}%
  \BibitemOpen
  \bibfield  {author} {\bibinfo {author} {\bibfnamefont {K.}~\bibnamefont
  {Wan}}\ and\ \bibinfo {author} {\bibfnamefont {I.}~\bibnamefont {Kim}},\
  }\href@noop {} {\enquote {\bibinfo {title} {Fast digital methods for
  adiabatic state preparation},}\ } (\bibinfo {year} {2020}),\ \Eprint
  {http://arxiv.org/abs/2004.04164} {arXiv:2004.04164} \BibitemShut {NoStop}%
\bibitem [{\citenamefont {Wiebe}\ \emph {et~al.}(2010)\citenamefont {Wiebe},
  \citenamefont {Berry}, \citenamefont {H{\o}yer},\ and\ \citenamefont
  {Sanders}}]{Wiebe2008}%
  \BibitemOpen
  \bibfield  {author} {\bibinfo {author} {\bibfnamefont {N.}~\bibnamefont
  {Wiebe}}, \bibinfo {author} {\bibfnamefont {D.}~\bibnamefont {Berry}},
  \bibinfo {author} {\bibfnamefont {P.}~\bibnamefont {H{\o}yer}}, \ and\
  \bibinfo {author} {\bibfnamefont {B.~C.}\ \bibnamefont {Sanders}},\ }\href
  {\doibase 10.1088/1751-8113/43/6/065203} {\bibfield  {journal} {\bibinfo
  {journal} {Journal of Physics A: Mathematical and Theoretical}\ }\textbf
  {\bibinfo {volume} {43}},\ \bibinfo {pages} {065203} (\bibinfo {year}
  {2010})}\BibitemShut {NoStop}%
\bibitem [{\citenamefont {Lemieux}\ \emph
  {et~al.}(2020{\natexlab{b}})\citenamefont {Lemieux}, \citenamefont
  {Duclos-Cianci}, \citenamefont {S{\'{e}}n{\'{e}}chal},\ and\ \citenamefont
  {Poulin}}]{Lemieux2020ResourceComputer}%
  \BibitemOpen
  \bibfield  {author} {\bibinfo {author} {\bibfnamefont {J.}~\bibnamefont
  {Lemieux}}, \bibinfo {author} {\bibfnamefont {G.}~\bibnamefont
  {Duclos-Cianci}}, \bibinfo {author} {\bibfnamefont {D.}~\bibnamefont
  {S{\'{e}}n{\'{e}}chal}}, \ and\ \bibinfo {author} {\bibfnamefont
  {D.}~\bibnamefont {Poulin}},\ }\href {http://arxiv.org/abs/2006.04650}
  {\bibfield  {journal} {\bibinfo  {journal} {arXiv:2006.04650}\ } (\bibinfo
  {year} {2020}{\natexlab{b}})}\BibitemShut {NoStop}%
\bibitem [{\citenamefont {Chiang}\ \emph {et~al.}(2014)\citenamefont {Chiang},
  \citenamefont {Xu},\ and\ \citenamefont {Somma}}]{Chiang2014}%
  \BibitemOpen
  \bibfield  {author} {\bibinfo {author} {\bibfnamefont {H.-T.}\ \bibnamefont
  {Chiang}}, \bibinfo {author} {\bibfnamefont {G.}~\bibnamefont {Xu}}, \ and\
  \bibinfo {author} {\bibfnamefont {R.~D.}\ \bibnamefont {Somma}},\ }\href
  {\doibase 10.1103/physreva.89.012314} {\bibfield  {journal} {\bibinfo
  {journal} {Physical Review A}\ }\textbf {\bibinfo {volume} {89}},\ \bibinfo
  {pages} {012314} (\bibinfo {year} {2014})}\BibitemShut {NoStop}%
\bibitem [{\citenamefont {Kaiser}\ and\ \citenamefont
  {Schafer}(1980)}]{Kaiser}%
  \BibitemOpen
  \bibfield  {author} {\bibinfo {author} {\bibfnamefont {J.}~\bibnamefont
  {Kaiser}}\ and\ \bibinfo {author} {\bibfnamefont {R.}~\bibnamefont
  {Schafer}},\ }\href {\doibase 10.1109/tassp.1980.1163349} {\bibfield
  {journal} {\bibinfo  {journal} {{IEEE} Transactions on Acoustics, Speech, and
  Signal Processing}\ }\textbf {\bibinfo {volume} {28}},\ \bibinfo {pages}
  {105} (\bibinfo {year} {1980})}\BibitemShut {NoStop}%
\bibitem [{\citenamefont {Berry}\ \emph {et~al.}(2018)\citenamefont {Berry},
  \citenamefont {Kieferov{\'{a}}}, \citenamefont {Scherer}, \citenamefont
  {Sanders}, \citenamefont {Low}, \citenamefont {Wiebe}, \citenamefont
  {Gidney},\ and\ \citenamefont {Babbush}}]{BerryNPJ18}%
  \BibitemOpen
  \bibfield  {author} {\bibinfo {author} {\bibfnamefont {D.~W.}\ \bibnamefont
  {Berry}}, \bibinfo {author} {\bibfnamefont {M.}~\bibnamefont
  {Kieferov{\'{a}}}}, \bibinfo {author} {\bibfnamefont {A.}~\bibnamefont
  {Scherer}}, \bibinfo {author} {\bibfnamefont {Y.~R.}\ \bibnamefont
  {Sanders}}, \bibinfo {author} {\bibfnamefont {G.~H.}\ \bibnamefont {Low}},
  \bibinfo {author} {\bibfnamefont {N.}~\bibnamefont {Wiebe}}, \bibinfo
  {author} {\bibfnamefont {C.}~\bibnamefont {Gidney}}, \ and\ \bibinfo {author}
  {\bibfnamefont {R.}~\bibnamefont {Babbush}},\ }\href
  {https://doi.org/10.1038/s41534-018-0071-5} {\bibfield  {journal} {\bibinfo
  {journal} {npj Quantum Information}\ }\textbf {\bibinfo {volume} {4}},\
  \bibinfo {pages} {22} (\bibinfo {year} {2018})}\BibitemShut {NoStop}%
\bibitem [{\citenamefont {Poulin}\ \emph {et~al.}(2018)\citenamefont {Poulin},
  \citenamefont {Kitaev}, \citenamefont {Steiger}, \citenamefont {Hastings},\
  and\ \citenamefont {Troyer}}]{PoulinPRL18}%
  \BibitemOpen
  \bibfield  {author} {\bibinfo {author} {\bibfnamefont {D.}~\bibnamefont
  {Poulin}}, \bibinfo {author} {\bibfnamefont {A.}~\bibnamefont {Kitaev}},
  \bibinfo {author} {\bibfnamefont {D.~S.}\ \bibnamefont {Steiger}}, \bibinfo
  {author} {\bibfnamefont {M.~B.}\ \bibnamefont {Hastings}}, \ and\ \bibinfo
  {author} {\bibfnamefont {M.}~\bibnamefont {Troyer}},\ }\href
  {https://doi.org/10.1103/physrevlett.121.010501} {\bibfield  {journal}
  {\bibinfo  {journal} {Physical Review Letters}\ }\textbf {\bibinfo {volume}
  {121}},\ \bibinfo {pages} {010501} (\bibinfo {year} {2018})}\BibitemShut
  {NoStop}%
\bibitem [{\citenamefont {H\"aner}\ \emph {et~al.}(2018)\citenamefont
  {H\"aner}, \citenamefont {Roetteler},\ and\ \citenamefont
  {Svore}}]{hner2018optimizing}%
  \BibitemOpen
  \bibfield  {author} {\bibinfo {author} {\bibfnamefont {T.}~\bibnamefont
  {H\"aner}}, \bibinfo {author} {\bibfnamefont {M.}~\bibnamefont {Roetteler}},
  \ and\ \bibinfo {author} {\bibfnamefont {K.~M.}\ \bibnamefont {Svore}},\
  }\href@noop {} {\enquote {\bibinfo {title} {Optimizing quantum circuits for
  arithmetic},}\ } (\bibinfo {year} {2018}),\ \Eprint
  {http://arxiv.org/abs/1805.12445} {arXiv:1805.12445} \BibitemShut {NoStop}%
\bibitem [{\citenamefont {Berry}\ \emph {et~al.}(2014)\citenamefont {Berry},
  \citenamefont {Childs}, \citenamefont {Cleve}, \citenamefont {Kothari},\ and\
  \citenamefont {Somma}}]{BerrySTOC14}%
  \BibitemOpen
  \bibfield  {author} {\bibinfo {author} {\bibfnamefont {D.~W.}\ \bibnamefont
  {Berry}}, \bibinfo {author} {\bibfnamefont {A.~M.}\ \bibnamefont {Childs}},
  \bibinfo {author} {\bibfnamefont {R.}~\bibnamefont {Cleve}}, \bibinfo
  {author} {\bibfnamefont {R.}~\bibnamefont {Kothari}}, \ and\ \bibinfo
  {author} {\bibfnamefont {R.~D.}\ \bibnamefont {Somma}},\ }in\ \href
  {http://doi.acm.org/10.1145/2591796.2591854} {\emph {\bibinfo {booktitle}
  {Proceedings of the 46th Annual ACM Symposium on Theory of Computing}}},\
  \bibinfo {series and number} {STOC '14}\ (\bibinfo  {publisher} {ACM},\
  \bibinfo {address} {New York, NY, USA},\ \bibinfo {year} {2014})\ pp.\
  \bibinfo {pages} {283--292}\BibitemShut {NoStop}%
\bibitem [{\citenamefont {Bravyi}\ and\ \citenamefont
  {Kitaev}(1998)}]{Bravyi1998}%
  \BibitemOpen
  \bibfield  {author} {\bibinfo {author} {\bibfnamefont {S.~B.}\ \bibnamefont
  {Bravyi}}\ and\ \bibinfo {author} {\bibfnamefont {A.~Y.}\ \bibnamefont
  {Kitaev}},\ }\href@noop {} {\enquote {\bibinfo {title} {Quantum codes on a
  lattice with boundary},}\ } (\bibinfo {year} {1998}),\ \Eprint
  {http://arxiv.org/abs/quant-ph/9811052} {arXiv:quant-ph/9811052} \BibitemShut
  {NoStop}%
\bibitem [{\citenamefont {Dennis}\ \emph {et~al.}(2002)\citenamefont {Dennis},
  \citenamefont {Kitaev}, \citenamefont {Landahl},\ and\ \citenamefont
  {Preskill}}]{Dennis2002}%
  \BibitemOpen
  \bibfield  {author} {\bibinfo {author} {\bibfnamefont {E.}~\bibnamefont
  {Dennis}}, \bibinfo {author} {\bibfnamefont {A.}~\bibnamefont {Kitaev}},
  \bibinfo {author} {\bibfnamefont {A.}~\bibnamefont {Landahl}}, \ and\
  \bibinfo {author} {\bibfnamefont {J.}~\bibnamefont {Preskill}},\ }\href
  {\doibase 10.1063/1.1499754} {\bibfield  {journal} {\bibinfo  {journal}
  {Journal of Mathematical Physics}\ }\textbf {\bibinfo {volume} {43}},\
  \bibinfo {pages} {4452} (\bibinfo {year} {2002})}\BibitemShut {NoStop}%
\bibitem [{\citenamefont {Raussendorf}\ and\ \citenamefont
  {Harrington}(2007)}]{Raussendorf2007}%
  \BibitemOpen
  \bibfield  {author} {\bibinfo {author} {\bibfnamefont {R.}~\bibnamefont
  {Raussendorf}}\ and\ \bibinfo {author} {\bibfnamefont {J.}~\bibnamefont
  {Harrington}},\ }\href {\doibase 10.1103/physrevlett.98.190504} {\bibfield
  {journal} {\bibinfo  {journal} {Physical Review Letters}\ }\textbf {\bibinfo
  {volume} {98}},\ \bibinfo {pages} {190504} (\bibinfo {year}
  {2007})}\BibitemShut {NoStop}%
\bibitem [{\citenamefont {Fowler}\ and\ \citenamefont
  {Gidney}(2018)}]{Fowler2018}%
  \BibitemOpen
  \bibfield  {author} {\bibinfo {author} {\bibfnamefont {A.~G.}\ \bibnamefont
  {Fowler}}\ and\ \bibinfo {author} {\bibfnamefont {C.}~\bibnamefont
  {Gidney}},\ }\href@noop {} {\enquote {\bibinfo {title} {Low overhead quantum
  computation using lattice surgery},}\ } (\bibinfo {year} {2018}),\ \Eprint
  {http://arxiv.org/abs/1808.06709} {arXiv:1808.06709} \BibitemShut {NoStop}%
\bibitem [{\citenamefont {Jones}\ \emph {et~al.}(2012)\citenamefont {Jones},
  \citenamefont {Whitfield}, \citenamefont {McMahon}, \citenamefont {Yung},
  \citenamefont {Meter}, \citenamefont {Aspuru-Guzik},\ and\ \citenamefont
  {Yamamoto}}]{Jones2012}%
  \BibitemOpen
  \bibfield  {author} {\bibinfo {author} {\bibfnamefont {N.~C.}\ \bibnamefont
  {Jones}}, \bibinfo {author} {\bibfnamefont {J.~D.}\ \bibnamefont
  {Whitfield}}, \bibinfo {author} {\bibfnamefont {P.~L.}\ \bibnamefont
  {McMahon}}, \bibinfo {author} {\bibfnamefont {M.-H.}\ \bibnamefont {Yung}},
  \bibinfo {author} {\bibfnamefont {R.~V.}\ \bibnamefont {Meter}}, \bibinfo
  {author} {\bibfnamefont {A.}~\bibnamefont {Aspuru-Guzik}}, \ and\ \bibinfo
  {author} {\bibfnamefont {Y.}~\bibnamefont {Yamamoto}},\ }\href {\doibase
  10.1088/1367-2630/14/11/115023} {\bibfield  {journal} {\bibinfo  {journal}
  {New Journal of Physics}\ }\textbf {\bibinfo {volume} {14}},\ \bibinfo
  {pages} {115023} (\bibinfo {year} {2012})}\BibitemShut {NoStop}%
\bibitem [{\citenamefont {Eastin}(2013)}]{Eastin2013}%
  \BibitemOpen
  \bibfield  {author} {\bibinfo {author} {\bibfnamefont {B.}~\bibnamefont
  {Eastin}},\ }\href {\doibase 10.1103/physreva.87.032321} {\bibfield
  {journal} {\bibinfo  {journal} {Physical Review A}\ }\textbf {\bibinfo
  {volume} {87}},\ \bibinfo {pages} {032321} (\bibinfo {year}
  {2013})}\BibitemShut {NoStop}%
\bibitem [{\citenamefont {Fowler}(2013)}]{Fowler2013}%
  \BibitemOpen
  \bibfield  {author} {\bibinfo {author} {\bibfnamefont {A.~G.}\ \bibnamefont
  {Fowler}},\ }\href@noop {} {\enquote {\bibinfo {title} {Optimal complexity
  correction of correlated errors in the surface code},}\ } (\bibinfo {year}
  {2013}),\ \Eprint {http://arxiv.org/abs/1310.0863} {arXiv:1310.0863}
  \BibitemShut {NoStop}%
\bibitem [{\citenamefont {Isakov}\ \emph {et~al.}(2015)\citenamefont {Isakov},
  \citenamefont {Zintchenko}, \citenamefont {R{\o}nnow},\ and\ \citenamefont
  {Troyer}}]{Isakov2015}%
  \BibitemOpen
  \bibfield  {author} {\bibinfo {author} {\bibfnamefont {S.~V.}\ \bibnamefont
  {Isakov}}, \bibinfo {author} {\bibfnamefont {I.}~\bibnamefont {Zintchenko}},
  \bibinfo {author} {\bibfnamefont {T.}~\bibnamefont {R{\o}nnow}}, \ and\
  \bibinfo {author} {\bibfnamefont {M.}~\bibnamefont {Troyer}},\ }\href
  {\doibase 10.1016/j.cpc.2015.02.015} {\bibfield  {journal} {\bibinfo
  {journal} {Computer Physics Communications}\ }\textbf {\bibinfo {volume}
  {192}},\ \bibinfo {pages} {265} (\bibinfo {year} {2015})}\BibitemShut
  {NoStop}%
\bibitem [{\citenamefont {Isakov}(2020)}]{SergeiEmail}%
  \BibitemOpen
  \bibfield  {author} {\bibinfo {author} {\bibfnamefont {S.~V.}\ \bibnamefont
  {Isakov}},\ }\href@noop {} {}\bibinfo {howpublished} {personal communication
  about simulated annealing code in \cite{Isakov2015}.} (\bibinfo {year}
  {2020})\BibitemShut {NoStop}%
\bibitem [{\citenamefont {Wouk}(1965)}]{WOUK}%
  \BibitemOpen
  \bibfield  {author} {\bibinfo {author} {\bibfnamefont {A.}~\bibnamefont
  {Wouk}},\ }\href {\doibase https://doi.org/10.1016/0022-247X(65)90073-9}
  {\bibfield  {journal} {\bibinfo  {journal} {Journal of Mathematical Analysis
  and Applications}\ }\textbf {\bibinfo {volume} {11}},\ \bibinfo {pages} {131
  } (\bibinfo {year} {1965})}\BibitemShut {NoStop}%
\bibitem [{\citenamefont {Hastings}(2018{\natexlab{b}})}]{Hastings2018b}%
  \BibitemOpen
  \bibfield  {author} {\bibinfo {author} {\bibfnamefont {M.~B.}\ \bibnamefont
  {Hastings}},\ }\href@noop {} {\enquote {\bibinfo {title} {Weaker assumptions
  for the short path optimization algorithm},}\ } (\bibinfo {year}
  {2018}{\natexlab{b}}),\ \Eprint {http://arxiv.org/abs/1807.03758}
  {arXiv:1807.03758} \BibitemShut {NoStop}%
\end{thebibliography}%


%

\appendix

\section{Addition for controlled rotations}\label{app:rotations}

Here we give more details on how to perform phase rotations using the method from \cite{Kitaev2002,GidneyAdder}. 
Prior to the simulation the following state is prepared
\begin{equation}
\ket{\phi} = \frac{1}{\sqrt{2^\gradbits}} \sum_{k=0}^{2^\gradbits - 1} e^{-2 \pi i k / 2^\gradbits} \ket{k}.
\end{equation}
This state is a tensor product of the form
\begin{equation}
\ket{\phi} = \frac{1}{\sqrt{2^\gradbits}} \bigotimes_{j=1}^{\gradbits} \left( \ket{0} + e^{-2 \pi i / 2^j} \ket{1}\right).
\end{equation}
It can be prepared using standard techniques for performing rotations on qubits.
To obtain overall error $\epsilon$, each rotation should be performed with error $\epsilon/\gradbits$, which has complexity $\mathcal{O}(\log(\gradbits/\epsilon))$ \cite{Bocharov2015},
giving overall complexity $\mathcal{O}(\gradbits\log(\gradbits/\epsilon))$ to prepare this state.
Because this state only need be prepared once, this complexity is negligible compared to the complexities in other parts of the algorithm.

Adding a value $\ell$ into this register gives
\begin{equation}
\frac{1}{\sqrt{2^\gradbits}} \sum_{k=0}^{2^\gradbits - 1} e^{-2 \pi i k / 2^\gradbits} \ket{k+\ell} = \frac{1}{\sqrt{2^\gradbits}} \sum_{k=0}^{2^\gradbits - 1} e^{-2 \pi i (k-\ell) / 2^\gradbits} \ket{k} = e^{2 \pi i \ell / 2^\gradbits}\ket{\phi}.
\end{equation}
This is why the addition yields a phase factor.
Moreover, the value of $\ell$ can be stored in a quantum register, in order to make this a controlled phase.
In order to make a controlled rotation on a qubit, we can perform the addition controlled by this qubit.
Then one would obtain
\begin{equation}
(\mu \ket{0} + \nu\ket{1})\ket{\ell} \ket{\phi} \mapsto (\mu \ket{0} + e^{2 \pi i \ell / 2^\gradbits}\nu\ket{1})\ket{\ell} \ket{\phi}.
\end{equation}
This approach is somewhat inefficient, because controlled addition has twice the complexity of addition.

Instead we can use the trick described in \sec{direnegor}, which enables a qubit to control whether addition or subtraction is performed with only Clifford gates.
The qubit simply needs to control \textsc{cnot}s on the target system before and after the addition.
Then we would obtain
\begin{equation}
(e^{-2 \pi i \ell / 2^\gradbits}\mu \ket{0} + e^{2 \pi i \ell / 2^\gradbits}\nu\ket{1})\ket{\ell} \ket{\phi}.
\end{equation}
This procedure therefore enables us to perform the rotation $e^{-2\pi i \ell Z/2^\gradbits}$ with $\gradbits-2$ Toffolis. This approach is far more efficient than techniques based on rotation synthesis with T gates when the rotation angle is given in a quantum register, because those techniques would need a separate rotation controlled on each bit.
When the rotation angle is given classically, this technique is slightly less efficient than rotation synthesis with T gates as in \cite{Bocharov2015},
because Toffolis have a cost equivalent to two T gates in magic state distillation \cite{Gidney2019}.
On the other hand, rotation angles that are integer multiples of $2\pi/2^\gradbits$ can be performed exactly, up to the accuracy of synthesizing the resource state $\ket{\phi}$.

To obtain a rotation that performs the mapping
\begin{equation}
    \ket{0} \mapsto \cos(2\pi\ell/2^\gradbits)  \ket{0} + \sin(2\pi\ell/2^\gradbits)  \ket{1},
\end{equation}
one can simply perform the operations
$S H e^{-2\pi i \ell Z/2^\gradbits} H$.
Here the Hadamard $H$ and $S$ gates are Clifford gates, so this gives the state preparation with the only Toffoli cost in synthesizing the $Z$-rotation.

\begin{figure}[!tbh]
\centering
  \resizebox{.65\linewidth}{!}{\includegraphics{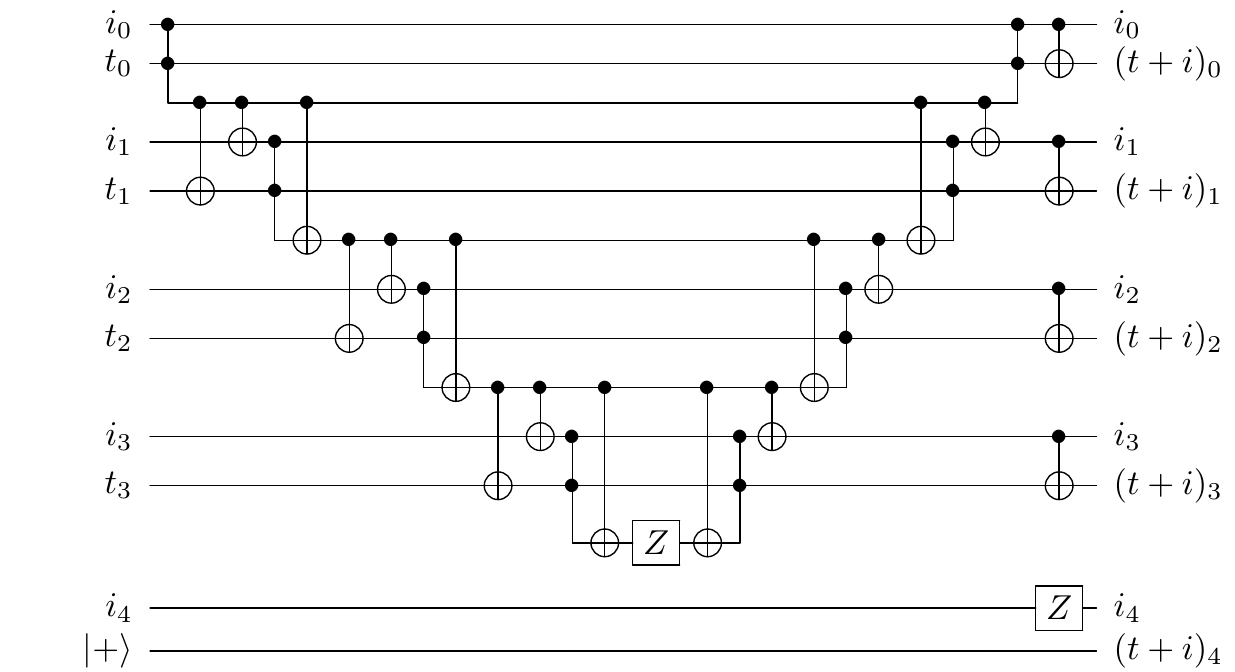}} \caption[Addition circuit when most significant target qubit is in plus state.]
  {\label{fig:adderp1}
    A circuit to perform addition on 5 qubits modulo $2^5$ when the most significant target qubit is in a $\ket{+}$ state.}

  \vspace{5mm}
\centering
  \resizebox{.65\linewidth}{!}{\includegraphics{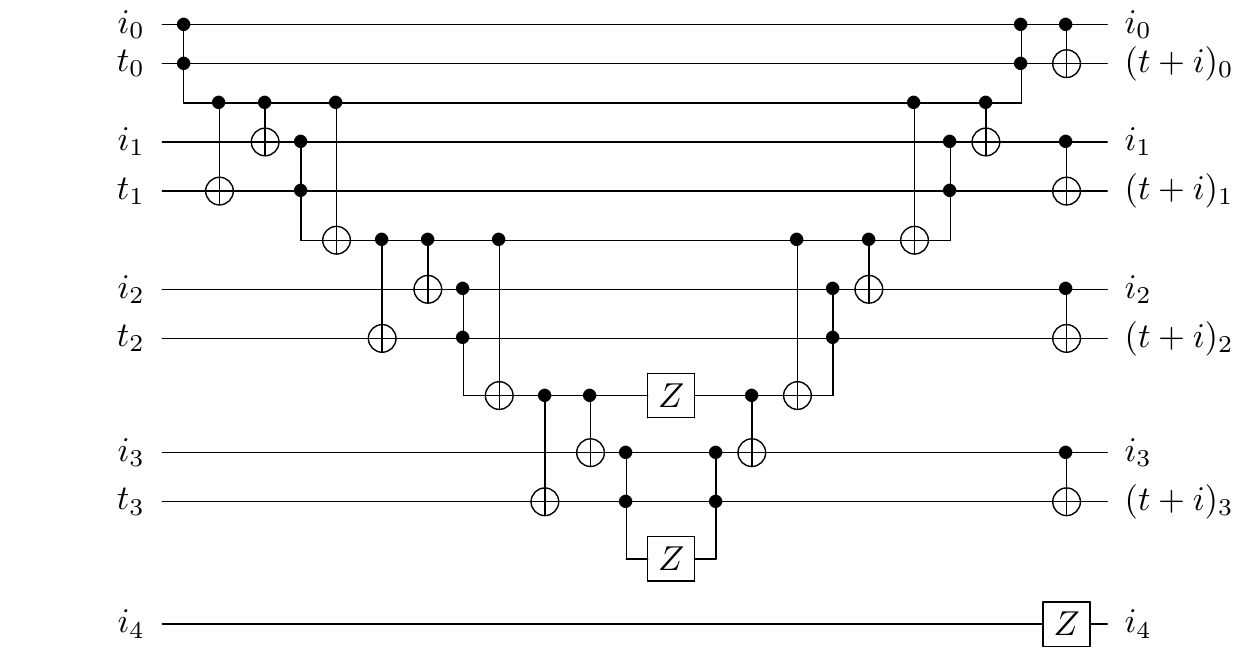}} \caption[Simplification of plus-state addition circuit to eliminate CNOT.]
  {\label{fig:adderp2} A simplification of~\fig{adderp1} to eliminate the \textsc{cnot}s on the last carry qubit. The $\ket{+}$ state is omitted here because it is not acted upon.}

  \vspace{5mm}
\centering
  \resizebox{.65\linewidth}{!}{\includegraphics{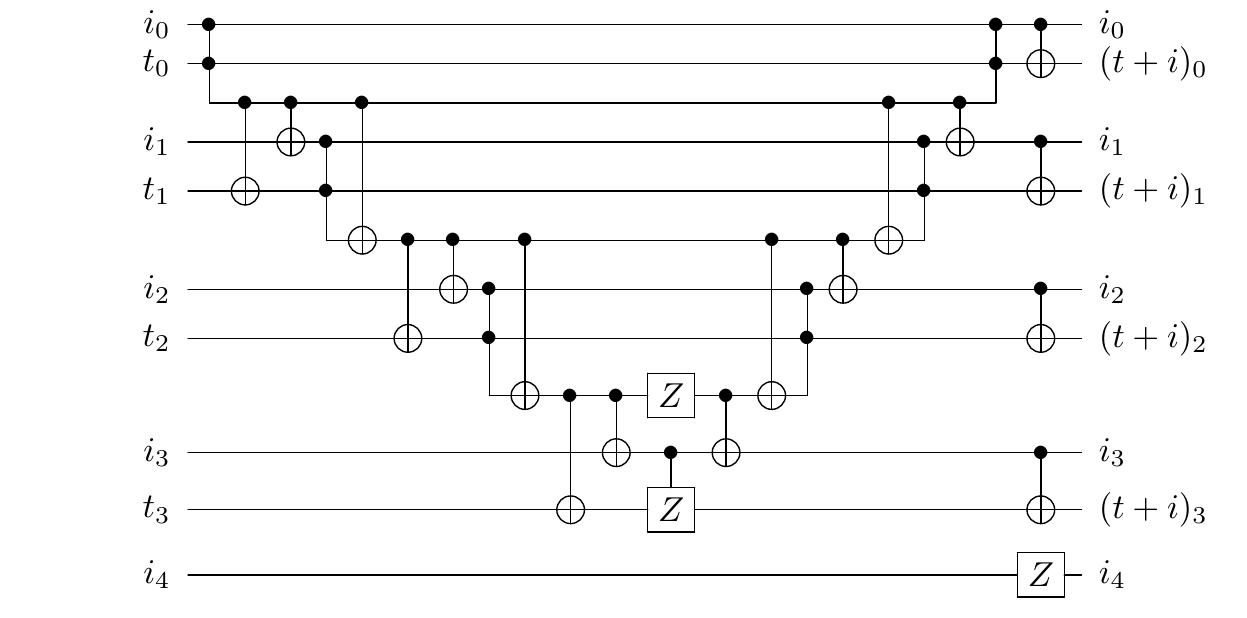}} \caption[Further simplification of plus-state adder to eliminate last carry qubit.]
  {\label{fig:adderp3}
    A simplification of \fig{adderp2} where the last carry qubit is eliminated entirely and the Toffoli is replaced with a controlled phase.}
\end{figure}

The complexity of performing the addition is only $\gradbits-2$ rather than $\gradbits-1$, as would normally be the case for addition of $\gradbits$-bit numbers (modulo $2^\gradbits$).
The reason is that the most significant qubit of $\ket{\phi}$ is in a $\ket{+}$ state, so \textsc{not} gates on this qubit can be replaced with phase gates, and this qubit can be discarded.
Doing that yields the circuit shown in \fig{adderp1}.
The Toffoli is not immediately saved, but the \textsc{cnot}s and $Z$ gate on the final carry qubit can be replaced with two $Z$ gates as shown in \fig{adderp2}.
Then the Toffoli used on the final carry qubit can simply be replaced with a controlled phase, as shown in \fig{adderp3}.
The resulting complexity is $\gradbits-2$ Toffolis.
If the angle to be rotated by is given as a classical variable, then the cost is further reduced to $\gradbits-3$ Toffolis, because addition of a classical number takes one fewer Toffoli.
This means that $\gradbits=4$, which would give a T gate, takes one Toffoli.

\begin{figure}[tbh]
\includegraphics[width=0.45\textwidth]{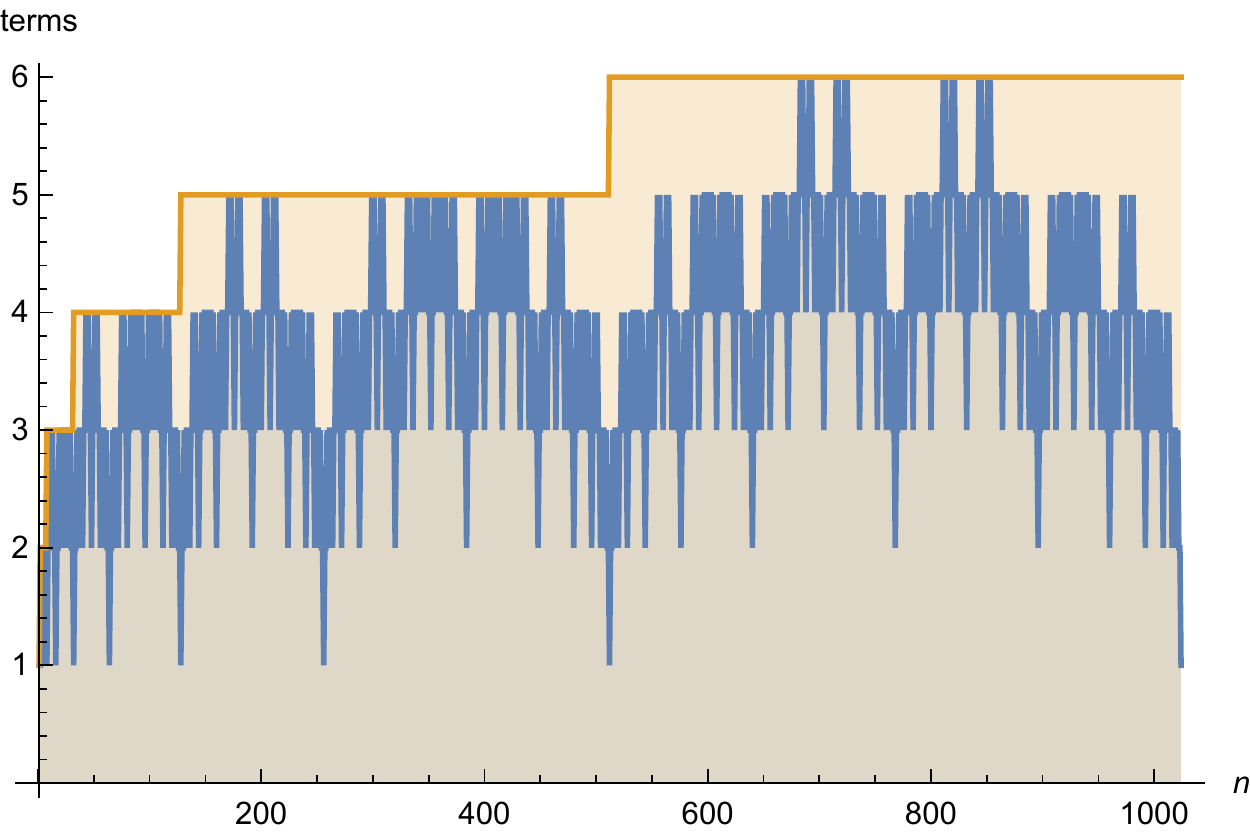}
\caption[Writing integers as sums of powers of 2 when we allow additions and subtractions.]{\label{fig:bits}
    In orange is the number of powers of $2$ needed to give integer $m$, when we allow additions and subtractions.
    The formula $\lceil (n+1)/2 \rceil$ is shown in orange, where the number of bits required to represent $m$ is $n=\lceil \log (m+1)\rceil$.
    }
\end{figure}

Next we consider the case that we need to multiply an integer $k$ with $b$ bits by a constant $\tilde{\gamma}$ to give the phase.
Given that $\tilde{\gamma}$ is represented on $n$ bits, we can write $\tilde{\gamma}$ as a sum of no more than $\lceil(n+1)/2\rceil$ powers of two, with positive and negative signs.
This formula is checked in \fig{bits}.
To prove the formula, assume that it is true for numbers with $\le n_0$ bits, and consider a number $m$ with $n=n_0+2$ bits (so the most significant bit must be a $1$).
There are then three cases to consider.
\begin{enumerate}
    \item For $m<(3/4)2^n$, we find that $m-2^{n-1}<2^{n-2}$, so $m-2^{n-1}$ has no more than $n-2=n_0$ bits, and so can be written as a sum of at most $\lceil(n_0+1)/2\rceil$ powers of two.
That means $m$ can be written as a sum of at most $\lceil(n_0+1)/2\rceil+1=\lceil(n+1)/2\rceil$ powers of two.
\item For $m>(3/4)2^n$, we have $2^n-m< 2^{n-2}$, so $2^n-m$ has no more than $n-2=n_0$ bits, and can be written as a sum of $\le \lceil(n_0+1)/2\rceil$ powers of two.
Since $m$ can be written as $2^n$ minus $2^n-m$, it can be written as at most $\lceil(n_0+1)/2\rceil+1=\lceil(n+1)/2\rceil$ powers of two.
\item The last case is that where $m=(3/4)2^n$, so $m=2^{n-1}+2^{n-2}$.
Since $n=n_0+2\ge 2$, $\lceil(n+1)/2\rceil\ge 2$, so again $m$ is written as a sum of at most $\lceil(n+1)/2\rceil$ powers of two.
\end{enumerate}

Since we have checked that the formula is true for small numbers of bits in \fig{bits}, the formula is true for all $n$ by induction.
To perform the multiplication, we will take each term in the sum for $\tilde{\gamma}$, and add or subtract a bit-shifted copy of $k$ to the phase gradient state.
We have no more than $(n+2)/2$ additions/subtractions, each of which is into the phase gradient state with $\gradbits$ bits, which gives cost no more than $(\gradbits-2)(n+2)/2$.

The error due to omitted bits in the multiplication (those omitted in bit-shifting $k$) can be bounded as follows.
First, note that the error for the additions is entirely in underestimating the product, since we are omitting digits.
For the subtractions the error is in overestimating the product.
Therefore, to obtain the maximum error we need to consider the case with entirely additions, since the subtractions would cancel the error.
For each addition the error is upper bounded by $2\pi/2^\gradbits$, because we omit adding bits that would correspond to phase shifts of $2\pi/2^{\gradbits+1}$, $2\pi/2^{\gradbits+2}$, and so forth.
That means the upper bound on the error from $(n+2)/2$ additions is $(n+2)\pi/2^\gradbits$.
To make the error in the multiplication no larger than $\epsilon$ we should take
\begin{equation}
    \gradbits = \lceil \log[(n+2)\pi/\epsilon]\rceil = \log (n/\epsilon) + \order{1}.
\end{equation}

\section{Discretizing adiabatic state preparation with qubitization}

\label{app:adiabatic_walk}

Here we place bounds on the error for the method of adiabatic evolution from \sec{heuwalk}.
For any fixed value of $r$ we can choose an adiabatic path between an initial Hamiltonian and a final Hamiltonian.  The accuracy of the adiabatic approximation depends strongly on how quickly we traverse this path, so it is customary to introduce a dimensionless time $s=t/T$ which allows us to easily change the speed without altering the shape of the path.

Using Trotter-Suzuki formulas for time-ordered operator exponentials we have that
\begin{align}
    \left\|\mathcal{T}e^{-iT\int_{s}^{s+1/r} H_{\rm eff}(s) \mathrm{d}s}-e^{-iH_{\rm eff}(s+1/2r) T/r}\right\| \in \mathcal{O}\left(\frac{ \max_s\|\partial_s^2 H_{\rm eff}(s)\|T+ \max_s\|\partial_s H_{\rm eff}(s)\| \|H_{\rm eff}(s)\|T^2}{r^3} \right).
\end{align}
However, the Hamiltonian $H_{\rm eff}$ for the time evolution operator in this case is not known except in terms of its action on the space containing the instantaneous eigenvectors of $H$.  In order to use this result, we need to bound the derivatives acting on the entire space.  In order to find an asymptotic bound on these derivatives we define,
\begin{equation}
    H_{\rm eff}(s) = \frac{ir}{4}\ln\left(W_r^4(s) \right)=\frac{ir}{4}\ln\left(\left(\left(I -2 I\otimes \ket{L(r,s)}\!\bra{L(r,s)}\right) \sel\right)^4 \right).
\end{equation}
It is then clear from the unitarity of $W_r(s)$ that for any $|s|\in O(1)$, if we choose the principal logarithm for $H_{\rm eff}$ then $\|H_{\rm eff}(s)\|\in \order{1}$.
The derivatives of the Hamiltonian are more involved to estimate.  

\subsection{Derivatives of matrix logarithms of unitary matrices}

In order to compute the derivatives of the effective Hamiltonian, we will need to compute the derivatives of the logarithm function.  Such an analysis is usually based on differentiating the Mercator series for the matrix logarithm; however, the Mercator series of $\log(A)$ does not converge for $\|A\|\ge 1$.  For greater generality we will use an integral representation for the matrix logarithm $\log(A)$ from \cite{WOUK},
\begin{equation}
    \log(A) = \int_0^1 \mathrm{d} t \, (A-\openone)[t(A-\openone)+\openone]^{-1} . \label{eq:logdef}
\end{equation}
This representation converges unless there is a real non-positive eigenvalue. For the case where $A$ is unitary, this requirement prohibits matrices that have any eigenvalues equal to precisely $-1$.
Next, defining $V:=[t(A-\openone)+\openone]^{-1}$, and in turn $(A-\openone)= t^{-1}(V^{-1} - \openone)$ this expression simplifies to
\begin{align}
\log(A) &= \int_0^1 \mathrm{d} t \, (A-\openone)V \nonumber\\
&=\int_0^1 \mathrm{d} t \, \frac 1t \left(\openone-V\right).
\end{align}
Next, note that for any invertible matrix-valued function $A(s)$ we have from the product rule that
\begin{equation}
    \partial_s (A(s) A^{-1}(s)) = 0 \Rightarrow \partial_s (A^{-1}(s)) = -A^{-1}(s) \dot{A}(s) A^{-1}(s) . \label{eq:invdiff}
\end{equation}
Using $\partial_s V = -tV\dot{A}V$ we get
\begin{equation}
 \partial_s  \log(A(s)) = \int_0^1 \mathrm{d} t \, V\dot{A}V.\label{eq:1deriv}
\end{equation}
Taking the derivative of~\eqref{eq:1deriv} gives
\begin{align}
\partial_s^2 \log(A(s)) &=  \int_0^1 \mathrm{d} t \, \left(\dot{V}\dot{A}V + V\ddot{A}V + V\dot{A}\dot{V}\right)\nonumber\\
&=\int_0^1 \mathrm{d} t \, \left(-tV\dot{A}V\dot{A}V + V\ddot{A}V -t V\dot{A}V\dot{A}V\right)\nonumber\\
&=\int_0^1 \mathrm{d} t \, V \left( \ddot{A}-2t \dot{A}V\dot{A} \right) V .\label{eq:2derivlog}
\end{align}
We can use the fact that $A$ is unitary to see that 
\begin{align}
    \|V\|^2 &=\left\|(t(A-\openone) +\openone)(t(A^\dagger-\openone) +\openone)\right\|^{-1} \nonumber\\
    &=\left\|(t^2 +(t-1)^2)\openone + (A+A^\dagger)\left(t -t^2 \right)\right\|^{-1} .
\end{align}
In the case where $A$ is close to the identity, if the absolute values of the phases of the eigenvalues are no greater than $\Gamma$, then
\begin{align}
    \|\partial_s \log(A(s))\| &\le \frac \Gamma{\sin \Gamma} \|\dot{A}\| \, , \label{eq:Aderiv1} \\
   \|\partial^2_s \log(A(s))\| &\le \frac \Gamma{\sin \Gamma} \|\ddot{A}\| + \frac 1{\cos^2(\Gamma/2)}\|\dot{A}\|^2 \, .\label{eq:Aderiv2}
\end{align}
Next we consider $W_r^2(t)$ in the case where the Hamiltonian is a linear combination of self-inverse unitaries so $\sel'^2=\openone$, which is the case for all Hamiltonians considered here.
Expanding it out we have
\begin{align}
    W_r^2(s) &= (\openone-2 \openone\otimes \ket{L(s,r)}\!\bra{L(s,r)}) \sel'(\openone -2 \openone\otimes \ket{L(s,r)}\!\bra{L(s,r)}) \sel'\nn
    &= (\openone-2 \openone\otimes \ket{L(s,r)}\!\bra{L(t,r)})(\openone -2\, \sel' \ket{L(s,r)}\!\bra{L(s,r)}\sel')\nn
    &= \openone-2 \openone\otimes \ket{L(t,r)}\!\bra{L(t,r)} -2\, \sel' \ket{L(s,r)}\!\bra{L(s,r)}\sel' \nn
    &\quad \quad +4\ket{L(s,r)}\!\bra{L(s,r)}\sel'\ket{L(s,r)}\!\bra{L(s,r)}\sel' .
\end{align}
Using
\begin{equation}
    \ket{L(s,r)} = \sum_{k}\sqrt{\frac{\lambda_k(s)}{\lambda(s)r}}\ket{k}\ket{00}+ \sqrt{\frac{r-1}{2r}}\ket{0}\left(\ket{10} + \ket{11}\right) ,
\end{equation}
we have
\begin{equation}
    \bra{L(s,r)}\sel'\ket{L(s,r)} = \frac{H(s)}{\lambda(s) r}.
\end{equation}
Then squaring again gives
\begin{align}
    W_r^4(s)
    &= \openone+4 \openone\otimes \ket{L(s,r)}\!\bra{L(s,r)} +4\, \sel' \ket{L(s,r)}\!\bra{L(s,r)}\sel'  + 16 \ket{L(s,r)}\left( \frac{H(s)}{\lambda(s) r}\right)^3\bra{L(s,r)}\sel' \nn
    &\quad -4 \openone\otimes \ket{L(s,r)}\!\bra{L(s,r)} -4\, \sel' \ket{L(s,r)}\!\bra{L(s,r)}\sel'  +8 \ket{L(s,r)}\left( \frac{H(s)}{\lambda(s) r}\right)\bra{L(s,r)}\sel' \nn
    &\quad +4\ket{L(s,r)}\left( \frac{H(s)}{\lambda(s) r}\right)\bra{L(s,r)}\sel'+4\,\sel'\ket{L(s,r)}\left( \frac{H(s)}{\lambda(s) r}\right)\bra{L(s,r)}\nn
    &\quad -8\ket{L(s,r)}\left( \frac{H(s)}{\lambda(s) r}\right)\bra{L(s,r)}\sel' -8\ket{L(s,r)}\left( \frac{H(s)}{\lambda(s) r}\right)^2\bra{L(s,r)}\nn
    &\quad -8\,\sel'\ket{L(s,r)}\left( \frac{H(s)}{\lambda(s) r}\right)^2\bra{L(s,r)}\sel'-8\ket{L(s,r)}\left( \frac{H(s)}{\lambda(s) r}\right)\bra{L(s,r)}\sel' \nn
    &= \openone+ 16 \ket{L(s,r)}\left( \frac{H(s)}{\lambda(s) r}\right)^3\bra{L(s,r)}\sel' -8\ket{L(s,r)}\left( \frac{H(s)}{\lambda(s) r}\right)^2\bra{L(s,r)}\nn
    &\quad-8\,\sel'\ket{L(s,r)}\left( \frac{H(s)}{\lambda(s) r}\right)^2\bra{L(s,r)}\sel'\nn
    &\quad -4\ket{L(s,r)}\left( \frac{H(s)}{\lambda(s) r}\right)\bra{L(s,r)}\sel'+4\,\sel'\ket{L(s,r)}\left( \frac{H(s)}{\lambda(s) r}\right)\bra{L(s,r)}\nn
    &= \openone +4\left[\sel',\ket{L(s,r)}\left( \frac{H(s)}{\lambda(s) r}\right)\bra{L(s,r)}\right] + \order{\frac{1}{r^2}} . \label{eq:qubiterate4}
\end{align}
This means that $\|W_r^4(s)-\openone\|\le 8/r+16/r^2+16/r^3$, so for $r\gtrsim 5.7$, $W_r^4$ does not have negative real eigenvalues, and our expression for the matrix logarithm holds.  This also implies that
\begin{align}
\|\log(W_r^4(s))\| &\le \int_0^1 \mathrm{d} t \, \|(A-\openone)V\|\le 8/r +\order{1/r^2} \label{eq:logWrbd}.
\end{align}

Next, under these assumptions we can use~\eqref{eq:Aderiv1} and~\eqref{eq:qubiterate4} to show that (neglecting terms of $\order{r^{-2}}$ which are negligible for large $r$ and using $\|H\|/\lambda \le 1$)
\begin{align}
    \|\partial_s \log(W_r^4(s))\| &\in \mathcal{O}\left(\| \partial_s W_r^4(s) \|  \right) \nn
    &\subseteq \mathcal{O}\left(\left\| \ket{\dot{L}(s,r)}\right\| \left\|\frac{H(s)}{\lambda(s) r}\right\| + \left\|\frac{\partial}{\partial s}\frac{H(s)}{\lambda(s) r}\right\|  \right)\nn
    &\subseteq \mathcal{O}\left(\frac{\left\| \ket{\dot{L}(s,r)} \right\|}{r} + \frac{|\dot{\lambda}| + \|\dot{H}\|}{\lambda r}  \right) .\label{eq:logderiv}
\end{align}
We observe from~\eqref{eq:L} and the definition of the Euclidean norm it follows that if the Hamiltonian is chosen to be independent of $r$ then
\begin{align}
\|\ket{\dot{L}}\|&\in \mathcal{O}\left(\sqrt{\sum_k \left|\frac{\partial}{\partial s} \sqrt{\frac{\lambda_k}{\lambda r}}\right|^2} \right) \subseteq \order{1/\sqrt{r}}.\label{eq:Lderiv}
\end{align}
Thus neglecting terms of order $\order{r^{-3/2}}$ we find from substituting this expression into~\eqref{eq:logderiv} that
\begin{align}
    \|\partial_s \log(W_r^4(s))\| &\in \mathcal{O}\left(\frac{|\dot{\lambda}| + \|\dot{H}\|}{\lambda r}  \right) . \label{eq:log1deriv}
\end{align}

It further follows from~\eqref{eq:Aderiv2} and~\eqref{eq:qubiterate4} that the second derivative of the matrix logarithm obeys (neglecting terms order $\order{r^{-3/2}}$ and higher)
\begin{align}
     \|\partial_s^2 \log(W_r^4(s))\| &\in \mathcal{O} \left( \|\partial_s^2 W_r^4(s)) \| + \|\partial_s W_r^4(s)) \|^2 \right) \nn
     &=\order{\|\partial_s^2 W_r^4(s)) \|}\nn
     &\subseteq\mathcal{O}\left( \frac{\|\ket{\ddot{L}(s,r)}\|}{r} + \left\|\frac{\partial^2}{\partial s^2} \frac{H}{\lambda r} \right\| + \|\ket{\dot{L}}\| \left\|\frac{\partial}{\partial s} \frac{H}{\lambda r} \right\| \right)\nn
     &\subseteq\mathcal{O}\left( \frac{\|\ket{\ddot{L}(s,r)}\|}{r} + \left\|\frac{\partial^2}{\partial s^2} \frac{H}{\lambda r} \right\| + \frac{1}{r^{3/2}} \right)\nn
     &\subseteq\mathcal{O}\left( \frac{\|\ket{\ddot{L}(s,r)}\|}{r} + \left( \frac{\|\ddot{H}\| + (\|\dot{H}\|+|\dot{\lambda}|)\left(\frac{|\dot{\lambda}|}{\lambda}\right)+|\ddot{\lambda}| }{\lambda r} \right)  \right) .
\end{align}
Note that in the above derivation terms of the form $\|\ket{\dot{L}(s,r)}\| \|\partial_s H/(\lambda r)\|$ are dropped because they are $\order{r^{-3/2}}$.
Again, if the Hamiltonian is chosen to be independent of $r$, then
\begin{equation}
    \|\ket{\ddot{L}(s,r)}\| \in \mathcal{O}\left(\sqrt{\sum_k \left|\frac{\partial^2}{\partial s^2} \sqrt{\frac{\lambda_k}{\lambda r}}\right|^2} \right) \subseteq \order{\frac{1}{\sqrt{r}}},
\end{equation}
which implies that, neglecting terms of $\order{r^{-3/2}}$ and higher
\begin{equation}
    \|\partial_s^2 \log(W_r^4(s))\| \in \mathcal{O} \left(  \frac{\|\ddot{H}\| + (\|\dot{H}\|+|\dot{\lambda}|)\left(\frac{|\dot{\lambda}|}{\lambda}\right)+|\ddot{\lambda}| }{\lambda r} \right).  \label{eq:log2deriv}
\end{equation}

Next in we bound the error that arises from approximating the time-ordered operator exponential by the exponential of the effective Hamiltonian evaluated at the midpoint. 
From the analysis of the midpoint rule for integration, we intuitively expect that the error should scale as $\mathcal{O}(1/r^3)$; however, such analysis cannot be directly applied here because of the fact that the derivatives of the Hamiltonian need not commute with the Hamiltonian.  It can be seen by performing a Taylor series expansion of the effective Hamiltonian to second order and substituting the result into the Dyson series that
\begin{align}
    \left\|\mathcal{T}e^{-iT\int_{s}^{s+4/r} H_{\rm eff}(s') \mathrm{d}s'}-e^{-i\frac{4T}{r}H_{\rm eff}(s+2/r) }\right\| \in \mathcal{O}\left( \max_s\frac{\|\partial_s^2 H_{\rm eff}(s)\| T}{r^3}+ \max_s \frac{\|\partial_s H_{\rm eff}(s)\|\|H_{\rm eff}(s)\|T^2}{r^3}\right). \label{eq:midpoint}
\end{align}
We then can bound the scaling of the error in the midpoint approximation by substituting \eqref{eq:log1deriv} and~\eqref{eq:log2deriv} into~\eqref{eq:midpoint} and noting from~\eqref{eq:logWrbd} that $\|H_{\rm eff}(s)\| = (r/4)\|\log(W_r^4(s))\|\in \mathcal{O}(1)$ to find
\begin{align}
    &\left\|\mathcal{T}e^{-iT\int_{s}^{s+4/r} H_{\rm eff}(s') \mathrm{d}s' }-e^{-i\frac{4T}{r}H_{\rm eff}(s+2/r) }\right\| \nonumber\\
    &\qquad= \mathcal{O}\left(\frac{ \|\partial_s^2 W_r^4(s)\| T+ \|\partial_s W_r^4(s)\|T^2}{r^2} \right) \nn
    &\qquad= \mathcal{O}\left(\frac{ \max_s\left( \|\ddot{H}\| + (\|\dot{H}\|+|\dot{\lambda}|)\left(\frac{|\dot{\lambda}|}{\lambda}\right)+|\ddot{\lambda}|\right)T+ \max_s\left(|\dot{\lambda}| + \|\dot{H}\| \right)T^2}{\lambda r^3} \right).
\end{align}
Since errors are sub-additive the error in performing a simulation from $s=0$ to $s=1$ is at most $\order{r}$ times the error given above.  This results in the following bound on the scaling of the value of  $r$  that suffices to guarantee simulation error at most $\epsilon$
\begin{equation}
    r\in \mathcal{O}\left( \sqrt{\frac{ \max_s\left( \|\ddot{H}\| + (\|\dot{H}\| + |\dot{\lambda}|)\left(\frac{|\dot{\lambda}|}{\lambda}\right)+|\ddot{\lambda}|\right)T+ \max_s\left(|\dot{\lambda}| + \|\dot{H}\| \right)T^2}{\lambda \epsilon}} \right).
\end{equation}

The adiabatic theorem then implies that, under reasonable assumptions about the derivatives of the Hamiltonian~\cite{elgart2012note} (specifically that the Hamiltonian is Gevrey class $G^\alpha$ for $\alpha\ge 1$), the value of $T$ needed to achieve error $\epsilon$, given that the minimum eigenvalue gap for the effective Hamiltonian is ${\gap}_{\rm eff}$, scales  at most as 
\begin{equation}
    T\in \widetilde{\mathcal{O}}\left(\frac{\max_s\|\dot{H}_{\rm eff}(s)\|}{{\gap}_{\rm eff}^2 \epsilon} \right) \subseteq \widetilde{\mathcal{O}} \left(\frac{\max_s \left( \|\dot{H}(s)\|+ |\dot{\lambda}|\right)}{\lambda {\gap}_{\rm eff}^2 \epsilon} \right).
\end{equation}
This implies that if $\lambda \in \Omega(1)$ and $\gap_{\rm eff} \in o(1)$
\begin{align}
    r&\in \widetilde{\mathcal{O}}\left( \frac{1}{\epsilon^{3/2}}\sqrt{\frac{ \max_s\left( \|\ddot{H}\| + (\|\dot{H}\|+|\dot{\lambda}|)\left(\frac{|\dot{\lambda}|}{\lambda}\right)+|\ddot{\lambda}|\right)\max_s\left(|\dot{\lambda}| + \|\dot{H}\| \right)}{{\gap}_{\rm eff}^2\lambda^2}+ \frac{\max_s\left(|\dot{\lambda}| + \|\dot{H}\| \right)^3}{{\gap}_{\rm eff}^4\lambda^3 }} \right)\nn
    &\subseteq \widetilde{\mathcal{O}}\left( \frac{1}{\epsilon^{3/2}}\sqrt{\frac{ \max_s\left( \|\ddot{H}\| +|\ddot{\lambda}|\right)\max_s\left(|\dot{\lambda}| + \|\dot{H}\| \right)}{{\gap}_{\rm eff}^2\lambda^2}+ \frac{\max_s\left(|\dot{\lambda}| + \|\dot{H}\| \right)^3}{{\gap}_{\rm eff}^4\lambda^3 }} \right).
\end{align}
If the Hamiltonian $H$ is maximum rank then the spectral gap of the effective Hamiltonian is on the order of ${\gap}_{\rm eff} \in \Omega(\min(\gap, \min_k |E_k|) / \lambda)$ where $\gap$ is the minimum spectral gap of the Hamiltonian $H$.  The minimum over energy comes from the fact that the eigenvalues of $W_r$ in the set $\{\pm 1, \pm i\}$ are mapped to $1$, which can lead to degeneracies in the effective Hamiltonian that were absent in the original Hamiltonian.  Thus the final scaling that we obtain is
\begin{equation}
    r\in \widetilde{\mathcal{O}}\left( \frac{1}{\epsilon^{3/2}}\sqrt{\frac{ \max_s\left( \|\ddot{H}\| +|\ddot{\lambda}|\right)\max_s\left(|\dot{\lambda}| + \|\dot{H}\| \right)}{\min({\gap},\min_k |E_k|)^2}+ \frac{\lambda \max_s\left(|\dot{\lambda}| + \|\dot{H}\| \right)^3}{\min({\gap},\min_k |E_k|)^4 }} \right).
\end{equation}

This confirms that by taking the number of steps sufficiently large that we can force the diabatic error to become arbitrarily small.  Thus we can use the walk operator in place of a Trotterized sequence for adiabatic state preparation and in turn as a heuristic that will converge to the global optima given a large enough $r$.  It should be noted, however, that the bounds used in this analysis are extremely loose and if a quantitatively correct estimate of the scaling is desired then many of the simplifications used above can be eschewed at the price of increasing the complexity of the expression.

Note that in practice, the adiabatic paths can be chosen such that the second derivative of the Hamiltonian is zero and similarly we can choose paths such that $\lambda$ is constant by absorbing it into the definition of the evolution time for each infinitesimal step.  However, we give the above expression for generality. Higher order versions of this can also be derived using time-dependent Trotter-Suzuki formulas~\cite{Wiebe2008}.

\section{In-place binary to unary conversion}\label{app:b2u}

\begin{figure}
\centering
\begin{minipage}[b]{.4\textwidth}
\Qcircuit @C=.7em @R=.7em {
&&\lstick{\text{(a)}\qquad k_{0}} &\qw  &\qw       &\qw       &\qswap    &\targ&\targ    &\qw       &\qw      &\targ    &\qw       &\qw       &\qw       &\qw      &\qw      &\qw      &\targ&\qw    \\
&&\lstick{k_{1}} &\qw  &\qw       &\qswap    &\qswap\qwx&\qw  &\ctrl{-1}&\qswap    &\qw      &\qw      &\qswap    &\qw       &\qw       &\qw      &\qw      &\qw      &\qw&\qw      \\
&&\lstick{k_{2}} &\qw  &\qswap    &\qswap\qwx&\qw       &\qw  &\qw      &\ctrl{-1} &\targ    &\ctrl{-2}&\qw       &\qswap    &\qw       &\qw      &\qw      &\qw      &\qw&\qw      \\
&&  &\qw  &\qw\qwx   &\qw       &\qw       &\qw  &\qw      &\qswap\qwx&\ctrl{-1}&\qw      &\qw       &\qw       &\qswap    &\qw      &\qw      &\qw      &\qw&\qw      \\
&&  &\qw  &\qswap\qwx&\qw       &\qw       &\qw  &\qw      &\qw       &\qw      &\qw      &\ctrl{-3} &\ctrl{-2} &\ctrl{-1} &\targ    &\targ    &\targ    &\ctrl{-4}&\qw\\
&&  &\qw  &\qw       &\qw       &\qw       &\qw  &\qw      &\qw       &\qw      &\qw      &\qswap\qwx&\qw\qwx   &\qw\qwx   &\ctrl{-1}&\qw      &\qw      &\qw&\qw      \\
&&  &\qw  &\qw       &\qw       &\qw       &\qw  &\qw      &\qw       &\qw      &\qw      &\qw       &\qswap\qwx&\qw\qwx   &\qw      &\ctrl{-2}&\qw      &\qw&\qw      \\
&&  &\qw  &\qw       &\qw       &\qw       &\qw  &\qw      &\qw       &\qw      &\qw      &\qw       &\qw       &\qswap\qwx&\qw      &\qw      &\ctrl{-3}&\qw&\qw \gategroup{1}{8}{1}{8}{.7em}{--}\gategroup{1}{7}{2}{9}{.7em}{--} \gategroup{1}{6}{4}{12}{.7em}{--}
 }
\end{minipage}\hspace{2cm}
\begin{minipage}[b]{.4\textwidth}
\Qcircuit @C=.7em @R=.7em {
\lstick{\text{(b)}\qquad k_0}    &\qw&\qw            &\qw         &\multigate{3}{\text{B2U}^{N/2}}&\qw       &\qw       &\qw       &\qw          &\qw      &\qw          &\targ    &\qw      \\
\lstick{k_{1,\ldots,n-2}} &\qw&{/}\qw^{n-2}   &\qw         &\ghost{B2U^{N/2}}                &\qswap    &\qw       &\qw       &\qw          &\qw      &\qw          &\qw      &\qw      \\
\lstick{k_{n-1}}&\qw&\qw            &\qswap      &\ghost{B2U^{N/2}}                &\qw       &\qswap    &\qw       &\qw          &\qw      &\qw          &\qw      &\qw      \\
                &\qw&{/}\qw^{N/2-n}  &\qw\qwx     &\ghost{B2U^{N/2}}                &\qw       &\qw       &\qswap    &\qw          &\qw      &\qw          &\qw      &\qw      \\
                &\qw&\qw            &\qswap\qwx  &\qw                             &\ctrl{-3} &\ctrl{-2} &\ctrl{-1} &\targ        &\targ    &\targ        &\ctrl{-4}&\qw      \\
                &\qw&{/}\qw^{n-2}   &\qw         &\qw                             &\qswap\qwx&\qw\qwx   &\qw\qwx   &\ctrl{-1}^{>0}&\qw\qwx &\qw\qwx      &\qw      &\qw      \\
                &\qw&\qw            &\qw         &\qw                             &\qw       &\qswap\qwx&\qw\qwx   &\qw          &\ctrl{-2}&\qw\qwx      &\qw      &\qw      \\
                &\qw&{/}\qw^{N/2-n}&\qw         &\qw                             &\qw       &\qw       &\qswap\qwx&\qw          &\qw     &\ctrl{-3}^{>0}&\qw      &\qw      \\
}
\end{minipage}
\caption{\label{fig:binary_to_unary}
A depiction of the binary-to-unary circuit mapping an $n$-bit binary number
to an $N$-bit ($N = 2^n$) unary encoding of the input.
In (a) we have the specific example where $N = 8$
(i.e.~$\operatorname{B2U}^{8}$). In (b) we have the circuit ($\operatorname{B2U}^N$)
defined recursively (in terms of $\operatorname{B2U}^{N/2}$),
In (b) the controlled-\textsc{swap} symbols are used to represent
many controlled-\textsc{swap}s, one for each qubit in the relevant
registers. The symbol ``$>\!0$'' signifies that
the multi-\textsc{cnot} is activated on any state other
than the state of all zeros, which can be implemented with a
cascade of \textsc{cnot}s because the input is
promised to have at most one non-zero qubit.
The labelled rails in both circuit diagrams refer to the bits of
the binary-encoded input $k$, and the unlabelled inputs are fresh ancillae.}
\end{figure}
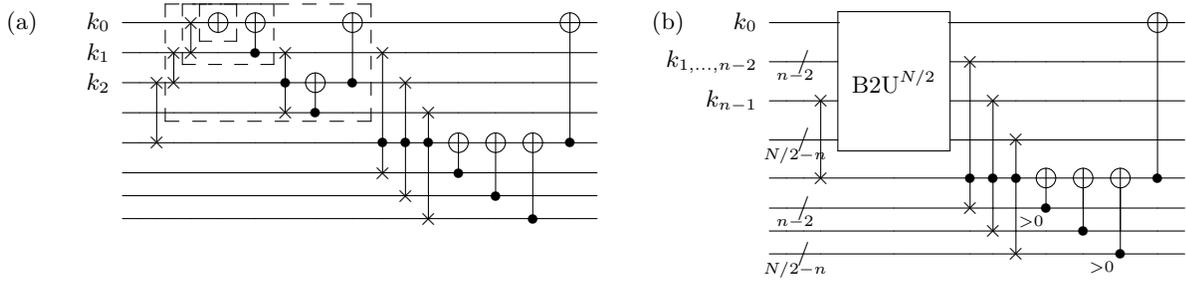

Here we present a quantum circuit ($\textsc{B2U}^N$) for converting a binary-encoded integer $k$ ($0 \leq k < N$) into one-hot unary on $N$ bits.
Recall that the one-hot unary encoding should have $k$ encoded as
$\ket{0}^{\otimes (k-1)}\ket{1}\ket{0}^{\otimes (N-k)}$.
An overview of the circuit is depicted in \fig{binary_to_unary}
in the special case that $N$ is a power of two.
First we sketch a proof that the circuit is correct.
Then we explain how to generalize the circuit to the case where $N$
is not a power of two. Finally, we count the non-Clifford gates needed
to perform our binary-to-unary conversion circuit.

We give a sketch of a proof that the circuit is correct for
$N$ a power of two.
Our proof works by induction and we begin by explaining the trivial
case $N = 1$. In this case, the output can only be the $1$-bit,
one-hot unary encoding of $0$ and hence the output should be
a single qubit, $\ket{1}$. The only input to the circuit is an ancilla initialized to $\ket{0}$ and so we can perform $\text{B2U}^1$ with a single
$\textsc{not}$ gate.

Now that we have explained the trivial case, we next explain our
recursion and why it works (see \figa{binary_to_unary}{(b)}). The idea is as follows.
First, we apply $\text{B2U}^{N/2}$ to $k' := k - 2^{n-1} k_{n-1}$,
where $k_{n-1}$ is the most significant bit of $k$.
This input is simply the last $n-1$ bits of $k$ and 
the output of $\text{B2U}^{N/2}$ is $N/2$ qubits.
Then, controlled on $k_{n-1}$, we swap bits $1$ through $N/2 - 1$
(counting from zero) of the output of $\text{B2U}^{N/2}$ with
$N/2 - 1$ ancilla qubits initialized to $0$. Note that this step does not
execute a controlled-\textsc{swap} on position $0$ of the
one-hot unary encoding of $k'$.
Having performed these controlled-\textsc{swap}s, we next wish
to erase qubit $N/2$ if $k > N/2$. We do this by performing
$N/2 - 1$ \textsc{cnot} gates targeted on the qubit at position
$N/2$ and controlled by each of the qubits
at positions above $N/2$. Finally, we have to resolve
the special cases where $k$ is $N/2$ or $0$. We do
this with one more \textsc{cnot}, with qubit
$N/2$ as the control and qubit $0$ as the target.

Having given an explanation of our recursive
construction, we next explain how to prove that the recursion works.
We consider three distinct cases.
\begin{enumerate}
  \item If $k < N/2$,
  we have $k_{n-1} = 0$ and hence none of the 
  controlled-\textsc{swap}s or \textsc{cnot}s
  will do anything. This is correct behaviour because the one-hot
  unary encoding of $k$ will be the one-hot unary encoding of $k'$
  with $N/2$ ancilla qubits appended to it.

  \item If $k = N/2$,
  the controlled-\textsc{swap}s will again do nothing but this is
  now because they are swapping pairs of identical qubits in the $\ket{0}$ state.
  The \textsc{cnot}s targeted on the qubit at position $N/2$
  will also do nothing because the control qubits are $0$. The final
  \textsc{cnot} will then erase the $1$ encoded in position $0$ of the output of $\textsc{B2U}^{N/2}$, which is there
  because the input was $k' = 0$.
  
  \item If $k > N/2$,
  the controlled-\textsc{swap}s will swap the one-hot unary encoding
  of $k'$ into the final $N/2$ qubits of the output register.
  The \textsc{cnot}s targeted on qubit $N/2$ then erase that qubit, leaving the correct unary encoding
  of $k$. The final \textsc{cnot} does nothing, as the
  control qubit was erased.
\end{enumerate}

The proof sketch demonstrates that our recursive binary-to-unary
circuit works when $N$ is a power of two. Next we explain how to
modify the circuit when $N$ is not a power of two. If $N$ is
not a power of two, define $n := \lceil \log N \rceil$ and
$N' = 2^n$. Apply $\text{B2U}^{N'/2}$ to the least significant
$n-1$ bits of $k$. Then perform the controlled-\textsc{swap}s
and \textsc{cnot}s involving the remaining 
$N - N'/2$ ancilla qubits, removing any operations that would 
involve deleted qubits. For $N = 7$, for example,
we would delete the bottom rail from \figa{binary_to_unary}{(a)} as well as the controlled-\textsc{swap} 
and the \textsc{cnot} involving that final rail.
To see that this works, observe that the circuit would
also work if we performed $\text{B2U}^{N'}$ and then remove the
final $N' - N$ qubits, which are guaranteed to be zero.
Our construction simply eliminates unnecessary gates from
$\text{B2U}^{N'}$.

Our final task is to count the number of non-Clifford gates
needed by our $\text{B2U}^N$ circuit. The only non-Clifford
gates are the controlled-\textsc{swap} operations, which can be
executed with a single Toffoli gate and two \textsc{cnot}s. We prove that the number of
controlled-\textsc{swap} gates is
\begin{equation}
\label{eq:B2U_cost}
    C_N := N - \lceil \log N \rceil - 1.
\end{equation}
First, it is clear that $C_1 = 0$ as required. Next,
it is clear from \figa{binary_to_unary}{(b)} that
$C_{N'} = N'/2 - 1 + C_{N'/2}$. Based on our analysis above,
$C_N = C_{N'} - (N' - N)$ and hence
\begin{equation}
\label{eq:B2U_cost_recursive}
    C_N = N - N'/2 - 1 + C_{N'/2}.
\end{equation}
Next assume \eq{B2U_cost} is true for some particular value $N'/2$. Then by substitution in \eq{B2U_cost_recursive},
\begin{align}
  C_N &= N - N'/2 - 1 + N'/2 - \lceil \log N'/2 \rceil - 1 \nonumber \\
      &= N - n - 1 = N - \lceil \log N \rceil - 1
\end{align}
thus satisfying \eq{B2U_cost} for $N$ as required. Therefore by induction \eq{B2U_cost} is correct for all $N$.

\section{Cost of multiplication}\label{app:multcost}
\label{app:multiplication}

As the multiplication operation is a major
contributor to the overall complexity of our algorithms, we need
to be quite careful in our analysis of the operation. We also
frequently require only low-precision arithmetic, meaning that 
we can make our multiplications less accurate and therefore
computationally cheaper. This Appendix presents our algorithms for
performing four variations of the multiplication task, with modifications
to be used when one of the inputs is given classically rather than quantumly.

Our strategy is to use schoolbook multiplication.
In schoolbook multiplication, the product
$\mulAB := \inA \times \inB$ is calculated by writing
$\inA = \sum_\ell 2^\ell \inA_\ell$ with $\inA_\ell \in \{0, 1\}$
and then calculating the sum $\mulAB = \sum_\ell 2^\ell \inA_\ell \inB$.
This reduces the task of multiplication to two very simple
multiplications and the task of adding a list of numbers.
The two multiplications are simple because multiplication
by a power of two can be accomplished by an appropriate bit-shift
operation, and multiplying by a single bit can be accomplished
by using that bit as a control for the addition operation.
That is, we perform that part of the addition if and only if the
control bit is one.

We begin in~\app{multiplication/uses} by reviewing the parts of the main text
where we need to multiply two numbers together.
In \app{multiplication/add_constant} we explain how to add a constant value to a quantum register,
which is used separately in \alg{directenergyLtermQUBO}
but is also used through the rest of this appendix in order to
multiply a quantum variable to a classical constant.
We then explain the simplest variant of multiplication
in~\app{multiplication/integer_integer},
where we must multiply two integers together.
We then explain the remaining variants
by modifying the integer-integer multiplication algorithm as appropriate.
In~\app{multiplication/integer_real}, we explain the case where
we multiply an integer to a real number.
In~\app{multiplication/real_real}, we explain the case where we
multiply a real number to another real number.
Finally, in~\app{multiplication/real_square}, we explain the case
where we calculate the square of a given real number.
In all cases we indicate how the algorithm is to be modified when one of
the inputs is classically specified.

\subsection{Uses of multiplication in this paper}
\label{app:multiplication/uses}

In \sec{phase_oracle} we need to multiply a quantum register by a classical constant $\tilde{\gamma}$ or $\gamma$ to obtain the phase to apply.
The multiplication is performed directly into the phase gradient state, so we cannot use the savings where the multiplication result is placed in an initially zero register.
The fastest method seems to be to write the classical constant as a sum of powers of 2 with plus and minus signs.

In \sec{functions} we consider QROM for interpolation of functions, and we need to multiply the input register by the slope.
In that case, both registers are quantum.
The input register is given to $\bdiff$ bits, and the goal is to give the approximation to the function to $\bsmooth$ bits.
This may require giving the slope to $\bsmooth+\order{\log\bsmooth}$ bits, or $\bsmooth+\bfun+\order{\log\bsmooth}$ bits in the case of the arcsine.
For Szegedy walks, we need to take the square of a quantum register, and need to multiply a quantum register by a constant.

\subsection{Methods for addition}
\label{app:multiplication/add_constant}

\begin{figure}[t]
\centering
  \resizebox{.7\linewidth}{!}{\includegraphics{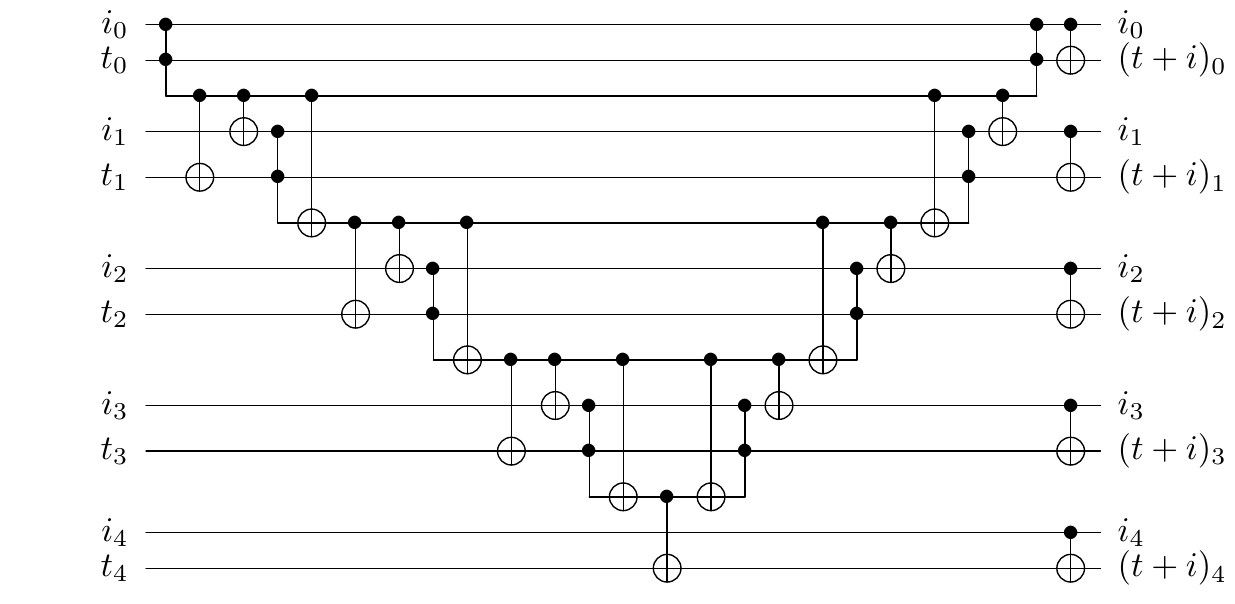}}
  \caption[Addition circuit]{\label{fig:adder}
    A circuit to perform addition on 5 qubits modulo $2^5$ from \cite{GidneyAdder}.}
\end{figure}

When adding a classically given constant to a quantum register, it is possible to save the qubits that would be used to store this classical constant.
Consider the quantum circuit for addition following \cite{GidneyAdder}, as shown in \fig{adder} where $i$ is the classically given integer and $t$ is the quantum register.
For this diagram we use the convention of \cite{GidneyAdder} where a Toffoli with a target known to be initially zeroed is shown with a $\llcorner$ for the target.
That is the first operation on the left in \fig{adder}.
The Toffolis with targets that are known to be zero afterwards are shown with $\lrcorner$ for the target.
These may be performed with measurements and Cliffords so do not add to the non-Clifford cost.

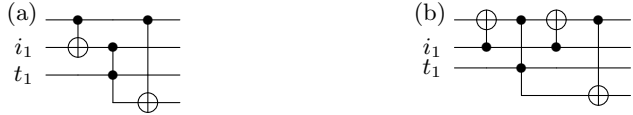
\begin{figure}[t]
\centerline{
(a)~\Qcircuit @C=0.9em @R=.6em {
& \ctrl{1} & \qw& \ctrl{3} &  \qw  \\
\lstick{i_1} & \targ & \ctrl{2}& \qw & \qw  \\
\lstick{t_1} & \qw & \control\qw& \qw & \qw \\
& & & \targ & \qw  \\
}
\hspace{3cm} (b)~\Qcircuit @C=0.9em @R=.6em {
 & \targ & \ctrl{3} & \targ & \ctrl{3} & \qw   \\
\lstick{i_1} & \ctrl{-1} & \qw & \ctrl{-1} & \qw &  \qw   \\
\lstick{t_1}  & \qw & \control\qw & \qw & \qw & \qw  \\
& & & \qw &  \targ & \qw \\
}}\caption[Adder circuit optimization]{\label{fig:adder1}(a) The component of the adder circuit where the qubit containing classical data is the target of a \textsc{cnot}.
(b) The circuit may be rewritten so the value on the second qubit is never changed.}
\end{figure}

\begin{figure}[t]
\centering
  \resizebox{.8\linewidth}{!}{\includegraphics{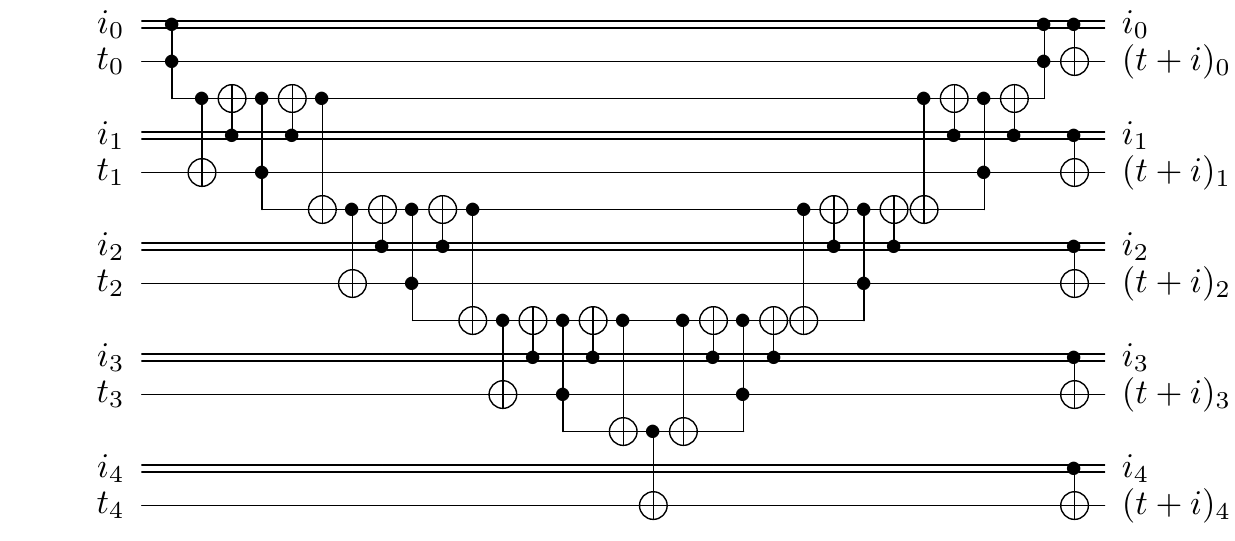}}
  \caption[Modified adder circuit with the optimization]{\label{fig:adder2}
    A circuit to perform addition on 5 qubits modulo $2^5$ such that the $i_j$ registers are only used as controls.
    Because they are only used as controls, if the number $i$ is given classically the addition can be performed entirely using classical control, without using any ancillas to store $i$.}
\end{figure}

The circuit for the adder contains a subsection where a \textsc{cnot} gate is performed on a qubit of $i$, say $i_1$, as shown in \figa{adder1}{(a)}.
The state after the \textsc{cnot} can alternatively be obtained on the control by switching the control and target for the \textsc{cnot}.
Then for the following Toffoli where $i_1$ would be the control, we switch the control to the carry register at the top.
After that the carry register needs to be used as a control where it should take its original value, so we need another \textsc{cnot} to undo the first.
The resulting section of the circuit is as shown in \figa{adder1}{(b)}.
Replacing all these sections of the circuit in this way, we obtain an addition circuit as shown in \fig{adder2}.
This adder only uses the $i_j$ registers as controls.
Since these registers have classically known values, all controls by these qubits may be replaced with classical controls, and these qubits need not be used.
This also reduces the Toffoli cost by 1, because the first Toffoli is replaced with a \textsc{cnot}.
The Toffoli cost is therefore the number of bits minus $2$.
The number of ancillas needed is the number of bits minus $1$.

\subsection{Multiplying two integers}
\label{app:multiplication/integer_integer}
In this variant of the multiplication task, we are to multiply
the $\dimA$-bit integer $\inA$ to the $\dimB$-bit integer $\inB$.
These integers are encoded into quantum registers \texttt{A} and \texttt{B},
respectively. Thus our task is to prepare a $(\dimA + \dimB)$-qubit register
\texttt{out} as follows:
\begin{equation}
    \ket{\inA}_\texttt{A} \ket{\inB}_\texttt{B} \ket{0}_\texttt{out}
    \mapsto
    \ket{\inA}_\texttt{A} \ket{\inB}_\texttt{B}
    \ket{\mulAB := \inA \times \inB}_\texttt{out}.
\end{equation}
We explain how to perform this multiplication using schoolbook
multiplication and the Gidney adder~\cite{GidneyAdder},
and we explain how to reduce the computational cost
if one of the inputs is presented to us classically rather than quantumly.

We now explain the schoolbook multiplication algorithm in some detail.
Let the bits of $\inA$ and $\inB$ be denoted as follows:
\begin{equation}
    \inA := \sum_{\ell = 1}^\dimA 2^{\dimA-\ell} \inA_\ell;\quad
    \inB := \sum_{\ell = 1}^\dimB 2^{\dimB-\ell} \inB_\ell;\quad
    \inA_\ell, \inB_\ell \in \{0, 1\}.
\end{equation}
Thus $\inB_\dimB$ refers to the least-significant bit of $\inB$.
Our procedure is then as follows.
\begin{enumerate}
    \item Controlled on the final qubit of \texttt{B}, copy all the qubits of
    \texttt{A} into the final $\dimA$ bits of \texttt{out}. \newline
    \emph{Result:} $\ket{0}_\texttt{out} \mapsto
    \ket{\inB_\dimB \inA}_\texttt{out}$. \newline
    \emph{Cost:} $\dimA$ Toffolis.
    
   \item For each $\ell = \dimB-1, \ldots, 1$, add $2^\ell$ times the value of \texttt{A} to \texttt{out} in place, controlled on the $(\dimB-\ell)^\text{th}$ qubit of \texttt{B}.
        This can be done by using the control to copy the $\dimA$ bits to an ancilla, and adding this ancilla.
        The ancilla can be erased with no Toffoli cost.
        We add \texttt{A} to \texttt{out} with the final
        $\ell$ qubits of \texttt{out} ignored. Note that the number of nonzero bits
        will always be no greater than $\dimA$. \newline
        \emph{Result:} $\ket{\xi}_\texttt{out} \mapsto
        \ket{\xi + 2^\ell \inB_{\dimB-\ell} \inA}_\texttt{out}$,
        where $\xi$ is the integer encoded in \texttt{out} before this step. \newline
        \emph{Cost:} $2\dimA$ Toffolis.
\end{enumerate}
The total number of Toffolis is $2 \dimA \dimB - \dimA$, and the total number of
temporary ancilla qubits needed is $2\dimA-1$ since we are copying $\dimA$ qubits out to an ancilla as well as using $\dimA-1$ temporary qubits in the addition.

We now consider how the cost of the algorithm can be reduced when one of the
inputs is presented classically, rather than quantumly.
The effect on the algorithm is different depending on whether $\inA$ or $\inB$
is known classically. In the case that $\inA$ is known classically,
we can replace all the Toffolis in the copy operation in step 1 with CNOTs or identity gates,
depending on whether the relevant bit of $\inA$ is $1$ or $0$. More interestingly,
each addition step would involve adding a known constant rather than
an unknown variable to be read from a quantum register during computation.
The effect on computational complexity depends on the classical constant $\inA$;
in particular, on the largest power of two that divides $\inA$.
In the worst case ($\inA \mod 2 = 1$), we save one Toffoli per addition step.
In the best case ($\inA = 0$), we have zero computational cost because we are
multiplying by zero and we know we are multiplying by zero.

In the case that $\inB$ is presented to us classically rather than quantumly,
we can make the addition controlled by performing the (non-controlled) addition circuit in the case of $1$, or doing nothing when the classical control is $0$. The number of quantum-to-quantum additions
would thus depend on the number of non-zero classical bits -- the greater the
Hamming weight of the classical input, the greater the number of additions to be
performed. Note that this is distinct from the cost of performing classical-to-quantum
multiplication when $\inA$ is the classical variable, in which case the complexity 
is determined by the number of zeros on the far right of the number.

The case where $\inB$ is given classically is more relevant for this paper.
We are unlikely to be dealing with classically specified integers that are a
multiple of a large power of two. On the other hand, we will frequently have some
information about the Hamming weight $\sum_\ell \inB_\ell$ of the classically
known number $\inB$. Each addition costs at most $\dimA$ Toffolis, we perform
$\sum_\ell \inB_\ell \leq \dimB$ such additions, and no other operations require
Toffoli gates. We therefore have a total Toffoli cost of at most $\dimA \dimB$,
which can be replaced with $\dimA \sum_\ell \inB_\ell$ if we can assume knowledge
of the Hamming weight of $\inB$. Thus we save a factor of $2$ if one of the inputs
is classical.

In the following subsections, we explain how to modify the above procedure for
variants of the multiplication task. These variants have at least one of the
inputs being a real number between $0$ and $1$, rather than an integer.
Thus the task is not to calculate the multiplication exactly, as this would involve
infinitely many bits for real numbers. Instead, we truncate the binary expansions
of real numbers to ensure that an error threshold is achieved.

\subsection{Multiplying an integer to a real number}
\label{app:multiplication/integer_real}

Now we consider a variant of the multiplication task where one of the inputs
is a real number between zero and one. For reasons that become clear below,
we specify $\inA$ to be the real number and $\inB$ to be the integer. We assume that
the real number is defined to infinitely many digits and that our task is
to approximate $\mulAB := \inA \times \inB$ to within an error tolerance $\mulError$.
Thus our task is to calculate some $\tilde\mulAB$ such that
$\abs{\mulAB - \tilde\mulAB} < \mulError$. That is to say, we are to prepare
a new quantum register \texttt{out} as follows:
\begin{equation}
    \ket{\inA}_\texttt{A} \ket{\inB}_\texttt{B} \ket{0}_\texttt{out} \mapsto
    \ket{\inA}_\texttt{A} \ket{\inB}_\texttt{B} \ket{\tilde\mulAB}_\texttt{out}.
\end{equation}
Here we are free to choose the number of bits for the register \texttt{A} and hence
the number of bits for the register \texttt{out}.
This choice will naturally depend on the error tolerance $\mulError$.

We begin by specifying symbols for the bits of the inputs $\inA$ and $\inB$.
Note that the indexing differs somewhat from the previous section. We define
\begin{equation}
    \inA := \sum_{\ell = 1}^\infty \inA_\ell/2^\ell;\quad
    \inB := \sum_{\ell = 1}^\dimB 2^{\dimB-\ell} \inB_\ell;\quad
    \inA_\ell, \inB_\ell \in \{0, 1\}.
\end{equation}
We then select an integer $\dimA \geq \dimB$ (presuming $\mulError < 1$)
that will count the number of bits of the input $\inA$ we plan to use.
We use $\dimA -1$ bits of $\inA$.
We explain our plan by first representing the ideal product as
\begin{equation}
\label{eq:mult_integer_real}
  \mulAB =
  \left(\begin{array}{clccccccccccc|ccc}
      & \inB_{    1    } & \times 
        & \inA_{        1        } & \inA_{        2        } 
        & \inA_{        3        } &         \cdots
        & \inA_{    \dimB - 2    } & \inA_{    \dimB - 1    } & .
        & \inA_{    \dimB        } &         \cdots      
        & \inA_{    \dimA - 1    } & \inA_{      \dimA      } 
        & \cdots \\
    + & \inB_{    2    } & \times
        &                      & \inA_{    1    } 
        & \inA_{        2        } &           \cdots
        & \inA_{    \dimB - 3    } & \inA_{    \dimB - 2    } & .
        & \inA_{    \dimB - 1    } &           \cdots      
        & \inA_{    \dimA - 2    } & \inA_{    \dimA - 1    } 
        & \cdots \\
    + & \inB_{    3    } & \times 
        &                          &                        
        & \inA_{        1        } &           \cdots
        & \inA_{    \dimB - 4    } & \inA_{    \dimB - 3    } & .
        & \inA_{    \dimB - 2    } &           \cdots      
        & \inA_{    \dimA - 3    } & \inA_{    \dimA - 2    } 
        & \cdots \\
      & & \vdots & & & & & & & & & & & \\
    + & \inB_{\dimB - 1} & \times 
        &                          &                        
        &                          &                 
        &                          & \inA_{        1        } & .
        & \inA_{        2        } &           \cdots      
        & \inA_{\dimA - \dimB + 1} & \inA_{\dimA - \dimB + 2} 
        & \cdots \\
     + & \inB_{ \dimB  } & \times 
        &                          &                        
        &                          &                 
        &                          &              0           & .
        & \inA_{        1        } &           \cdots      
        & \inA_{  \dimA - \dimB  } & \inA_{\dimA - \dimB + 1} 
        & \cdots
  \end{array}\right),
\end{equation}
where the vertical line denotes where we truncate the binary expansion
of $\inA$. We thus calculate
\begin{equation}
  \tilde\mulAB_\dimA := \sum_{\ell = 1}^\dimB
  \inB_\ell \lfloor \inA 2^{\dimA - \ell} \rfloor  2^{\dimB-\dimA};
  \quad  \mulAB - \tilde\mulAB_\dimA \leq \dimB 2^{\dimB - \dimA}.
\end{equation}
To ensure that the error tolerance $\mulError$ is achieved,
we should choose $\dimA > \dimB + \log (\dimB/\mulError)$.
We therefore choose 
\begin{equation}
    \dimA = \dimB +\lceil \log (\dimB/\mulError) \rceil.
\end{equation}

We follow a similar strategy to that described in \app{multiplication/integer_integer},
meaning that we are to perform a sequence of controlled additions.
We work bottom to top in \eq{mult_integer_real}.

We start with the bottom line by copying $\dimA - \dimB$ bits into the output register, with Toffoli cost $\dimA - \dimB$.
After that the number of Toffolis is twice the number of bits.
The total number of Toffolis is then
\begin{align}
(\dimA - \dimB) + 2(\dimA - \dimB + 1) + \ldots + 2(\dimA - 1)
&= \dimA - \dimB + \sum_{\ell = \dimA - \dimB+1}^{\dimA - 1} \ell \nn
&= \dimA (2\dimB - 1)-\dimB^2 \nn
&=(\dimB +\lceil \log (\dimB/\mulError) \rceil)(2\dimB - 1)-\dimB^2 \nn
&= \dimB^2+(2\dimB-1) \lceil \log (\dimB/\mulError) \rceil -\dimB.
\end{align}
Hence the Toffoli cost of multiplying a real number to an integer
on a quantum computer is no more than
\begin{equation}
    \dimB^2+(2\dimB-1) \lceil \log (\dimB/\mulError) \rceil -\dimB,
\end{equation}
where $\dimB$ is the number of bits used to specify the integer
and $\mulError$ is the allowable error in the overall multiplication.
The algorithm requires that the real number is specified to
$\dimA = \lceil \dimB \log(\dimB/\mulError) \rceil$ bits and uses $\dimA-1$ ancilla qubits.

\subsection{Multiplying two different real numbers}
\label{app:multiplication/real_real}

In this subsection we consider the task where we are to multiply
two real numbers $\inA$ ($0 \leq \inA < 1$) and $\inB$ ($0 \leq \inA < 1$).
Our task is to calculate an approximation $\tilde\mulAB$ to
$\mulAB := \inA \times \inB$ such that $\abs{\mulAB - \tilde\mulAB} < \mulError$,
where $\mulError > 0$ is some given error tolerance.
That is to say, we are to prepare a new quantum register \texttt{out}
as follows:
\begin{equation}
  \ket{\inA}_\texttt{A} \ket{\inB}_\texttt{B} \ket{0}_\texttt{out}
  \mapsto
  \ket{\inA}_\texttt{A} \ket{\inB}_\texttt{B} \ket{\tilde\mulAB}_\texttt{out}.
\end{equation}
We are free to choose the number of qubits in each of the registers
\texttt{A}, \texttt{B}, and \texttt{out} to ensure that the output
encodes a value for $\tilde\mulAB$ that approximates $\mulAB$ to within
the error tolerance $\mulError$. We begin by discussing these choices of
register size, starting with the size of \texttt{A} and \texttt{B}.

To explain our choice for the numbers of qubits for registers \texttt{A} and
\texttt{B}, we begin by introducing notation for the inputs $\inA$ and $\inB$.
As before, we define the bits of the inputs according to the equations
\begin{equation}
    \inA := \sum_{\ell = 1}^\infty \inA_\ell/2^\ell;\quad
    \inB := \sum_{\ell = 1}^\infty \inB_\ell/2^\ell;\quad
    \inA_\ell, \inB_\ell \in \{0, 1\}.
\end{equation}
This suggests our strategy for calculating $\mulAB$. As before, we have
\begin{equation}
\label{eq:mult_real_real}
  \mulAB =
  \left(\begin{array}{clccccccccccc|ccc}
      & \inB_{    1    } & \times & .0
        & \inA_{        1        } & \inA_{        2        } 
        & \inA_{        3        } &         \cdots
        & \inA_{    \dimB - 2    } & \inA_{    \dimB - 1    }
        & \inA_{    \dimB        } &         \cdots      
        & \inA_{    \dimA - 1    } & \inA_{      \dimA      } 
        & \cdots \\
    + & \inB_{    2    } & \times & .0
        &            0         & \inA_{    1    } 
        & \inA_{        2        } &           \cdots
        & \inA_{    \dimB - 3    } & \inA_{    \dimB - 2    }
        & \inA_{    \dimB - 1    } &           \cdots      
        & \inA_{    \dimA - 2    } & \inA_{    \dimA - 1    } 
        & \cdots \\
    + & \inB_{    3    } & \times & .0
        &            0           &              0         
        & \inA_{        1        } &           \cdots
        & \inA_{    \dimB - 4    } & \inA_{    \dimB - 3    }
        & \inA_{    \dimB - 2    } &           \cdots      
        & \inA_{    \dimA - 3    } & \inA_{    \dimA - 2    } 
        & \cdots \\
      & & \vdots & & & & & & & & & & & & \\
    + & \inB_{\dimB - 1} & \times & .0 
        &            0             &              0         
        &            0             &           \cdots
        &            0             & \inA_{        1        }
        & \inA_{        2        } &           \cdots      
        & \inA_{\dimA - \dimB + 1} & \inA_{  \dimA - \dimB  } 
        & \cdots \\ \hline
    + & \inB_{  \dimB  } & \times & .0 
        &            0             &              0         
        &            0             &           \cdots
        &            0             &              0           
        & \inA_{        1        } &           \cdots      
        & \inA_{  \dimA - \dimB  } & \inA_{\dimA - \dimB - 1} 
        & \cdots \\
      & & \vdots & & & & & & & & & & & &
  \end{array}\right),
\end{equation}
where solid lines indicate where we truncate the calculation in order to 
produce the approximation $\tilde\mulAB$ instead of $\mulAB$.
Here we have assumed that $\dimA \geq \dimB$; if $\dimA < \dimB$,
our repeated addition procedure would involve several additions by zero.
The repeated addition strategy has a Toffoli cost of
\begin{align}
  (\dimA - \dimB + 1) + 2(\dimA - \dimB + 2) + \ldots + 2(\dimA - 1)
  &= (\dimA - \dimB + 1) + 2\sum_{\ell=\dimA - \dimB + 2}^{\dimA - 1} \ell \nn
  &= \dimA (2\dimB - 3) - (\dimB - 1)^2.
\end{align}
It seems reasonable to set $\dimA = \dimB$, and numerical evidence
indicates that this choice makes the optimal tradeoff between computational
complexity and error tolerance.
Setting $\dimAB := \dimA = \dimB$, the Toffoli cost is simply
$\dimAB^2-\dimAB - 1$.

\begin{figure}
    \centering
    \includegraphics[width=.5\textwidth]{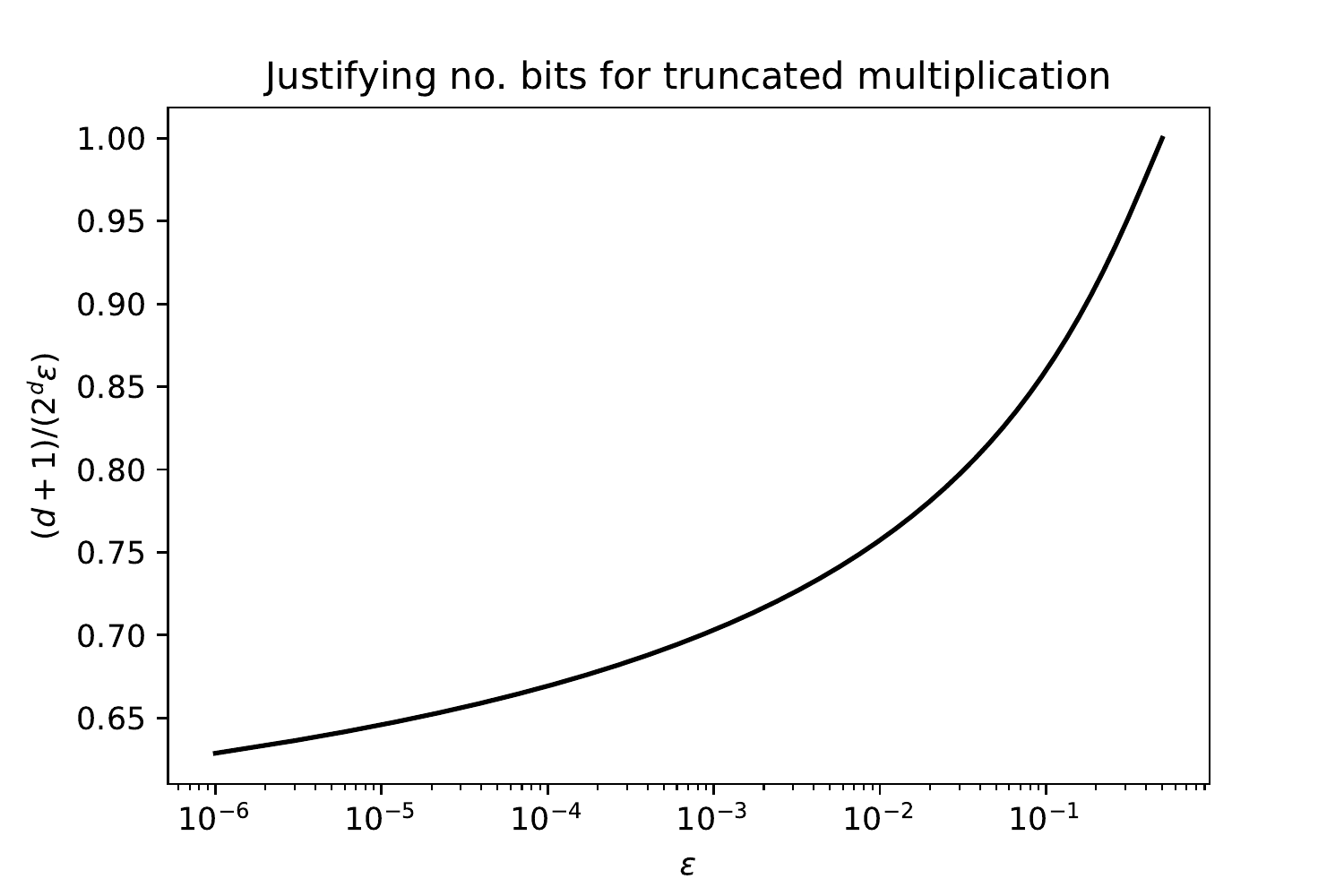}
    \caption[Justification for number of bits used when multiplying
    one real number to another.]{Numerical justification for our choice of
    $\dimAB$ in \eq{no_qubits_mult_real_real}.
    We plot
    the ratio of $(\dimAB + 1)/2^\dimAB$ to the error bound $\mulError$.
    A value less than $1$ ensures that our choice of $\dimAB$ yields a
    multiplication result whose error is less than $\mulError$.}
    \label{fig:trunc_mult}
\end{figure}

We now consider the error of the sum in
\eq{mult_real_real}.
There
\begin{equation}
  \tilde\mulAB =  \sum_{\substack{n,m=1 \\ n+m \leq \dimAB}}^\infty \inA_n \inB_m 2^{-(n+m)} ,
\end{equation}
so
\begin{align}
  \mulAB - \tilde\mulAB 
  = \sum_{\substack{n,m=1 \\ n+m > \dimAB}}^\infty \inA_n \inB_m 2^{-(n+m)}
  \leq \sum_{\substack{n,m=1 \\ n+m > \dimAB}}^\infty  2^{-(n+m)}
  = \frac{\dimAB+1}{2^{\dimAB}}.
\end{align}
We can ensure that the error of the approximation $\tilde\mulAB$ is within tolerance $\mulError$ by setting
\begin{equation}
    \frac{\dimAB + 1}{2^\dimAB} \leq \mulError.
\end{equation}
Though the above could be solved exactly using a Lambert-W function, it is
satisfied with
\begin{equation}
\label{eq:no_qubits_mult_real_real}
    \dimAB = 1 + \log(1/\mulError) + \log(1 + \log(1/\mulError)) .
\end{equation}
\fig{trunc_mult} justifies
this choice by depicting the value of $\frac{\dimAB + 1}{2^\dimAB \mulError}$
as a function of $\mulError$ with $d$ chosen as per
\eq{no_qubits_mult_real_real}.
To choose $\dimAB$, we take the ceiling of this expression, because $\dimAB$ must be chosen to be an integer.

Hence our strategy for multiplying two real numbers uses
\begin{equation}
 \dimAB^2-\dimAB - 1 =
  \log^2 (1/\mulError) + 2\log (1/\mulError) \log \log (1/\mulError)+ \order{\log (1/\mulError)}
\end{equation}
Toffoli gates to achieve an output with error less than $\mulError$.
The ancilla cost is $\dimAB-1$ bits for a copy of the bits of $\inA$ for the controlled addition, and another $\dimAB-1$ bits for the addition itself.
Thus the ancilla cost is
\begin{equation}
2\log (1/\mulError)+ \order{\log\log (1/\mulError)}.
\end{equation}

\subsection{Squaring a real number}
\label{app:multiplication/real_square}

We are given a quantum register \texttt{A} with a real number $\inA$ that satisfies $0 \leq \inA < 1$.
Our task is to calculate
an approximation $\tilde\mulAB$ of $\mulAB := \inA^2$ such that
$\abs{\mulAB - \tilde\mulAB} < \mulError$, where $\mulError$ is given
($0 < \mulError < 1$). That is to say, we are to prepare a new quantum register 
\texttt{out} as follows:
\begin{equation}
    \ket{\inA}_\texttt{A} \ket{0}_\texttt{out} \mapsto
    \ket{\inA}_\texttt{A} \ket{\tilde\mulAB}_\texttt{out}.
\end{equation}
We will include $\dimAB$ bits in the sum, so the sum can be expressed as
\begin{equation}
\tilde\mulAB = \sum_{\substack{n,m=1 \\ n+m\le \dimAB}}^\infty \inA_n \inA_m 2^{-(n+m)}.
\end{equation}
We take advantage of symmetry to rewrite the sum as
\begin{align}\label{eq:symmsum}
\tilde\mulAB &= 2\sum_{\substack{n,m=1 \\ n+m\le \dimAB,n>m}}^\infty \inA_n \inA_m 2^{-(n+m)} + \sum_{n=1}^{{\lfloor \dimAB/2 \rfloor}} \inA_n 2^{-2n} \nn
&= 2\sum_{n=1}^{\dimAB-1} \inA_n \sum_{m=1}^{\min(n-1,\dimAB-n)} \inA_m 2^{-(n+m)} + \sum_{n=1}^{{\lfloor \dimAB/2 \rfloor}} \inA_n 2^{-2n}.
\end{align}
The first term in this sum contains the parts where $n>m$, and is multiplied by 2 because those parts with $n<m$ are equal by symmetry.
The second term is that for $n=m$.
This sum is more efficient, because only about half as many terms appear.
Now the term in the second sum for $n=\lfloor \dimAB/2 \rfloor$ is half the size of any of the other terms, so it is convenient to omit it.

The form of the sum can be shown, for the odd example $\dimAB=15$,
\begin{equation}
\label{eq:ex15}
  \tilde\mulAB =
  \left(\begin{array}{clccccccccccccccc}
      & \inA_1 & \times & .0
        & \blu \inA_1  &    0    &    0    &    0    &    0    &    0    &    0    &    0    &    0    &    0    &    0    &    0    &    0    \\
    + & \inA_2 & \times & .0
        & \inA_1  &    0    & \blu \inA_2  &    0    &    0    &    0    &    0    &    0    &    0    &    0    &    0    &    0    &    0    \\
    + & \inA_3 & \times & .0 
        &    0    & \inA_1  & \inA_2  &    0    & \blu \inA_3  &    0    &    0    &    0    &    0    &    0    &    0    &    0    &    0    \\
     + & \inA_4 & \times & .0  
        &    0    &    0    & \inA_1  & \inA_2  & \inA_3  &    0    & \blu \inA_4  &    0    &    0    &    0    &    0    &    0    &    0    \\
        + & \inA_5 & \times & .0  
        &    0    &    0    &    0    & \inA_1  & \inA_2  & \inA_3  & \inA_4  &    0    & \blu \inA_5  &    0    &    0    &    0    &    0    \\
    + & \inA_6 & \times & .0  
        &    0    &    0    &    0    &    0    & \inA_1  & \inA_2  & \inA_3  & \inA_4  & \inA_5  &    0    & \blu \inA_6  &    0    &    0    \\
    + & \inA_7 & \times & .0  
        &    0    &    0    &    0    &    0    &    0    & \inA_1  & \inA_2  & \inA_3  & \inA_4  & \inA_5  & \inA_6  &    0    & \blu \inA_7     \\ \hline
    + &  \inA_8 & \times & .0  
        &    0    &    0    &    0    &    0    &    0    &    0    & \inA_1  & \inA_2  & \inA_3  & \inA_4  & \inA_5  & \inA_6  & \inA_7   \\
    + & \inA_9 & \times & .0  
        &    0    &    0    &    0    &    0    &    0    &    0    &    0    & \inA_1  & \inA_2  & \inA_3  & \inA_4  & \inA_5   & \inA_6   \\
    + & \inA_{10} & \times & .0  
        &    0    &    0    &    0    &    0    &    0    &    0    &    0    &    0    & \inA_1  & \inA_2  & \inA_3  & \inA_4   & \inA_5   \\
    + & \inA_{11} & \times & .0  
        &    0    &    0    &    0    &    0    &    0    &    0    &    0    &    0    &    0    & \inA_1  & \inA_2  & \inA_3   & \inA_4   \\
    + & \inA_{12} & \times & .0  
        &    0    &    0    &    0    &    0    &    0    &    0    &    0    &    0    &    0    &    0    & \inA_1  & \inA_2  & \inA_3    \\
     + & \inA_{13} & \times & .0  
        &    0    &    0    &    0    &    0    &    0    &    0    &    0    &    0    &    0    &    0    &    0    & \inA_1  & \inA_2    \\
    + & \inA_{14} & \times & .0  
        &    0    &    0    &    0    &    0    &    0    &    0    &    0    &    0    &    0    &    0    &    0    &    0    & \inA_1    \\
    \end{array}\right).
\end{equation}
Here we have shown terms from the second sum from \eq{symmsum} in blue.
In the case where $\dimAB$ is odd, $\lfloor \dimAB/2 \rfloor=(\dimAB-1)/2$, and we can write the sum as
\begin{equation}\label{eq:odd}
\tilde\mulAB = \sum_{n=1}^{(\dimAB-1)/2} \inA_n \left( 2^{-2n} + 2\sum_{m=1}^{n-1} \inA_n \inA_m 2^{-(n+m)} \right)
+2\sum_{n=(\dimAB+1)/2}^{\dimAB-1} \inA_n \sum_{m=1}^{\dimAB-n} \inA_m 2^{-(n+m)}.
\end{equation}
The first sum in \eq{odd} corresponds to the part above the horizontal line in \eq{ex15}, and the second sum in \eq{odd} corresponds to the part below the line.
To compute \eq{odd}, we start at the least significant digit, and move to the most significant digit (corresponding to moving from the bottom row to the top row in \eq{ex16}), as
\begin{equation}
\tilde\mulAB = \sum_{n=\dimAB-1}^{(\dimAB+1)/2} \inA_n \sum_{m=1}^{\dimAB-n} \inA_m 2^{-(n+m-1)} + \sum_{n=(\dimAB-1)/2}^1 \inA_n \left( 2^{-2n} + \sum_{m=1}^{n-1} \inA_n \inA_m 2^{-(n+m-1)} \right).
\end{equation}
To compute the sum we start with $n=\dimAB-1$, and copy the value $\inA_{\dimAB-1} \inA_1$ into the output at position $\dimAB-1$ (to initialize the output as $\inA_{\dimAB-1} \inA_1 2^{-(\dimAB-1)}$) with Toffoli cost 1.
Next, we use $\inA_{\dimAB-2}$ to control addition of $\inA_1 2^{-(\dimAB-2)}+\inA_2 2^{-(\dimAB-1)}$ into the output.
This controlled addition has cost $2\times 2$ because it is for 2 bits.
At step $j=\dimAB-n\le (\dimAB-1)/2$, the cost of controlled addition of $j$ bits is $2j$.
The cost of that part is therefore
\begin{equation}
1+\sum_{j=2}^{(\dimAB-1)/2} 2j = (\dimAB^2-5)/4.
\end{equation}
For the remaining steps with $n=(\dimAB-1)/2$ to 2, we have a cost of $n-1$ to produce the $n-1$ values of $\inA_n \inA_m$, and there are $n+1$ bits that need to be added into the output.
That includes the bit for $\inA_n 2^{-2n}$ which is spaced by one bit from the remaining bits for $\inA_n \inA_m$.
That gives cost $(n-1)+(n+1)=2n$.
For $n=1$, we just have a cost of one Toffoli to add in the single bit (without a control, because it is just $\inA_n$).
That gives the same cost as the first half, for a total cost of
\begin{equation}
\dimAB^2/2-5/2.
\end{equation}

In the case where $\dimAB$ is even, $\lfloor \dimAB/2 \rfloor=\dimAB/2$, and we can write the sum as
\begin{equation}\label{eq:even}
\tilde\mulAB = \sum_{n=1}^{\dimAB/2-1} \inA_n \left( 2^{-2n} + 2\sum_{m=1}^{n-1} \inA_n \inA_m 2^{-(n+m)} \right)
+ 2\inA_{\dimAB/2}\sum_{m=1}^{\dimAB/2-1} \inA_m 2^{-(\dimAB/2+m)}
+2\sum_{n=\dimAB/2+1}^{\dimAB} \inA_n \sum_{m=1}^{d+1-n} \inA_m 2^{-(n+m)}.
\end{equation}
The form of the sum for an even example $\dimAB=16$ is
\begin{equation}
\label{eq:ex16}
  \tilde\mulAB =
  \left(\begin{array}{clcccccccccccccccc}
      & \inA_1 & \times & .0
        & \blu \inA_1  &    0    &    0    &    0    &    0    &    0    &    0    &    0    &    0    &    0    &    0    &    0    &    0    &    0    \\
    + & \inA_2 & \times & .0
        & \inA_1  &    0    & \blu \inA_2  &    0    &    0    &    0    &    0    &    0    &    0    &    0    &    0    &    0    &    0    &    0    \\
    + & \inA_3 & \times & .0 
        &    0    & \inA_1  & \inA_2  &    0    & \blu \inA_3  &    0    &    0    &    0    &    0    &    0    &    0    &    0    &    0    &    0    \\
     + & \inA_4 & \times & .0  
        &    0    &    0    & \inA_1  & \inA_2  & \inA_3  &    0    & \blu \inA_4  &    0    &    0    &    0    &    0    &    0    &    0    &    0    \\
        + & \inA_5 & \times & .0  
        &    0    &    0    &    0    & \inA_1  & \inA_2  & \inA_3  & \inA_4  &    0    & \blu \inA_5  &    0    &    0    &    0    &    0    &    0    \\
    + & \inA_6 & \times & .0  
        &    0    &    0    &    0    &    0    & \inA_1  & \inA_2  & \inA_3  & \inA_4  & \inA_5  &    0    & \blu \inA_6  &    0    &    0    &    0    \\
    + & \inA_7 & \times & .0  
        &    0    &    0    &    0    &    0    &    0    & \inA_1  & \inA_2  & \inA_3  & \inA_4  & \inA_5  & \inA_6  &    0    & \blu \inA_7  &    0    \\ \hline 
    + &  \inA_8 & \times & .0  
        &    0    &    0    &    0    &    0    &    0    &    0    & \inA_1  & \inA_2  & \inA_3  & \inA_4  & \inA_5  & \inA_6  & \inA_7  &    0  \\ \hline
    + & \inA_9 & \times & .0  
        &    0    &    0    &    0    &    0    &    0    &    0    &    0    & \inA_1  & \inA_2  & \inA_3  & \inA_4  & \inA_5   & \inA_6 & \inA_7  \\
    + & \inA_{10} & \times & .0  
        &    0    &    0    &    0    &    0    &    0    &    0    &    0    &    0    & \inA_1  & \inA_2  & \inA_3  & \inA_4   & \inA_5 & \inA_6  \\
    + & \inA_{11} & \times & .0  
        &    0    &    0    &    0    &    0    &    0    &    0    &    0    &    0    &    0    & \inA_1  & \inA_2  & \inA_3   & \inA_4 & \inA_5  \\
    + & \inA_{12} & \times & .0  
        &    0    &    0    &    0    &    0    &    0    &    0    &    0    &    0    &    0    &    0    & \inA_1  & \inA_2  & \inA_3  & \inA_4  \\
     + & \inA_{13} & \times & .0  
        &    0    &    0    &    0    &    0    &    0    &    0    &    0    &    0    &    0    &    0    &    0    & \inA_1  & \inA_2  & \inA_3  \\
    + & \inA_{14} & \times & .0  
        &    0    &    0    &    0    &    0    &    0    &    0    &    0    &    0    &    0    &    0    &    0    &    0    & \inA_1  & \inA_2  \\
    + & \inA_{15} & \times & .0  
        &    0    &    0    &    0    &    0    &    0    &    0    &    0    &    0    &    0    &    0    &    0    &    0    &    0    & \inA_1  \\
    \end{array}\right).
\end{equation}
Again we have shown terms from the second sum from \eq{symmsum} in blue.
The first sum in \eq{even} corresponds to the part above the first horizontal line in \eq{ex16}, the second sum in \eq{even} corresponds to the part between the two lines in \eq{ex16}, and the third sum in \eq{even} corresponds to the part below the second horizontal line in \eq{ex16}.
To compute \eq{even}, we again start at the least significant digit, and move to the most significant digit, as
\begin{equation}\label{eq:evsqsum}
\tilde\mulAB = \sum_{n=\dimAB}^{\dimAB/2+1} \inA_n \sum_{m=1}^{d+1-n} \inA_m 2^{-(n+m-1)}
+ \inA_{\dimAB/2}\sum_{m=1}^{\dimAB/2-1} \inA_m 2^{-(\dimAB/2+m-1)}
+ \sum_{n=\dimAB/2-1}^{1} \inA_n \left( 2^{-2n} + \sum_{m=1}^{n-1} \inA_n\inA_m 2^{-(n+m-1)} \right) .
\end{equation}
For the costing of the additions, we have the same costing for the first sum in \eq{evsqsum} as in the odd case, giving cost
\begin{equation}
1+\sum_{j=2}^{\dimAB/2-1} 2j = (\dimAB^2-2\dimAB-4)/4.
\end{equation}
The middle sum in \eq{evsqsum} has cost $\dimAB-2$, then the final sum has cost $2n$ for $n=2$ to $\dimAB/2-1$, and cost $1$ for $n=1$, giving the same cost as the first sum.
That gives a total complexity
\begin{equation}
\dimAB^2/2-4.
\end{equation}
Thus in both the odd and even cases the complexity is less than $\dimAB^2/2$.

To estimate the error, we have
\begin{align}
\mulAB-\tilde\mulAB &= 2\sum_{\substack{n,m=1 \\ n+m>\dimAB,n>m}}^\infty \inA_n \inA_m 2^{-(n+m)} + \sum_{n=\lfloor \dimAB/2 \rfloor}^{\infty} \inA_n 2^{-2n} \nn
&\le 
2\sum_{\substack{n,m=1 \\ n+m>\dimAB,n>m}}^\infty 2^{-(n+m)} + \sum_{n=\lfloor \dimAB/2 \rfloor}^{\infty} 2^{-2n} \nn
&= \sum_{\ell=\dimAB+1}^{\infty} \lfloor(\ell-1)/2\rfloor 2^{-\ell} + \frac 43 2^{-2\lfloor \dimAB/2 \rfloor} .
\end{align}
In the case of even $\dimAB$ we get
\begin{equation}
\frac 12 \dimAB \, 2^{-\dimAB} + \frac 13 2^{-\dimAB}  + \frac 43 2^{-\dimAB} = \frac 12 \dimAB \, 2^{-\dimAB} + \frac 53 2^{-\dimAB},
\end{equation}
and in the case of odd $\dimAB$ we get
\begin{equation}
\frac 12 \dimAB \, 2^{-\dimAB} + \frac 16 2^{-\dimAB}  + \frac 83 2^{-\dimAB} = \frac 12 \dimAB \, 2^{-\dimAB} + \frac {17}6 2^{-\dimAB}.
\end{equation}
We find that we can limit the error to $\epsilon$ using
\begin{equation}
    \dimAB = \lceil \log(1/\epsilon)+\log(11/3+\log(1/\epsilon)) \rceil.
\end{equation}
The Toffoli cost of squaring is then (regardless of whether $\dimAB$ is odd or even)
\begin{equation}
   \dimAB^2/2 = \frac 12 \log^2 (1/\mulError) + \log (1/\mulError) \log \log (1/\mulError)+ \order{\log (1/\mulError)}.
\end{equation}

The ancilla cost in the case where $d$ is even has a maximum of $\dimAB-2$.
When $n=\dimAB/2$ (corresponding to the part between the two horizontal lines in \eq{ex16}), there are $\dimAB/2-1$ bits to add in a controlled way, so there are $\dimAB/2-1$ bits for the copy and another $\dimAB/2-1$ bits for the addition itself.
When $n=\dimAB/2-1$, there are $\dimAB/2-2$ bits to add in a controlled way, and a range of $\dimAB/2$ bits to add in which give an ancilla cost of $\dimAB/2$.
In both these cases, the ancilla cost is $\dimAB-2$.
When $\dimAB$ is odd, the ancilla cost is $\dimAB-1$.
When $n=\lceil\dimAB/2\rceil$ (the part just below the horizontal line in \eq{ex15}), there are $\lceil\dimAB/2\rceil-1$ bits to add in a controlled way, which takes $2(\lceil\dimAB/2\rceil-1)=\dimAB-1$ ancillas for $\dimAB$ odd.
When $n=\lceil\dimAB/2\rceil-1$, there are $\lceil\dimAB/2\rceil-2$ bits to add in a controlled way, and a range of $\lceil\dimAB/2\rceil$ bits to add, for a total ancilla cost of $\dimAB-1$.
Hence the ancilla cost of squaring is only half that for multiplication, and is
\begin{equation}
   \log (1/\mulError) + \order{\log\log (1/\mulError)}.
\end{equation}

\section{Other approaches to Hamiltonian evolution based optimization}
\label{app:other_evolution_optimization_approaches}

Here we outline two other approaches in the literature to optimization based on
Hamiltonian evolution. We consider ``shortest path'' optimization in
\app{shortest_path} and we consider quantum-enhanced population transfer
in \app{pop_transfer}. In both cases, we review the techniques and explain how the
algorithmic primitives we develop in this paper could be applied
in each approach.

\subsection{Heuristic variant of the shortest path algorithm}
\label{app:shortest_path}

Hastings' ``shortest path algorithm'' \cite{Hastings2018} (SPA) is an interesting approach to quantum optimization that is also based on time evolution under a cost function with some non-commuting driver Hamiltonian. Perhaps the most intriguing property of the SPA is that Hastings was able to rigorously show that the SPA gives a super-Grover (i.e., better than quadratic) speedup for certain classical optimization problems -- e.g., for an arbitrary instance of the problem MAX-2-LIN2 (which is a problem very closely related to QUBO) \cite{Hastings2018b}. The results of \cite{Hastings2018} and \cite{Hastings2018b} also rigorously (and in some cases, empirically) show similar speedups under a variety of assumptions about related problems.

The SPA essentially involves applying amplitude amplification to a variant of the adiabatic algorithm which uses the time-dependent Hamiltonian
\begin{equation}
\label{eq:shortest_h}
H(s) = C + s \, B \,\left(\frac{\sum_{p} X_p}{N}\right)^K
\end{equation}
where $C$ is the diagonal cost function of interest. Here, $K$ is a positive integer and $B$ is a scalar, and in order to rigorously show super-Grover speedups, both are chosen carefully based on properties of $C$.
In Algorithm 1 of \cite{Hastings2018}, the system is initialized in $\ket{+}^{\otimes N}$ with $s=1$ and then the transverse field is adiabatically turned off. Then, one computes the energy $C$ in a quantum register, and the idea is to apply amplitude amplification using this state preparation in order to amplify outcomes for which the computed energy is below some target threshold. In order to simplify analysis of the algorithm Hastings proposes to use a measurement based scheme similar to the Zeno approach described in \sec{zeno}. For the cases considered in \cite{Hastings2018} this combination reduces to the following very simple algorithm (Algorithm 3 of \cite{Hastings2018}) on which amplitude amplification is applied:

\begin{enumerate}
    \item Initialize the system in the state $\ket{\psi} = \ket{+}^{\otimes N}$.
    \item Perform phase estimation on $\ket{\psi}$ under the Hamiltonian $H(1)$ defined in \eq{shortest_h}. If the energy is greater than a threshold $E_{\rm threshold}$, terminate the algorithm and return failure to the amplitude amplification flag.
    \item If the previous step has succeeded, use a direct energy oracle to measure the energy of the state into a quantum register.
    If the energy is equal to $E_0$, 
    return success to the amplitude amplification flag (else return failure).
\end{enumerate}
The algorithm is to use amplitude amplification to boost the flag bit to near unit success. The work of \cite{Hastings2018} points out that the algorithm could work either by using a quantum walk such as qubitization, or with time evolution.

We note that it is possible to simplify the implementation of this algorithm with a technique that will also marginally improve performance (by increasing the success probability by an exponentially small factor). Our modification is to suggest that one proceed to step 3 regardless of whether or not step 2 succeeds. In doing this, we see that because the result of the phase estimation measurement is never used, we don't actually need the ancilla or controls involved in phase estimation. Instead, we can follow similar logic to \cite{Boixo2009a} to see that the procedure becomes equivalent to performing time evolution (or applying a quantum walk) for randomly chosen duration. We can choose the probability distribution to suppress phase measurement errors as large as the energy gap, as described in \sec{zeno}.

To explain the effect of this approach in a different way, consider writing the initial state as
\begin{equation}
\ket{\psi} = \sum_j \braket{\psi_{j,1}}{\psi} \ket{\psi_{j,1}},
\end{equation}
where $\ket{\psi_{j,1}}$ are the eigenstates of $H(1)$.
Then the evolution for time $t$ gives
\begin{equation}
\ket{\psi} = \sum_j \braket{\psi_{j,1}}{\psi} e^{-iE_{j,1}t} \ket{\psi_{j,1}}.
\end{equation}
The squared overlap with the desired solution state $\ket{\psi_{0,0}}$ is
\begin{equation}
    p_{\rm succ}(t) = \sum_{j,k} \braket{\psi_{j,1}}{\psi} \braket{\psi}{\psi_{k,1}} e^{-i(E_{j,1}-E_{k,1})t}\braket{\psi_{0,0}}{\psi_{j,1}}\braket{\psi_{k,1}}{\psi_{0,0}}.
\end{equation}
This expression corresponds to the probability of measuring the solution state after the evolution.
If we average over $t$ with probability $p_{\rm time}(t)$, then we have
\begin{equation}
    p_{\rm succ} = \sum_{j,k} \braket{\psi_{j,1}}{\psi} \braket{\psi}{\psi_{k,1}} \tilde p_{\rm time}(E_{j,1}-E_{k,1})\braket{\psi_{0,0}}{\psi_{j,1}}\braket{\psi_{k,1}}{\psi_{0,0}},
\end{equation}
where
\begin{equation}
    \tilde p_{\rm time}(E_{j,1}-E_{k,1}) = \int dt\, p_{\rm time}(t) e^{-i(E_{j,1}-E_{k,1})t}.
\end{equation}
Thus $\tilde p_{\rm time}$ corresponds to a Fourier transform of $p_{\rm time}$.
If $p_{\rm time}$ is chosen such that its Fourier transform goes to zero before the minimum energy gap, then $\tilde p_{\rm time}(E_{j,1}-E_{k,1})$ is nonzero only for $j=k$.
That is equivalent to having a measurement of phase with zero probability of error as large as the energy gap.
Then the average probability is
\begin{equation}
    p_{\rm succ} = \sum_{j} |\braket{\psi_{j,1}}{\psi}|^2 |\braket{\psi_{0,0}}{\psi_{j,1}}|^2 \ge |\braket{\psi_{0,1}}{\psi}|^2 |\braket{\psi_{0,0}}{\psi_{0,1}}|^2.
\end{equation}
Thus the average probability of success is at least as large as $|\braket{\psi_{0,1}}{\psi}|^2 |\braket{\psi_{0,0}}{\psi_{0,1}}|^2$ as given by Hastings' approach.
A minor drawback as compared to Hastings' approach is that only a single time is used, so if it happens that this time gave $p_{\rm succ}(t)$ significantly smaller than average the amplitude amplification would not give the solution.

The original motivation for the SPA seems to be primarily to produce an algorithm where a rigorous analysis can be performed, and so it is debatable whether one would actually want to try to use the algorithm heuristically rather than via some other approach. If did one want to use this approach heuristically there are many ways that could be accomplished; for instance, by choosing $E_{\rm target}$, $B$ and $K$ heuristically and then resolving to use a fixed number of rounds of amplitude amplification. Note that in the variant we have described it is no longer necessary to have an $E_{\rm threshold}$, although one will still need to choose the precision to which one performs phase estimation. One can see that such a heuristic variant of this algorithm could be implemented by using either our Hamiltonian walk or Trotter step oracles for the evolution, followed by our direct energy oracles for computing the amplitude amplification target.

\subsection{Quantum enhanced population transfer}
\label{app:pop_transfer}

Another heuristic algorithm for optimization which has been proposed is the quantum enhanced population transfer (QEPT) method of \cite{Smelyanskiy2018,Kechedzhi2018}. Unlike quantum heuristics which begin in a uniform superposition state, QEPT proposes to use quantum dynamics to evolve from one low energy solution of an optimization problem to other low energy solutions of similar energy. The idea is motivated by the search of configuration space in the classically non-ergodic phase associated with hard optimization problems. Such energy landscapes contain an extensive number of local minima separated by large Hamming distances. Algorithms relying on classical dynamics satisfying the detailed balance condition, such as simulated annealing, tend to get trapped at local minima of these landscapes. Thus, one could alternatively apply classical simulated annealing until the algorithm becomes trapped, then apply QEPT starting from that state, then again apply simulated annealing starting from the QEPT solutions, and so on.

Specifically, the context studied in \cite{Smelyanskiy2018,Kechedzhi2018} is as follows. Consider a cost function $C$ on $N$ qubits and bitstring $x$ with energy $E_x$ (so that $C \ket{x} = E_x \ket{x}$). The problem solved by QEPT is to produce another bitstring $y$ within a small energy window $E_y \in [E_x -\delta/2, E_y + \delta/2]$ such that the Hamming distance $d_{x, y}$ between $x$ and $y$ is $\mathcal{O}(N)$. In the presence of a spin glass type energy landscape finding such states $y$ using a classical algorithm takes exponential resources. The QEPT procedures suggests solving the above computational task as follows.
\begin{enumerate}
    \item Prepare the system in the initial state $\ket{x}$.
    \item Turn on a transverse field Hamiltonian $\sum_{k=1}^N X_i$ up to some optimal field strength $B_\perp = \mathcal{O}(\|C\| / N)$ with ramp up time polynomial in $N$.
    \item Evolve for time $T$ under the fixed Hamiltonian 
\begin{equation}
H = C + B_\perp \sum_{k=1}^N X_k.
\end{equation}
\item Measure in the computational basis and check the classical energy of the observed state.
\end{enumerate}
In general, we would expect that $T$ will scale exponentially in order for the procedure to succeed with fixed probability. However, for the worst case scenario when there are $M$ states with energy $-1$ and $2^N - M$ states of energy $0$, the work of \cite{Smelyanskiy2018} was able to show that this procedure succeeds with high probability for $T = {\cal O}(\sqrt{2^N / M})$, which is the same as the Grover scaling. However, unlike Grover, this protocol does not require any fine tuning of the transverse field or computation time. The procedure has also been shown empirically to produce similar results for the random energy model (where each bit string has a totally random energy).

The suggestion to use this algorithm heuristically is simply to choose $T$, as well as the accuracy with which we implement the time evolution, heuristically. Like with the adiabatic algorithm, this will essentially correspond to the number of steps that we take in either a product formula, or quantum walk approach to simulating the evolution. We propose that when using the quantum walk form of the algorithm, one not use signal processing and instead perform population transfer directly on the quantum walk. We note that since the norm of the problem and driver Hamiltonians are similar in magnitude, there would be no advantage to performing simulation in the interaction picture and so an approach based on qubitization is likely the best LCU style algorithm for QEPT.

\end{document}